\documentclass[12pt,preprint]{aastex}

\shorttitle{Warm H$_2$ in SINGS galaxies}
\shortauthors{Roussel et al.}

\begin{document}

\title{Warm molecular hydrogen in the Spitzer SINGS galaxy sample}

\author{H. Roussel\altaffilmark{1},
G. Helou\altaffilmark{2},
D.J. Hollenbach\altaffilmark{3},
B.T. Draine\altaffilmark{4},
J.D. Smith\altaffilmark{5},
L. Armus\altaffilmark{2},
E. Schinnerer\altaffilmark{1},
F. Walter\altaffilmark{1},
C.W. Engelbracht\altaffilmark{5},
M.D. Thornley\altaffilmark{6},
R.C. Kennicutt\altaffilmark{7,5},
D. Calzetti\altaffilmark{8},
D.A. Dale\altaffilmark{9},
E.J. Murphy\altaffilmark{10},
C. Bot\altaffilmark{2}
}

\email{roussel@mpia-hd.mpg.de}

\altaffiltext{1}{Max-Planck-Institut f\"ur Astronomie, Heidelberg, 69117 Germany}
\altaffiltext{2}{California Institute of Technology, Pasadena, CA 91125}
\altaffiltext{3}{NASA Ames Research Center, Moffett Field, CA 94035}
\altaffiltext{4}{Princeton University, Princeton, NJ 08544}
\altaffiltext{5}{Steward Observatory, University of Arizona, Tucson, AZ 85721}
\altaffiltext{6}{Bucknell University, Lewisburg, PA 17837}
\altaffiltext{7}{IoA, University of Cambridge, UK}
\altaffiltext{8}{University of Massachusetts, Amherst, MA 01003}
\altaffiltext{9}{University of Wyoming, Laramie, WY 82071}
\altaffiltext{10}{Yale University, New Haven, CT 06520}

\begin{abstract}
Results on the properties of warm molecular hydrogen in 57
normal galaxies are derived from measurements of H$_2$ rotational transitions
in the mid-infrared, obtained as part of the Spitzer Infrared Nearby Galaxies Survey
(SINGS). This study extends previous extragalactic surveys of emission lines of H$_2$,
the most abundant constituent of the molecular interstellar medium, to fainter
and more common systems ($L_{\rm FIR} = 10^7$ to $6 \times 10^{10}$\,L$_{\sun}$)
of all morphological and nuclear types.
In our sensitive integral-field observations covering kiloparsec-scale areas,
the 17\,$\mu$m S(1) transition is securely detected in the nuclear regions of
86\% of SINGS galaxies with stellar masses above $10^{9.5}$\,M$_{\sun}$.
The derived column densities of warm H$_2$ (with temperatures above $\sim 100$\,K),
even though averaged over large areas, are commensurate with values observed in
resolved photodissociation regions; the median of the sample is
$3 \times 10^{20}$\,cm$^{-2}$. They amount to a significant fraction
of the column densities of total molecular hydrogen, between 1\% and more than 30\%.
The power emitted in the sum of the three lowest-energy transitions is on average
30\% of the power emitted in the bright [Si{\small II}] cooling line (34.8\,$\mu$m),
and represents about $4 \times 10^{-4}$ of the total infrared power within the same
area for star-forming galaxies, which is consistent with excitation in
photodissociation regions. The fact that the H$_2$ line intensities scale tightly
with the emission in the aromatic bands, even though the average radiation field
intensity within the same area varies by a factor ten, can also be understood
if both tracers originate predominantly in photodissociation regions, either
dense or diffuse. A large fraction of the 25 targets classified as {\small LINER}s
or Seyferts, however, strongly depart from the rest of the sample, in having warmer
H$_2$ in the excited states, smaller mass fractions of H$_2$ in the warm phase,
and an excess of power emitted in the rotational transitions with respect to
aromatic bands, the total infrared emission and the [Si{\small II}] line.
We propose a threshold in H$_2$ to aromatic band power ratios, allowing the
identification of low-luminosity AGNs by an excess H$_2$ excitation.
A dominant contribution from shock heating is favored in these objects.
Finally, we detect in nearly half the star-forming targets, in particular
in low-density central regions of late-type galaxies, non-equilibrium ortho
to para ratios, consistent with the effects of pumping by far-ultraviolet
photons combined with incomplete ortho-para thermalization by collisions,
or possibly non-equilibrium photodissociation fronts advancing into cold gas.
\end{abstract}
 
\keywords{galaxies: ISM -- infrared: galaxies -- infrared: ISM -- ISM: lines and bands --
ISM: molecules -- surveys}

\section{Introduction}

Rotational transitions of molecular hydrogen, lying in the mid-infrared range
between 5 and 30\,$\mu$m (Table~\ref{tab_lines}), provide measurements of the mass
and temperature distribution of the bulk of the warm molecular gas phase,
at temperatures of $\sim 100$ to 1000\,K.
By contrast, the transitions from higher vibrational levels in the
near-infrared, better studied because more easily observable from the ground,
arise from gas at apparent excitation temperatures of
more than 1000\,K located in a thin layer of molecular clouds,
hosting a negligible fraction (of the order of $10^{-6}$) of the total H$_2$ mass
\citep{Black76, Burton92}. The rotational lines are thus more appropriate
tracers of molecular gas being exposed to moderate heating, and arise from a much
larger volume fraction of the molecular clouds. They constitute one of the
most important coolants of warm molecular gas \citep{Neufeld93}.

H$_2$ emission lines have been detected in a wide array of sources, including
outflows from young stars \citep{Gautier76, Bally82}, photodissociation regions
\citep{Gatley87, Tanaka89}, planetary nebulae \citep{Treffers76, Beckwith78},
supernova remnants \citep{Treffers79, Burton89}, large regions at the centers of galaxies
\citep{Thompson78, Gatley84} and extranuclear large-scale shocks in galaxy collisions
\citep{Herbst90, Sugai97, Appleton06}. The possible excitation mechanisms are
accordingly varied. In normal
galaxies, the major excitation source is expected to be the far-ultraviolet
radiation of massive stars in photodissociation regions, with photon energies between
6 and 13.6\,eV \citep[and references therein]{Hollenbach97}. H$_2$ molecules can be
pumped by FUV photons into electronically-excited states, followed by fluorescence
and radiative cascade through the vibration-rotation levels of the ground electronic
state. Pure fluorescent spectra are produced only if the cascade is not significantly
altered by collisions with hydrogen atoms and molecules; if the critical densities
for collisional deexcitation are exceeded, a portion of the pump energy is converted
to heat by collisions, and the lowest rotational levels are populated by collisions
and thermalized. Pure fluorescence is thus much more likely in the vibrational transitions,
that have high critical densities, than in the pure rotational transitions considered
here, with critical densities below a few $10^3$\,cm$^{-3}$ for S(0) to S(3).
Additionally, FUV photons can be absorbed by dust grains, followed by the ejection
of photoelectrons that heat the gas. This also results in the thermal excitation
of the low-energy levels of H$_2$ by collisions with the warm gas. Besides the radiation
of massive stars, a second
important source of excitation is shocks, in molecular outflows, supernova remnants
or cloud collisions in a disturbed gravitational potential \citep{Shull78, Draine83}.
In addition to the above processes, X-rays produced in active nuclei or in supernova
remnant shocks can partially ionize and heat the gas over large column densities,
leading to H$_2$ excitation by collisions with hydrogen atoms and molecules, and
with fast electrons \citep{Lepp83, Draine91, Maloney96}.
Finally, H$_2$ molecules can be formed directly into excited states.

Surveys of molecular hydrogen line emission in galaxies have been so far mostly
restricted to starbursts, active galactic nuclei and ultraluminous systems,
and have been performed mostly in the near-infrared, targetting vibration-rotation
lines that arise from
upper levels with much higher excitation energies than the mid-infrared lines.
It has been speculated that the major source of H$_2$ heating in star-forming galactic
nuclei was shocks in supernova remnants, based on comparison of the luminosity of some
vibration-rotation H$_2$ lines with a limited number of Galactic templates and with
shock models \citep{Moorwood88, Mouri90}. However, scaling individual templates to
the integrated emission of galaxies has large inherent uncertainties,
and the near-infrared line ratios most often used to discriminate between thermal
and non-thermal emission are not always sufficient to distinguish between shocks and
fluorescent excitation followed by collisional deexcitation in high-density regions
\citep{Sternberg89}.
\citet{Puxley88} surveyed starburst galaxies in several vibration-rotation lines,
and found that the dominant excitation mechanism was pumping
by the far-ultraviolet radiation of massive stars, rather than collisional excitation.
\citet{Davies03} reached the same conclusion for a small sample of ultraluminous
galaxies, in which the first vibrational level is thermalized by high densities in
photodissociation regions.
Active nuclei ({\small LINER}s or Seyferts) can show an excess of H$_2$ emission relative
to hydrogen recombination lines and aromatic bands \citep[e.g.][]{Moorwood88, Larkin98},
but the exact nature of the additional source of excitation, namely X-ray excitation,
fluorescence induced by a non-thermal ultraviolet continuum, or shocks induced by
dynamical perturbations, is often unclear \citep[e.g.][]{Quillen99}. It is however
unlikely that significant H$_2$ emission could arise from interaction between
molecular clouds and jets from Seyfert nuclei \citep{Rotaciuc91, Knop01}.

The detection of a rotational line of H$_2$ was first reported by \citet{Beck79}
(the S(2) transition at 12.3\,$\mu$m in Orion) from observations at Las Campanas
Observatory. It was soon followed by many more ground-based detections, but the
majority of data on the rotational spectrum of H$_2$ were produced by the SWS
instrument on board ISO \citep[e.g.][]{Lutz00, Rigopoulou02}.
Furthermore, with previous infrared spectroscopic capabilities,
observations of normal galaxies have proven difficult due to sensitivity
limitations, so that our current knowledge is mainly extrapolated from studies of
very bright objects, maybe not representative of the general galaxy population.
The purpose of this paper is thus to extend previous work to fainter systems than
formerly accessible, and to characterize directly the generic properties of the warm
molecular hydrogen content of normal galaxies. The SINGS sample
\citep[Spitzer Infrared Nearby Galaxies Survey;][]{Kennicutt03}, covering a broad range
of infrared luminosities, morphologies and nuclear types, is ideally suited to such
a pursuit.

Studies of rotational lines alone, without information on vibrational levels, have
very limited diagnostic value concerning the source of excitation, because the
low critical densities of the rotational levels make it likely that they will be
thermalized most of the time, and thus cannot be used to distinguish between
the various heating mechanisms.
Because observations of vibration-rotation transitions in the near-infrared are
still scarce for normal galaxies, and because they are typically performed in
apertures that are not matched to our observations, we did not attempt to include
vibrational levels in our analysis. The characterization of excitation mechanisms
and physical conditions in the gas would greatly benefit from such
information, but would necessitate an additional dedicated survey.

The rotational lines are, however, energetically important and can characterize 
the temperature and density conditions of a large mass fraction of the
interstellar medium in galaxies, i.e. that consisting of warm molecular gas.
From a SWS survey of rotational lines in nearby starburst and Seyfert galaxies,
\citet{Valentijn96} and \citet{Rigopoulou02} obtained mass fractions of H$_2$ in
the warm phase of several percent. In ultraluminous galaxies observed with Spitzer,
\citet{Higdon06} derive much lower mass fractions of warm gas, but the fact that the
majority of their sample has only upper limits for the S(0) line makes it possible
that the temperatures are overestimated
(because computed from the S(1) to S(3) lines only, whenever S(0) is undetected)
and thus the masses of warm H$_2$ underestimated.

This paper presents observations of warm molecular hydrogen in nearby galaxies
obtained as part of SINGS \citep{Kennicutt03}. From these data, we present
quantifications of the temperatures and column densities of warm H$_2$ encountered
in kiloparsec-scale areas, mostly nuclear regions,
and a comparison of the power emitted in the rotational lines with those produced
by [Si{\small II}] at 34.8\,$\mu$m, which is the dominant cooling line
of normal galaxies in the mid-infrared range, and by dust.
We emphasize the different properties of star-forming regions and nuclei classified
as {\small LINER}s or Seyferts, and discuss their H$_2$ excitation mechanisms.
The data, analysis methods and observational results are described in Sections
\ref{data} to \ref{powers}.\footnote{For easier comparison to future observations
and models, ascii flux tables of all the measured quantities are available
upon e-mail request.}
The interpretation of the main findings is presented in Sections~\ref{excitation}
and \ref{syliner}, and a summary of the results and conclusions can be found
in Section~\ref{summary}.

\section{Data and measurements}
\label{data}

\subsection{Targets}

The SINGS sample \citep{Kennicutt03}, comprising 75 galaxies, is intended to be a
valuable representative set of local galaxies that
are not ultraluminous, and whose moderate distances ensure that the properties of the
interstellar medium can be studied at relatively small spatial scales (a few hundreds
of parsecs at the shortest wavelengths). Numerous sources with mild starbursts or
low-luminosity active nuclei are included. Of this sample, we excluded from the present
study the objects that were not observed in spectroscopic mode because of their very
low brightness (DDO\,154, Ho\,I, M81\,dwA, M81\,dwB),
or containing very little dust and nebular emission within the nuclear
area mapped by the high spectral resolution modules
(the quiescent ellipticals NGC\,584 and NGC\,1404, the quiescent dwarf galaxies DDO\,53,
DDO\,165, Ho\,IX and the asymmetric magellanic galaxies NGC\,4236, NGC\,5398, NGC\,5408
and IC\,4710). The dwarf galaxies IC\,2574 and NGC\,5474 were also rejected because they
lack observations in some of the spectroscopic modules. Of the two star-forming
dwarf galaxies with several extranuclear pointings, Ho\,II and NGC\,6822, we retained
only NGC\,6822 here; the regions within Ho\,II are indeed two faint to allow an
analysis of the H$_2$ excitation diagram, contrary to some regions within NGC\,6822.
Low-mass galaxies with extranuclear pointings will be discussed elsewhere.
NGC\,3034 (M\,82) was excluded due to the unavailability of nuclear spectroscopy from
SINGS, as well as NGC\,1377, which constitutes a galaxy class of its own very different
from the rest of the SINGS sample, and has been discussed separately \citep{Roussel06}.
The sample for H$_2$ measurements comprises 66 targets in 57 galaxies
(Table~\ref{tab_target}).
The pointings are centered either on the nuclear regions (for most targets) or on some
bright star-forming complexes (for a few dwarf galaxies and a spiral galaxy).
Diffuse regions within galactic disks are not covered by the present study.

\begin{figure}[!ht]
\vspace*{-1cm}
\resizebox{12cm}{!}{\rotatebox{90}{\plotone{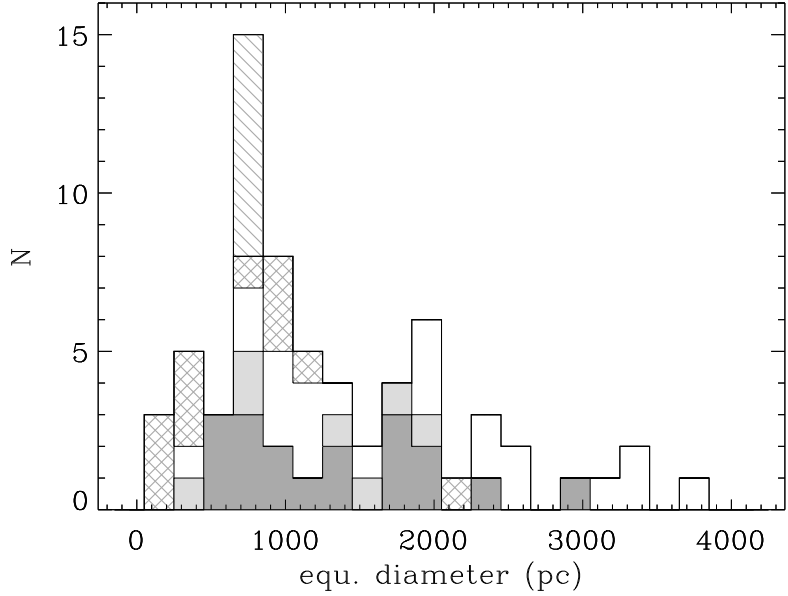}}}
\caption{Histogram of the equivalent diameters of the areas over which the line and continuum
fluxes were integrated. The apertures were defined by the intersection of the spectral maps
in the LH, SH and SL modules of the IRS instrument. The extranuclear regions in NGC\,5194
are shown by the hatched histogram, the regions within dwarfs by the cross-hatched histogram,
the {\small LINER} nuclei by the darker shade and the Sy nuclei by the lighter shade.
}
\label{fig:diameters}
\end{figure}

The aperture over which we extracted the spectra is the intersection of the various
areas covered by all four spectroscopic modules. The central position and solid
angle of this aperture, used to measure all the quantities presented in this paper (line
and continuum fluxes), is listed for each galaxy in Table~\ref{tab_target}. In practice,
the limiting size is that of the maps performed with the high-resolution modules,
which were enlarged in a few cases in order to cover the emission from a star-forming
circumnuclear ring. At the distances of the targets, the equivalent linear diameters
of the apertures range from 60\,pc to 3.8\,kpc (distribution shown in
Fig.~\ref{fig:diameters}), and the median is 900\,pc.
Although the apertures are in general small fractions of the optical extent of the
galaxies, the measurements are still averages over very large and complex areas.
It is expected that a large number of disconnected star formation sites, in addition
to the nucleus, contribute to the total emission.

\subsection{Broadband imaging}
\label{images}

To estimate flux densities of the dust continuum and of the aromatic bands
(also referred to as the emission from PAHs, or polycyclic aromatic hydrocarbons),
we used images in the 3.6\,$\mu$m and 7.9\,$\mu$m bands of the IRAC camera \citep{Fazio04},
and scan maps in the three bands of the MIPS instrument \citep{Rieke04} at effective
wavelengths of 24, 71 and 156\,$\mu$m.
Since in early-type galaxies photospheric emission can make an important contribution
to 7.9\,$\mu$m fluxes, we subtracted an estimate of this component in order to
obtain pure measurements of aromatic band emission. To this effect, we scaled
3.6\,$\mu$m fluxes, assumed to be dominated by stellar emission, as described
in \citet{Helou04}. The resulting flux densities are noted $F_{\rm 7.9\,dust}$.

The observing strategy and data reduction are
described by \citet{Kennicutt03}. The full width at half maximum of the point spread
function (PSF) is close to 2\arcsec\ at 7.9\,$\mu$m, 6\arcsec\ at 24\,$\mu$m,
18\arcsec\ at 71\,$\mu$m and 40\arcsec\ at 156\,$\mu$m. Flux calibration uncertainties
are of the order of 10\% in the IRAC bands, and 5\%, 10\% and 15\% in the MIPS 24, 71
and 156\,$\mu$m bands, respectively.
To correct for the effects of light scattering in IRAC arrays, we applied to flux
densities measured from IRAC maps corrective factors that are appropriate for the
photometry of extended sources within apertures of arbitrary size
\citep[derived by T. Jarrett and published by][]{Dale07}.
For our apertures, the correction factor at 7.9\,$\mu$m is of the order of 10\%.

\subsection{Spectroscopic data}
\label{spectra}

The targets were observed in mapping mode with the IRS instrument \citep{Houck04},
at low spectral resolution between 5 and 38\,$\mu$m, with the SL and LL slits
($\lambda / \Delta \lambda \approx 60$--130) and at high spectral resolution between
10 and 37\,$\mu$m, with the SH and LH slits ($\lambda / \Delta \lambda \approx 600$).
The observing strategy is described by \citet{Kennicutt03} and \citet{Smith04}.
The data were pre-processed with the S13 version of the Spitzer Science Center pipeline.
Pixels with an abnormal responsivity were masked, and spectral cubes were built
with the Cubism software \citep{Smith07a}. The flux calibration
was performed as described by \citet{Roussel06}. We checked the accuracy of this
procedure by systematically comparing broadband fluxes from imaging observations and
from spectra, and line fluxes from high and low spectral resolution spectra, for bright
lines that are minimally contaminated by broad aromatic features at low resolution
(but note that even if the flux calibrations of the different modules were in perfect
agreement, deviations would be expected from slight misalignment between the apertures).
We obtain $F_{24}{\rm (MIPS)} / F_{24}{\rm (LL)} = 1.01 \pm 0.04$ (for targets with
$F_{24} > 0.025$\,Jy within a diameter of about 50\arcsec),
$F_{7.9}{\rm (IRAC)} / F_{7.9}{\rm (SL)} = 0.99 \pm 0.05$ (for targets with
$F_{7.9} > 0.025$\,Jy within a diameter of about 30\arcsec\ and accurately determined
backgrounds in SL maps),
$F_{\rm [SiII]}{\rm (LL)} / F_{\rm [SiII]}{\rm (LH)} = 0.96 \pm 0.14$,
$F_{\rm [SIII]~34}{\rm (LL)} / F_{\rm [SIII]~34}{\rm (LH)} = 0.92 \pm 0.20$
and $F_{\rm [SIII]~19}{\rm (LL)} / F_{\rm [SIII]~19}{\rm (SH)} = 1.05 \pm 0.21$
(for targets with line fluxes above $6 \sigma$).

\subsection{Measurements}
\label{measurements}

The S(0) to S(3) rotational transitions of H$_2$ (Table~\ref{tab_lines}) were measured
for all targets. In addition, we measured the S(4) to S(7) transitions in three
galaxies in which these lines are bright enough to become detectable
at low spectral resolution (see Table~\ref{tab_flux}).

In high spectral resolution data, we defined errors from fluctuations of the
pseudo-continuum, which was fitted as an affine function of wavelength
($F_{\nu} = {\rm a} \lambda + {\rm b}$).
In SL data, errors at each wavelength were estimated from spatial fluctuations
of blank fields within the satellite spectral maps that are automatically obtained
when the source lies in the other half of the slit \citep[see][]{Smith04}.
Both the fluxes and the errors presented in Table~\ref{tab_flux} were then
added linearly for each point of the line profile above the pseudo-continuum.
The line profiles were constrained to have a width compatible with the spectral
resolution, since the latter is sufficiently low that no line is resolved.
Fig.~\ref{fig:lines} shows the line spectra for the representative galaxies
NGC\,1097, NGC\,6946, NGC\,7552, NGC\,1266, NGC\,4569 and NGC\,4579.

The S(1) line is usually the brightest.
Of the non-dwarf galaxies of the SINGS sample (with stellar masses estimated
as by \citet{Lee06} above $10^{9.5}$\,M$_{\sun}$), the nuclear regions of 86\%
are securely detected in the S(1) line, with fluxes above three times the
measured error. The other 14\% are either ellipticals of the {\small LINER}
type, or late-type spirals (Sc-Sd).

There are two galaxies in common between this sample and that of \citet{Rigopoulou02},
namely NGC\,7552 and NGC\,6946, the latter from the study of \citet{Valentijn96}.
For both, our aperture is larger than the beam of ISO-SWS, which covered an area
of 280 to 380 arcsec$^2$. For the lines that were detected with SWS, we obtain
fluxes that are higher by factors of 2.3 (S(1) in NGC\,6946), 5.6 (S(0) in NGC\,6946)
and 1.1 (S(1) and S(3) in NGC\,7552).
The exact placement of the ISO-SWS beam is not known. For NGC\,6946, given this
uncertainty, it is conceivable that the H$_2$ emission be twice as bright in our
800 arcsec$^2$ aperture as in the SWS aperture; but the S(0) line flux of
\citet{Valentijn96} is inconsistent with our data.

For this study, we estimate total infrared fluxes (TIR) between 3 and 1100\,$\mu$m,
defined as a linear combination of 24, 71 and 156\,$\mu$m flux densities.
The formula of \citet{Dale02} is used here, and we have checked that replacing
it with the more recent prescription by \citet{Draine07a} does not change the
following results in any appreciable way.
The infrared fluxes are measured within the same area as the other quantities for direct
comparison. The PSF width at 156\,$\mu$m is however much larger than the size of our
spectroscopic aperture, so that some extrapolation is needed. We first measure MIPS
fluxes within the larger aperture used to compare total infrared fluxes with line fluxes
measured in the LL module. Then,
we scale these fluxes by the ratio of $F_{24}$ measured in the small aperture to
$F_{24}$ measured in the larger aperture, which is equivalent to assuming that the
spectral energy distribution does not change from an area of $\approx 300$\,arcsec$^2$
to an area of $\approx 2000$\,arcsec$^2$. The associated errors are however expected
to be small compared with the dynamic range of the quantities discussed in
Section~\ref{powers}.
Simulations of the overestimation of the far-infrared fluxes caused by
the extrapolation, using a simple model of a point-source starburst (with
the spectral energy distribution of Mrk\,33) superposed on quasi-uniform
emission from low radiation field intensity regions (with the colors
of the central regions of NGC\,24 or NGC\,2403), indicate that the effect
should be in most cases of the order of 20\% (when the starburst and quiescent
components contribute equally at 156\,$\mu$m), and in extreme cases reach a
maximum of a factor 2 (when the quiescent component dominates).
\citet{Smith07b} reached a similar conclusion (see their Section 3.2).

\section{Excitation diagrams}

Excitation diagrams provide a convenient visualization of the distribution of level
populations and allow first constraints on the excitation mechanisms (thermal or
non-thermal) that can produce this distribution. They represent the column density
in the upper level of each observed transition $N_{\rm u}$, normalized by its statistical
weight $g_{\rm u}$, as a function of the upper level energy $E_{\rm u}$.
The flux of a transition can be written as
$F = h \nu~ A~ N_{\rm u}~ \Omega / (4 \pi)$, where $A$ is the spontaneous emission
probability, $h \nu$ is the transition energy and $\Omega$ is the beam solid angle.
In the assumption of local thermodynamic equilibrium, the total column density
$N_{\rm tot}$ can be derived from
$N_{\rm u} = g_{\rm u}~ N_{\rm tot}~ {\rm exp}(-E_{\rm u}~ /~ (kT))~ /~ Z(T)$, where
$g_{\rm u} = (2 I + 1)~ (2 J + 1)$ is the statistical weight (with the spin number
$I=0$ for even J or para transitions, and $I=1$ for odd J or ortho transitions),
and $Z(T) \sim 0.0247~ T~ /~ (1 - {\rm exp}(-6000\,{\rm K}~ /~ T))$ is the partition
function \citep{Herbst96}, valid for $T > 40$\,K.

The apparent excitation temperature can then be derived from each pair of transitions by:
\begin{equation}
kT = (E_{\rm u2} - E_{\rm u1})~ /~ {\rm ln}(N_{\rm u1} / N_{\rm u2} \times g_{\rm u2} / g_{\rm u1})
\end{equation}
with $N_{\rm u1} / N_{\rm u2} = F_1 / F_2 \times A_2 / A_1 \times \lambda_1 / \lambda_2$.
Since both radiative decay and collisions with H$_2$ change the rotational number $J$
by an even number, the ortho and para states are largely 
decoupled and should in principle be dealt with independently.

\subsection{Ortho-para thermalization and departures therefrom}

As emphasized by \citet{Burton92}, the lower rotational levels of H$_2$ will be in
collisional equilibrium over a wide range of conditions, because their critical
densities are low. Figure~\ref{fig:ncrit} shows the critical densities of all the
rotational transitions observable with the IRS instrument, as a function of temperature,
computed using the functional form for the collisional de-excitation rate coefficient
by H$_2$ given by \citet{Shull82} and the transition probabilities given by
\citet{Black76}.
The derived critical densities for each line are about an order of magnitude lower
than those for collisions with H computed by \citet{Mandy93}, the comparison being
made at 600\,K, since \citet{Mandy93} provide results only for high temperatures.

The integrated emission from warm H$_2$ in star-forming galaxies is likely to
come predominantly from the densest photodissociation regions (PDRs) within the beam,
with densities above $10^3$\,cm$^{-3}$ \citep{Burton92, Kaufman06},
in which case the lowest
rotational levels will be thermalized. Observations of starburst galaxies with ISO-SWS
\citep{Rigopoulou02} as well as ultraluminous galaxies with Spitzer-IRS \citep{Higdon06}
are indeed consistent with this expectation. At first sight, the same applies to
the galaxies studied here.

However, some of the excitation diagrams show departures from thermalization
of ortho levels with para levels,
in the sense that the apparent temperatures derived from each pair of transitions
of consecutive rotational number are not monotonic as a function of upper level energy.
Clear examples are NGC\,1266
($T{\rm (S0-S1)} = (201 \pm 45)$\,K, $T{\rm (S1-S2)} = (465 \pm 34)$\,K
and $T{\rm (S2-S3)} = (347 \pm 18)$\,K);
NGC\,4254 ($(162 \pm 9)$\,K, $(358 \pm 59)$\,K and $(259 \pm 38)$\,K);
and NGC\,4631 ($(127 \pm 8)$\,K, $(342 \pm 39)$\,K and $(268 \pm 25)$\,K).
Such deviations from thermalization can be explained by an ortho to para density
ratio in the excited states apparently different from the equilibrium value.
We have
\begin{eqnarray}
OPR~ & =~ \frac{OPR_{\rm \,high\,T}}{3}~ \frac{\sum_o (2 I_o + 1)~ (2 J_o + 1)~ {\rm exp}(-E_o~ /~ (kT))}{\sum_p (2 I_p + 1)~ (2 J_p + 1)~ {\rm exp}(-E_p~ /~ (kT))} \nonumber \\
     & =~ OPR_{\rm \,high\,T}~ \frac{\sum_o (2 J_o + 1)~ {\rm exp}(-E_o~ /~ (kT))}{\sum_p (2 J_p + 1)~ {\rm exp}(-E_p~ /~ (kT))}
\label{eq:opr}
\end{eqnarray}
where the subscripts $o$ and $p$ designate ortho and para levels respectively
($I_p = 0$ and $I_o = 1$).
$OPR_{\rm \,high\,T}$, equal to the actual ortho to para ratio ($OPR$) in the
high-temperature limit, expresses deviations from local thermodynamic equilibrium
(LTE) if it differs from three.
It may be called the effective nuclear spin degeneracy ratio,
but will hereafter be called the ortho to para ratio for convenience.
In LTE, $OPR \sim 2$ for $T \sim 100$\,K and $OPR \sim 3$ for $T > 200$\,K \citep{Burton92},
but $OPR_{\rm \,high\,T} = 3$ at all temperatures.
Although $OPR_{\rm \,high\,T} < 3$ may be inferred for the excited states ($J \geq 2$),
this does not imply that the ortho to para ratio of the bulk of the gas in the
$J = 1$ and $J = 0$ states be out of LTE.
In the following, LTE will refer more particularly to the equilibrium
between the ortho and para levels, and not of the ortho levels or para
levels separately. Extinction effects are discussed in Section~\ref{extinction}
and the interpretation of $OPR_{\rm \,high\,T}$ values is postponed to Section~\ref{excitation}.

To derive temperatures and column densities,
we first determine whether the excitation diagram is compatible or not with LTE
by inserting explicitly the factor $OPR_{\rm \,high\,T} / 3$ in the equations
for column densities of the ortho levels, and deriving temperatures from each pair of
consecutive transitions as a function of $OPR_{\rm \,high\,T}$, to verify whether these
conditions are satisfied:
$T{\rm (S0-S1)} \leq T{\rm (S0-S2)} \leq T{\rm (S1-S2)} \leq T{\rm (S1-S3)} \leq T{\rm (S2-S3)}$,
since in gas with a distribution of temperatures, ratios of transitions with low-energy
upper levels always probe lower excitation temperatures than ratios of transitions
with higher-energy upper levels.
$T{\rm (S0-S2)}$ and $T{\rm (S1-S3)}$ are independent of $OPR_{\rm \,high\,T}$ and determined
directly from the observed fluxes, but $T{\rm (S0-S1)}$, $T{\rm (S1-S2)}$ and $T{\rm (S2-S3)}$
depend on $OPR_{\rm \,high\,T}$. For each pair $(p,o)$ = (0,1), (2,1) and (2,3), we have:
\begin{equation}
k T({\rm S}_p-{\rm S}_o) = (E_{\rm u~o} - E_{\rm u~p})~ /~ \ln(OPR_{\rm \,high\,T} \times R)\,.
\end{equation}
with $R = F_p / F_o \times A_o / A_p \times \lambda_p / \lambda_o \times (2 J_o + 1)~/~(2 J_p + 1)$.
Figure~\ref{fig:diag_temp} shows the corresponding diagram for two galaxies.
In case the above condition on the temperatures is satisfied for $OPR_{\rm \,high\,T} = 3$,
as illustrated for NGC\,3198, we fix $OPR_{\rm \,high\,T} = 3$~; in the opposite case,
illustrated by NGC\,4631, we fit $OPR_{\rm \,high\,T}$ as explained below.
The excitation diagrams of all the galaxies, with fits overlaid, are shown in
Fig.~\ref{fig:diag_exc}.

\subsection{Temperatures and column densities}
\label{fits}

Since in all cases the excitation diagrams indicate that a single temperature does not fit
all the line fluxes, we assume that the H$_2$ emission is the sum of two discrete components
of different temperatures, which is enough to reproduce accurately the observed fluxes.
In the general case of $OPR_{\rm \,high\,T} = 3$, we perform a least-squares fit of the
excitation diagram to determine
the parameters of the two discrete components (the lower temperature $T_1$, the upper
temperature $T_2$, and their mass fraction) and the normalization by the total
column density. The results are listed in Table~\ref{tab_fit}.

When the gas is at a range of temperatures, it is in practice impossible to lift the
degeneracy between mass and temperature from the lowest-energy levels.
Since the column density has a very steep dependence on $T_1$, we adopt two different
procedures to fit the excitation diagrams and ascertain the amplitude of the
uncertainties caused by this degeneracy. In the first case, we constrain $T_1$ to
exceed the value for which the column density is 20\% higher than the nominal density
derived from $T{\rm (S0-S1)}$. In the second case, we leave $T_1$ unconstrained.
In the following, both approaches will be retained when discussing results that
depend on $T_1$.

For the results not to be biased by systematic sensitivity
differences at the wavelengths of the H$_2$ transitions, we also replace the measured
errors by a uniform weight. When $OPR_{\rm \,high\,T}$ is allowed to be fitted, we fix $T_2$ at
$1.3 \times T{\rm (S1-S3)}$ in the constrained-$T_1$ fits, which was chosen from
the median value of $T_2$ in galaxies with $OPR_{\rm \,high\,T} = 3$. In free-$T_1$ fits
with $OPR_{\rm \,high\,T} = 3$, the distribution of $T_2 / T{\rm (S1-S3)}$ is large, with a tail
of high values; therefore, $T_2$ is first fixed at the median value,
$1.14 \times T{\rm (S1-S3)}$, and then at $1.5 \times T{\rm (S1-S3)}$, to probe
the full range of most likely values.
Finally, when one flux is an upper limit, we fix both $T_1$ at $0.98 \times T{\rm (S0-S1)}$
(which increases the total column density by a maximum of $\sim 20$\% with respect to
that obtained with $T_1 = T{\rm (S0-S1)}$ but allows a small contribution from hotter
gas to the S(0) and S(1) lines), and $T_2$ as above.

For the three galaxies from which more transitions, up to S(7), could be measured,
the procedure is the same except that a third component has to be added. The
additional parameters are $T_3$ and the mass fraction of the second component,
and $T_2$ is fixed at 400\,K.

Several galaxies barely satisfy the criterion on temperatures to have
$OPR_{\rm \,high\,T} = 3$, with $T{\rm (S1-S2)} \geq 0.95 \times T{\rm (S1-S3)}$
and $T{\rm (S2-S3)} \leq 1.05 \times T{\rm (S1-S3)}$. When $T_1$ is constrained,
the quality of their fits can be improved by allowing $OPR_{\rm \,high\,T}$ to vary.
For these objects, we provide results with $OPR_{\rm \,high\,T} < 3$. Allowing
$OPR_{\rm \,high\,T}$ to be smaller than the equilibrium value has the indirect consequence
that the derived column densities are smaller. The amplitude of this effect is
indicated in Table~\ref{tab_fit}. Similarly, for NGC\,1705 and NGC\,4552, we provide
results with $OPR_{\rm \,high\,T} < 3$ and indicate the change in column density with
respect to $OPR_{\rm \,high\,T} = 3$, because although the S(2) transition being an upper
limit prevents any reliable determination of $OPR_{\rm \,high\,T}$, the $T_1$ temperatures
derived with $OPR_{\rm \,high\,T} = 3$ are the two lowest of the whole sample, raising
the suspicion that they might be artifacts of the constraint on $OPR_{\rm \,high\,T}$.
We also consider $OPR_{\rm \,high\,T} < 3$ more likely for these galaxies in view of the
dependence of $OPR_{\rm \,high\,T}$ on H$_2$ brightness, 
discussed later in Section~\ref{excitation}.

The median $T_1$ temperature is 154\,K when the fits are constrained (ranging between 97
and 300\,K); when no constraint is applied, the median $T_1$ is 118\,K with
$T_2 = 1.14 \times T{\rm (S1-S3)}$, and 161\,K with $T_2 = 1.5 \times T{\rm (S1-S3)}$.
The total column densities that we obtained, averaged over kiloparsec-scale regions
in galactic centers,
range between $10^{19}$ and $2 \times 10^{21}$\,cm$^{-2}$ (for constrained-$T_1$ fits),
or $2 \times 10^{22}$\,cm$^{-2}$ (for free-$T_1$ fits), and their medians are
respectively $3 \times 10^{20}$\,cm$^{-2}$ and 5--$6 \times 10^{20}$\,cm$^{-2}$
(Fig.~\ref{fig:coldens}).
This can be compared with typical column densities of resolved photodissociation regions
in the Milky Way. In the Orion Bar, column densities of H$_2$ warmer than 400\,K,
derived from rotational lines, lie between $10^{20}$ and $10^{21}$\,cm$^{-2}$
\citep{Parmar91, Allers05}.
Note that because the Orion Bar is observed nearly edge-on, an equivalent PDR
seen face-on would have lower column densities.
In NGC\,7023, \citet{Fuente99} derived a total column
density of $5 \times 10^{20}$\,cm$^{-2}$ for H$_2$ warmer than 300\,K.
Thus, if the H$_2$ emission in our targets comes from similar photodissociation
regions, they must occupy in general a very large fraction of the observing beam,
assuming that they do not overlap on the line of sight.

Figure~\ref{fig:coldens} also shows a clear dependence of the local (nuclear) column
density of warm H$_2$ on the total stellar mass of the host galaxy.
The stellar mass and the infrared luminosity being correlated for star-forming
galaxies, there is a similar dependence on far-infrared luminosities.
To first order, the column density of warm H$_2$ shows the same behavior as
tracers of molecular gas and star formation rate densities, which suggests
that the primary source of H$_2$ heating is the star formation activity in
non-AGN galaxies, and the nuclear regions respond to the global mass and luminosity.
{\small LINER} and Sy nuclei do not follow the correlation shown
by star-forming regions, and tend to have smaller column densities of warm H$_2$.
The differences in terms of energy output and excitation mechanisms will be studied
in more detail in Sections~\ref{powers}, \ref{excitation} and \ref{syliner}.
Since the few extranuclear regions and dwarf galaxies included in the sample
do not distinguish themselves from the other star-forming targets in any obvious
way, here and in the following, they are not discussed as separate categories.

\subsection{Optical depth toward H$_2$}
\label{extinction}

Consistent with the negligible optical depths inferred from the silicate absorption
bands at 10\,$\mu$m and 18\,$\mu$m in most SINGS galaxies \citep{Smith07b},
that support the modest values of nebular extinction derived from the Balmer
decrement \citep{Dale06}, we assume zero extinction both in the lines and in
the dust continuum for all the targets. In eight galactic centers among the
SINGS sample (included here),
\citet{Smith07b} obtained a better fit in their decomposition of the low
spectral resolution spectra by including a finite optical depth in the silicate
bands. We expect the warm H$_2$ component to suffer less extinction, on average,
than the warm dust continuum, because the two emission sources will not be cospatial
in general, and the regions of high optical depth will be confined to compact
regions, probably more concentrated than the regions participating in H$_2$ emission
\citep[see the striking example of NGC\,1377;][]{Roussel06}.
In particular, \citet{Higdon06} did not see any evidence for significant extinction
in the rotational H$_2$ lines of ultraluminous galaxies, although these objects
are expected to have much higher optical depths than the present sample.
In the absence of any quantitative constraint on the differential extinction
between the dust and H$_2$, we do not attempt to correct H$_2$ fluxes for extinction.

Using the extinction law of \citet{Moneti01}, valid for the Galactic center, we have
$A(9.7\,\mu{\rm m}) / A_{\rm V} = 0.15$, $A(28.2\,\mu{\rm m}) / A(9.7\,\mu{\rm m}) = 0.25$ and
$A(17.0\,\mu{\rm m}) / A(9.7\,\mu{\rm m}) = A(12.3\,\mu{\rm m}) / A(9.7\,\mu{\rm m}) = 0.46$\,.
Even assuming the same optical depth toward the warm molecular hydrogen
as toward the hot dust,
the extinction correction would not change significantly the derived column densities.
The extinction is modest at 10\,$\mu$m, and therefore negligible at 28\,$\mu$m,
the wavelength of the S(0) line which dominates the total column density determination.
Extinction effects would however depress the S(1) and S(3) line fluxes with respect
to S(0) and S(2), and could thus artificially lower the derived $OPR_{\rm \,high\,T}$.
In the following, we put lower limits to $OPR_{\rm \,high\,T}$ values, when less than 3,
derived for the eight galaxies with non-zero optical depth at 10\,$\mu$m.

NGC\,3198 is the sample galaxy with the highest optical depth in the silicate
feature according to \citet{Smith07b}, but its excitation diagram shows no sign of
attenuation of the S(1) and S(3) lines relative to the others, and is consistent with
$OPR_{\rm \,high\,T} = 3$ (Fig.~\ref{fig:diag_temp}). The second most obscured galaxy
of the present sample is NGC\,1266 (it also has the highest nebular extinction according
to \citet{Dale06}, $A_{\rm V} = 4.1$\,mag), for which we derive $OPR_{\rm \,high\,T} < 3$.
If this were due to optical depth effects, then the S(3) line at 9.7\,$\mu$m should
be more attenuated than the S(1) line at 17.0\,$\mu$m. Since this would be consistent
with the excitation diagram, we cannot exclude that the apparently low $OPR_{\rm \,high\,T}$
value be an extinction artifact in at least this galaxy.
The dissimilar behavior of the two galaxies in terms of differential
extinction between H$_2$ and the dust could then arise from different excitation
mechanisms and geometries: whereas in the nuclear regions of NGC\,3198, classified
as purely H{\small II}, the H$_2$ emission is presumably distributed over a large
volume, the H$_2$ emission in the {\small LINER} nucleus of NGC\,1266 may be much
more compact, and not produced by star formation processes (see Section~\ref{powers}).
For 13 galaxies with negligible silicate extinction in the spectral decomposition
performed by \citet{Smith07b}, the excitation diagrams do imply $OPR_{\rm \,high\,T} < 3$,
whether a constraint on the lower temperature $T_1$ is applied or not.
In addition, of the 6 galaxies found to have non-zero silicate extinction and
$OPR_{\rm \,high\,T} < 3$, three would require $\tau{\rm(H_2)} > \tau_{\rm sil}$ in order
to obtain $OPR_{\rm \,high\,T} = 3$ after extinction correction (by $\geq 25$\% for
NGC\,1266, by a factor $\geq 6$ for NGC\,4631 and by a factor $\geq 3.5$ for NGC\,5866).
The three others (NGC\,1482, 4536 and 6946) would require either
$\tau{\rm(H_2)} > \tau_{\rm sil}$, or very low $T_1$ temperatures ($\leq 100$\,K).
Since it is unlikely that the optical depth toward H$_2$ be higher than toward the
dust continuum, we conclude that our finding, discussed in Section~\ref{excitation},
is robust against extinction effects.

\section{Mass fraction in the warm phase}
\label{h2co}

In order to estimate the fraction of molecular hydrogen that is heated to temperatures
above $\sim 100$\,K, we searched the literature for observed intensities of the 2.6\,mm CO(1-0)
line within a beam comparable to the solid angle of our observations. Table~\ref{tab_mass}
summarizes the adopted data. The column density of cold H$_2$ as given here is derived
from CO velocity-integrated intensities on the main-beam temperature scale, assuming
a uniform conversion factor of CO(1-0) intensities to H$_2$ column densities of
$2.3 \times 10^{20}$\,cm$^{-2}$/(K\,km\,s$^{-1}$) \citep{Strong88}.
We derived aperture corrections to the CO intensities by projecting on a map both
the IRS beam and the CO beam. We did not use any deconvolution technique.
Whenever possible, a map from the BIMA SONG interferometric survey, including
the zero-spacing total intensity \citep{Helfer03}, was used.
Otherwise, we used instead the 7.9\,$\mu$m map and assumed the spatial distributions
of aromatic bands in emission and CO(1-0) line emission to be similar
at the large spatial scales corresponding to our apertures.
This can be justified qualitatively by the association of dust with molecular gas
and the Schmidt law (for a recent study of the spatially-resolved Schmidt law,
see Kennicutt et al. 2007, in preparation).
The applied
correction factors are listed in Table~\ref{tab_mass}. In some cases, there are
several available measurements all giving consistent estimates to within 30\%~;
the corresponding unused references are given within parentheses.

There are two major sources of uncertainty in this comparison. The first one is inherent
to the difficulty of matching the physical area covered by the IRS integral-field
measurements, from single-dish or aperture-synthesis measurements within
a different beam. The second dominant source of uncertainty comes
from the conversion factor of CO intensities to H$_2$ masses, assumed uniform here.
The result of \citet{Strong88} is derived from a comparison of Galactic $\gamma$-ray
emission with CO and H{\small I} emission. \citet{Dame01} obtained a consistent
conversion factor by extrapolating the gas-to-dust mass ratio measured from H{\small I}
and far-infrared emission, in areas devoid of CO emission, to molecular clouds.
Both methods provide an estimate of the total H$_2$ column density, including the
warm gas as well as the cold gas, for molecular clouds under similar average
physical conditions as Galactic clouds.
Note however that conversion factors both significantly lower and significantly higher
have been derived for normal galaxies. For instance, the recent study of \citet{Draine07b}
favors an average value of $4 \times 10^{20}$\,cm$^{-2}$/(K\,km\,s$^{-1}$), based
on global gas-to-dust mass ratios in the SINGS sample. In addition,
the ratio of H$_2$ column density to CO intensity can vary by at least a factor
two, depending on the physical conditions of the regions emitting in CO \citep{Maloney88},
even though our observing aperture is large enough to cover a large number of molecular
clouds and dilute some of the dispersion in their physical properties.
In particular, the conversion factor is expected to be lower for compact and actively
star-forming regions than for more diffuse and more quiescent regions.
We discount here variations due to metal abundance, since we could find CO measurements
for only two low-metallicity targets (NGC\,2915 and NGC\,6822\_A).

Figure~\ref{fig:frac_warm} shows the mass fraction of molecular hydrogen in the warm
phase ($T \geq T_1 \approx 100$\,K) as a function of the minimum temperature of the
warm component, as determined by the lowest-energy rotational H$_2$ lines.
The nuclei classified as star-forming have a relatively narrow range of lower
temperatures ($T_1 = 144 \pm 24$\,K for 31 nuclei, with or without CO data,
from the constrained fits). However, for nuclear regions classified as
{\small LINER}s or Seyferts, the spread in temperatures is higher
($T_1 = 180 \pm 45$\,K for 25 nuclei). No statistically-significant difference
exists between the 18 {\small LINER} and 7 Sy nuclei.

A clear anticorrelation exists between the two quantities plotted
(partly the result of the degeneracy between temperatures and column densities),
which remains intact
when restricting the sample to those galaxies for which we could find well-matched
CO data (i.e. with correction factors close to unity and with several consistent
measurements). The dynamic range in the warm gas mass fraction is much
higher than accounted for by the uncertainty on the total H$_2$ mass.
The uncertainty on the warm H$_2$ mass for individual objects is however extremely
large, owing to the degeneracy between $T_1$, often ill-constrained by the data,
and the column density. The example of NGC\,4579 is the most striking
(see Table~\ref{tab_fit}).
Since its rotational levels up to J=5 are close to thermal equilibrium (at a
single temperature of the order of 300-400\,K), such a component at 70\,K
as found in the free-$T_1$ fit is unlikely to be real.
Because the fits where $T_1$ is unconstrained allow mass fractions in the warm phase
that are sometimes unphysical (for example for NGC\,2976 and NGC\,4826),
we favor the constrained fits as more plausible, but emphasize that the
mass distribution at low temperatures is in general unconstrained.

In the case of constrained-$T_1$ fits,
it appears that for a small set of nuclear regions classified as {\small LINER}s
or Seyferts, the warm H$_2$ phase consists only of a very small fraction of the total
mass, but heated to higher temperatures than in regions classified as purely star-forming.
This behavior arises naturally if normal photodissociation region excitation is missing,
and if the hotter gas is located in a thin layer of molecular clouds, or has a small
filling factor.
In the case of free-$T_1$ fits, only NGC\,1316 (Fornax\,A) remains robustly in the part
of the diagram with high $T_1$ and mass fraction below 3\%. The average temperature
is however still higher for {\small LINER}s and Seyferts than for H{\small II}
nuclei, and the average mass fraction in the warm phase likewise lower.
The reason for this difference will be further discussed in Section~\ref{syliner},
addressing the excitation mechanisms.

\section{Comparison of the powers emitted by warm H$_2$, [Si{\small II}] and dust in star-forming regions}
\label{powers}

In order to empirically quantify the importance of the H$_2$ rotational lines in cooling
the interstellar medium of normal galaxies, and to put constraints on the possible
excitation mechanisms of H$_2$, discussed in more detail in Section~\ref{excitation},
we examine power ratios of H$_2$ to other tracers of the warm interstellar medium
extracted from the same observations. The results presented here are independent
of any fits to the excitation diagrams. Only the H{\small II} nuclei and complexes
are considered, {\small LINER} and Sy nuclei being separately discussed in
Section~\ref{syliner}.
Since the bulk of warm H$_2$, at the lowest rotational temperatures, emits mostly
in the S(0) to S(2) lines, whereas the S(3) line emission has a noticeably higher
contribution from hotter H$_2$, probably indicating more mixed excitation sources
(anticipating the discussion of excitation mechanisms, see
Section~\ref{temp_constraints}), we choose, as the most useful quantification of
H$_2$ rotational emission in star-forming targets, the sum of the S(0) to S(2) lines.

\subsection{Total infrared emission}

In photodissociation regions, almost all the far-ultraviolet power from massive stars
that does not escape is absorbed by dust and converted to infrared continuum radiation,
or is absorbed by H$_2$. Only a very small fraction of the power absorbed
by dust, of the order of 1\%, is converted to photoelectrons that heat
the gas, and emerges as infrared lines \citep{Tielens85}.
The dominant gas coolants are the [O{\small I}] and [C{\small II}] lines
at 63\,$\mu$m and 158\,$\mu$m, but mid-infrared lines, in particular [Si{\small II}]
at 34.8\,$\mu$m and the H$_2$ rotational lines, are also energetically significant.
Although the transition rate coefficients of H$_2$ are low
and the excitation energies relatively high, H$_2$ molecules are dominant in number.

The observed ratios of the power emitted in the sum of the S(0) to S(2) lines
to the total dust power emitted in the infrared (TIR; see Section~\ref{measurements})
range between $2.5 \times 10^{-4}$ and $7.5 \times 10^{-4}$ for nuclear regions that
are not classified as {\small LINER}s or Seyferts (Fig.~\ref{fig:frac_tir}a).
These ratios are in agreement with predictions of the photodissociation models of
\citet{Kaufman06} for a wide variety of radiation field intensities $G_0$ and hydrogen
densities $n$, but a relatively narrow range of $G_0 / n$ ratios, approximately
between 0.1 and 1 with $G_0$ in units of $1.6 \times 10^{-3}$\,erg\,s$^{-1}$\,cm$^{-2}$
and $n$ in units of cm$^{-3}$. Note that models predict the ratio of the H$_2$ line power
to the far-ultraviolet (FUV) power (for photon energies between 6 and 13.6\,eV), rather
than the total infrared power. Since the intrinsic FUV flux heating the photodissociation
regions is unknown, the comparison between observations and models is here made by assuming
an exact conversion of FUV photons to infrared photons.
The fraction of dust heating provided by non-FUV photons can however be
significant.
Allowing for this effect would reduce the derived $G_0 / n$ ratios.
The H$_2$ rotational line fluxes predicted by \citet{Kaufman06} are nearly an
order of magnitude higher than those from the older models of \citet{Burton92},
because of the inclusion of photoelectric heating by PAHs,
a better H$_2$ model, and a finer numerical grid near the region of H$_2$ emission.

The inferred $G_0 / n$ ratios are lower than the results of \citet{Malhotra01},
who derived the physical conditions of an ensemble of bright star-forming galaxies
from the [C{\small II}] and [O{\small I}] lines. They found $G_0 / n$ ratios between
about 0.5 and 6, i.e. on average 5 times higher than those indicated here by the
rotational H$_2$ lines.
A possible explanation is that H$_2$ emission comes from cooler and denser regions
than [C{\small II}] and [O{\small I}], 
because H$_2$ exists at higher optical depths inside the clouds than C$^+$ and O
\citep{Hollenbach97}.
The difference in physical conditions could thus merely reflect a different spatial origin.
Besides the different locations within PDRs, the two studies also deal with different
regions within galaxies: the targets of \citet{Malhotra01} were selected to have
most of their line emission encompassed by the ISO-LWS beam of 70\arcsec,
whereas our apertures usually cover small fractions of the line and dust emitting
areas.
Alternatively, the observations of \citet{Malhotra01} could reflect intrinsically
different physical conditions because their sample contains galaxies on average
brighter and more active than the sample used here.
Their far-infrared luminosities \citep[in the definition of][]{Helou88} range from
$6 \times 10^7$ to $8 \times 10^{11}$\,L$_{\sun}$, with a median of
$1.5 \times 10^{10}$\,L$_{\sun}$, whereas the far-infrared luminosities of the
present sample range from $10^7$ to $6 \times 10^{10}$\,L$_{\sun}$, with a median
of $3 \times 10^9$\,L$_{\sun}$.
The median $F_{60}/F_{100}$ ratio is also higher in the sample of \citet{Malhotra01}
(0.57) than in our sample (0.41), indicating higher radiation field intensities
on average.
The $G_0 / n$ ratios derived by \citet{Malhotra01} however do not display any
clear correlation with either infrared luminosity or color.
Only NGC\,1482 and NGC\,5713, included in both samples, allow a direct comparison
of model results (we discard the {\small LINER} NGC\,1266 because most of its H$_2$
emission is not produced by PDRs, as shown in Sect.~\ref{syliner}).
For both sources, the H$_2$ line fluxes indicate consistently $G_0 \sim 4000$
and $n \sim 1$--$2 \times 10^4$. For NGC\,1482, $G_0$ is in agreement with one of
the two models of \citet{Malhotra01}, but $n$ is at least four times higher.
For NGC\,5713, $G_0$ is two times higher that that of \citet{Malhotra01}, and
$n$ is at least six times higher.
In conclusion, we favor differences in spatial origin (both within PDRs and within
galaxies) as a likely cause for the different model results.

\subsection{[Si{\small II}] line emission}

Figure~\ref{fig:frac_tir}b shows the ratio of powers emitted in the H$_2$ rotational
lines and in the [Si{\small II}] line. The dispersion in the ratio is very similar
to that seen in Fig.~\ref{fig:frac_tir}a, and the [Si{\small II}] line alone emits
more power than the sum of the S(0) to S(3) transitions in H{\small II} nuclei.
The [Si{\small II}] line has indeed been found to be the brightest mid-infrared
cooling line and to scale tightly with the total infrared power both in nuclear
and extranuclear regions within the SINGS sample galaxies (Helou et al., in
preparation), with only a very slight dependence on the radiation field intensity.
We have on average $F{\rm (S0-S2)}/F{\rm ([Si{\small II}])} = 0.3$
(ranging between 0.15 and 0.5 for nuclei),
and $F{\rm ([Si{\small II}])}/TIR = 2 \times 10^{-3}$.
Using the [Si{\small II}] line as a substitute for the total dust emission
is advantageous because it is observed at about the same angular resolution
as the H$_2$ lines, whereas estimating the total infrared
power within these apertures requires a large extrapolation (because of the
large width of the point spread function at 70 and 160\,$\mu$m), making the
uncertainty on H$_2$/TIR relatively high.
The [Si{\small II}] power predicted by the photodissociation region model
of \citet{Kaufman06}, with the same physical conditions as above,
is however smaller than observed by a factor greater than 3,
which implies either that the majority
of [Si{\small II}] emission comes from H{\small II} regions in high-metallicity
nuclear regions, or that the fraction of silicon incorporated in dust grains
is smaller than 90\%.

Only the regions B and C in NGC\,6822 have significantly less [Si{\small II}] emission,
with respect to H$_2$ emission, than the nuclear regions of spiral galaxies.
Their H$_2$ emission is also slightly overluminous with respect to the
aromatic bands (Fig.~\ref{fig:frac_tir}c). This may not be entirely attributable to
a metallicity effect, decreasing the abundances
of PAHs and silicon, since region A (Hubble\,V) has normal flux ratios,
and oxygen abundances are quite uniform in NGC\,6822 \citep{Pagel80}. An alternative
explanation is that additional excitation of H$_2$ may be provided in regions B and
C, with respect to region A, by shocks in supernova remnants
(see the more general discussion in Sect.~\ref{agn_shock}).
To our knowledge, no independant evidence exists to test the existence of
shocks in these regions. \citet{Chandar00} obtained a normal H{\small II}
optical line spectrum at the center of NGC\,6822\_C, but since their beam
of 2.5 arcsec$^2$ is only about 1\% of ours, we cannot rule out shock excitation.
Finally, given the small distance of NGC\,6822, the regions covered by the
IRS aperture are less than 100\,pc in size. Greater fluctuations around
the average properties are thus not unexpected.
At present, we are unable to decide which scenario is the most likely.

\subsection{Aromatic bands}
\label{pah_power}

Figure~\ref{fig:frac_tir}c shows a remarkable constancy of the power ratio
of the H$_2$ rotational lines to the aromatic bands. Among the measured dust
and gas observables, PAH emission provides the tightest correlation with H$_2$.
Observations of photodissociation regions have shown that the emission from aromatic
band carriers and from fluorescently-excited H$_2$ just outside photoionized regions
are nearly cospatial, with H$_2$ sometimes seen to extend slightly deeper into molecular
clouds \citep{Sellgren90, Graham93, Tielens93, Brooks00, Habart03}.
Cospatiality might be expected since both species
can be excited by FUV photons. Aromatic band carriers can also be excited
by lower-energy photons in the ultraviolet and optical, but with smaller
absorption cross-sections \citep[see][]{Li01}, so that FUV photons will dominate
the excitation whenever massive stars are present. H$_2$ is however dissociated by
FUV photons between 11.3 and 13.6\,eV where it is not self-shielded, whereas
PAHs survive the absorption of these photons.
Therefore, in the case of relatively dense PDRs (associated with molecular clouds),
where collisional heating is expected to be the major origin of the H$_2$ rotational
lines, H$_2$ emission should peak at slightly higher optical depth than aromatic bands,
in the transition layer between atomic and molecular hydrogen, with $A_{\rm V} > 1$.
In addition, PAHs probably cannot be excited as deep into molecular clouds
as H$_2$, because at sufficiently high densities they will be coagulated
onto grain mantles on short timescales \citep{Boulanger90}.
If photodissociation regions dominate the excitation of H$_2$, as consistent with
the above results, a tight relation between aromatic band emission and
rotational H$_2$ emission can arise
only if the physical conditions in PDRs, especially the $G_0 / n$ ratio, are
relatively uniform, because H$_2$ fluxes and PAH fluxes depend in very different
ways on these two parameters. The condition of relatively constant $G_0 / n$ ratios
seems verified in the present sample at least for the average emission within
kiloparsec-scale regions (see above). Based on the modelling of [C{\small II}] and
[O{\small I}] emission, \citet{Malhotra01} proposed that a regulation of $G_0 / n$
might be achieved at the scale of individual PDRs by expanding H{\small II} regions
in pressure equilibrium with their surrounding PDRs.

A correlation was previously claimed by \citet{Mouri90}
between the 3.3\,$\mu$m band and the v=1-0 S(1) line at 2.12\,$\mu$m for a small
sample of starburst and Seyfert galaxies. The dominant source that they propose
for H$_2$ excitation, following \citet{Moorwood88}, is however not photodissociation
regions, but shocks in supernova remnants. Using the shock models of \citet{Kaufman96}
to estimate the sum of the S(0) to S(2) transitions (up to 6\%
of the mechanical power, assuming that its totality is dissipated
in molecular clouds), and the population synthesis model
of \citet{Leitherer99} to estimate both the total mechanical power and the FUV luminosity
from continuous star formation with a Salpeter initial mass function, shocks alone
are in principle able to produce a significant fraction of the observed H$_2$ emission,
but only if the efficiency of conversion of mechanical power into H$_2$ emission
is unrealistically high. The rotational line ratios are also inconsistent
with shock models, which predict higher temperatures ($T > 1000$\,K) except
for very low shock velocities (in which case the power fraction radiated away
by rotational H$_2$ lines is lower). If the collective rotational line emission
from shocks in supernova remnants is similar to that observed in individual objects
such as 3C\,391 and IC\,443 \citep{Reach02}, then this mechanism can provide only
a modest fraction of the total H$_2$ emission.
In addition, if H$_2$ emission came predominantly from supernova
remnants whereas aromatic bands arise mostly in photodissociation regions, the partial
deconnection between the two, both temporal and spatial, would manifest itself by a large
scatter in the observed relation between H$_2$ and PAH fluxes
for galaxies with diverse star formation histories, which is not observed.

More recently, \citet{Rigopoulou02} proposed a relation similar to that
presented by \citet{Mouri90}, between the 7.7\,$\mu$m aromatic band and
the rotational S(1) line in starburst galaxies.
Figure~\ref{fig:frac_tir}c not only confirms this result for lower-luminosity galaxies,
but also shows that the dispersion for the whole sample of star-forming nuclei is
very small, and much smaller with the aromatic bands than with the
24\,$\mu$m emission (Fig.~\ref{fig:frac_tir}d), which is dominated by the continuum
from transiently-heated very small grains, as well as from big grains in intense
radiation fields. The quantification of the average H$_2$ to dust power ratios
and their dispersions is given in the caption of Fig.~\ref{fig:frac_tir}.
The energy coupling between aromatic band carriers and H$_2$ strongly suggests
that both are excited predominantly in photodissociation regions,
although they may not come from the exact same layers (at the same optical depths
within the clouds). We present further analysis in the next section.

A similar correlation, with a similarly small dispersion, was observed between
the [C{\small II}] line and aromatic band emission \citep{Helou01}. This relation
suggests that aromatic band carriers are the source of a major part of gas
heating in photodissociation regions, via the photoelectric effect, at least
at modest radiation field intensities, since [C{\small II}] emission is the
dominant cooling channel in this case.
In the narrow range of physical conditions that seem to apply if the emission from
H{\small II} nuclei is interpreted in the framework of photodissociation region models
(a dynamic range in $G_0 / n$ of only a factor 10), then the same link between aromatic
band carriers and H$_2$ would follow if H$_2$ were heated in relatively dense
photodissociation regions by the PAHs. Our results, however, suggest that in
nearly half the star-forming targets, the dominant excitation mechanism of the
rotational levels may be fluorescence in low-density regions, so that
ortho-para thermalization is not achieved by collisions
(see Sect.~\ref{excitation}). If the lines are fluorescently excited,
the cause underlying the tight relation between H$_2$ and aromatic band emission
may be that both are proportional to the incident far-ultraviolet flux which excited
them.

\subsection{Cirrus clouds versus PDRs with high radiation field intensities}

The tight association between H$_2$ emission and aromatic bands (Fig.~\ref{fig:frac_tir}c)
may be surprising if one assumes that a significant fraction of aromatic band
emission arises from diffuse, mostly atomic regions with low radiation field
intensities. The infrared emission of such clouds is often termed
cirrus \citep{Low84, Terebey86}. If this were the case, then the
scaling of PAH flux with H$_2$ flux could be explained only if a constant fraction of
the total FUV flux escaped PDRs and were absorbed in the more diffuse
interstellar medium. We stress that we adopt here the definition of photodissociation
regions stated by \citet{Hollenbach97}: these are not restricted to the interfaces
between bright H{\small II} regions and dense molecular clouds, but
apply more generally to all the neutral interstellar medium illuminated by
FUV photons (with energies between 6 and 13.6\,eV).
Figure~\ref{fig:frac_pah_frac_highU} demonstrates the great
difficulty of making the idea of an important contribution from the cirrus medium
consistent with the data. It shows the flux ratios of
H$_2$ to PAH emission on the one hand, and 24\,$\mu$m to PAH emission on the other
hand, as a function of $P_{24} \sim \nu_{24} F_{24} / (\nu_{71} F_{71} + \nu_{156} F_{156})$,
estimated within the spectroscopic apertures. The quantity $P_{24}$ is closely related
to $f_{U>100} = f(L_{\rm dust}~; U_{\rm rad} > 100)$, derived from the modelling of
the global spectral energy distributions by \citet{Draine07b}, which is the fraction
of the total dust luminosity emitted by regions with radiation field intensities
$U_{\rm rad}$ higher than 100 times the local average value. The dust luminosity
fraction of cirrus clouds,
$\sim f_{U<10}$, can be evaluated as $1 - c~ f_{U>100}$, if one assumes that
$f_{U>100}$ and $f_{U>10}$ are in constant proportion to each other.
Fig.~\ref{fig:frac_pah_frac_highU} shows that while the $F_{24}$/PAH ratio rises
by about one order of magnitude, the H$_2$/PAH ratio is invariant as a function of
$P_{24}$ or $f_{U>100}$.
The results obtained by replacing $P_{24}$ with $F_{71}/F_{156}$, which has the
same physical significance as $F_{60}/F_{100}$, a more traditional indicator
of the relative importance of H{\small II} regions and cirrus clouds
\citep{Helou86}, are identical.

We conclude that aromatic bands are mostly associated with photodissociation
regions (illuminated by FUV photons able to provide H$_2$ excitation).
In addition, since PAHs are excited not only by FUV photons but also by
low-energy photons, the observed constancy of the H$_2$ to PAH ratio imposes
some restrictions on possible variations of the radiation field hardness.
Assuming that cirrus clouds, i.e. PDRs with low radiation field intensities,
receive appreciably softer radiation than PDRs with high radiation field
intensities, it would be difficult to understand how both types of regions
could produce similar H$_2$/PAH ratios. As a corollary, the hypothesis that
cirrus clouds could make a large contribution to H$_2$ and PAH emission in
our targets, although not definitely ruled out, is not favored.
Note that the situation may be different in more quiescent parts of galaxies,
not probed by the present sample, and deserves further investigation.
Measurement of H$_2$ line fluxes in quiescent regions is however challenging,
because they depend steeply on the $G_0$ and $n$ parameters.

The above does not preclude a large portion of the H$_2$ and PAH emission
to originate in relatively diffuse molecular gas. Estimates of the optical
depth of the $^{12}$CO(1-0) line, over large areas of the Galaxy, indicate
that the total molecular medium comprises a substantial diffuse component
\citep{Polk88}. We will see in the next section that our data, for a portion
of the targets, do support an important contribution from low-density
PDRs to the total warm H$_2$ emission.

\section{Excitation mechanisms in star-forming regions}
\label{excitation}

In sources with purely stellar activity, H$_2$ emission is expected to arise in
varying proportions from
two main energy sources: the far-ultraviolet radiation of OB stars illuminating PDRs;
and shocks in supernova remnants or other sources, providing collisional heating.
In the first case, the excitation can be thermal, by collisions with gas heated
by photoelectrons, or by inelastic collisions with H$_2$ molecules pumped by FUV
photons. The excitation mechanism can also be fluorescence, followed by radiative
cascade to the ground vibrational state. Heating by supernova remnants is unlikely
to be dominant for two reasons, as we have seen in Sect.~\ref{pah_power}. First,
at the low observed temperatures dominating the warm H$_2$ mass, heating would have
to be provided by slow shocks, which are not efficient enough to compete with PDR
excitation. Second, variations in star formation histories within the sample,
which are shown by Moustakas et al. (2007, in prep.) to be very large (from
population synthesis fitting to optical spectra), would produce more scatter
than observed in the H$_2$/PAH ratio. We conclude that H$_2$ is
heated predominantly by PDRs, and this interpretation is supported both by
energetics arguments and by the close association with aromatic band emission
(Sect.~\ref{powers}). In this section, we focus on
additional constraints on the physical conditions and excitation mechanisms
in PDRs (thermal or fluorescent) from the line ratios and excitation diagrams.

\subsection{Constraints from the temperature distribution}
\label{temp_constraints}

We have seen that the H$_2$ to far-infrared ratios are consistent with values
of $G_0 / n$, the ratio of the average radiation field intensity to the hydrogen
density in PDRs, between about 0.1 and 1. In principle, separate constraints on
$G_0$ and on $n$ can be obtained from the temperature distribution reflected in
the H$_2$ line ratios \citep{Kaufman06}.
Given the complexity of the surveyed regions, however, the interpretation of
the line ratios by comparison with models of a single PDR is severely limited.
First, the emission from many distinct PDRs, presumably showing a large range
of physical conditions, is averaged within the beam. Second, even though the
total emission in the sum of the S(0) to S(3) lines is probably dominated by
PDRs, shocks also related to the star formation activity must be present,
driven in particular by supernova remnants, protostellar outflows, or turbulence
dissipation \citep{Falgarone05}. Since these shocks are characterized by higher
rotational temperatures than PDRs, they would contribute mostly to the S(3) line
\citep{Kaufman96}, and in view of the observed line ratios, negligibly to the
lower-lying transitions.

In both cases, the superposition of PDRs of different conditions or of shocks
induces a spread in temperatures, and as a consequence these are not reproduced
by single PDR models. We compared the line ratios of the star-forming targets
with $OPR_{\rm \,high\,T} = 3$ to the predictions of \citet{Kaufman06}. Although
the ranges in $G_0$ and $n$ derived from S(2)/S(0) have a broad overlap with
those derived from S(1)/S(0), the S(3)/S(1) ratios are inconsistent, indicating
more intense radiation fields and higher densities. The least biased tracer of
physical conditions in the bulk of the PDRs is thus the S(1)/S(0) ratio (for
galaxies with $OPR_{\rm \,high\,T} = 3$). It suggests that the average $G_0$
varies between about 100 and 5000 (in units of
$1.6 \times 10^{-3}$\,erg\,s$^{-1}$\,cm$^{-2}$)
and the average $n$ between about 500 and $10^4$\,cm$^{-3}$.

\subsection{Fluorescent excitation}

It is surprising that in several objects,
the S(0) to S(3) transitions are not thermalized, as indicated by deviations
from an apparent $OPR_{\rm \,high\,T}$ of 3. $OPR$ values that
are different from the equilibrium value at the temperature of the H$_2$ gas
arise naturally from fluorescent excitation, if the gas density is lower than
the critical density
for ortho-para equilibration by collisions with H and H$^+$ \citep{Sternberg99}.
This is because the ultraviolet
absorption lines have a greater optical depth in the ortho states than in the
para states, so that the ortho states at a given depth are pumped less.
The apparent $OPR_{\rm \,high\,T}$ values only apply to the states excited by
FUV pumping, and do not imply that the true $OPR_{\rm \,high\,T}$ value
(for the J=1 and J=0 states, where most of the gas resides)
be different from 3. For a total $OPR_{\rm \,high\,T}$ of 3, the $OPR_{\rm \,high\,T}$
of the excited levels is predicted to be close to $\sqrt{3} = 1.7$.
A thorough review of the phenomenon of selective excitation
and its implications for the interpretation of excitation diagrams was provided by
\citet{Sternberg99}. Note that radiative decay in the electronic ground state
and most collisional (de)excitations
always occur at fixed spin number and thus preserve
the ortho or para state. Conversion from one state to the other can be accomplished
by H$_2$ dissociation followed by reformation on dust grains, or by reactions
with protons and hydrogen atoms in the gas phase.

Deviations from thermalization had previously remained unseen for rotational
lines of extragalactic sources \citep{Rigopoulou02, Higdon06}.
In the frame of PDR excitation, values $OPR_{\rm \,high\,T} < 3$ may be interpreted
as arising in FUV-pumped gas with sufficiently low densities to prevent ortho-para
equilibration by collisions. However, because this process depends on unknown timescales
for gas heating, cooling, dissociation and reformation, we are unable to quantify
in a simple way the implied density conditions.
Alternatively, it is conceivable that the emission comes from initially cold gas,
that has been heated by slow shocks and has not had time to
reach the equilibrium value of $OPR$ \citep{Timmermann98, Wilgenbus00}.
For H{\small II} nuclei, however, we have seen that photodissociation region
excitation is more likely than shock heating (Sect.~\ref{powers}).

Another possibility is that in a fraction of the PDRs within the beam, the photodissociation
front is advancing into cold gas (in LTE at time $t_1$ with $OPR(t_1) \ll 3$ and
$OPR_{\rm \,high\,T}(t_1) = 3$), and the recently-heated gas has not yet had enough time
to reach LTE at time $t_2$ (with $OPR(t_2) \sim OPR(t1)$ and $OPR_{\rm \,high\,T}(t_2) < 3$
according to Equ.~\ref{eq:opr}). This is the interpretation
favored by \citet{Fuente99} and \citet{Habart03} to explain the non-LTE ortho to para ratios
observed in the PDRs of NGC\,7023 and $\rho$ Ophiuchi, respectively. In this scenario, the
observed portion of the interstellar medium would have to contain a much larger fraction
of non-equilibrium PDRs in targets with low $OPR_{\rm \,high\,T}$ than in targets with
$OPR_{\rm \,high\,T} = 3$, and the underlying reason would be unclear. It is also unknown
whether the timescales involved in ortho-para equilibration in our sources are long enough
for this scenario to be viable.

Figure~\ref{fig:opr_sb} shows the derived $OPR_{\rm \,high\,T}$ values (fixed to 3 whenever
the temperatures derived from each pair of adjacent transitions were compatible
with this assumption) as a function of the total brightness of the S(0) to S(2)
lines. We have here included {\small LINER}s and Seyferts because they do not
display a different behavior in this diagram. Under the hypothesis of photodissociation
region excitation, and assuming first that sites of star formation occupy
a constant fraction of the observing beam, sources with low H$_2$ brightnesses
should consist of regions with both low densities and low radiation field intensities,
while sources with the highest H$_2$ brightnesses should include a greater fraction
of high-density, high-radiation regions. In this simplified view, low values of
$OPR_{\rm \,high\,T}$, indicating that H$_2$ is not thermalized by collisions, could be
obtained only in the low-brightness sources, as seen generally in Fig.~\ref{fig:opr_sb}.
Variations in the beam filling factor by sites of star formation would then induce
a horizontal scatter, which is indeed very large. Fig.~\ref{fig:opr_sb} also shows
$OPR_{\rm \,high\,T}$ as a function of the average surface brightness in the 24\,$\mu$m band.
The latter quantity incorporates a significant contribution from H{\small II} regions,
and should be dominated by variations in radiation field intensity, rather than variations
in gas density. The fact that the horizontal spread is larger in this diagram than
in the diagram involving the H$_2$ brightness supports our tentative interpretation
in terms of density effects.
In addition, $OPR_{\rm \,high\,T}$ does not show any variation as a function of $F_{71}/F_{156}$,
and only a weak tendency to increase with the quantity $P_{24}$ discussed in the
previous section. Both these quantities are indicators of the average radiation
field intensity;
they are observed to be generally correlated with the gas density, but only weakly.
Although the data do not allow us to truly estimate average densities in photodissociation
regions, they suggest that in a substantial number of the observed nuclear regions,
the emission can be dominated by low density gas,
relative to well-studied Galactic PDRs and starburst galaxies. New modelling is required
to quantify the conditions under which rotational lines indicate $OPR_{\rm \,high\,T} < 3$
in PDRs.

The H$_2$ line ratios, compared with the PDR model of \citet{Kaufman06}, do not indicate
lower gas densities, on average, in galaxies with $OPR_{\rm \,high\,T} < 3$ than in
galaxies with $OPR_{\rm \,high\,T} = 3$. However, the densities estimated in this
way are averages within the whole beam, and the densest and warmest regions have a
greater weight, because they are more luminous in H$_2$; if a large spread in densities
exists, with both dense clumps and diffuse PDRs, an increase in the diffuse
fraction may not be easily detectable in the average density, while still leaving
an imprint on $OPR_{\rm \,high\,T}$. We note that the proton density should also
play an important role, since ortho-para conversion is effected by collisions
with H and H$^+$. Whether galaxies with $OPR_{\rm \,high\,T} < 3$ have lower
proton densities in molecular clouds, due to reduced ionization by cosmic rays,
is in principle testable with radio continuum observations
of synchrotron radiation from cosmic ray electrons. The number of sample
galaxies with adequate data, at sufficiently high angular resolution, is however
too small to apply this test.

We have seen in Section~\ref{extinction} that extinction effects are unlikely
to modify our results in a statistical sense. The fitted $OPR_{\rm \,high\,T}$ values
are correlated with the brightness of the dust emission,
in the aromatic bands, the 24\,$\mu$m continuum and total infrared emission (not
shown here but similar to Fig.~\ref{fig:opr_sb}), which is a further indirect argument
against extinction effects being responsible for low $OPR_{\rm \,high\,T}$ values.
If extinction played a significant role, then low $OPR_{\rm \,high\,T}$ values
would be seen preferentially in bright and compact regions, which is not the case.

The galaxies for which we derive the lowest $OPR_{\rm \,high\,T}$ values (and among the lowest
H$_2$ surface brightnesses, regardless of the constraint on the $T_1$ temperature) are
NGC\,337, NGC\,1705 ($OPR_{\rm \,high\,T}$ is a lower limit), NGC\,2915, NGC\,4552
($OPR_{\rm \,high\,T}$ is a lower limit) and NGC\,7793. The first three are dwarf galaxies,
NGC\,7793 a very late-type spiral and NGC\,4552 is a small elliptical galaxy classified
as {\small LINER}, with the smallest infrared brightness of the whole sample.
Since smaller gas densities are expected in general in dwarf galaxies than in the
central regions of massive galaxies, except in blue compact dwarfs such as Mrk\,33,
this finding is consistent with our interpretation of small $OPR_{\rm \,high\,T}$ values
in terms of low density. We may also remark that in NGC\,1705, which is a starburst,
no ultracompact H{\small II} region was detected in the radio \citep{Johnson03};
thus no conflict exists with the hypothesis that the photodissociation regions
in NGC\,1705 have low densities.

\section{Excitation mechanisms in {\small LINER} and Seyfert nuclei}
\label{syliner}

A large number of the galaxies classified as {\small LINER}s or Seyferts deviate
significantly from the relations discussed in Section~\ref{powers}, in having a
strong excess of H$_2$ emission with respect to all the other tracers used here,
not only aromatic bands but also [Si{\small II}], the 24\,$\mu$m flux and the total
infrared emission (Fig.~\ref{fig:frac_tir}),
arguing for an alternative excitation mechanism in these
galaxies. The average and dispersion of each power ratio are given separately
for H{\small II} nuclei and for {\small LINER} and Sy nuclei in the figure caption.
Given these results, the quantities by which the two categories are most clearly
separated are the H$_2$ to aromatic band ratio and the H$_2$ to [Si{\small II}] ratio.
In particular, nuclear regions with $F{\rm (S0-S2)}/F_{\rm 7.9\,dust} > 10^{-1.94}$
are likely to be of the {\small LINER} or Sy type at the 99\% confidence level.

We thus define an excess of H$_2$ emission, with respect to H{\small II} nuclei,
based on the observed relation with aromatic bands. We choose as the maximal H$_2$
power associated purely with star formation the quantity
$10^{-1.94} \times L_{\rm 7.9\,dust}$.
Aromatic band carriers are thought to be destroyed in intense radiation field environments,
such as H{\small II} region cores \citep{Giard94} and ionized regions around Seyfert nuclei
\citep{Desert88, Voit92}. They would
however survive where H$_2$ is not dissociated, so that an enhancement of the H$_2$/PAH
ratio is more likely caused by a genuine excess of H$_2$ emission, in conditions where
PAHs are not excited, rather than normal H$_2$ emission occurring where PAHs would
have been destroyed.
Furthermore, H$_2$ emission is seen in excess not only with respect to aromatic bands,
but also with respect to [Si{\small II}] and other dust tracers (24\,$\mu$m and
total infrared emission).
[Si{\small II}] emission may be depressed because the ionization state of silicon
becomes higher, but the dust continuum cannot be suppressed like aromatic bands.
Our empirical quantification of the H$_2$ excess is intended to extract the part of
H$_2$ emission that cannot originate in photodissociation regions. The excitation
mechanism that is required to account for this excess, while not exciting PAHs, is
either X-ray irradiation or shock heating.

We discuss {\small LINER} and Sy nuclei as a single category, because no detectable
difference exists in their H$_2$ properties. This may be expected for several reasons:
small-number statistics; the fact that the classification of low-luminosity AGNs
can be ambiguous as it depends on the aperture size in particular; and the fact
that the source of H$_2$ excitation might not be directly linked to the nuclear
activity, as the results discussed below suggest.

\subsection{Heating by X-rays from an active galactic nucleus}

The idea that a nuclear X-ray source may modify the chemistry and excitation of the
surrounding molecular clouds through sufficiently large column densities as to
offer a convenient way to identify active nuclei hidden by dust has recently
received much attention. Models predict, in particular, unusual ratios of tracers
of dense molecular gas \citep{Lepp96, Meijerink05}, consistent with observations
of NGC\,1068 \citep{Usero04}. X-ray excitation would also manifest itself in the
properties of H$_2$, the most abundant molecule. To test the hypothesis that the
additional excitation in galaxies showing a significant excess of H$_2$ emission
with respect to aromatic bands (Figure~\ref{fig:frac_tir}c) is predominantly
produced by nuclear X-rays, we have compiled estimated X-ray fluxes
in the 2-10\,keV band obtained from Chandra observations. The data and references
are summarized in Table~\ref{tab_xrays}. The H$_2$ excess is shown as a function
of the X-ray luminosity of the nucleus in Fig.~\ref{fig:xrays}. Here, galaxies
with no H$_2$ excess according to our definition are shown below the dashed line.

The spread in Figure~\ref{fig:xrays} is very large.
Since the available X-ray measurements
do not isolate the hard X-ray component, but include soft
emission, a substantial part can be thermal emission from supernova remnants,
as opposed to power-law emission from the activity related to a central supermassive
black hole. Many H{\small II} nuclei with no H$_2$ excess indeed have X-ray luminosities
that are comparable to those of {\small LINER} and Seyfert nuclei. On the other hand,
some galaxies, in particular NGC\,3627, NGC\,4569 and NGC\,5866, are very luminous
in H$_2$ but have only modest X-ray to H$_2$ luminosity ratios compared with
other H$_2$-excess galaxies.
We shall assume a power-law spectrum with a standard photon index of $-1.8$,
as adopted by \citet{Ho01}, to extrapolate the total X-ray luminosity to lower
photon energies. Under this assumption, the luminosity between 0.2 and 10\,keV
is only two times the luminosity between 2 and 10\,keV compiled here.
Up to 10\% of the intrinsic 1-10\,keV luminosity may emerge in the sum of all H$_2$
transitions \citep{Lepp83}, whereas in our comparison we have summed only the three
rotational lines S(0) to S(2). Therefore, pure X-ray excitation of the excess H$_2$
emission should not be energetically possible even in objects such as NGC\,4579 and
NGC\,5195, which have relatively low H$_2$ to X-ray luminosity ratios
in Fig.~\ref{fig:xrays}. Even if the intrinsic emission in soft X-rays
were underestimated by a large factor, X-ray excitation would still be
unlikely to dominate in most cases.
The apparent trend of increasing excess H$_2$ luminosities with increasing
X-ray luminosities in Fig.~\ref{fig:xrays} does not imply a direct excitation of
H$_2$ by X-rays, since it could also be understood in the frame of multi-phase
shocks, produced by supernova remnants or by starburst winds (see below).
From near-infrared line ratios, \citet{Davies05} also reached the conclusion that X-rays
do not contribute significantly to the excitation of H$_2$ in a small sample of
active galactic nuclei, within regions smaller than those sampled here (about 100\,pc).

Supernova remnants are another important source of X-rays. More than half their
mechanical energy can be converted to X-rays, mostly with energies below $\sim 500$\,eV
\citep{Draine91}, but the X-ray power is converted to H$_2$ emission in the S(0) to S(2)
lines with low efficiencies, less than $10^{-3}$ \citep{Draine90}. Therefore, in view
of the estimates presented in Fig.~\ref{fig:snshocks} (see the next Section for details),
heating of H$_2$ by X-rays from supernova remnants is completely negligible.

\subsection{Heating by shocks}
\label{agn_shock}

For galaxies such as NGC\,3627 and NGC\,4569, with very high H$_2$ to X-ray luminosity
ratios, an efficient mechanism has to be invoked to account for the H$_2$ brightness.
It has been shown recently that galactic shocks can convert a very large fraction
of the kinetic energy into rotational H$_2$ emission without producing a lot of
X-rays \citep{Appleton06}. We thus propose that large-scale shocks play
a major role. NGC\,3627 is an interacting galaxy in the Leo Triplet, characterized
by severe morphological and kinematical distortions and, in the center, a massive
molecular gas concentration in the form of an asymmetric bar-like structure
\citep{Zhang93}, with a peak in the stellar velocity dispersion that is shifted by
about 3\arcsec\ from the nucleus, and probably strong gas inflow into the nucleus
\citep{Afanasiev05}. NGC\,4569 is a starburst {\small LINER} \citep{Maoz98} having
recently produced a large number of supernova explosions, triggering the expansion
of X-ray, H$\alpha$ \citep{Tschoke01} and synchrotron emission lobes \citep{Chyzy06},
and also has a circumnuclear ring of molecular gas with strong non-circular motions
\citep{Nakanishi05}. The situation could be similar to NGC\,4945, where H$_2$ emission
is seen to follow the innermost part of a starburst outflow, with an extent of
$\sim 200$\,pc \citep{Moorwood96}.
These two examples thus support the idea of a dominant excitation by shocks, triggered
by dynamical perturbations and by a starburst wind, respectively.

In order to test the possibility that the excess H$_2$ emission be caused by
supernova remnant shocks, we estimated the required heating efficiencies in the
following way. We first computed the star formation rate required to account for the
excess H$_2$ emission, assuming constant star formation, if all the
mechanical power is converted to the power emitted by the sum of the S(0) to S(2)
lines \citep[using the Starburst99 population synthesis model of][]{Leitherer99}.
We then compared this with the star formation rate estimated from the 24\,$\mu$m
luminosity, using the calibration of \citet{Wu05} for the same initial mass function.
The ratio of the two rates gives an order-of-magnitude estimate of the H$_2$ heating
efficiency $\eta$ that is needed if supernova shocks are invoked as the dominant heating
mechanism, and is shown in Fig.~\ref{fig:snshocks} as a function of the excess H$_2$
luminosity. In this simplified computation, the H$_2$ excess in the star-forming
regions B and C of NGC\,6822 can easily be attributed to supernova remnant shocks,
as they would imply efficiencies of at most 3\%. Besides these regions, supernova
remnants are also a sufficient heating source of the excess H$_2$ for at least six
of the {\small LINER} and Sy nuclei. In NGC\,5195, in particular, the necessary
heating efficiency (including the fraction of the total mechanical power injected
and absorbed in molecular clouds) is very small. Galaxies such as NGC\,4450
and NGC\,4579, on the other hand, unambiguously require much more power than
available in supernova remnants. The excitation source could be shocks triggered by
cloud collisions induced by
gravitational perturbations, maybe in combination with X-rays.

NGC\,1316 (Fornax\,A) stands out in Fig.~\ref{fig:frac_warm} as having the highest
$T_1$ temperature ($T_1 > 280$\,K) and the lowest mass fraction in warm phase ($< 1$\%),
and the lowest levels up to J=4 are characterized by a single temperature.
In this particular case,
H$_2$ may be heated by fast shocks caused by the powerful jet that has produced
large-scale radio lobes, and that is observed in the form of knots in the inner
kiloparsec, which may be a signature of interaction with the interstellar medium,
i.e. shocks \citep{Geldzahler84}. Excitation by far-ultraviolet radiation and X-rays
or by slow shocks would produce a large quantity of H$_2$ at temperatures
between 100 and 300\,K \citep{Burton92, Maloney96}, which is ruled out by the data.
NGC\,1316 is the only galaxy of the sample for which the excess H$_2$ may be heated
by an AGN jet.

Of the 16 H$_2$-excess galaxies, all have excitation diagrams consistent with
$OPR_{\rm \,high\,T} = 3$ (at least in the free-$T_1$ fits),
except NGC\,1266, NGC\,5866 and NGC\,4125.
NGC\,1266 may be unusual simply by virtue of its significant optical depth,
as discussed in Section~\ref{extinction}. This predominance of $OPR_{\rm \,high\,T} = 3$,
combined with more elevated temperatures than in H{\small II} nuclei
on average (Fig.~\ref{fig:frac_warm}) would be consistent with the excess H$_2$
emission originating in shocks where ortho-para equilibration is fast. It remains
however difficult to test this idea, and to identify the source of these shocks.

\section{Summary and conclusions}
\label{summary}

We present the measurements and results of a survey of the four lowest-energy rotational
transitions of H$_2$, S(0) to S(3), in a local sample of 57 galaxies, from the SINGS program.
For three galaxies in this sample, higher-energy transitions, up to S(7), could be measured.
Characterizing the amount and physical conditions of the warm molecular hydrogen phase
traced by these lines is of prime interest, because molecular hydrogen represents a major
mass fraction of the interstellar medium of normal galaxies, and the warm phase itself
(gas heated to temperatures of $\approx 100$ to 1000\,K) can constitute a substantial
fraction of the total H$_2$.

The emission is measured over areas of median size 0.9\,kpc, thus including a large number
of distinct star formation sites and molecular clouds. The sample comprises mostly nuclear
regions (in 47 massive galaxies and 9 dwarf galaxies), of which 45\% are optically
classified as {\small LINER} or Seyfert, as well as 10 extranuclear star-formation
complexes within a dwarf galaxy and a spiral.
With respect to earlier studies of molecular hydrogen emission in galaxies, and
particularly rotational lines, which had focussed on very bright systems (nearby
starbursts and AGNs, as well as ultraluminous galaxies), this paper provides results
on the average properties of warm H$_2$ of relatively faint systems, more representative
of the general population of galaxies.

Perhaps the most significant observational results (detailed below) are:
(1) the tight correlation of the powers emitted by the sum of the S(0) to S(2)
lines and by aromatic bands, and the fact that the $F$(S0-S2)/PAH ratio
is insensitive to the marked variations in average radiation field intensities
existing in our sample; (2) the existence of non-equilibrium ortho to para
ratios in the rotational levels, that are weakly correlated with the
surface brightness of the H$_2$ lines. These results call for further modelling
in order to be better understood.

{\bf Masses and column densities:}
The total masses of warm H$_2$ within our apertures range between $10^5$\,M$_{\sun}$
and close to $3 \times 10^8$\,M$_{\sun}$ in galaxy nuclear regions. In star formation
complexes of nearby dwarf galaxies (NGC\,2915 and NGC\,6822), we probe warm H$_2$ masses
down to a few $10^3$\,M$_{\sun}$ within equivalent diameters of 60 to 250\,pc.
The mass densities range between 0.2 and 30\,M$_{\sun}$\,pc$^{-2}$.
The column densities that we derive are on average of the same order of magnitude
as the column densities observed in individual Galactic photodissociation regions.
For systems in which H$_2$ is predominantly excited by PDRs,
assuming they have similar characteristics to the Orion Bar,
this implies that they fill most of the observing beam.

{\bf H$_2$ mass fraction in the warm phase ($\mathbf T \geq 100$\,K):}
Under a conservative assumption about the distribution of temperatures, we find
that the warm H$_2$ gas makes up between 1\% and more than 30\% of the total H$_2$.
For star-forming galaxies, the median mass fraction in the warm phase is 10\%.
The column density nevertheless has a steep inverse dependence on the temperature,
and we cannot rule out that the unconstrained cool H$_2$
component ($70\,{\rm K} < T < 100$\,K) might in some cases dominate the S(0) and
S(1) emission and account for most of the H$_2$ mass.

{\bf PDR excitation in star-forming regions:}
In H{\small II} nuclei, we observe a remarkably narrow range of H$_2$ to aromatic
band flux ratio.
This result argues for photodissociation regions providing most of the power used
for H$_2$ excitation, since aromatic bands are known to arise predominantly from
these regions,
as defined by \citet{Hollenbach97}, that include all the neutral interstellar
medium illuminated by FUV photons.
Two main excitation mechanisms can be at work simultaneously in
PDRs: pumping by FUV photons, followed by fluorescent decay to the ground electronic
state, and collisions with hydrogen atoms and molecules heated by photoelectrons
or FUV pumping.
Comparison with the predictions of PDR models for the ratio of H$_2$ to FUV luminosity
indicates a narrow range of average physical conditions, $G_0 / n$ between 0.1 and 1
with the radiation field intensity $G_0$ in units of
$1.6 \times 10^{-3}$\,erg\,s$^{-1}$\,cm$^{-2}$ and the hydrogen nucleus density
$n$ in units of cm$^{-3}$. The sum of the S(0) to S(2) transitions represents
between $2.5 \times 10^{-4}$ and $7.5 \times 10^{-4}$ of the total infrared power,
and on average 30\% of the [Si{\small II}] line power.
The observed temperatures suggest that the average $G_0$ varies between about
100 and 5000, and the average $n$ between about 500 and $10^4$\,cm$^{-3}$.
This seems to imply that H$_2$ rotational line emission comes mostly
from molecular clouds illuminated by OB associations.
We have seen however that the estimator of these parameters is biased
(see Section~\ref{temp_constraints}),
and a non-negligible contribution from less dense regions with less intense
radiation fields is not excluded.

{\bf Evidence for fluorescence in star-forming regions:}
Previous surveys of rotational lines in galaxies had not revealed any departure
from thermalization of H$_2$, which is a consequence of the relatively low
critical densities of the lower rotational transitions. By contrast, we find that
nearly half the targets in our sample deviate significantly
from local thermodynamic equilibrium in having apparent ortho to para ratios
$OPR_{\rm \,high\,T}$ lower than the equilibrium value of three.
We have seen that this result cannot be an artifact caused by extinction effects.
Low values of $OPR_{\rm \,high\,T}$ may thus be interpreted as evidence of fluorescent
excitation, which naturally leads to low ortho to para ratios in the excited states,
occurring in regions of sufficiently low density that
ortho-para equilibration by collisions is incomplete.
The fraction of relatively diffuse molecular gas in normal
galaxies could thus be far from negligible.
In order to test this idea, it would be desirable to obtain independent estimates
of the average gas density in PDRs. In the absence of any robust estimate of this
quantity, we used the surface brightness of the sum of the S(0) to S(2) transitions
as a tracer of the coupled variations of $n$ and $G_0$ to show that the data are
compatible with an interpretation of low apparent $OPR_{\rm \,high\,T}$ values in terms
of low density PDRs. In particular, the lowest values occur preferentially in very
late type galaxies, that may have a more
diffuse interstellar medium than earlier-type galaxies.
In the present sample, close to half the star-forming targets have low $OPR_{\rm \,high\,T}$
values. We infer that fluorescence can be the predominant excitation mechanism of rotational
H$_2$ lines in normal star-forming galaxies. In more active galaxies, however, collisional
excitation in photodissociation regions is likely to overtake fluorescence, even though
the latter still contributes to gas heating.
Alternatively, low $OPR_{\rm \,high\,T}$ values may be caused by non-equilibrium PDRs,
in which initially cold gas, recently reached by the photodissociation front and heated,
has not had enough time to adjust its ortho to para ratio. A disadvantage of this scenario
is that it does not explain why PDRs would have systematically different properties
in targets with $OPR_{\rm \,high\,T} < 3$ and in those with $OPR_{\rm \,high\,T} = 3$.
We thus emphasize that the cause of non-equilibrium ortho to para ratios in the
rotational levels is not well understood at present.

{\bf Differences between H{\small II} and {\small LINER}/Sy nuclei:}
Despite our large observing beam implying that the emission from the immediate vicinity
of the nucleus is diluted in the emission from extended areas decoupled from nuclear
activity, a large fraction of nuclei classified as {\small LINER} or Seyfert distinguish
themselves from purely star-forming nuclei in several ways. In a statistical sense,
the temperatures of the warm H$_2$ phase are slightly higher, and as a corollary the
mass fractions of warm to total H$_2$ are lower
(with a median of 4\% instead of 10\% for H{\small II} nuclei).
The correlation between H$_2$ and
aromatic band emission observed in H{\small II} nuclei also breaks down in
{\small LINER} and Sy nuclei. A large number of them have excess emission in the
H$_2$ lines, with respect to aromatic bands and to [Si{\small II}], which
in general is the brightest cooling line in the mid-infrared range (with possible
contributions from both PDRs and H{\small II} regions as well as X-ray irradiated
gas), and, to a lesser degree, with respect to the total infrared emission.
The fact that less contrast between the different nuclear categories is seen
in the H$_2$/TIR ratio may partly be due to the fact that estimating the 70\,$\mu$m
and 160\,$\mu$m fluxes within our small apertures requires a large extrapolation.

{\bf Threshold for nonstellar excitation:}
We propose that nuclear ratios $F{\rm (S0-S2)}/F_{\rm 7.9\,dust} > 10^{-1.94}$,
with observables defined as in our study, are indicative of the {\small LINER}
and Sy categories. It should however be kept in mind that some sources belonging
to these classes are indistinguishable from H{\small II} nuclei. This could be
thought of as mainly a distance and beam dilution effect; but the H$_2$/PAH ratio
behaves contrary to this expectation, since in the present sample it shows a
correlation with distance, rather than an anti-correlation.
None of the quantities derived in this paper shows any dependence on distance,
with this exception of the H$_2$ excess in {\small LINER} and Sy nuclei.
The fact that it tends to increase with distance could be a selection bias,
but it also suggests that the H$_2$ excess is in general spatially extended.
We may also remark that for galaxies like NGC\,1377, whose dust emission is
interpreted as dominated by an extremely young and opaque starburst \citep{Roussel06},
the above criterion to select {\small LINER} and Sy nuclei will not be applicable.

{\bf Shock excitation in {\small LINER}/Sy nuclei:}
We interpret the differences in H$_2$/PAH ratios as a genuine excess of H$_2$ emission
(and not a deficit of the other tracers while preserving H$_2$), i.e. as requiring
at least one additional mechanism to excite H$_2$ molecules with respect to PDR heating.
Excitation by nuclear X-rays seems implausible, as models predict much lower heating
efficiencies than what would be necessary to account for the estimated H$_2$ excess,
compared with X-ray luminosities derived from Chandra observations.
We thus favor excess heating by large-scale shocks, caused either by the collective
effect of supernov\ae\ in an aging starburst, or by dynamical perturbations.
An order-of-magnitude estimate suggests that supernova remnant shocks can easily
account for the H$_2$ excess of a fraction of the {\small LINER} and Sy nuclei
(for example NGC\,5195), but do not provide enough mechanical power for the galaxies
with the highest H$_2$/PAH ratios.
For the latter, shocks triggered by dynamical perturbations are the best candidate
to supply the excess H$_2$ heating.
In one case, it is conceivable that the excess H$_2$ emission may be produced
by the interaction of a nuclear jet with the interstellar medium, namely in NGC\,1316
(Fornax\,A). This target has the warmest H$_2$ of the whole sample, with no evidence
for a cool ($T < 300$\,K) component, the lowest mass fraction in the warm phase,
and is remarkable for its large-scale radio lobes.

{\bf Ortho-para thermalization in {\small LINER}/Sy nuclei:}
Consistent with the hypothesis that the additional H$_2$ may be caused by shocks
in which ortho-para equilibration is fast, the excitation diagrams of most galaxies
with excess H$_2$ emission are consistent with ortho-para
thermalization ($OPR_{\rm \,high\,T} = 3$), with three exceptions: NGC\,1266,
where we have seen that the apparent $OPR_{\rm \,high\,T} < 3$ may be a result of high
optical depth toward H$_2$, NGC\,5866 and NGC\,4125.
Note that the latter two galaxies are observed edge-on, so that $OPR_{\rm \,high\,T} < 3$
could also be an extinction artefact; this would however imply that the optical depth derived
from the silicate bands be underestimated by large factors (see Sect.~\ref{extinction}).
If excess H$_2$ emission originated from shocks with $OPR_{\rm \,high\,T} < 3$ in the
latter two galaxies, it would imply that the gas was initially cold and had not had
time to reach equilibrium.

The rotational H$_2$ lines are most often fainter than the forbidden lines
in the same wavelength range, in particular [Si{\small II}], [Ne{\small II}],
[Ne{\small III}] and [S{\small III}] (with no example of an H$_2$-dominated line
spectrum in the present sample), but are among the dominant coolants of
molecular gas and provide important constraints on the excitation of the warm
molecular
interstellar medium, where a large mass fraction of the gas resides in normal
galaxies. The results presented here are assumed to be representative of
moderate-luminosity galaxies of all types, and can serve as a comparison point
for future studies of distant galaxies.

\acknowledgements
Support for this work, part of the Spitzer Space Telescope Legacy Science
Program, was provided by NASA through an award issued by the Jet Propulsion
Laboratory, California Institute of Technology under NASA contract 1407.
We thank Adam Leroy for a useful discussion.

\clearpage

\begin{deluxetable}{llrlrr}
\tablecaption{Observed H$_2$ lines.\tablenotemark{a}
\label{tab_lines}
}
\tablehead{
transition & short    & rest $\lambda$ & spectral & E$_{\rm u}$/k & A \\
v=0        & notation & ($\mu$m)       & order    & (K)           & (10$^{-11}$\,s$^{-1}$)\\
}
\startdata
J=2-0 & S(0) & 28.219 & LH\,14 &  510 & 2.95 \\
J=3-1 & S(1) & 17.035 & SH\,12 & 1015 & 47.6 \\
J=4-2 & S(2) & 12.279 & SH\,17 & 1681 & 275. \\
J=5-3 & S(3) &  9.665 & SL\,1  & 2503 & 980. \\
J=6-4 & S(4) &  8.025 & SL\,1  & 3473 & 2640. \\
J=7-5 & S(5) &  6.910 & SL\,2  & 4585 & 5880. \\
J=8-6 & S(6) &  6.109 & SL\,2  & 5828 & 11400. \\
J=9-7 & S(7) &  5.511 & SL\,2  & 7196 & 20000. \\
\enddata
\tablenotetext{a}{The rotational upper level energies were computed from the molecular
constants given by \citet{Huber79} and the transition probabilities are from \citet{Black76}.}
\end{deluxetable}

\begin{deluxetable}{lrlcl}
\tabletypesize{\footnotesize}
\tablecaption{Targets.\tablenotemark{a}
\label{tab_target}
}
\tablehead{
galaxy & D     & class & center (J2000)   & solid angle \\
~      & (Mpc) & ~     & RA~~~~~~~~~~~DEC & (arcsec$^2$) \\
}
\startdata
N24   &  8.2 & dwarf & 00 09 56.31 $-$24 57 52.0 & 294. \\
N337  & 24.7 & nuc   & 00 59 49.99 $-$07 34 42.5 & 289. \\
N628  & 11.4 & nuc   & 01 36 41.72 +15 46 59.4 & 300. \\
N855  &  9.6 & dwarf & 02 14 03.64 +27 52 40.6 & 288. \\
N925  & 10.1 & nuc   & 02 27 17.06 +33 34 43.8 & 287. \\
N1097 & 16.9 & nuc   & 02 46 18.77 $-$30 16 30.0 & 809. \\
N1266 & 31.3 & liner & 03 16 00.68 $-$02 25 38.8 & 287. \\
N1291 &  9.7 & liner & 03 17 18.50 $-$41 06 27.7 & 287. \\
N1316 & 26.3 & liner & 03 22 41.61 $-$37 12 28.7 & 287. \\
N1482 & 22.0 & nuc   & 03 54 38.68 $-$20 30 08.5 & 798. \\
N1512 & 10.4 & nuc   & 04 03 53.96 $-$43 20 55.6 & 831. \\
N1566 & 18.0 & sy    & 04 20 00.53 $-$54 56 17.4 & 306. \\
N1705 &  5.8 & dwarf & 04 54 13.26 $-$53 21 39.2 & 802. \\
N2403 &  3.5 & nuc   & 07 36 50.25 +65 36 04.6 & 287. \\
N2798 & 24.7 & nuc   & 09 17 22.99 +42 00 00.8 & 293. \\
N2841 &  9.8 & liner & 09 22 02.75 +50 58 35.5 & 291. \\
N2915 &  2.7 & dwarf & 09 26 11.59 $-$76 37 34.2 & 291. \\
N2976 &  3.5 & dwarf & 09 47 15.63 +67 55 00.3 & 289. \\
N3031 &  3.5 & sy    & 09 55 33.50 +69 03 55.8 & 296. \\
N3049 & 19.6 & nuc   & 09 54 49.66 +09 16 19.2 & 320. \\
N3184 &  8.6 & nuc   & 10 18 17.03 +41 25 28.3 & 287. \\
N3190 & 17.4 & liner & 10 18 05.71 +21 49 56.8 & 304. \\
N3198 &  9.8 & nuc   & 10 19 55.10 +45 32 59.6 & 287. \\
N3265 & 20.0 & nuc   & 10 31 06.85 +28 47 49.1 & 301. \\
Mrk33 & 21.7 & dwarf & 10 32 32.02 +54 24 05.4 & 315. \\
N3351 &  9.3 & nuc   & 10 43 57.83 +11 42 10.9 & 761. \\
N3521 &  9.0 & liner & 11 05 48.51 $-$00 02 10.1 & 285. \\
N3621 &  6.2 & liner & 11 18 16.62 $-$32 48 49.9 & 291. \\
N3627 &  8.9 & sy    & 11 20 15.04 +12 59 31.0 & 301. \\
N3773 & 12.9 & dwarf & 11 38 12.91 +12 06 43.0 & 285. \\
N3938 & 12.2 & nuc   & 11 52 49.45 +44 07 15.0 & 292. \\
N4125 & 21.4 & liner & 12 08 05.73 +65 10 27.9 & 294. \\
N4254 & 20.0 & nuc   & 12 18 49.71 +14 25 01.4 & 299. \\
N4321 & 20.0 & nuc   & 12 22 55.08 +15 49 18.2 & 795. \\
N4450 & 20.0 & liner & 12 28 29.51 +17 05 04.9 & 285. \\
N4536 & 25.0 & nuc   & 12 34 27.22 +02 11 14.3 & 777. \\
N4552 & 20.0 & liner & 12 35 39.71 +12 33 21.9 & 285. \\
N4559 & 11.6 & nuc   & 12 35 57.79 +27 57 36.7 & 295. \\
N4569 & 20.0 & liner & 12 36 49.91 +13 09 46.8 & 287. \\
N4579 & 20.0 & sy    & 12 37 43.70 +11 49 07.1 & 295. \\
N4594 & 13.7 & liner & 12 39 59.58 $-$11 37 22.3 & 289. \\
N4625 &  9.5 & dwarf & 12 41 52.65 +41 16 26.5 & 294. \\
N4631 &  9.0 & nuc   & 12 42 07.84 +32 32 34.8 & 282. \\
N4725 & 17.1 & sy    & 12 50 26.71 +25 30 03.2 & 294. \\
N4736 &  5.3 & liner & 12 50 53.23 +41 07 13.6 & 298. \\
N4826 &  5.6 & liner & 12 56 43.79 +21 41 00.7 & 292. \\
N5033 & 13.3 & sy    & 13 13 27.67 +36 35 38.2 & 311. \\
N5055 &  8.2 & liner & 13 15 49.50 +42 01 46.6 & 313. \\
N5194 &  8.2 & sy    & 13 29 52.94 +47 11 44.2 & 330. \\
N5194\_A\tablenotemark{b}
      &  8.2 & hii   & 13 29 49.54 +47 13 28.6 & 256. \\
N5194\_B\tablenotemark{b}
      &  8.2 & hii   & 13 30 01.69 +47 12 50.9 & 271. \\
N5194\_C\tablenotemark{b}
      &  8.2 & hii   & 13 30 00.02 +47 11 11.9 & 352. \\
N5194\_D\tablenotemark{b}
      &  8.2 & hii   & 13 30 02.76 +47 09 53.0 & 338. \\
N5194\_E\tablenotemark{b}
      &  8.2 & hii   & 13 29 56.84 +47 10 45.9 & 234. \\
N5194\_F\tablenotemark{b}
      &  8.2 & hii   & 13 29 52.93 +47 12 38.7 & 297. \\
N5194\_G\tablenotemark{b}
      &  8.2 & hii   & 13 29 43.99 +47 10 20.7 & 285. \\
N5195 &  8.2 & liner & 13 29 59.80 +47 16 00.1 & 323. \\
N5713 & 26.6 & nuc   & 14 40 11.57 $-$00 17 19.1 & 292. \\
N5866 & 12.5 & liner & 15 06 29.58 +55 45 46.7 & 286. \\
N6822\_A\tablenotemark{c}
      &  0.6 & dwarf & 19 44 52.84 $-$14 43 09.8 & 308. \\
N6822\_B\tablenotemark{c}
      &  0.6 & dwarf & 19 44 50.57 $-$14 52 49.0 & 288. \\
N6822\_C\tablenotemark{c}
      &  0.6 & dwarf & 19 44 48.67 $-$14 52 26.5 & 304. \\
N6946 &  5.5 & nuc   & 20 34 51.79 +60 09 12.6 & 803. \\
N7331 & 15.7 & liner & 22 37 04.11 +34 24 57.5 & 263. \\
N7552 & 22.3 & nuc   & 23 16 10.56 $-$42 35 04.0 & 722. \\
N7793 &  3.2 & dwarf & 23 57 49.79 $-$32 35 28.4 & 296. \\
\enddata
\tablenotetext{a}{
Distances are from \citet{Kennicutt03}.
Nuclear classifications derived from optical spectroscopy, published
by \citet{Smith07b}, were modified for the following targets.
N1097, N1512, N4321 and N7552: Since our aperture includes a bright
star-forming ring, which dominates the dust and line emission, we adopt
the H{\small II} class instead of the {\small LINER} class.
N2841, N4552, N4569, N4594, N4826 and N5195: The Sy class was
changed to {\small LINER} \citep{Ho97}.
N3198 and N3938: The {\small LINER} class was changed to H{\small II} \citep{Ho97}.
The dwarf galaxy class is here arbitrarily defined by a total stellar mass,
estimated as by \citet{Lee06}, below $10^{9.7}$\,M$_{\sun}$.
}
\tablenotetext{b}{A total of 11 extranuclear regions were observed in high-resolution
spectroscopy in NGC\,5194, from which we included 7 in our sample.}
\tablenotetext{c}{Several locations were observed in high-resolution spectroscopy
in this galaxy. We present here results for three among the brightest regions at
7.9 and 24\,$\mu$m. N6822\_A corresponds to the H{\small II} region Hubble\,V
\citep{Hubble25}, and N6822\_C to the H{\small II} region K$\gamma$ \citep{Kinman79}.}
\end{deluxetable}

\begin{deluxetable}{lrrrrrrrr}
\tabletypesize{\footnotesize}
\rotate
\tablecaption{Fluxes.\tablenotemark{a}
\label{tab_flux}
}
\tablehead{
galaxy & S(0) & S(1) & S(2) & S(3) & [SiII]            & F$_{\rm 7.9\,dust}$ & F$_{24}$ & TIR \\
~      & \multicolumn{5}{c}{(10$^{-18}$\,W\,m$^{-2}$)} & (Jy)               & (Jy)     & (10$^{-15}$\,W\,m$^{-2}$) \\
}
\startdata
N24   &     $6.6 \pm 2.6$ &    $2.2 \pm 2.4$ &    $3.2 \pm 2.0$ &   $10.3 \pm 6.4$ &     $36.7 \pm 3.6$ & 0.010 & 0.010 &     $18. \pm 3.$ \\
N337  &    $11.6 \pm 3.8$ &   $21.0 \pm 4.9$ &   $15.4 \pm 9.9$ &    $9.7 \pm 5.2$ &    $186.1 \pm 7.0$ & 0.048 & 0.087 &     $89. \pm 9.$ \\
N628  &     $7.9 \pm 2.5$ &   $13.7 \pm 5.3$ &    $0.0 \pm 3.0$ &    $5.7 \pm 8.6$ &     $47.8 \pm 2.8$ & 0.026 & 0.026 &     $39. \pm 3.$ \\
N855  &     $2.9 \pm 0.9$ &   $11.2 \pm 6.5$ &    $0.0 \pm 5.1$ &   $22.4 \pm 7.2$ &     $70.2 \pm 3.8$ & 0.017 & 0.039 &     $53. \pm 7.$ \\
N925  &     $3.8 \pm 1.7$ &    $7.8 \pm 2.9$ &    $3.7 \pm 2.1$ &    $0.0 \pm 8.8$ &     $81.0 \pm 4.5$ & 0.021 & 0.022 &     $37. \pm 4.$ \\
N1097 &  $213.1 \pm 42.6$ & $726.1 \pm 43.3$ & $293.6 \pm 29.4$ & $423.0 \pm 23.4$ & $4921.1 \pm 121.9$ & 1.139 & 3.763 & $2214. \pm 227.$ \\
N1266\tablenotemark{b}
      &    $16.4 \pm 8.5$ &  $148.5 \pm 6.6$ &  $121.8 \pm 7.1$ & $189.8 \pm 11.6$ &   $153.5 \pm 41.5$ & 0.048 & 0.579 &   $438. \pm 30.$ \\
N1291 &     $3.7 \pm 1.8$ &   $29.8 \pm 4.8$ &   $13.9 \pm 8.2$ &  $31.6 \pm 10.0$ &     $75.1 \pm 5.8$ & 0.013 & 0.049 &     $60. \pm 7.$ \\
N1316 &     $1.5 \pm 0.8$ &   $36.0 \pm 6.1$ &   $18.4 \pm 5.9$ &   $84.0 \pm 9.9$ &    $102.4 \pm 6.4$ & 0.025 & 0.094 &   $105. \pm 11.$ \\
N1482 &  $106.8 \pm 35.1$ & $424.0 \pm 53.4$ & $184.0 \pm 22.8$ & $207.5 \pm 26.9$ &  $3954.0 \pm 75.1$ & 1.163 & 2.974 & $1752. \pm 154.$ \\
N1512 &    $25.9 \pm 2.9$ &   $83.6 \pm 8.9$ &  $27.2 \pm 10.4$ &  $48.3 \pm 13.6$ &   $390.7 \pm 13.9$ & 0.091 & 0.150 &   $203. \pm 24.$ \\
N1566 &    $24.0 \pm 2.2$ &  $129.5 \pm 7.1$ &   $55.3 \pm 4.8$ &  $91.6 \pm 17.6$ &    $101.9 \pm 7.8$ & 0.098 & 0.237 &   $235. \pm 19.$ \\
N1705 &     $5.6 \pm 3.9$ &    $2.2 \pm 1.3$ &   $0.0 \pm 10.3$ &   $11.6 \pm 4.0$ &    $86.3 \pm 12.0$ & 0.007 & 0.028 &     $40. \pm 5.$ \\
N2403 &     $5.5 \pm 1.5$ &    $6.6 \pm 3.1$ &    $2.0 \pm 2.0$ &    $4.6 \pm 8.2$ &     $82.0 \pm 5.9$ & 0.029 & 0.022 &     $35. \pm 2.$ \\
N2798 &   $43.2 \pm 25.3$ & $207.8 \pm 22.1$ &   $90.7 \pm 9.2$ & $105.2 \pm 12.1$ &   $840.4 \pm 42.2$ & 0.354 & 1.723 &   $936. \pm 67.$ \\
N2841 &     $3.1 \pm 1.4$ &    $9.6 \pm 2.5$ &    $4.1 \pm 4.5$ &  $28.0 \pm 10.8$ &     $81.1 \pm 5.5$ & 0.010 & 0.027 &     $48. \pm 4.$ \\
N2915 &     $0.8 \pm 0.6$ &    $1.8 \pm 2.1$ &    $3.8 \pm 1.8$ &    $0.0 \pm 8.0$ &     $40.1 \pm 3.7$ & 0.005 & 0.019 &     $25. \pm 3.$ \\
N2976 &     $8.9 \pm 2.4$ &   $15.7 \pm 3.0$ &    $4.9 \pm 3.5$ &    $5.2 \pm 4.2$ &     $66.1 \pm 5.4$ & 0.026 & 0.055 &     $77. \pm 6.$ \\
N3031 &     $6.3 \pm 3.6$ &   $46.1 \pm 6.1$ &   $26.1 \pm 8.1$ &   $65.3 \pm 8.0$ &    $217.9 \pm 9.7$ & 0.081 & 0.337 &   $234. \pm 17.$ \\
N3049 &     $6.4 \pm 2.3$ &   $23.7 \pm 3.8$ &   $11.7 \pm 6.5$ &   $11.1 \pm 5.2$ &   $197.1 \pm 11.9$ & 0.051 & 0.258 &   $141. \pm 13.$ \\
N3184 &     $9.8 \pm 2.3$ &   $22.6 \pm 2.9$ &    $6.3 \pm 3.5$ &   $0.0 \pm 15.9$ &    $105.0 \pm 4.2$ & 0.031 & 0.082 &   $103. \pm 13.$ \\
N3190 &    $18.1 \pm 2.9$ &   $75.3 \pm 8.0$ &   $20.9 \pm 6.4$ &  $71.6 \pm 12.6$ &     $67.1 \pm 7.3$ & 0.051 & 0.068 &   $137. \pm 23.$ \\
N3198 &    $13.2 \pm 4.9$ &   $32.1 \pm 5.4$ &   $10.2 \pm 3.5$ &   $18.7 \pm 8.5$ &     $96.5 \pm 6.6$ & 0.059 & 0.260 &   $200. \pm 20.$ \\
N3265 &     $9.3 \pm 3.2$ &   $30.6 \pm 3.5$ &   $11.4 \pm 7.9$ &   $15.3 \pm 7.4$ &    $159.1 \pm 7.7$ & 0.062 & 0.193 &    $119. \pm 9.$ \\
Mrk33 &    $13.9 \pm 5.6$ &   $29.3 \pm 3.7$ &   $11.5 \pm 5.3$ &   $45.7 \pm 8.6$ &   $245.4 \pm 10.9$ & 0.076 & 0.583 &   $239. \pm 13.$ \\
N3351 &   $62.8 \pm 10.5$ & $218.4 \pm 20.6$ &  $88.3 \pm 10.1$ & $169.8 \pm 19.9$ &  $1858.9 \pm 35.7$ & 0.366 & 1.385 &   $917. \pm 93.$ \\
N3521 &    $17.8 \pm 3.6$ &   $31.2 \pm 6.9$ &   $12.3 \pm 4.0$ &   $20.5 \pm 8.8$ &    $157.5 \pm 4.6$ & 0.101 & 0.106 &   $129. \pm 10.$ \\
N3621 &    $16.7 \pm 2.1$ &   $38.3 \pm 3.2$ &   $17.1 \pm 4.7$ &   $49.0 \pm 8.4$ &    $150.3 \pm 7.5$ & 0.075 & 0.065 &     $88. \pm 6.$ \\
N3627 &    $31.2 \pm 3.9$ & $318.6 \pm 12.2$ &  $141.7 \pm 5.2$ & $209.1 \pm 12.6$ &    $162.1 \pm 9.3$ & 0.190 & 0.406 &   $518. \pm 31.$ \\
N3773 &     $4.8 \pm 3.4$ &    $7.3 \pm 3.2$ &    $3.7 \pm 2.5$ &    $8.3 \pm 8.6$ &    $106.9 \pm 6.1$ & 0.024 & 0.087 &     $67. \pm 7.$ \\
N3938 &     $8.0 \pm 1.2$ &   $13.4 \pm 3.3$ &    $4.2 \pm 3.2$ &    $0.0 \pm 8.0$ &     $57.7 \pm 4.2$ & 0.027 & 0.030 &     $51. \pm 5.$ \\
N4125 &     $2.6 \pm 1.6$ &   $17.6 \pm 4.7$ &   $12.6 \pm 6.4$ &   $18.0 \pm 8.8$ &     $46.8 \pm 5.9$ & 0.005 & 0.015 &     $21. \pm 3.$ \\
N4254 &    $17.6 \pm 1.3$ &   $87.3 \pm 9.5$ &   $46.6 \pm 9.2$ &   $32.5 \pm 8.8$ &   $409.0 \pm 10.6$ & 0.160 & 0.203 &   $231. \pm 23.$ \\
N4321 &   $78.2 \pm 11.7$ & $267.5 \pm 22.8$ & $132.1 \pm 24.6$ & $164.5 \pm 16.2$ &  $1507.3 \pm 27.2$ & 0.392 & 0.730 &   $778. \pm 90.$ \\
N4450 &    $10.3 \pm 1.6$ &   $91.4 \pm 5.2$ &   $33.7 \pm 9.5$ &  $89.0 \pm 11.2$ &     $29.6 \pm 3.1$ & 0.011 & 0.028 &     $55. \pm 9.$ \\
N4536 &  $108.7 \pm 32.3$ & $411.6 \pm 40.5$ & $175.6 \pm 23.3$ & $216.8 \pm 20.7$ &  $3207.4 \pm 70.0$ & 0.813 & 2.401 & $1400. \pm 118.$ \\
N4552 &     $1.8 \pm 1.4$ &    $1.1 \pm 1.5$ &    $0.0 \pm 3.9$ &    $9.1 \pm 4.6$ &     $13.4 \pm 3.9$ & 0.004 & 0.017 &      $6. \pm 1.$ \\
N4559 &    $13.6 \pm 2.1$ &   $17.9 \pm 3.5$ &    $4.0 \pm 1.5$ &  $23.4 \pm 13.7$ &     $81.1 \pm 3.5$ & 0.034 & 0.030 &     $47. \pm 5.$ \\
N4569\tablenotemark{c}
      &    $41.6 \pm 6.9$ &  $316.1 \pm 8.7$ &  $152.4 \pm 5.2$ & $309.4 \pm 18.7$ &   $288.5 \pm 17.6$ & 0.128 & 0.475 &   $365. \pm 40.$ \\
N4579\tablenotemark{d}
      &     $8.9 \pm 2.3$ &  $164.1 \pm 5.2$ & $105.8 \pm 12.4$ & $256.3 \pm 10.0$ &   $175.3 \pm 12.2$ & 0.032 & 0.135 &   $160. \pm 18.$ \\
N4594 &     $1.9 \pm 2.3$ &    $6.4 \pm 3.8$ &    $4.2 \pm 2.7$ &    $4.6 \pm 4.0$ &    $126.1 \pm 7.3$ & 0.017 & 0.068 &     $70. \pm 8.$ \\
N4625 &     $8.0 \pm 1.6$ &    $9.6 \pm 1.9$ &    $5.4 \pm 4.3$ &   $5.0 \pm 11.4$ &     $67.5 \pm 4.8$ & 0.024 & 0.024 &     $33. \pm 5.$ \\
N4631 &    $58.2 \pm 6.4$ & $123.6 \pm 16.1$ &   $60.4 \pm 5.6$ &   $46.8 \pm 9.1$ &  $1350.4 \pm 22.7$ & 0.338 & 0.661 &   $636. \pm 61.$ \\
N4725 &    $11.7 \pm 1.7$ &   $37.1 \pm 5.1$ &   $19.9 \pm 8.1$ &   $38.0 \pm 9.0$ &     $24.9 \pm 3.2$ & 0.011 & 0.035 &     $56. \pm 9.$ \\
N4736 &    $33.2 \pm 6.8$ & $250.8 \pm 16.4$ & $100.1 \pm 10.4$ & $221.7 \pm 13.6$ &   $219.5 \pm 21.4$ & 0.293 & 0.636 &   $678. \pm 50.$ \\
N4826 &    $71.3 \pm 6.5$ & $345.3 \pm 15.8$ & $150.9 \pm 10.9$ & $210.6 \pm 10.3$ &   $927.6 \pm 15.0$ & 0.343 & 0.612 &   $756. \pm 92.$ \\
N5033 &    $36.6 \pm 3.5$ & $182.0 \pm 10.4$ &   $63.5 \pm 3.1$ & $126.9 \pm 19.1$ &   $530.9 \pm 10.0$ & 0.177 & 0.239 &   $311. \pm 45.$ \\
N5055 &    $44.0 \pm 2.4$ &  $158.0 \pm 9.1$ &   $52.0 \pm 6.7$ &  $80.2 \pm 10.6$ &    $295.2 \pm 9.0$ & 0.148 & 0.166 &   $300. \pm 37.$ \\
N5194 &    $17.7 \pm 2.8$ &  $134.4 \pm 9.1$ &   $75.7 \pm 4.7$ & $187.3 \pm 17.7$ &   $634.9 \pm 12.2$ & 0.168 & 0.291 &   $351. \pm 39.$ \\
N5194\_A &  $5.8 \pm 1.3$ &   $16.4 \pm 2.7$ &    $4.3 \pm 2.6$ &    $5.4 \pm 1.8$ &     $27.8 \pm 3.0$ & 0.030 & 0.036 &     $53. \pm 2.$ \\
N5194\_B & $18.1 \pm 2.2$ &   $43.1 \pm 8.2$ &   $12.0 \pm 4.3$ &  $18.1 \pm 10.2$ &    $339.7 \pm 8.5$ & 0.094 & 0.193 &   $201. \pm 13.$ \\
N5194\_C & $23.2 \pm 2.2$ &   $43.4 \pm 6.7$ &    $5.7 \pm 4.5$ &   $14.1 \pm 4.6$ &    $223.6 \pm 6.8$ & 0.079 & 0.115 &    $155. \pm 7.$ \\
N5194\_D & $14.7 \pm 2.8$ &   $30.3 \pm 6.5$ &    $3.9 \pm 3.1$ &    $8.4 \pm 3.0$ &    $222.9 \pm 6.2$ & 0.058 & 0.118 &   $130. \pm 14.$ \\
N5194\_E & $22.7 \pm 1.7$ &   $47.9 \pm 6.0$ &   $19.2 \pm 7.1$ &   $23.3 \pm 3.1$ &    $222.2 \pm 4.8$ & 0.072 & 0.089 &    $116. \pm 8.$ \\
N5194\_F & $19.4 \pm 2.2$ &   $39.2 \pm 5.2$ &   $15.4 \pm 9.1$ &  $21.6 \pm 14.0$ &    $271.7 \pm 5.8$ & 0.075 & 0.110 &    $132. \pm 4.$ \\
N5194\_G & $22.2 \pm 3.0$ &   $49.6 \pm 5.4$ &   $23.6 \pm 2.7$ &  $27.7 \pm 11.2$ &    $425.6 \pm 8.5$ & 0.093 & 0.192 &   $173. \pm 15.$ \\
N5195 &   $54.2 \pm 20.9$ & $310.4 \pm 12.3$ &  $127.7 \pm 6.5$ & $275.0 \pm 17.2$ &   $127.8 \pm 19.6$ & 0.271 & 0.835 &   $600. \pm 55.$ \\
N5713 &    $27.4 \pm 3.6$ & $151.6 \pm 14.3$ &   $50.9 \pm 7.9$ &   $85.6 \pm 7.0$ &   $798.7 \pm 14.8$ & 0.268 & 0.738 &   $500. \pm 54.$ \\
N5866 &    $15.5 \pm 1.5$ &   $90.0 \pm 5.5$ &   $39.1 \pm 7.0$ &   $37.2 \pm 6.5$ &     $78.8 \pm 3.9$ & 0.038 & 0.052 &   $169. \pm 26.$ \\
N6822\_A &  $8.1 \pm 4.3$ &   $21.1 \pm 4.3$ &   $16.0 \pm 4.1$ &   $25.9 \pm 3.8$ &   $187.1 \pm 11.2$ & 0.036 & 0.517 &   $283. \pm 24.$ \\
N6822\_B &  $7.8 \pm 1.3$ &   $11.2 \pm 3.6$ &    $8.8 \pm 2.8$ &   $10.9 \pm 5.1$ &     $19.0 \pm 2.8$ & 0.010 & 0.037 &     $44. \pm 5.$ \\
N6822\_C &  $4.4 \pm 1.9$ &    $8.0 \pm 5.2$ &    $6.2 \pm 5.4$ &    $8.0 \pm 2.9$ &     $16.0 \pm 3.6$ & 0.009 & 0.100 &     $76. \pm 7.$ \\
N6946 &  $227.7 \pm 61.9$ & $636.9 \pm 33.5$ & $271.6 \pm 17.1$ & $286.3 \pm 29.3$ &  $3203.9 \pm 91.9$ & 1.314 & 4.234 & $2354. \pm 193.$ \\
N7331 &    $16.4 \pm 1.6$ &   $49.6 \pm 5.0$ &   $13.8 \pm 1.9$ &   $23.4 \pm 8.3$ &    $213.9 \pm 4.8$ & 0.074 & 0.091 &   $125. \pm 10.$ \\
N7552 & $257.8 \pm 106.9$ & $565.3 \pm 42.6$ & $245.9 \pm 22.5$ & $315.9 \pm 14.8$ & $3619.9 \pm 165.2$ & 1.495 & 7.840 & $3237. \pm 219.$ \\
N7793 &     $7.0 \pm 1.7$ &   $15.0 \pm 3.2$ &   $11.7 \pm 4.6$ &    $5.3 \pm 3.7$ &     $79.7 \pm 4.3$ & 0.041 & 0.046 &     $71. \pm 5.$ \\
\enddata
\tablenotetext{a}{See Sections~\ref{images} and \ref{measurements}.
For NGC\,1266, NGC\,4569 and NGC\,4579, it was possible to measure higher-level
transitions of H$_2$, which are provided in the following notes.}
\tablenotetext{b}{The S(4) to S(7) line fluxes are respectively: $(102.7 \pm 17.4)$, $(241.9 \pm 19.9)$,
$> 18.5$ and $(192.2 \pm 28.6) \times 10^{-18}$\,W\,m$^{-2}$. The S(6) transition was not
accurately measurable because it was observed at low spectral resolution on the blue shoulder
of a bright 6.2\,$\mu$m aromatic band.}
\tablenotetext{c}{The S(4) to S(7) line fluxes are respectively: $(52.6 \pm 27.7)$, $< 252.6$,
$< 61.3$ and $(50.3 \pm 24.2) \times 10^{-18}$\,W\,m$^{-2}$. The S(5) and S(6) transitions are
upper limits because S(5) is contaminated by the [Ar{\small II}] line at 6.985\,$\mu$m and S(6)
is diluted in the blue shoulder of the 6.2\,$\mu$m aromatic band.}
\tablenotetext{d}{The S(4) to S(7) line fluxes are respectively: $(48.7 \pm 10.7)$, $< 324.9$,
$< 147.2$ and $(83.3 \pm 37.5) \times 10^{-18}$\,W\,m$^{-2}$. The same remark as for NGC\,4569
applies.}
\end{deluxetable}

\begin{deluxetable}{lrrrrrrr}
\tabletypesize{\footnotesize}
\tablecaption{Derived temperatures, column densities and $OPR_{\rm \,high\,T}$.\tablenotemark{a}
\label{tab_fit}
}
\tablehead{
galaxy & $T_1$ & $f_1$ & $T_2$ & $f_2$ & $OPR_{\rm \,high\,T}$ & $N_{\rm tot}(T > T_1)$       & $N_{\rm tot}(OPR=3)$ \\
~      & (K)   & ~     & (K)   & ~     & ~                     & ($10^{20}$\,mol.\,cm$^{-2}$) & ~$/N_{\rm tot}(OPR<3)$ \\
}
\startdata
N24   &    97. & 0.99941 &  769. & 0.00059 & $1.49 \pm 0.91$ &  4.232 & 1.75 \\
~     &    78. & 0.99980 &  675. & 0.00020 & $1.90 \pm 0.57$ & 15.210 & 1.46 \\
~     &    90. & 0.99971 &  888. & 0.00029 & $1.45 \pm 0.37$ &  7.314 & 1.73 \\
N337  &   160. & 0.97793 &  402. & 0.02207 & $1.02 \pm 0.15$ &  1.745 & 2.13 \\
~     &   128. & 0.97803 &  352. & 0.02197 & $1.21 \pm 0.25$ &  3.218 & 0.97 \\
~     &   165. & 0.98711 &  464. & 0.01289 & $0.86 \pm 0.14$ &  1.824 & 2.19 \\
N628  &   119. & 0.99678 &  393. & 0.00322 & 3               &  2.598 & ~    \\
N855  &   146. & 0.98750 &  578. & 0.01250 & 3               &  0.508 & ~    \\
N925  &   123. & 0.99473 &  494. & 0.00527 & 3               &  1.050 & ~    \\
~     &    99. & 0.99648 &  421. & 0.00352 & 3               &  2.780 & ~    \\
N1097 &   157. & 0.98027 &  422. & 0.01973 & $2.03 \pm 0.36$ & 11.837 & 1.29 \\
~     &    94. & 0.99033 &  335. & 0.00967 & 3               & 62.925 & ~    \\
N1266\tablenotemark{b}
      &   247. & 0.83691 &  400. & 0.15619 & $1.79 \pm 0.29$ &  1.231 & 1.19 \\
~     &   204. & 0.82588 &  400. & 0.16935 & $1.80 \pm 0.18$ &  1.594 & 0.84 \\
N1291 &   184. & 0.95420 &  484. & 0.04580 & 3               &  0.415 & ~    \\
~     &   141. & 0.95010 &  414. & 0.04990 & 3               &  0.762 & ~    \\
N1316 &   300. & 0.96436 & 1456. & 0.03564 & 3               &  0.114 & ~    \\
~     & $> 278.$ & $> 0.93750$ & $> 900.$ & $< 0.06250$ & 3  &  0.121 & ~    \\
N1482 &   171. & 0.96943 &  407. & 0.03057 & $1.83 \pm 0.25$ &  5.098 & 1.38 \\
~     &   137. & 0.96699 &  357. & 0.03301 & $2.12 \pm 0.43$ &  8.848 & 1.49 \\
~     &   175. & 0.98252 &  469. & 0.01748 & $1.58 \pm 0.25$ &  5.378 & 1.40 \\
N1512 &   138. & 0.98887 &  415. & 0.01113 & 3               &  1.883 & ~    \\
~     &   118. & 0.98955 &  375. & 0.01045 & 3               &  3.338 & ~    \\
N1566 &   177. & 0.96436 &  441. & 0.03564 & $2.24 \pm 0.37$ &  2.749 & 1.19 \\
~     &   115. & 0.97305 &  359. & 0.02695 & 3               &  8.517 & ~    \\
N1705 &   127. & 0.99453 &  811. & 0.00547 & $0.49 \pm 0.29$ &  0.494 & 3.78 \\
~     &    92. & 0.99844 &  711. & 0.00156 & $0.65 \pm 0.14$ &  2.050 & 0.95 \\
~     &   114. & 0.99727 &  936. & 0.00273 & $0.49 \pm 0.11$ &  0.876 & 2.33 \\
N2403 &   109. & 0.99844 &  449. & 0.00156 & 3               &  2.486 & ~    \\
~     &    99. & 0.99873 &  419. & 0.00127 & 3               &  4.138 & ~    \\
N2798 &   179. & 0.96182 &  409. & 0.03818 & $1.92 \pm 0.23$ &  5.169 & 1.32 \\
~     &   144. & 0.95518 &  359. & 0.04482 & $2.17 \pm 0.43$ &  8.460 & 1.16 \\
~     &   184. & 0.97812 &  472. & 0.02188 & $1.67 \pm 0.27$ &  5.419 & 1.27 \\
N2841 &   137. & 0.99678 & 1030. & 0.00322 & 3               &  0.709 & ~    \\
~     &   132. & 0.99707 &  970. & 0.00293 & 3               &  0.844 & ~    \\
N2915 &   169. & 0.97090 &  763. & 0.02910 & $1.05 \pm 0.47$ &  0.096 & 1.74 \\
~     &   116. & 0.98838 &  669. & 0.01162 & $1.30 \pm 0.28$ &  0.305 & 0.55 \\
~     &   151. & 0.98350 &  880. & 0.01650 & $1.00 \pm 0.20$ &  0.150 & 1.13 \\
N2976 &   134. & 0.99199 &  376. & 0.00801 & $1.84 \pm 0.30$ &  2.095 & 1.43 \\
~     &    86. & 0.99619 &  300. & 0.00381 & 3               & 12.046 & ~    \\
N3031 &   178. & 0.95908 &  523. & 0.04092 & 3               &  0.700 & ~    \\
~     &   126. & 0.96387 &  433. & 0.03613 & 3               &  1.699 & ~    \\
N3049 &   174. & 0.96611 &  403. & 0.03389 & $1.60 \pm 0.20$ &  0.740 & 1.50 \\
~     &   141. & 0.96113 &  353. & 0.03887 & $1.83 \pm 0.36$ &  1.238 & 1.17 \\
~     &   179. & 0.98076 &  465. & 0.01924 & $1.39 \pm 0.22$ &  0.773 & 1.55 \\
N3184 &   127. & 0.99502 &  440. & 0.00498 & 3               &  2.631 & ~    \\
~     &   118. & 0.99648 &  449. & 0.00352 & 3               &  3.805 & ~    \\
N3190 &   148. & 0.99355 &  574. & 0.00645 & 3               &  3.286 & ~    \\
~     &   142. & 0.99385 &  554. & 0.00615 & 3               &  3.883 & ~    \\
N3198 &   128. & 0.99316 &  415. & 0.00684 & 3               &  3.438 & ~    \\
~     &   111. & 0.99385 &  375. & 0.00615 & 3               &  6.283 & ~    \\
N3265 &   156. & 0.98105 &  408. & 0.01895 & $2.02 \pm 0.35$ &  1.421 & 1.30 \\
~     &    95. & 0.98975 &  325. & 0.01025 & 3               &  7.006 & ~    \\
Mrk33 &   124. & 0.99736 &  624. & 0.00264 & 3               &  3.844 & ~    \\
~     &   115. & 0.99775 &  584. & 0.00225 & 3               &  5.478 & ~    \\
N3351 &   141. & 0.98633 &  437. & 0.01367 & 3               &  4.575 & ~    \\
~     &   114. & 0.98867 &  387. & 0.01133 & 3               &  9.893 & ~    \\
N3521 &   118. & 0.99531 &  414. & 0.00469 & 3               &  5.715 & ~    \\
~     &    91. & 0.99687 &  354. & 0.00313 & 3               & 18.837 & ~    \\
N3621 &   126. & 0.99541 &  532. & 0.00459 & 3               &  4.490 & ~    \\
~     &   110. & 0.99609 &  472. & 0.00391 & 3               &  8.195 & ~    \\
N3627 &   220. & 0.91836 &  433. & 0.08164 & $2.42 \pm 0.19$ &  2.672 & 1.12 \\
~     &   124. & 0.90469 &  344. & 0.09531 & 3               &  7.324 & ~    \\
N3773 &   115. & 0.99678 &  480. & 0.00322 & 3               &  1.671 & ~    \\
~     &    90. & 0.99805 &  410. & 0.00195 & 3               &  5.446 & ~    \\
N3938 &   118. & 0.99629 &  425. & 0.00371 & 3               &  2.603 & ~    \\
~     &   102. & 0.99697 &  377. & 0.00303 & 3               &  5.116 & ~    \\
N4125 &   211. & 0.92803 &  481. & 0.07197 & $1.75 \pm 0.24$ &  0.238 & 1.35 \\
~     &   163. & 0.91826 &  422. & 0.08174 & $1.90 \pm 0.37$ &  0.377 & 1.15 \\
~     &   212. & 0.95625 &  555. & 0.04375 & $1.56 \pm 0.25$ &  0.262 & 1.35 \\
N4254 &   200. & 0.94111 &  384. & 0.05889 & $1.49 \pm 0.11$ &  1.763 & 1.53 \\
~     &   167. & 0.90293 &  337. & 0.09707 & $1.59 \pm 0.29$ &  2.368 & 1.14 \\
~     &   206. & 0.96982 &  444. & 0.03018 & $1.36 \pm 0.22$ &  1.812 & 1.45 \\
N4321 &   164. & 0.97480 &  427. & 0.02520 & $1.76 \pm 0.29$ &  4.001 & 1.41 \\
~     &   131. & 0.97627 &  374. & 0.02373 & $2.07 \pm 0.43$ &  7.511 & 1.27 \\
~     &   168. & 0.98477 &  493. & 0.01523 & $1.49 \pm 0.25$ &  4.258 & 1.46 \\
N4450 &   190. & 0.96992 &  536. & 0.03008 & 3               &  1.178 & ~    \\
~     &   172. & 0.96152 &  476. & 0.03848 & 3               &  1.467 & ~    \\
N4536 &   166. & 0.97344 &  413. & 0.02656 & $1.90 \pm 0.28$ &  5.602 & 1.35 \\
~     &   133. & 0.97334 &  362. & 0.02666 & $2.23 \pm 0.45$ & 10.170 & 1.35 \\
~     &   170. & 0.98437 &  476. & 0.01563 & $1.62 \pm 0.27$ &  5.908 & 1.40 \\
N4552 &   126. & 0.99463 &  986. & 0.00537 & $0.84 \pm 0.56$ &  0.449 & 2.61 \\
~     &    89. & 0.99873 &  865. & 0.00127 & $1.09 \pm 0.24$ &  2.156 & 0.54 \\
~     &   108. & 0.99766 & 1138. & 0.00234 & $0.87 \pm 0.19$ &  0.959 & 1.30 \\
N4559 &   111. & 0.99951 &  775. & 0.00049 & 3               &  6.202 & ~    \\
~     &   109. & 0.99961 &  825. & 0.00039 & 3               &  6.967 & ~    \\
N4569\tablenotemark{c}
      &   184. & 0.91348 &  400. & 0.08564 & 3               &  4.062 & ~    \\
~     &   138. & 0.95226 &  400. & 0.04755 & 3               &  9.055 & ~    \\
N4579\tablenotemark{d}
      &   286. & 0.55439 &  400. & 0.43999 & $2.80 \pm 0.56$ &  0.534 & 1.00 \\
~     &    68. & 0.98551 &  400. & 0.01439 & 3               & 21.592 & ~    \\
N4594 &   175. & 0.96621 &  441. & 0.03379 & $1.49 \pm 0.23$ &  0.232 & 1.57 \\
~     &   138. & 0.96709 &  387. & 0.03291 & $1.71 \pm 0.35$ &  0.421 & 1.06 \\
~     &   178. & 0.97969 &  509. & 0.02031 & $1.28 \pm 0.21$ &  0.251 & 1.64 \\
N4625 &   133. & 0.99268 &  411. & 0.00732 & $1.27 \pm 0.28$ &  1.856 & 1.85 \\
~     &   107. & 0.99453 &  361. & 0.00547 & $1.65 \pm 0.35$ &  4.144 & 1.20 \\
~     &   136. & 0.99551 &  475. & 0.00449 & $1.05 \pm 0.20$ &  1.964 & 2.38 \\
N4631 &   154. & 0.98174 &  386. & 0.01826 & $1.36 \pm 0.19$ &  9.852 & 1.75 \\
~     &   125. & 0.98174 &  338. & 0.01826 & $1.64 \pm 0.32$ & 18.343 & 0.96 \\
~     &   159. & 0.98955 &  445. & 0.01045 & $1.14 \pm 0.19$ & 10.160 & 2.01 \\
N4725 &   155. & 0.98223 &  481. & 0.01777 & $1.97 \pm 0.50$ &  1.814 & 1.31 \\
~     &    88. & 0.99482 &  381. & 0.00518 & 3               & 13.562 & ~    \\
N4736 &   180. & 0.96455 &  478. & 0.03545 & 3               &  3.822 & ~    \\
~     &   151. & 0.95732 &  418. & 0.04268 & 3               &  5.791 & ~    \\
N4826 &   176. & 0.96523 &  426. & 0.03477 & $2.05 \pm 0.30$ &  8.766 & 1.27 \\
~     &    72. & 0.99570 &  328. & 0.00430 & 3               & 223.50 & ~    \\
N5033 &   157. & 0.97900 &  438. & 0.02100 & 3               &  5.255 & ~    \\
~     &   134. & 0.97920 &  398. & 0.02080 & 3               &  8.435 & ~    \\
N5055 &   142. & 0.98242 &  386. & 0.01758 & 3               &  7.586 & ~    \\
~     &   114. & 0.98506 &  346. & 0.01494 & 3               & 16.336 & ~    \\
N5194 &   180. & 0.95732 &  521. & 0.04268 & 3               &  1.753 & ~    \\
~     &   127. & 0.96123 &  431. & 0.03877 & 3               &  4.171 & ~    \\
N5194\_A & 133. & 0.98672 & 349. & 0.01328 & 3               &  1.469 & ~    \\
~        & 111. & 0.98848 & 319. & 0.01152 & 3               &  2.951 & ~    \\
N5194\_B & 128. & 0.99150 & 373. & 0.00850 & 3               &  4.933 & ~    \\
~        & 109. & 0.99287 & 343. & 0.00713 & 3               &  9.421 & ~    \\
N5194\_C & 120. & 0.99873 & 458. & 0.00127 & 3               &  6.824 & ~    \\
~        & 118. & 0.99873 & 448. & 0.00127 & 3               &  7.448 & ~    \\
N5194\_D & 123. & 0.99824 & 429. & 0.00176 & 3               &  4.144 & ~    \\
~        & 121. & 0.99824 & 419. & 0.00176 & 3               &  4.572 & ~    \\
N5194\_E & 142. & 0.98896 & 406. & 0.01104 & $1.79 \pm 0.33$ &  5.653 & 1.44 \\
~        &  63. & 0.99922 & 313. & 0.00078 & 3               &  235.5 & ~    \\
N5194\_F & 137. & 0.99102 & 417. & 0.00898 & $1.91 \pm 0.40$ &  4.127 & 1.37 \\
~        &  85. & 0.99648 & 331. & 0.00352 & 3               & 26.208 & ~    \\
N5194\_G & 148. & 0.98594 & 418. & 0.01406 & $1.64 \pm 0.31$ &  4.041 & 1.53 \\
~        & 118. & 0.98857 & 367. & 0.01143 & $2.03 \pm 0.43$ &  8.457 & 1.21 \\
~        & 151. & 0.99160 & 483. & 0.00840 & $1.37 \pm 0.24$ &  4.337 & 1.74 \\
N5195 &   164. & 0.97441 &  468. & 0.02559 & 3               &  6.769 & ~    \\
~     &   134. & 0.97305 &  408. & 0.02695 & 3               & 11.899 & ~    \\
N5713 &   162. & 0.97266 &  413. & 0.02734 & 3               &  3.901 & ~    \\
~     &   133. & 0.96875 &  363. & 0.03125 & 3               &  6.684 & ~    \\
N5866 &   194. & 0.94678 &  393. & 0.05322 & $1.88 \pm 0.15$ &  1.676 & 1.33 \\
~     &   160. & 0.92070 &  344. & 0.07930 & $2.03 \pm 0.37$ &  2.379 & 1.10 \\
~     &   200. & 0.97158 &  453. & 0.02842 & $1.68 \pm 0.27$ &  1.731 & 1.25 \\
N6822\_A & 158. & 0.98027 & 504. & 0.01973 & $1.53 \pm 0.42$ &  1.094 & 1.56 \\
~     &   120. & 0.98721 &  442. & 0.01279 & $1.89 \pm 0.41$ &  2.561 & 1.31 \\
~     &   156. & 0.98799 &  581. & 0.01201 & $1.33 \pm 0.25$ &  1.306 & 1.76 \\
N6822\_B & 140. & 0.99043 & 476. & 0.00957 & $1.29 \pm 0.35$ &  1.609 & 1.82 \\
~     &   109. & 0.99404 &  417. & 0.00596 & $1.63 \pm 0.36$ &  3.946 & 1.30 \\
~     &   139. & 0.99414 &  549. & 0.00586 & $1.09 \pm 0.21$ &  1.867 & 2.00 \\
N6822\_C & 148. & 0.98604 & 478. & 0.01396 & $1.33 \pm 0.35$ &  0.716 & 1.77 \\
~        & 115. & 0.99092 & 419. & 0.00908 & $1.67 \pm 0.34$ &  1.690 & 1.04 \\
~        & 148. & 0.99150 & 552. & 0.00850 & $1.14 \pm 0.21$ &  0.830 & 2.19 \\
N6946 &   156. & 0.98066 &  399. & 0.01934 & $1.70 \pm 0.26$ & 13.119 & 1.46 \\
~     &   126. & 0.98105 &  350. & 0.01895 & $2.04 \pm 0.41$ & 24.494 & 1.09 \\
~     &   161. & 0.98857 &  461. & 0.01143 & $1.43 \pm 0.24$ & 13.603 & 1.62 \\
N7331 &   136. & 0.99043 &  401. & 0.00957 & 3               &  3.976 & ~    \\
~     &   118. & 0.99023 &  361. & 0.00977 & 3               &  6.658 & ~    \\
N7552 &   144. & 0.98818 &  418. & 0.01182 & $1.77 \pm 0.35$ & 20.019 & 1.43 \\
~     &   115. & 0.99053 &  367. & 0.00947 & $2.21 \pm 0.46$ & 42.249 & 1.12 \\
~     &   147. & 0.99277 &  483. & 0.00723 & $1.47 \pm 0.26$ & 21.235 & 1.67 \\
N7793 &   176. & 0.96455 &  381. & 0.03545 & $0.91 \pm 0.09$ &  0.869 & 2.29 \\
~     &   144. & 0.95400 &  334. & 0.04600 & $1.02 \pm 0.20$ &  1.368 & 1.50 \\
~     &   181. & 0.98096 &  440. & 0.01904 & $0.79 \pm 0.13$ &  0.900 & 2.62 \\
\enddata
\tablenotetext{a}{
The parameters $f_{\rm i}$ are the mass fractions of the discrete
components at the temperatures $T_{\rm i}$.
For each galaxy, the first line gives the results of the fits where $T_1$
is constrained in order not to overestimate the column density, and the second (third)
line where $T_1$ is unrestricted. For galaxies where the S(2) flux is an upper limit,
only constrained-$T_1$ fits were performed.
When $OPR_{\rm \,high\,T} < 3$, $T_2$ is fixed, and the free-$T_1$ fit results are then
provided for two
different values of $T_2$: $1.14 \times T$(S1-S3) (second line) and $1.5 \times T$(S1-S3)
(third line). See Section~\ref{fits} for explanations.}
\tablenotetext{b}{The third component is characterized by $T_3 = 1415$\,K and $f_3 = 0.00690$
(constrained-$T_1$ fit) or $T_3 = 1455$\,K and $f_3 = 0.00477$ (free-$T_1$ fit).}
\tablenotetext{c}{$T_3 = 1139$\,K and $f_3 = 0.00088$ (constrained-$T_1$ fit),
or $T_3 = 1319$\,K and $f_3 = 0.00019$ (free-$T_1$ fit).}
\tablenotetext{d}{
$T_3 = 1324$\,K and $f_3 = 0.00561$ (constrained-$T_1$ fit), or $T_3 = 1414$\,K and
$f_3 = 0.00010$ (free-$T_1$ fit).}
\end{deluxetable}

\begin{deluxetable}{lrrrrrll}
\tabletypesize{\footnotesize}
\rotate
\tablecaption{Masses in the warm and cold phases.\tablenotemark{a}
\label{tab_mass}
}
\tablehead{
galaxy & $M(T > T_1)$         & $N_{\rm tot}$(cold H$_2$)    & $M(T > T_1)$            & CO beam      & corr. & map & ref CO \\
~      & ($10^6$\,M$_{\sun}$) & ($10^{20}$\,mol.\,cm$^{-2}$) & ~$/~ M({\rm cold H}_2)$ & (arcsec$^2$) & ~     & ~   & ~ \\
}
\startdata
N24   &   3.13 & ~        & ~        & ~     & ~    & ~    & ~ \\
N337  &  11.61 &   9.4    & 0.18     & 1521. & 1.52 & IRAC & E96 \\
N628  &   3.79 &  10.5    & 0.25     & 627.  & 1.04 & BIMA & S05 (B93, Y95) \\
N855  &   0.50 &   2.3    & 0.22     & 415.  & 1.24 & IRAC & W95 \\
N925  &   1.15 & $< 18.7$ & $> 0.06$ & 386.  & ~    & ~    & S05 \\
N1097 & 103.01 & 183.     & 0.06     & 1590. & 1.74 & IRAC & Y95 (VV98, H93, K03) \\
N1266 &  12.97 &  ~       & ~        & ~     & ~    & ~    & ~ \\
N1291 &   0.42 &  11.7    & 0.04     & 1452. & 2.33 & IRAC & T91 \\
N1316 &   0.85 &  19.9    & 0.006    & 1452. & 2.55 & IRAC & H01 \\
N1482 &  73.91 & 136.     & 0.04     & 1521. & 1.76 & IRAC & E96 \\
N1512 &   6.34 &  ~       & ~        & ~     & ~    & ~    & ~ \\
N1566 &  10.22 &  ~       & ~        & ~     & ~    & ~    & ~ \\
N1705 &   0.50 &  ~       & ~        & ~     & ~    & ~    & ~ \\
N2403 &   0.33 &  23.8    & 0.10     & 855.  & 0.96 & BIMA & E96 \\
N2798 &  34.54 &  46.5    & 0.11     & 855.  & 2.50 & IRAC & E96 \\
N2841 &   0.74 & $< 27.4$ & $> 0.03$ & 370.  & ~    & ~    & S05 \\
N2915 & 0.0077 & $< 6.0$  & $> 0.02$ & 1452. & 2.04 & IRAC & A04 \\
N2976 &   0.28 &  10.6    & 0.20     & 452.  & 1.12 & BIMA & A04 (I05, Y95) \\
N3031 &   0.10 & $< 7.9$  & $> 0.09$ & 3068. & ~    & ~    & S05 \\
N3049 &   3.39 &  31.3    & 0.02     & 573.  & 1.48 & IRAC & C97 \\
N3184 &   2.09 &  40.1    & 0.07     & 441.  & 1.23 & BIMA & S05 (Y95, S93) \\
N3190 &  11.33 &  37.8    & 0.09     & 380.  & 1.14 & IRAC & L98 \\
N3198 &   3.54 &  33.1    & 0.10     & 415.  & 1.31 & IRAC & B93 \\
N3265 &   6.45 &  25.8    & 0.06     & 227.  & 0.76 & IRAC & G91 \\
Mrk33\tablenotemark{b}
      &  21.35 &  16.5    & 0.23     & 380.  & 1.15 & IRAC & S92 (A04) \\
N3351 &  11.30 &  95.1    & 0.05     & 606.  & 0.85 & BIMA & P97 (S05) \\
N3521 &   4.95 &  26.1    & 0.22     & 201.  & 0.68 & BIMA & N01 (Y95) \\
N3621 &   1.88 &  ~       & ~        & ~     & ~    & ~    & ~ \\
N3627 &   2.39 & 196.     & 0.01     & 271.  & 0.95 & BIMA & S05 (Y95) \\
N3773 &   2.98 & $< 13.5$ & $> 0.12$ & 2376. & 5.83 & IRAC & L05 \\
N3938 &   4.23 &  26.4    & 0.10     & 262.  & 0.96 & BIMA & S05 (Y95) \\
N4125 &   1.20 &   6.5    & 0.04     & 415.  & 1.13 & IRAC & W95 \\
N4254 &   7.86 &  81.8    & 0.02     & 201.  & 0.75 & S99  & N01 (K88) \\
N4321 &  47.83 & 133.     & 0.03     & 415.  & 0.74 & BIMA & B93 (S05, K96, H93) \\
N4450 &   5.04 &  16.4    & 0.07     & 131.  & 0.90 & BIMA & S05 \\
N4536 & 102.41 &  50.2    & 0.11     & 855.  & 1.08 & S03  & E96 \\
N4552 &   1.91 &  ~       & ~        & ~     & ~    & ~    & ~ \\
N4559\tablenotemark{c}
      &   9.29 &  13.9    & 0.45     & 355.  & 0.98 & BIMA & S05 \\
N4569 &  17.48 & 200.     & 0.02     & 118.  & 0.73 & BIMA & S05 (N01, K88, H93) \\
N4579 &   2.36 &  36.7    & 0.01     & 118.  & 0.42 & BIMA & S05 (K88) \\
N4594 &   0.47 & $< 6.2$  & $> 0.04$ & 346.  & ~    & ~    & B91 \\
N4625 &   1.85 &   9.2    & 0.20     & 452.  & 1.10 & IRAC & A04 (B03) \\
N4631 &   8.44 &  96.2    & 0.10     & 1590. & 1.76 & IRAC & Y95 \\
N4725 &   5.85 & $< 54.8$ & $> 0.03$ & 210.  & ~    & ~    & S05 \\
N4736 &   1.20 &  81.3    & 0.05     & 767.  & 1.37 & BIMA & S05 (S98, Y95) \\
N4826 &   3.00 & 250.     & 0.04     & 1337. & 1.81 & BIMA & S05 (N01, Y95) \\
N5033 &  10.86 &  96.0    & 0.05     & 855.  & 1.39 & BIMA & C01 (S05, N01, E96, A95) \\
N5055 &   5.98 & 122.     & 0.06     & 645.  & 1.23 & BIMA & S05 (M99, E96, A95, Y95, S93) \\
N5194 &   1.46 &  40.0    & 0.04     & 474.  & 0.98 & BIMA & S05 \\
N5194\_A & 0.95 &   ~     & ~        & ~     & ~    & ~    & ~ \\
N5194\_B & 3.37 &   ~     & ~        & ~     & ~    & ~    & ~ \\
N5194\_C & 6.07 &   ~     & ~        & ~     & ~    & ~    & ~ \\
N5194\_D & 3.53 &   ~     & ~        & ~     & ~    & ~    & ~ \\
N5194\_E & 3.33 &   ~     & ~        & ~     & ~    & ~    & ~ \\
N5194\_F & 3.07 &   ~     & ~        & ~     & ~    & ~    & ~ \\
N5194\_G & 2.91 &   ~     & ~        & ~     & ~    & ~    & ~ \\
N5195 &   5.51 & 110.     & 0.06     & 177.  & 0.58 & IRAC & K02 (S89) \\
N5713 &  30.27 &  50.8    & 0.08     & 1590. & 2.01 & IRAC & Y95 (Y03) \\
N5866 &   2.80 &  50.9    & 0.03     & 346.  & 1.02 & IRAC & W03 (T94) \\
N6822\_A\tablenotemark{b}
      & 0.0047 & 7.8   & 0.14     & 415.  & 1.26 & IRAC & I03 \\
N6822\_B & 0.0062 &  ~       & ~        & ~     & ~    & ~    & ~ \\
N6822\_C & 0.0030 &  ~       & ~        & ~     & ~    & ~    & ~ \\
N6946 &  11.95 & 486.     & 0.03     & 816.  & 0.95 & BIMA & S05 \\
N7331 &   9.69 &  56.0    & 0.07     & 346.  & 0.94 & BIMA & I99 (N01) \\
N7552 & 269.01 & 171.     & 0.12     & 1452. & 1.86 & IRAC & A95 (C92) \\
N7793 &   0.10 &  25.4    & 0.03     & 1452. & 1.87 & IRAC & I95 \\
\enddata
\tablenotetext{a}{We retain here results from
constrained-$T_1$ fits, giving lower masses in the warm phase. Whenever the excitation
diagrams are ambiguous regarding the value of $OPR_{\rm \,high\,T}$, we adopt the mass
of warm H$_2$ derived with $OPR_{\rm \,high\,T} < 3$, which is also smaller than the mass
derived with $OPR_{\rm \,high\,T} = 3$ (see Table~\ref{tab_fit}). References for CO
intensities are given in abbreviated form.
The quantity "corr." designates the factor that was applied to the CO brightness in
order to correct for the different apertures of the CO and H$_2$ observations,
and "map" is the image used to derive this factor
(see Section~\ref{h2co}).}
\tablenotetext{b}{Mrk\,33: The H$_2$ mass fraction in the warm phase may be severely overestimated
in this blue compact galaxy, because of the use of an inappropriate factor to
convert CO flux to total H$_2$ mass. \citet{Israel05} derive a factor about four
times higher than the standard factor used here, which would make the total H$_2$
mass also four times higher.
NGC\,6822\_A: In their detailed study of this region, \citet{Israel03} derive a CO
flux to total H$_2$ mass conversion factor twenty times higher than the standard factor,
so that the same remark as for Mrk\,33 applies.}
\tablenotetext{c}{NGC\,4559 has the highest mass fraction in the warm phase, even when
constraining $T_1$ to be near-maximal. The warm phase is dominated by gas at $\sim 100$\,K.
NGC\,4559 is a quiescent galaxy, as obvious from the low $F_{24}/F_{7.9}$ flux density
ratio within the IRS aperture. The galaxy has little molecular gas, and it is possible
that it is mostly diffuse, and thus more exposed to FUV radiation. The same remarks
may apply, to a lesser degree, to other galaxies of the sample such as NGC\,628.}
\end{deluxetable}

\begin{deluxetable}{lrll}
\tabletypesize{\footnotesize}
\tablecaption{Nuclear X-ray fluxes in the 2-10\,keV band and references.
\label{tab_xrays}
}
\tablehead{
galaxy & $F_{\rm X}$               & ref & note\tablenotemark{a} \\
~      & (10$^{-17}$ W\,m$^{ -2}$) & ~   & ~ \\
}
\startdata
N628  & 0.59     & K05  & 2 \\
N1097 & 173.     & N06  & 1 \\
N1291 & 7.45     & K05  & 2 \\
N1316 & 10.1     & KF03 & 2 \\
N2841 & 1.06     & HF01 & 1 \\
N3031 & 1020.    & HF01 & 1 \\
N3184 & 1.12     & K05  & 2 \\
N3627 & $< 0.32$ & HF01 & 1 \\
N4125 & 1.07     & S04  & 1 \\
N4321 & $< 1.25$ & HF01 & 1 \\
N4552 & 12.1     & SD05 & 1 \\
N4569 & 7.61     & HF01 & 1 \\
N4579 & 264.     & HF01 & 1 \\
N4594 & 120.     & HF01 & 1 \\
N4725 & 7.15     & HF01 & 1 \\
N4736 & 27.4     & SD05 & 1 \\
N4826 & $< 1.08$ & HF01 & 1 \\
N5033 & 123.     & HF01 & 1 \\
N5194 & 1300.    & S04  & 1 \\
N5195 & 4.45     & K05  & 2 \\
N5866 & 0.18     & SD05 & 1 \\
N7331 & 2.28     & S04  & 1 \\
\enddata
\tablenotetext{a}{1: Quantity given directly in the cited paper
(abbreviated notation as in the list of references). \\
2: Quantity computed as in \citet{Ho01} (HF01) from counts in the 0.2-8\,keV band.}
\end{deluxetable}

\clearpage

\begin{figure}[!ht]
\hspace*{-2.5cm}
\resizebox{5cm}{!}{\rotatebox{90}{\plotone{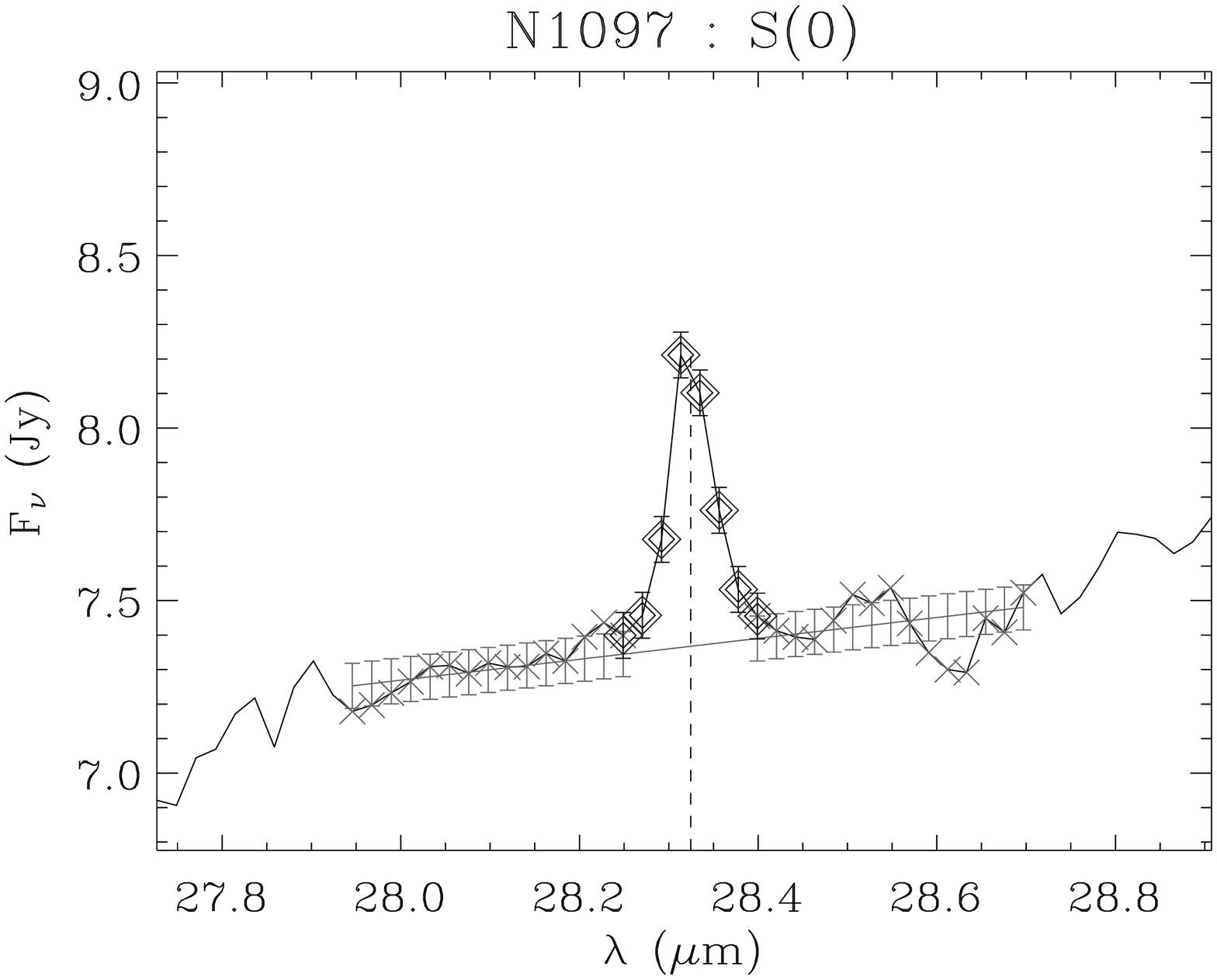}}}
\hspace*{-0.55cm}
\resizebox{5cm}{!}{\rotatebox{90}{\plotone{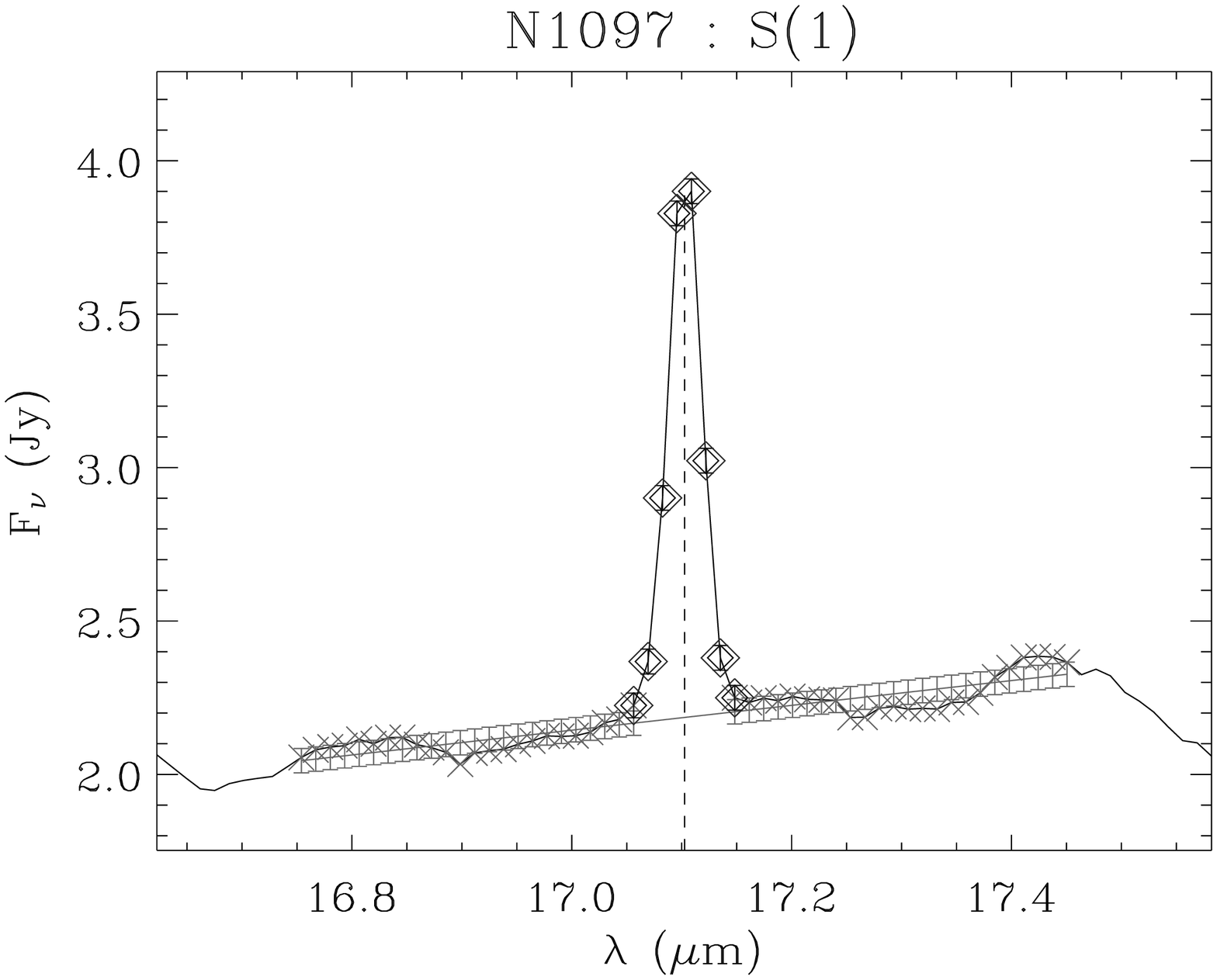}}}
\hspace*{-0.55cm}
\resizebox{5cm}{!}{\rotatebox{90}{\plotone{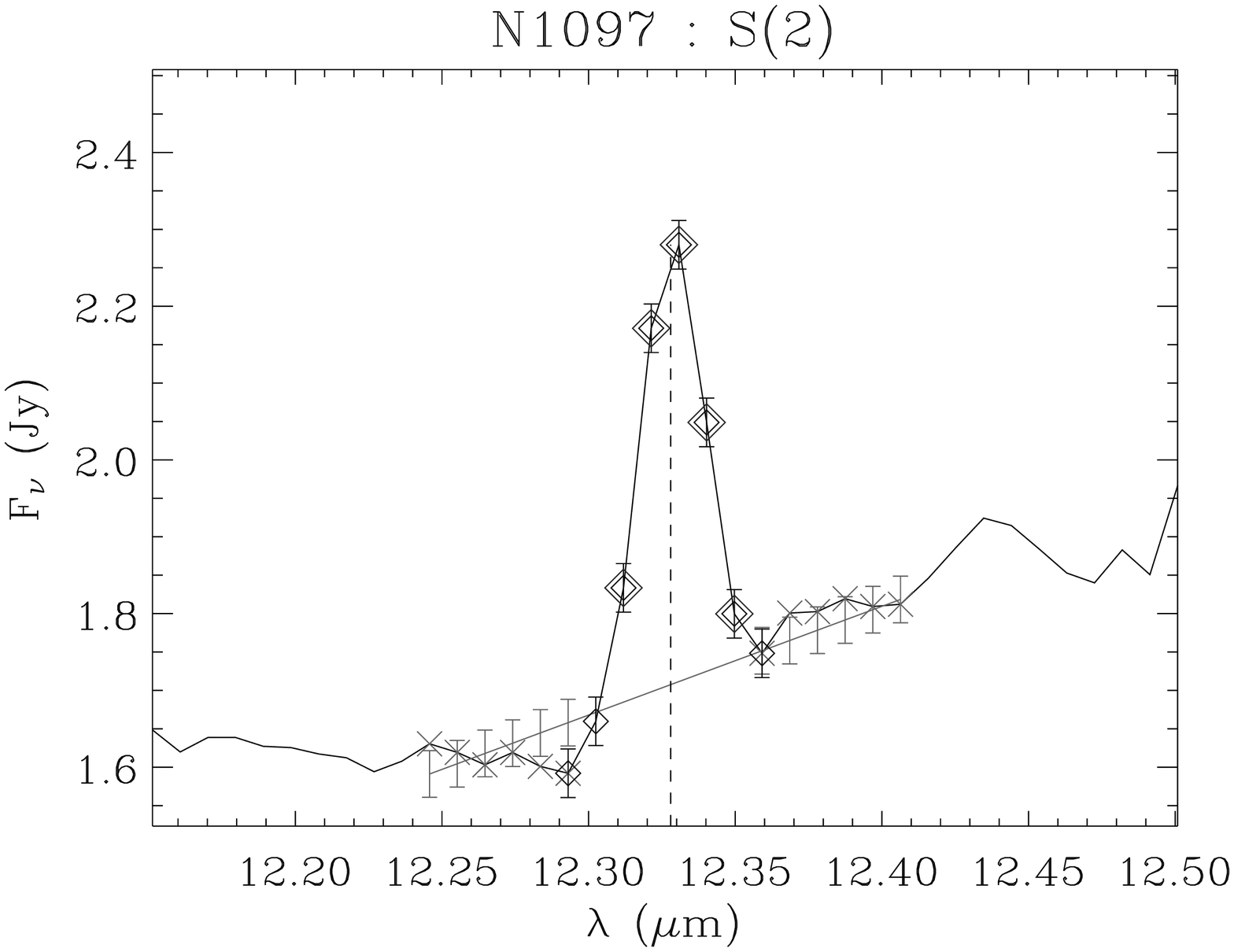}}}
\hspace*{-0.55cm}
\resizebox{5cm}{!}{\rotatebox{90}{\plotone{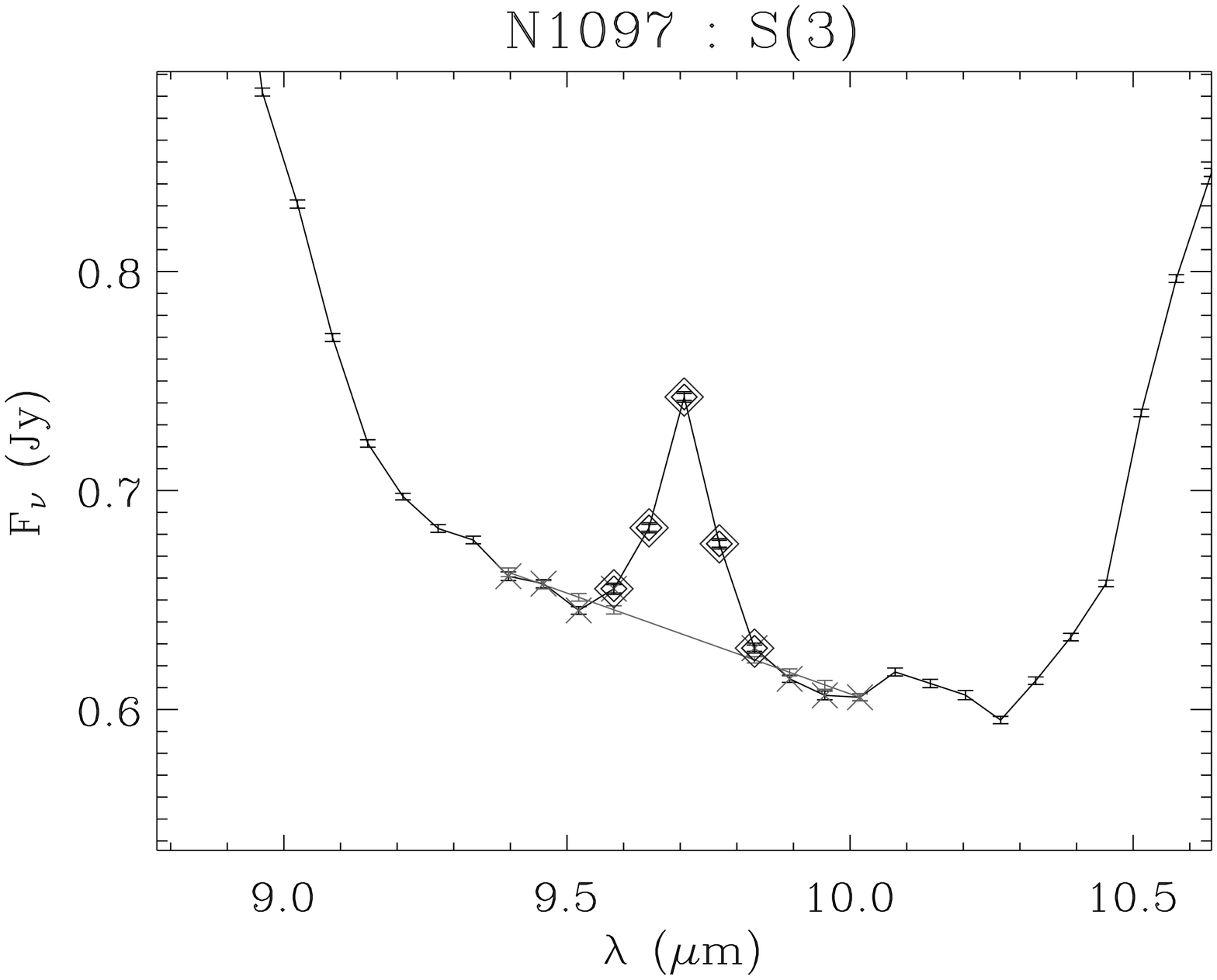}}} \\
\hspace*{-2.5cm}
\resizebox{5cm}{!}{\rotatebox{90}{\plotone{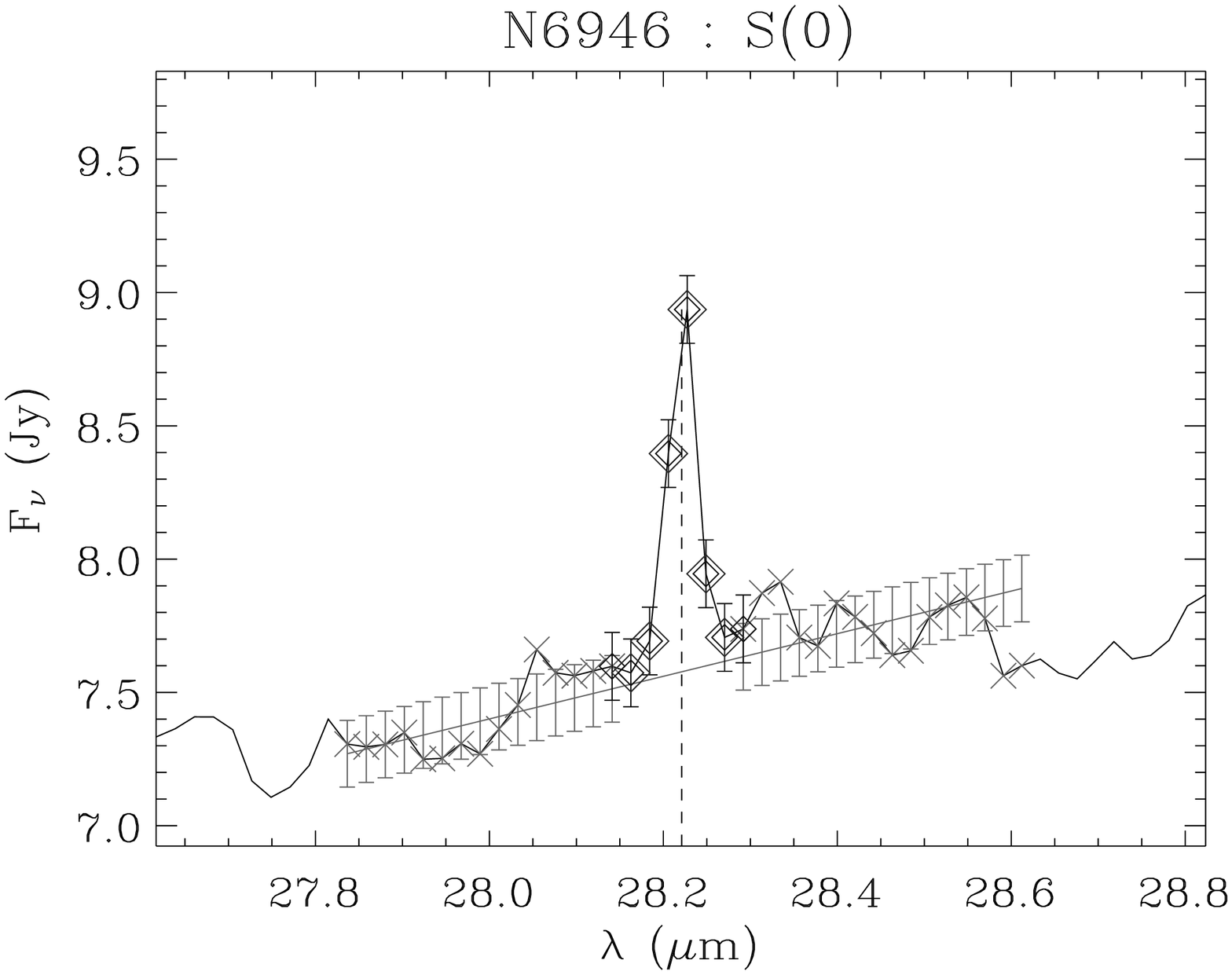}}}
\hspace*{-0.55cm}
\resizebox{5cm}{!}{\rotatebox{90}{\plotone{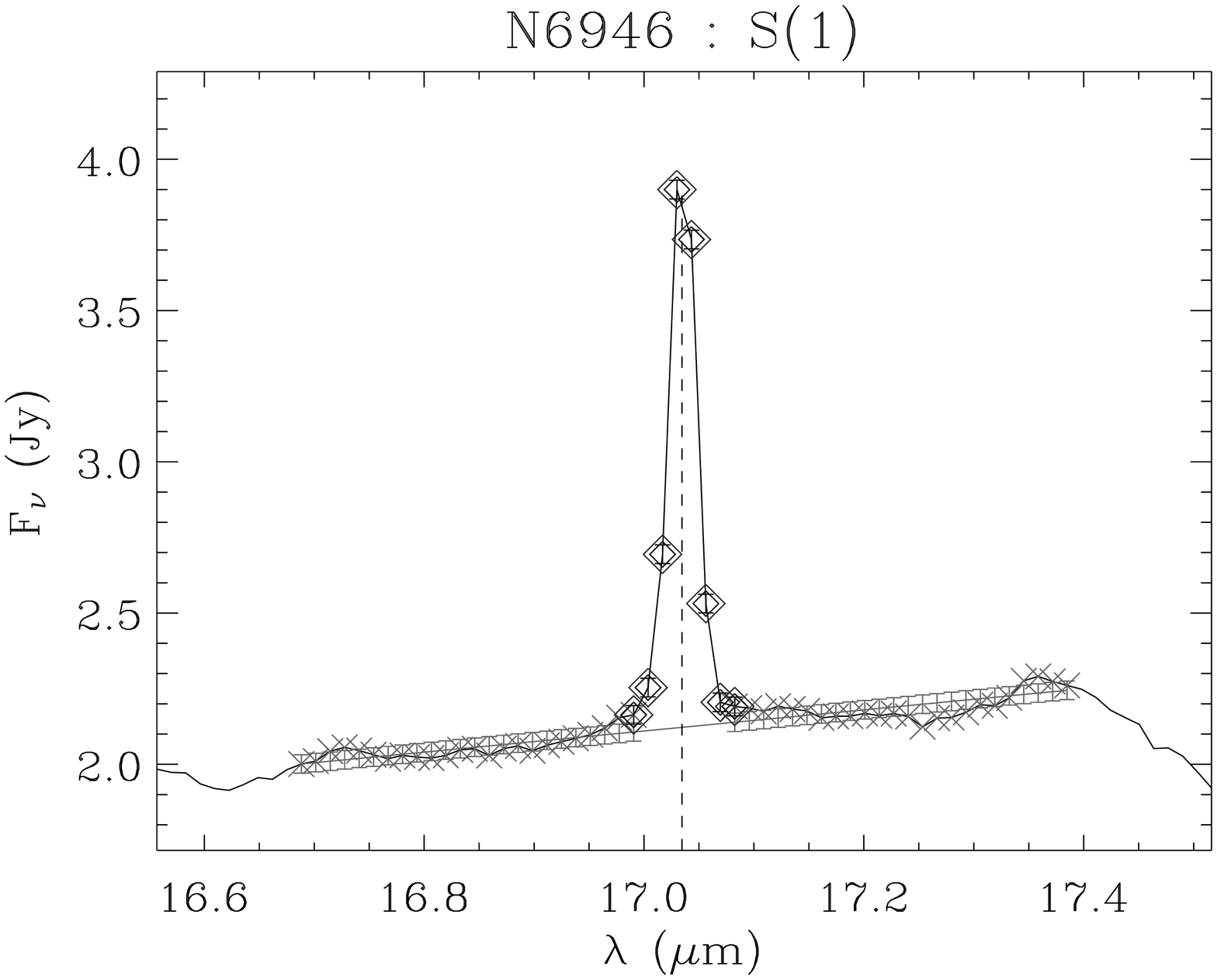}}}
\hspace*{-0.55cm}
\resizebox{5cm}{!}{\rotatebox{90}{\plotone{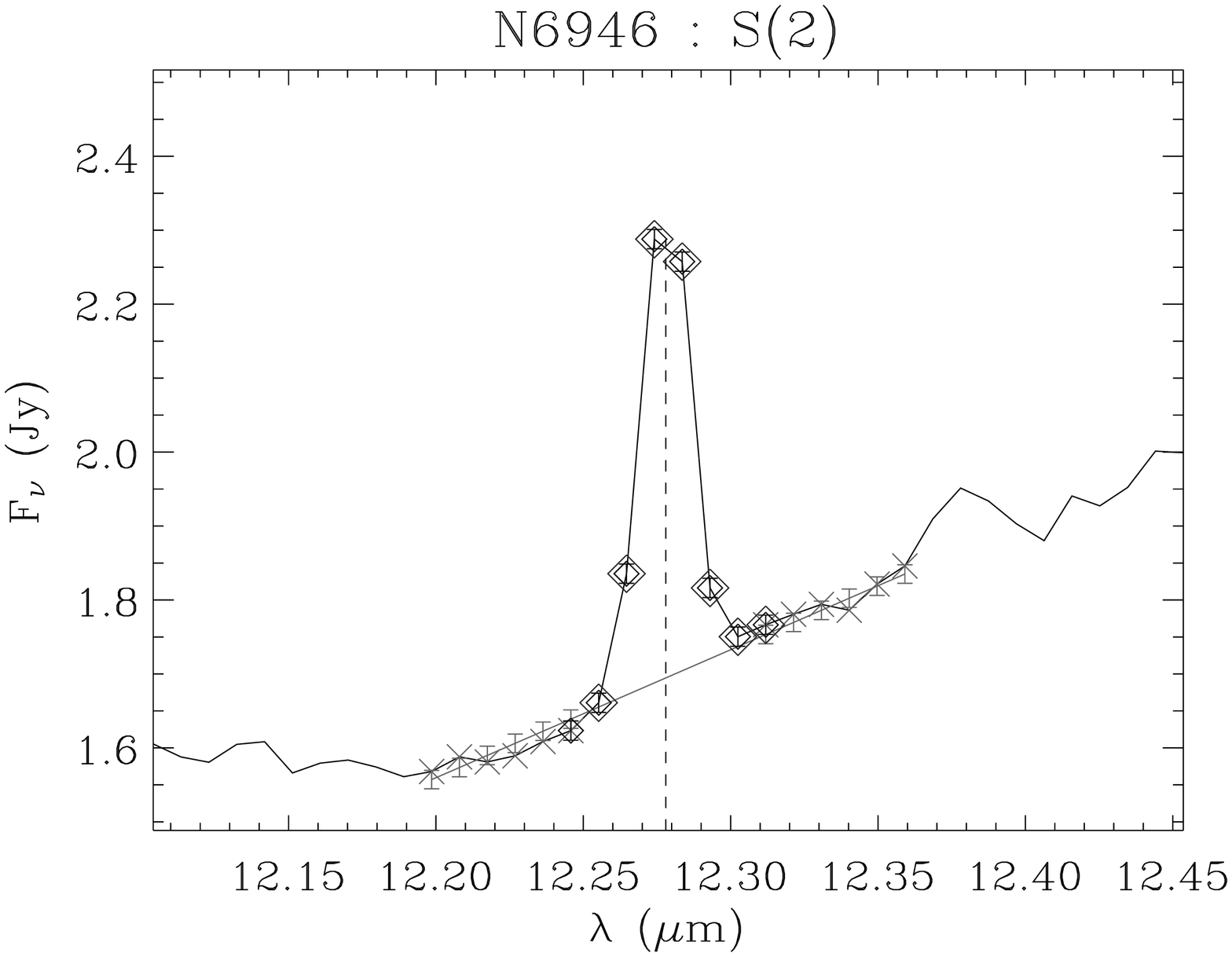}}}
\hspace*{-0.55cm}
\resizebox{5cm}{!}{\rotatebox{90}{\plotone{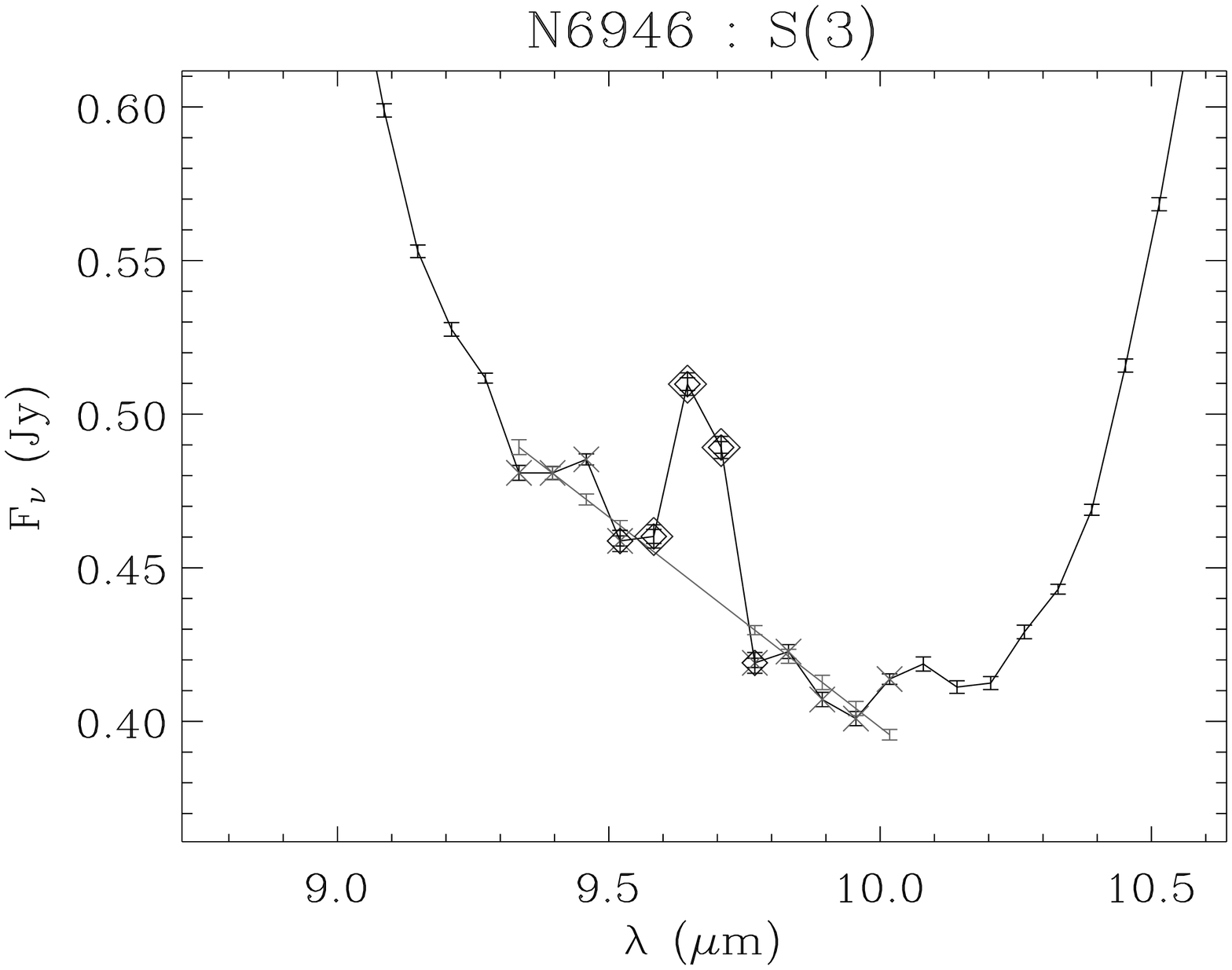}}} \\
\hspace*{-2.5cm}
\resizebox{5cm}{!}{\rotatebox{90}{\plotone{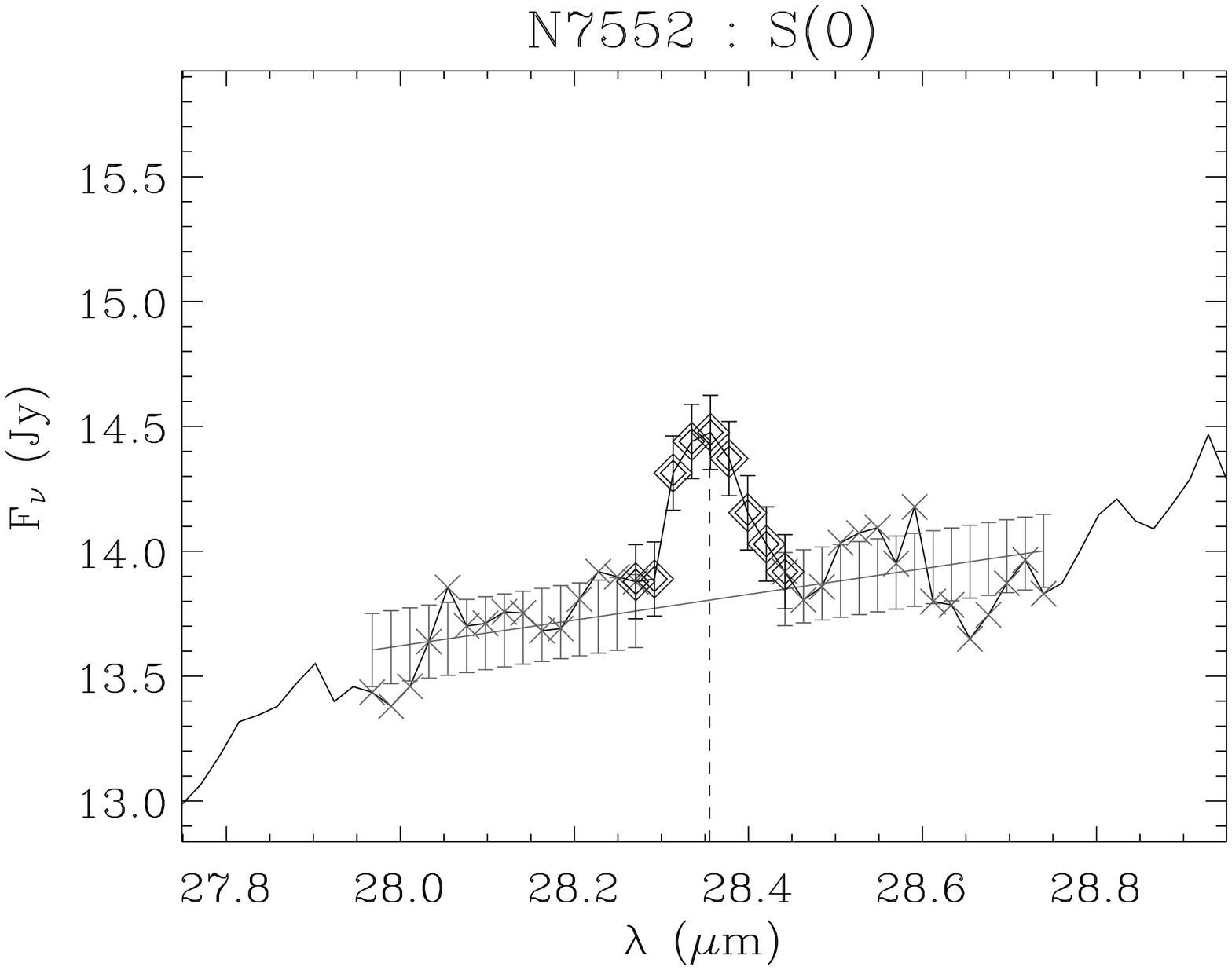}}}
\hspace*{-0.55cm}
\resizebox{5cm}{!}{\rotatebox{90}{\plotone{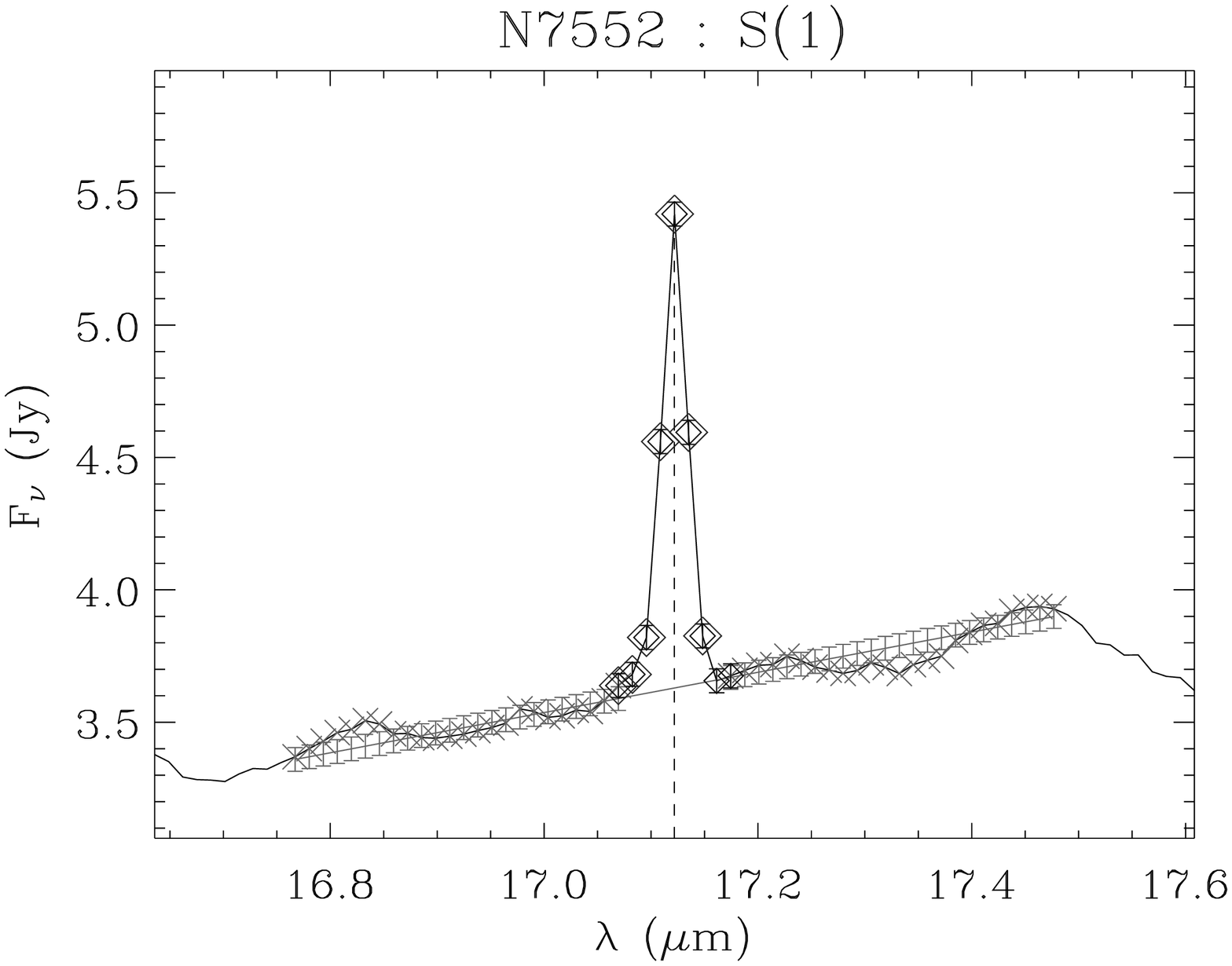}}}
\hspace*{-0.55cm}
\resizebox{5cm}{!}{\rotatebox{90}{\plotone{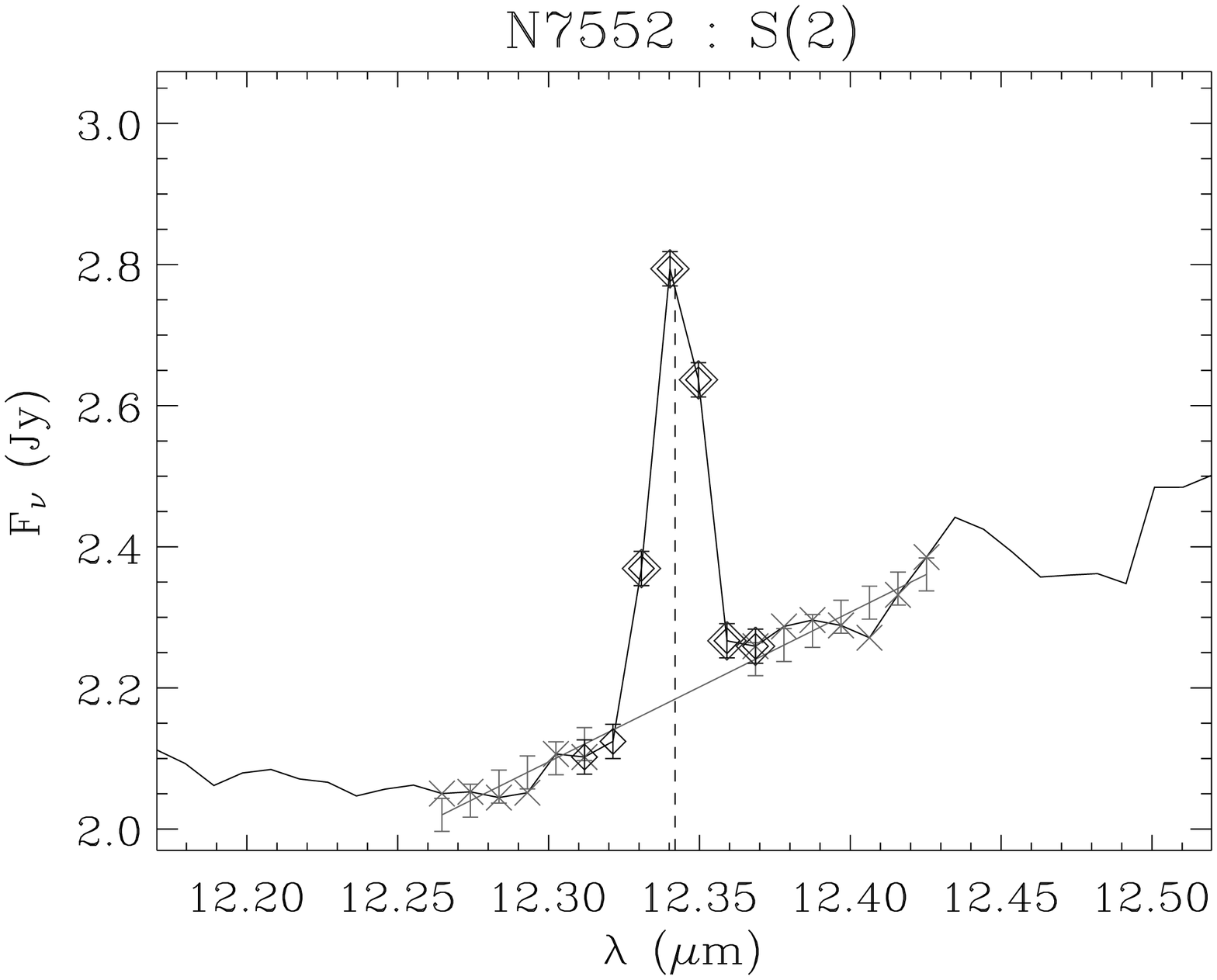}}}
\hspace*{-0.55cm}
\resizebox{5cm}{!}{\rotatebox{90}{\plotone{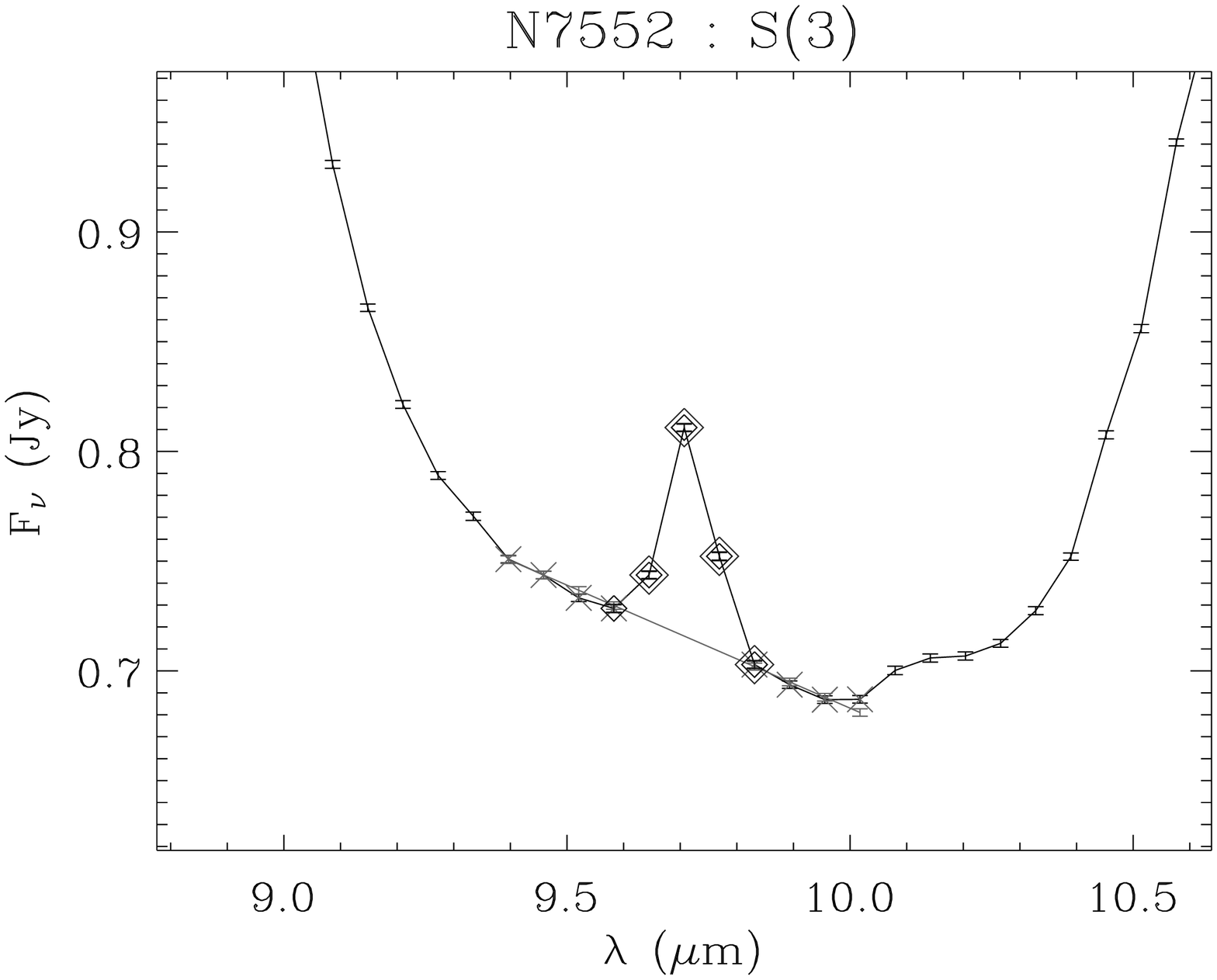}}} \\
\hspace*{-2.5cm}
\resizebox{5cm}{!}{\rotatebox{90}{\plotone{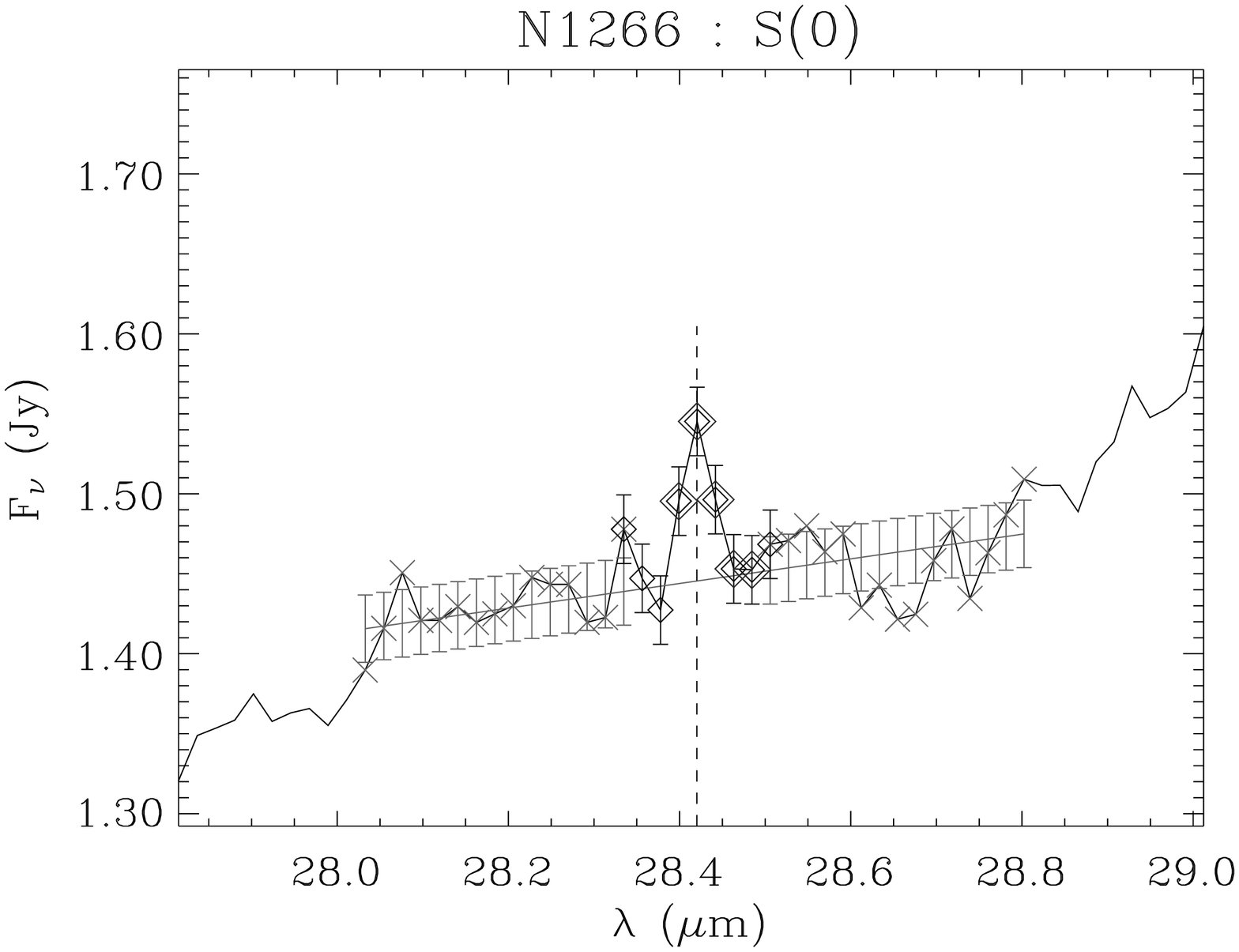}}}
\hspace*{-0.55cm}
\resizebox{5cm}{!}{\rotatebox{90}{\plotone{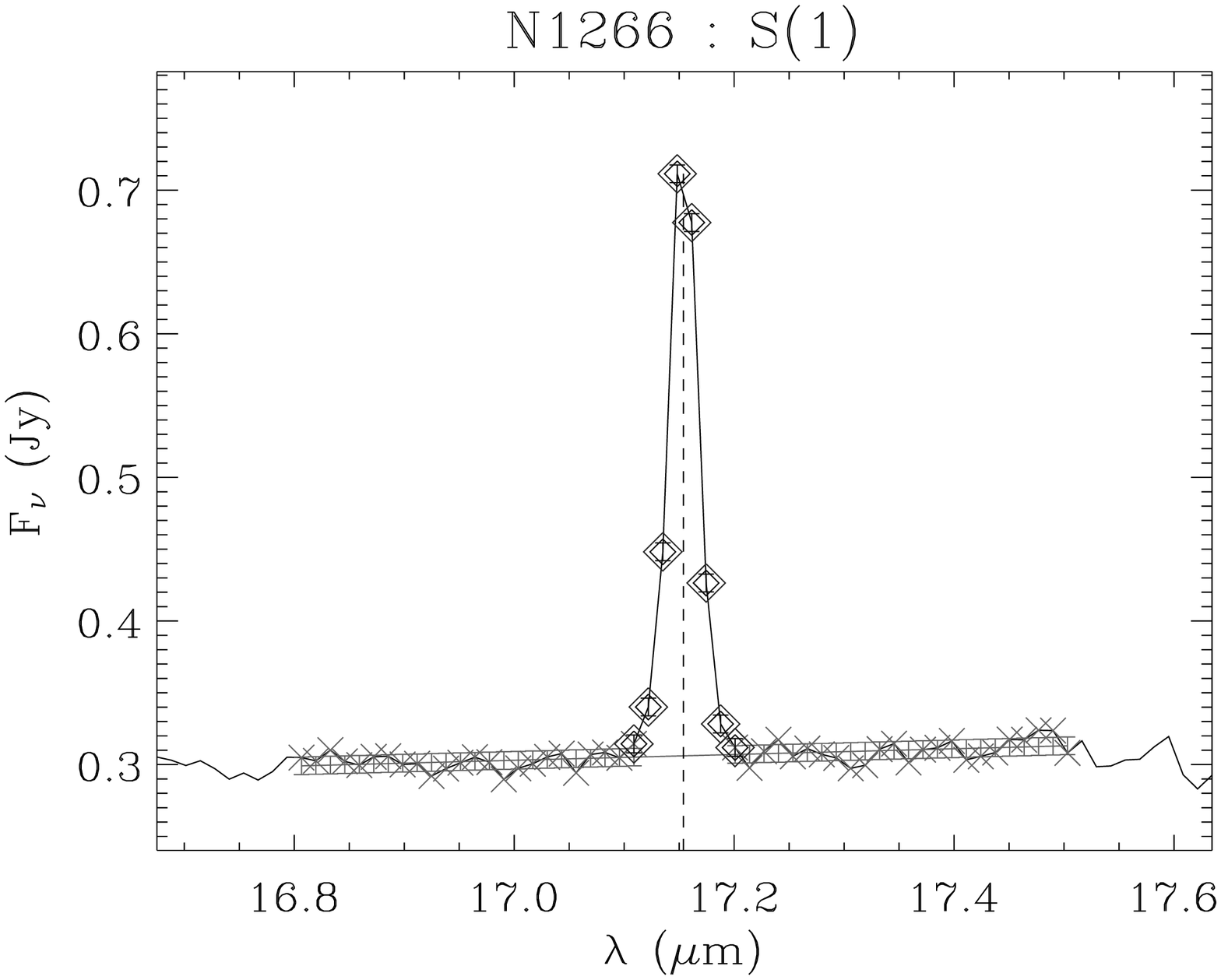}}}
\hspace*{-0.55cm}
\resizebox{5cm}{!}{\rotatebox{90}{\plotone{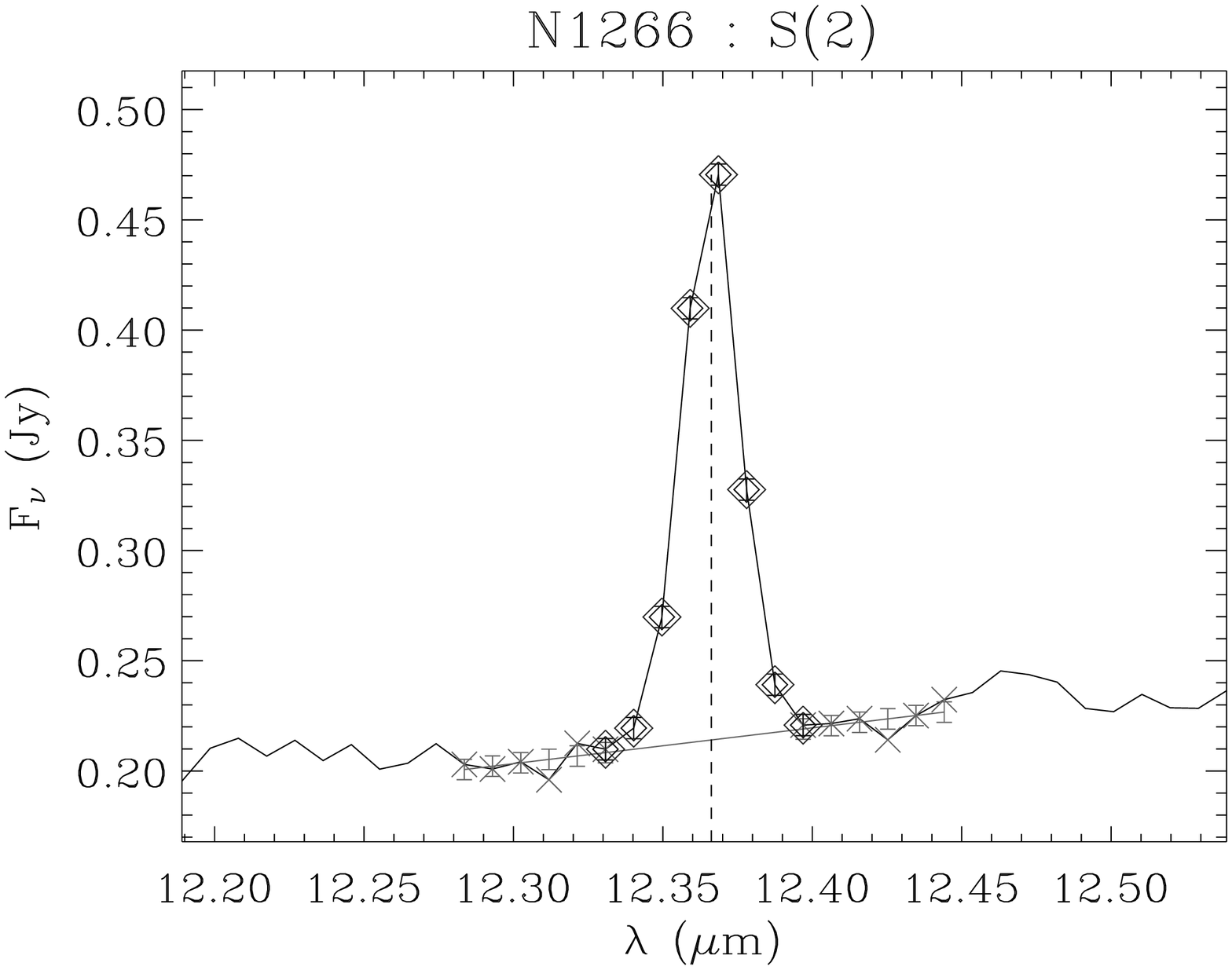}}}
\hspace*{-0.55cm}
\resizebox{5cm}{!}{\rotatebox{90}{\plotone{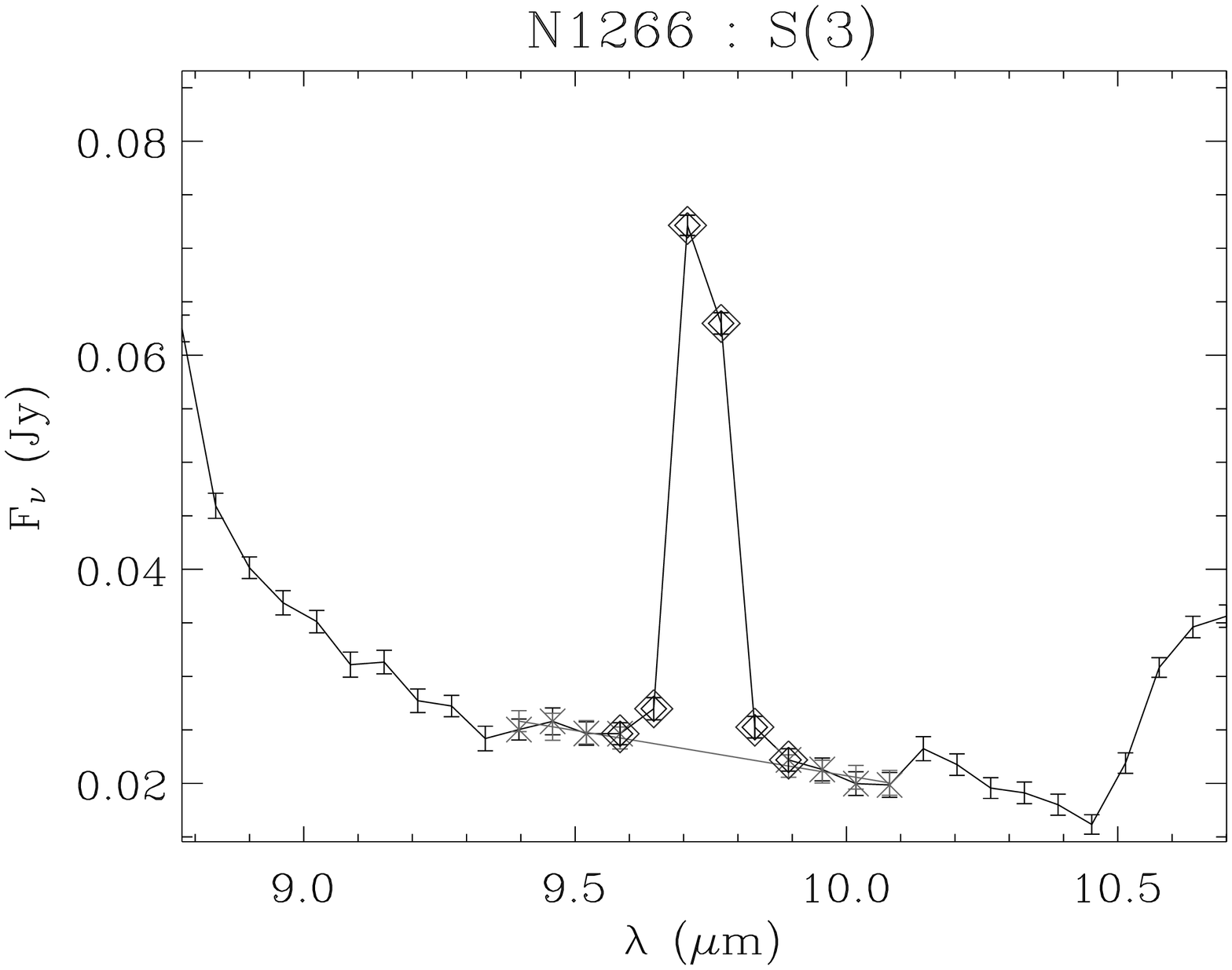}}} \\
\hspace*{-2.5cm}
\resizebox{5cm}{!}{\rotatebox{90}{\plotone{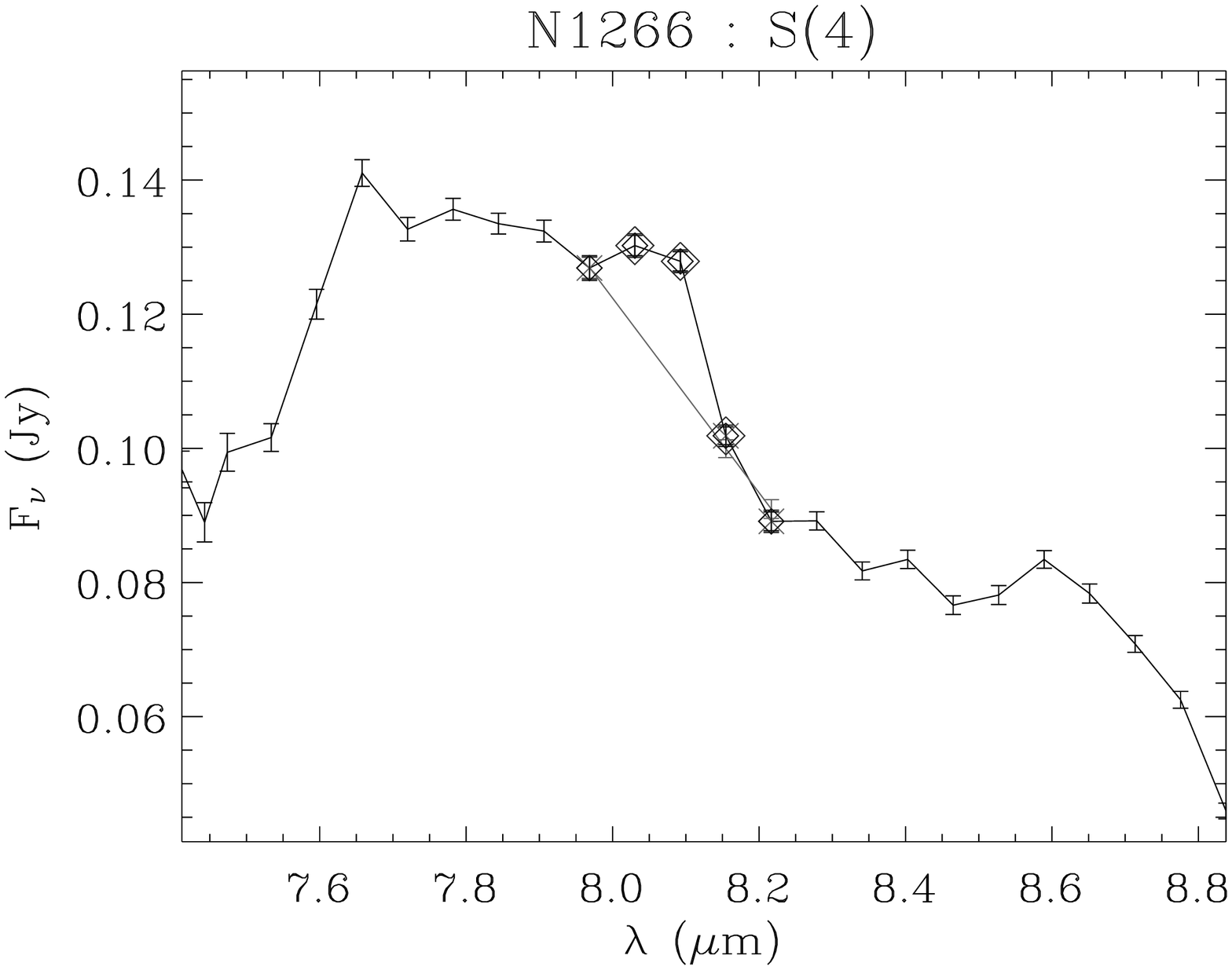}}}
\hspace*{-0.55cm}
\resizebox{5cm}{!}{\rotatebox{90}{\plotone{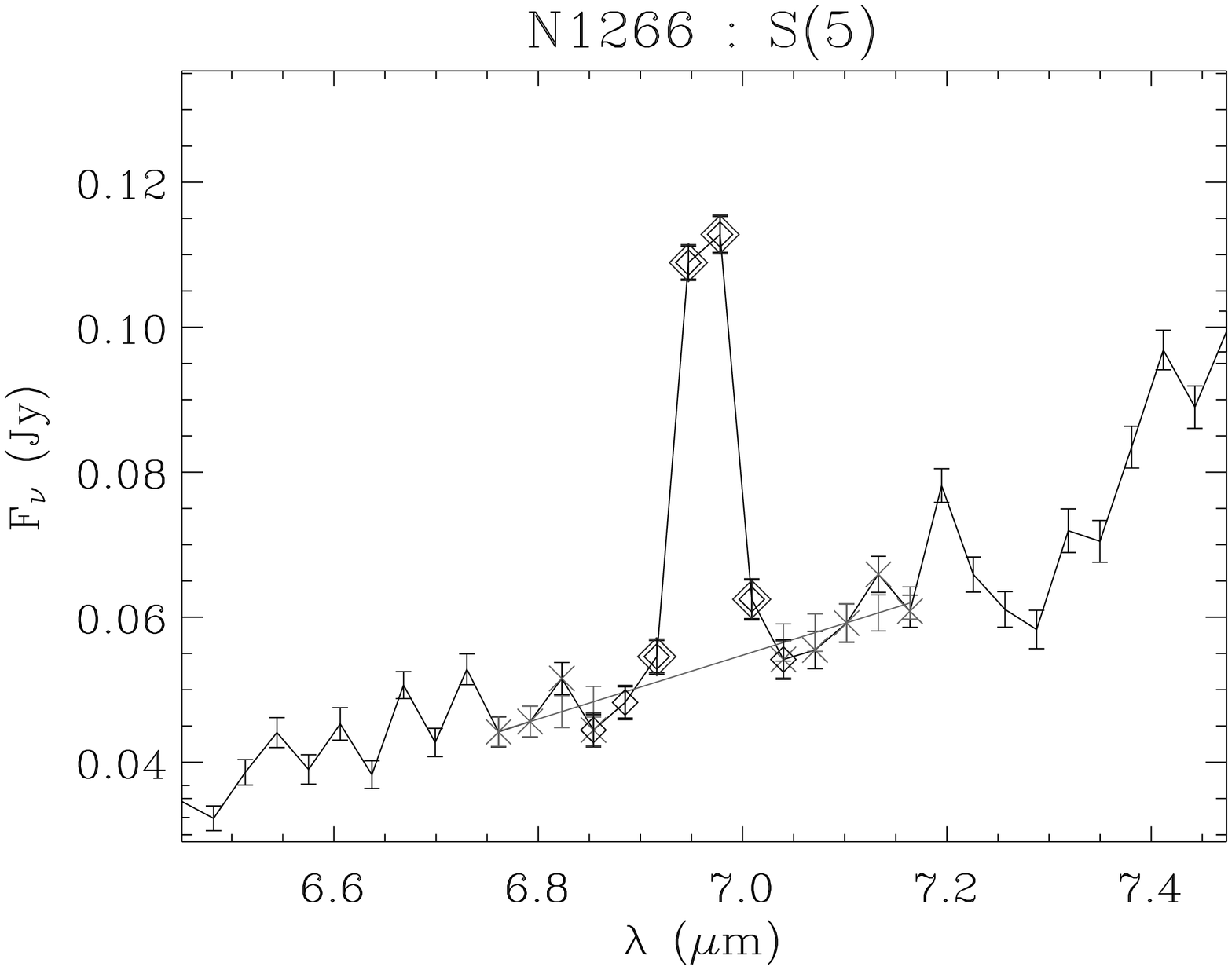}}}
\hspace*{-0.55cm}
\resizebox{5cm}{!}{\rotatebox{90}{\plotone{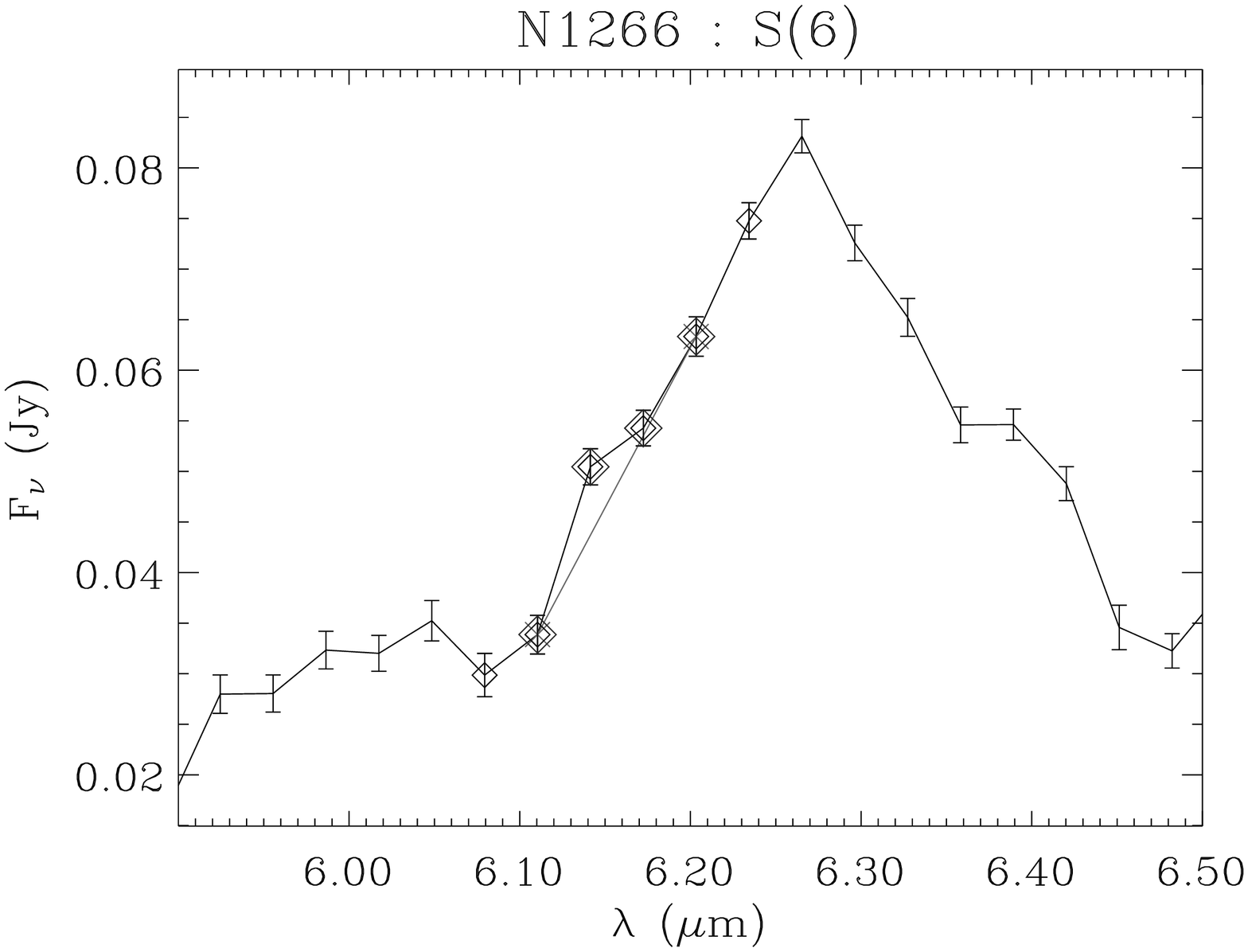}}}
\hspace*{-0.55cm}
\resizebox{5cm}{!}{\rotatebox{90}{\plotone{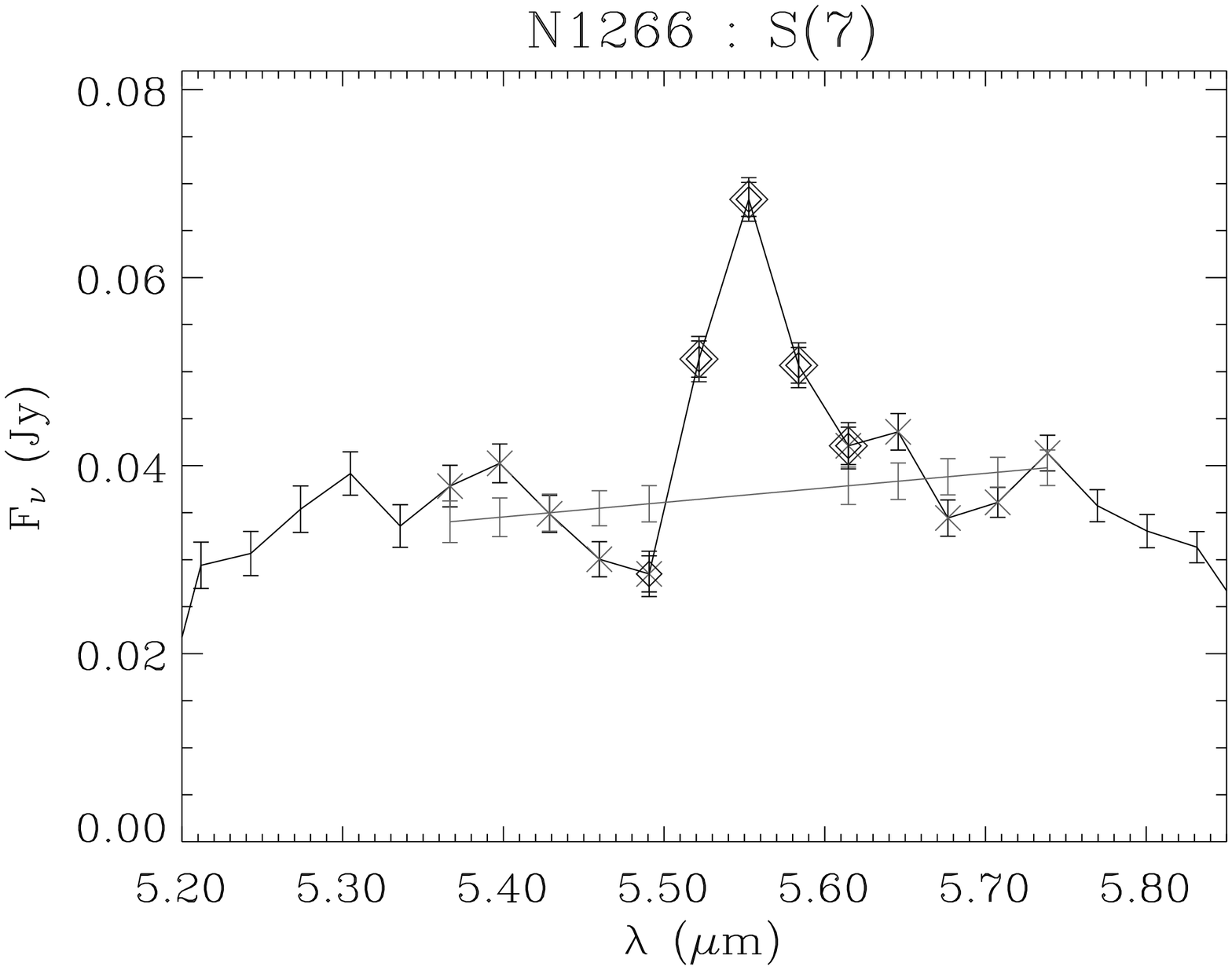}}}
\caption{Examples of H$_2$ line spectra: the circumnuclear starbursts NGC\,1097,
NGC\,6946 and NGC\,7552, and the three galaxies for which we could estimate the
fluxes of higher transitions than S(3), NGC\,1266, NGC\,4569 and NGC\,4579.
The straight line indicates the fitted pseudo-continuum, and the diamond symbols
show the wavelength range over which the line flux was integrated.}
\label{fig:lines}
\end{figure}

\addtocounter{figure}{-1}
\begin{figure}[!ht]
\hspace*{-2.5cm}
\resizebox{5cm}{!}{\rotatebox{90}{\plotone{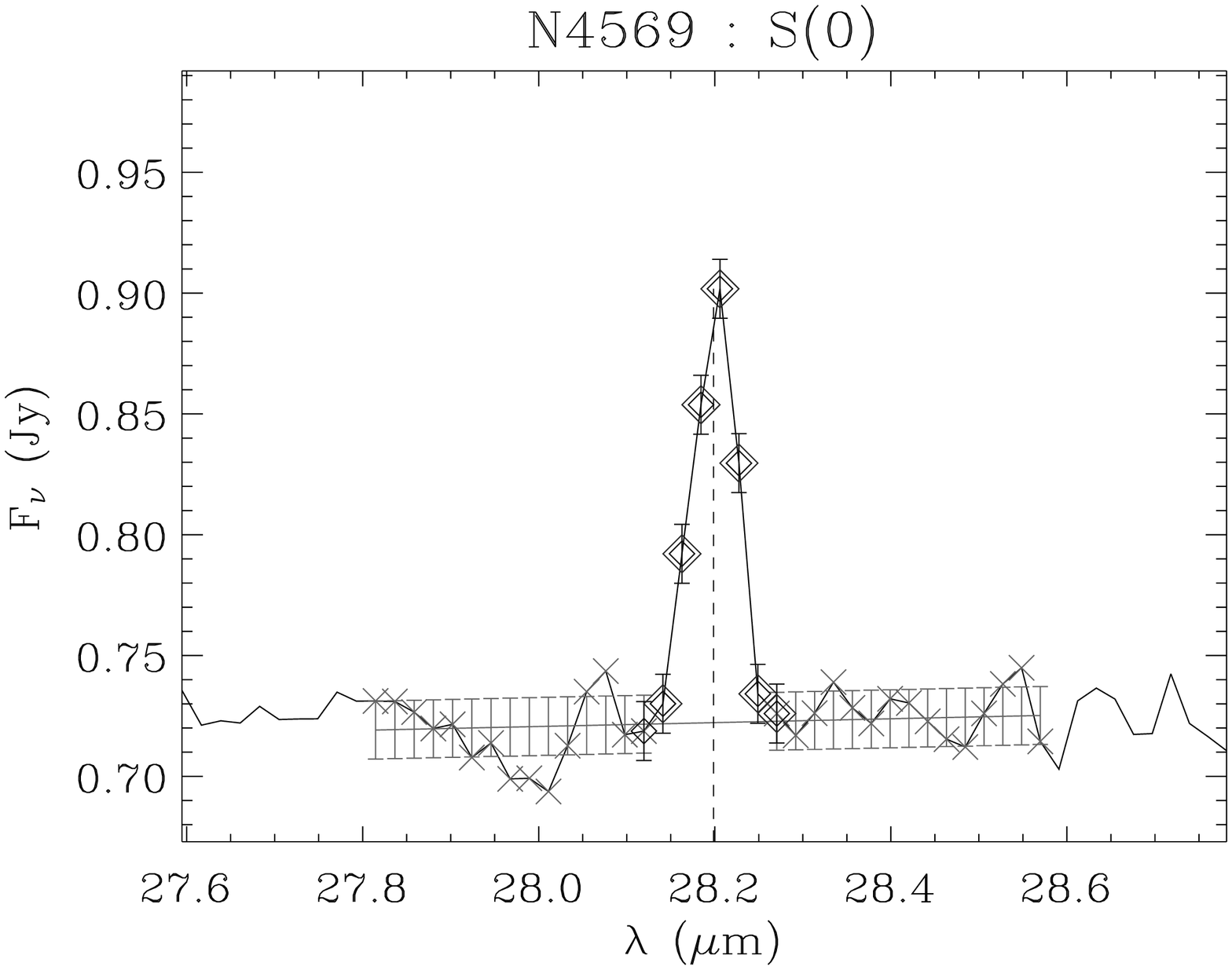}}}
\hspace*{-0.55cm}
\resizebox{5cm}{!}{\rotatebox{90}{\plotone{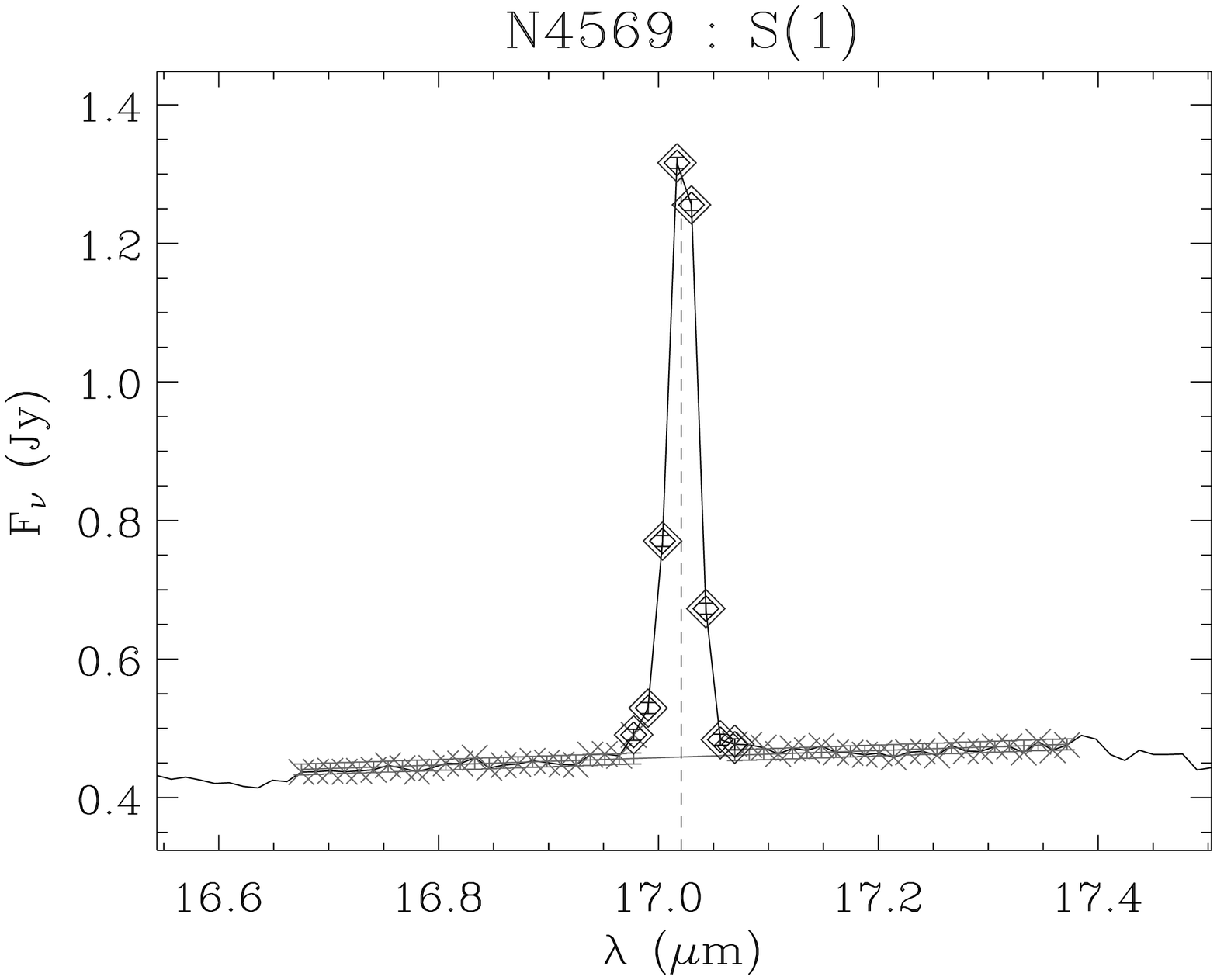}}}
\hspace*{-0.55cm}
\resizebox{5cm}{!}{\rotatebox{90}{\plotone{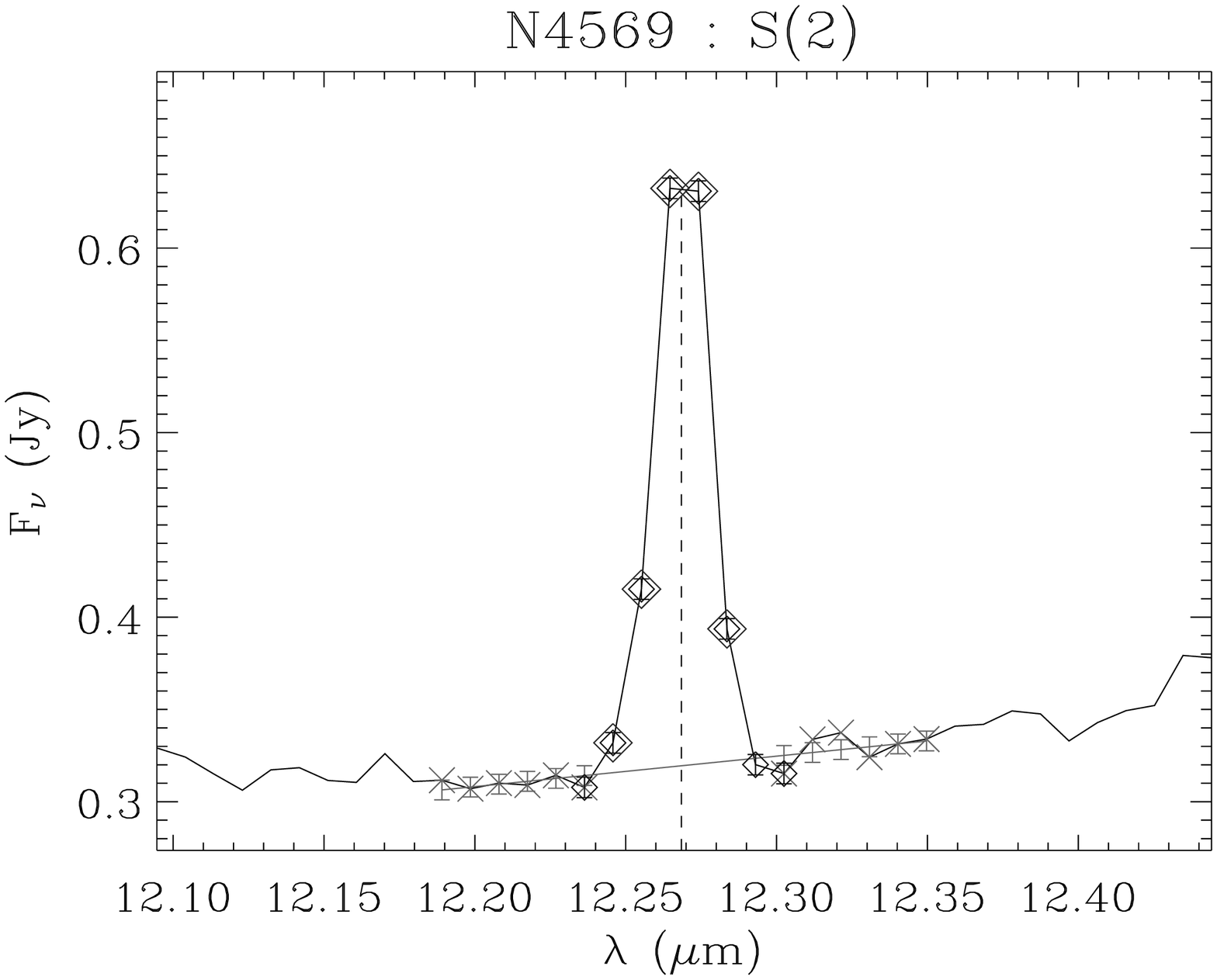}}}
\hspace*{-0.55cm}
\resizebox{5cm}{!}{\rotatebox{90}{\plotone{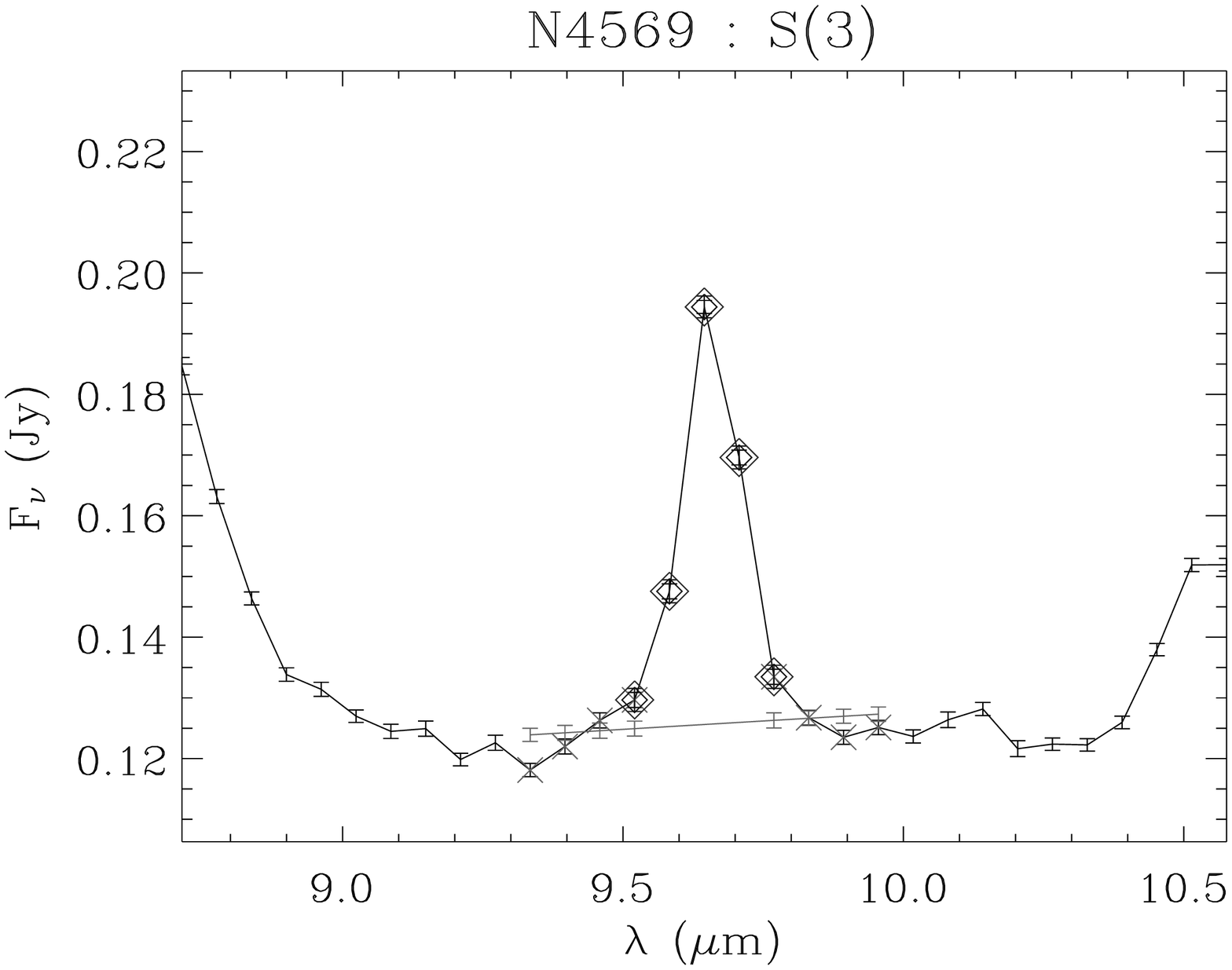}}} \\
\hspace*{-2.5cm}
\resizebox{5cm}{!}{\rotatebox{90}{\plotone{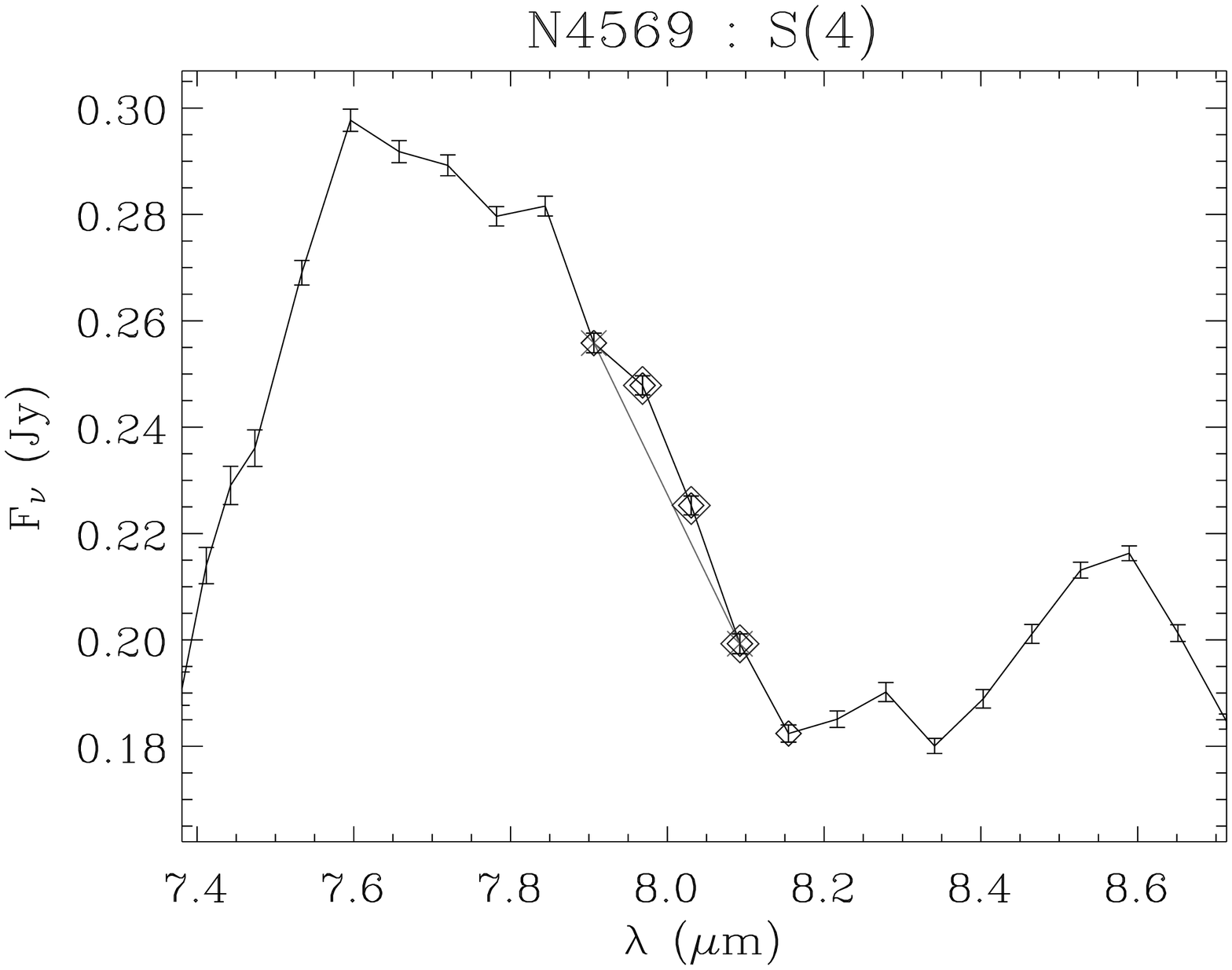}}}
\hspace*{-0.55cm}
\resizebox{5cm}{!}{\rotatebox{90}{\plotone{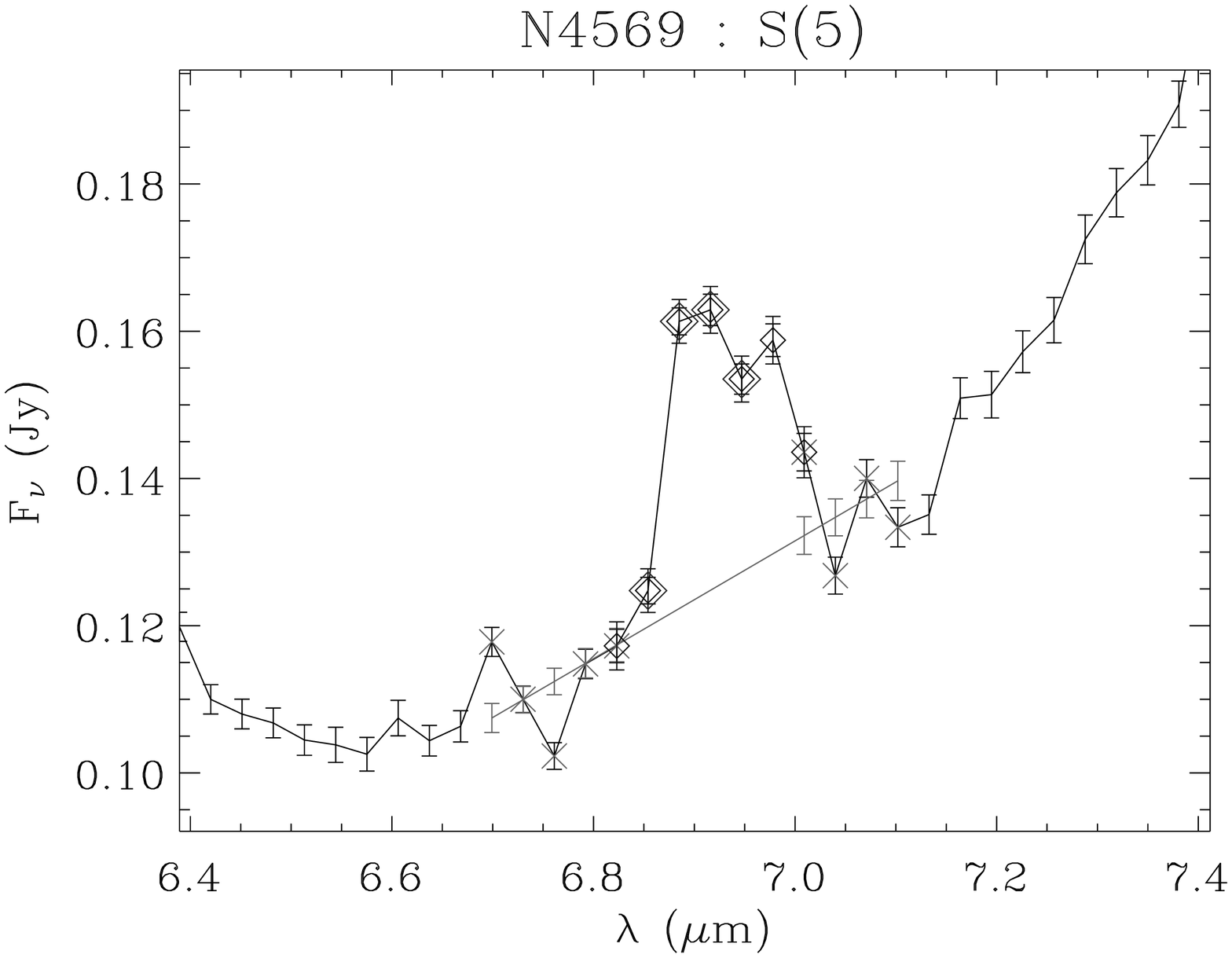}}}
\hspace*{-0.55cm}
\resizebox{5cm}{!}{\rotatebox{90}{\plotone{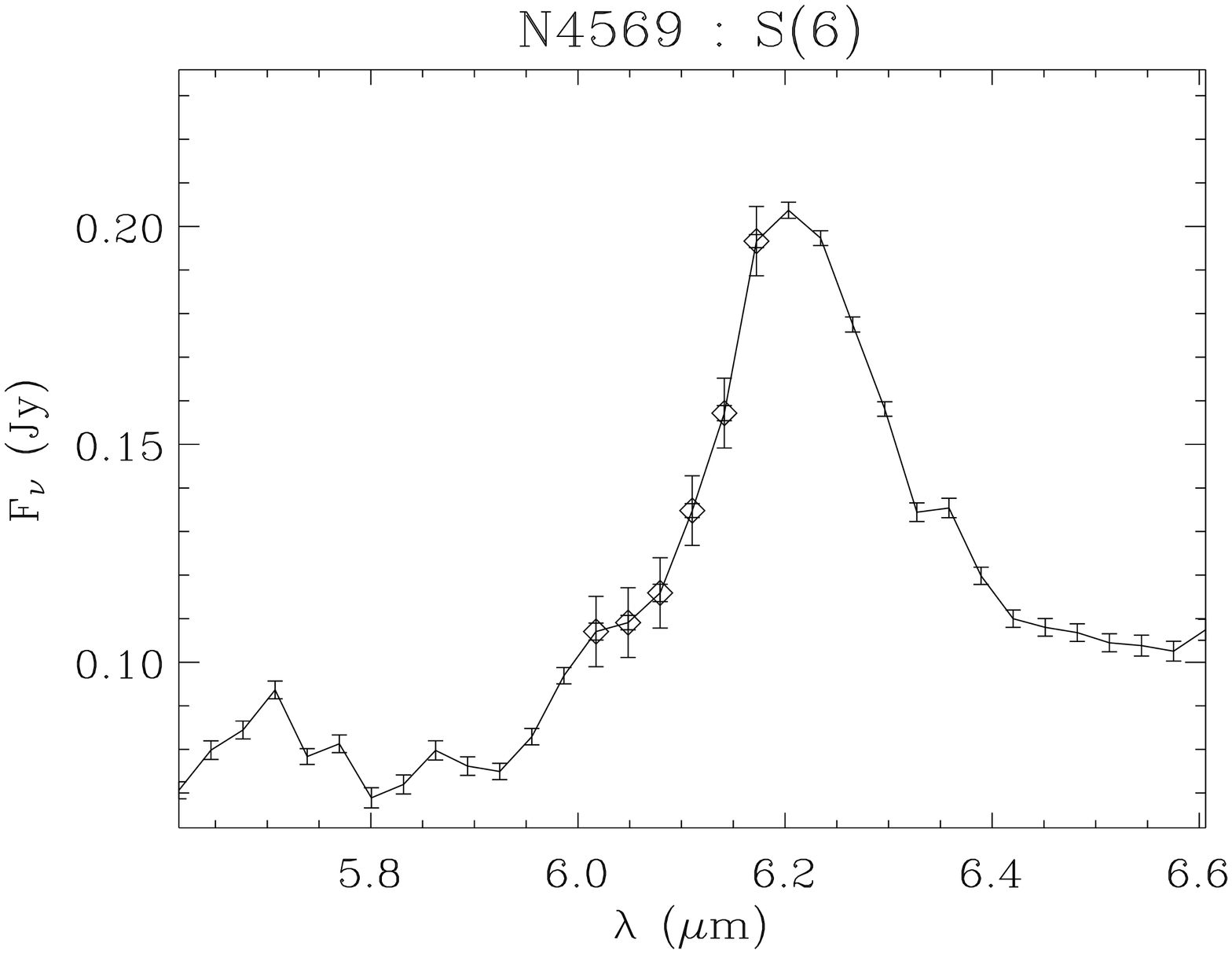}}}
\hspace*{-0.55cm}
\resizebox{5cm}{!}{\rotatebox{90}{\plotone{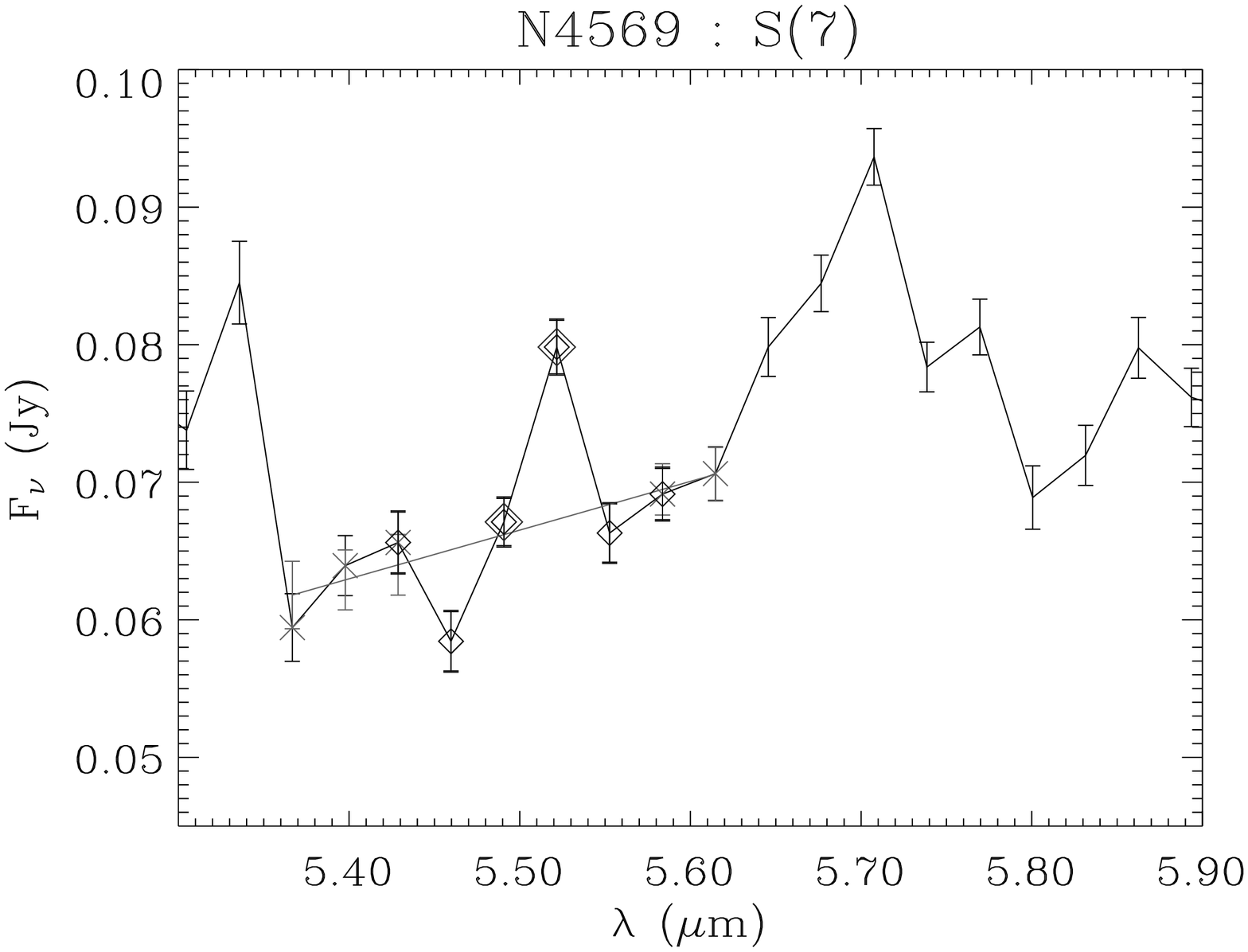}}} \\
\hspace*{-2.5cm}
\resizebox{5cm}{!}{\rotatebox{90}{\plotone{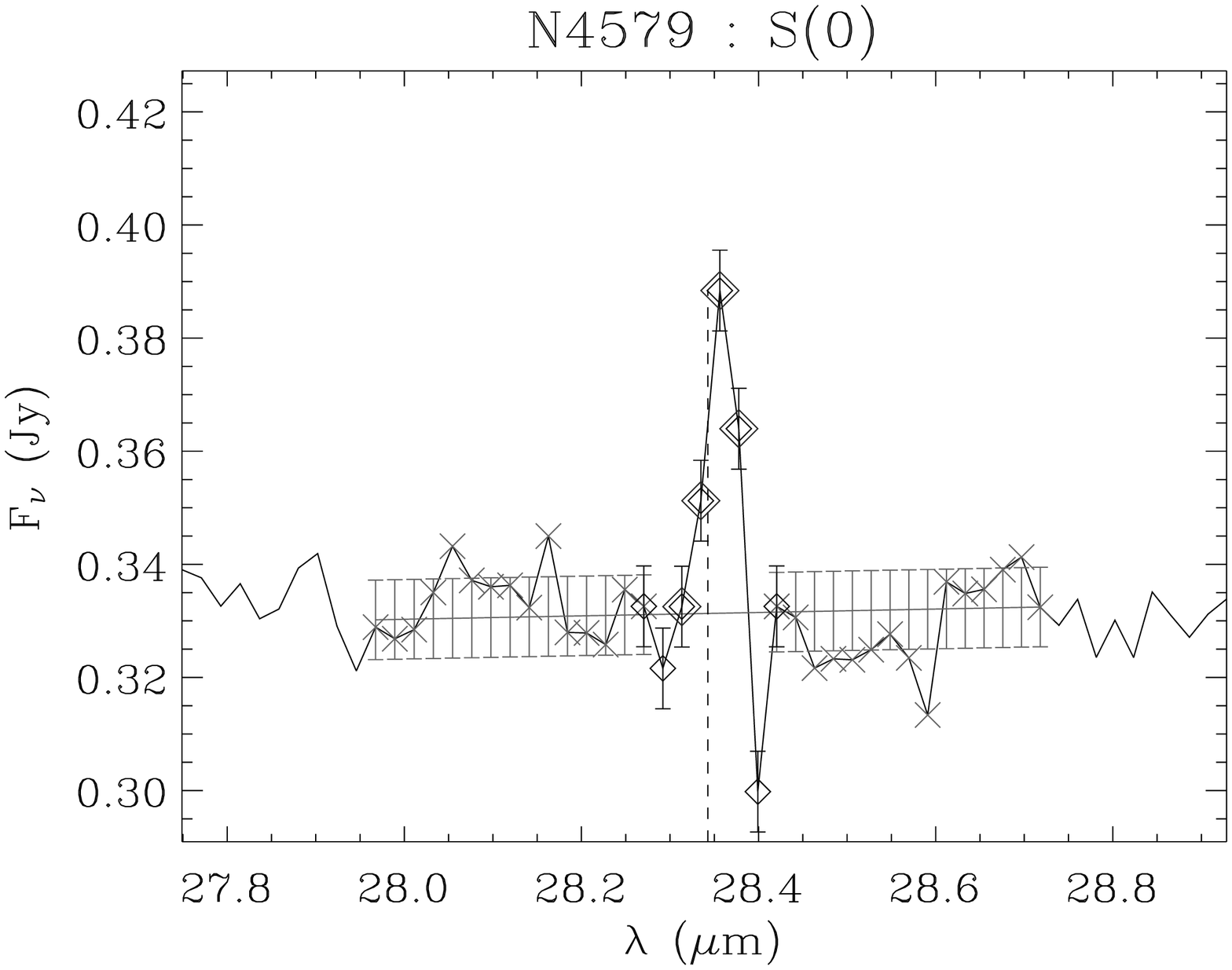}}}
\hspace*{-0.55cm}
\resizebox{5cm}{!}{\rotatebox{90}{\plotone{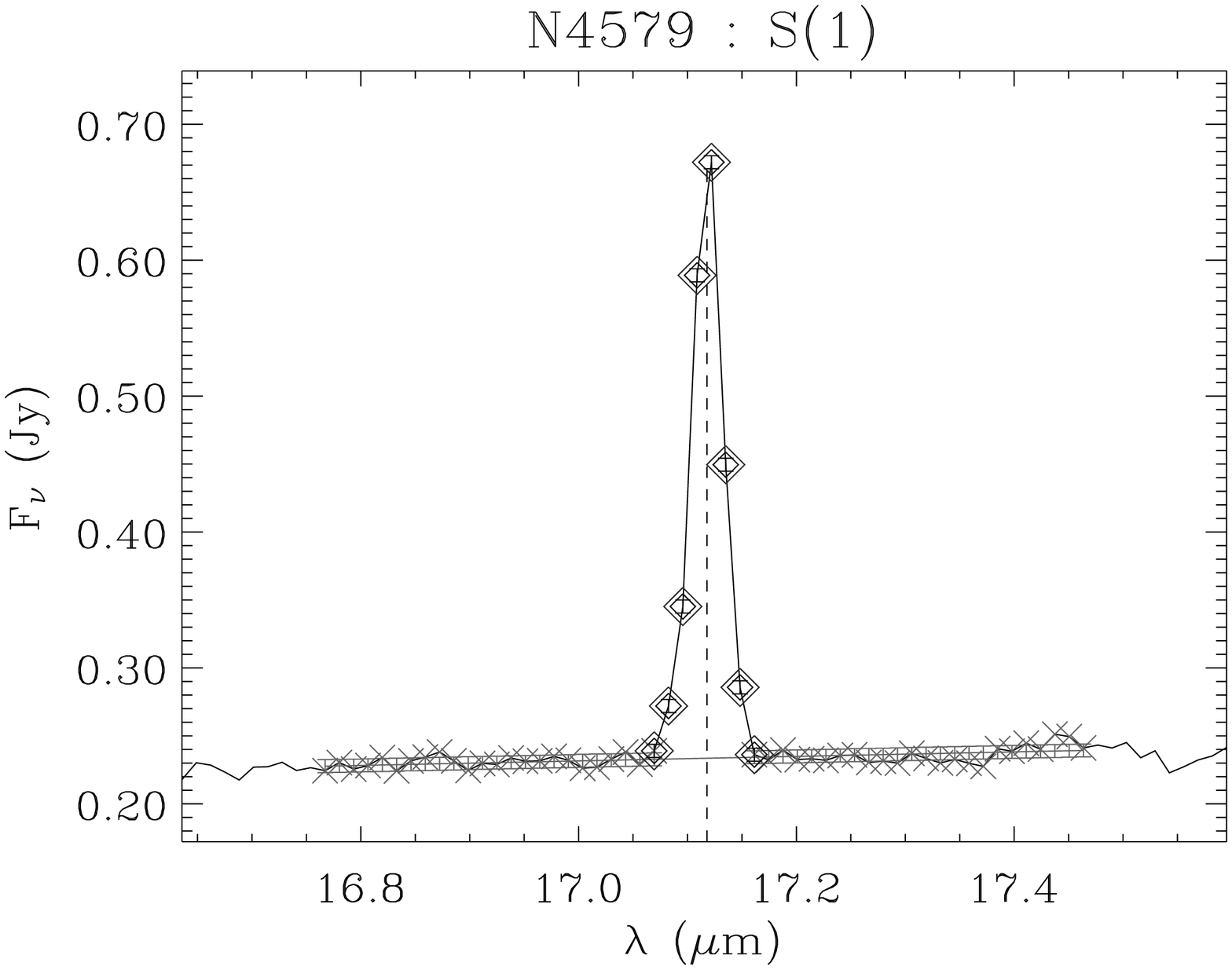}}}
\hspace*{-0.55cm}
\resizebox{5cm}{!}{\rotatebox{90}{\plotone{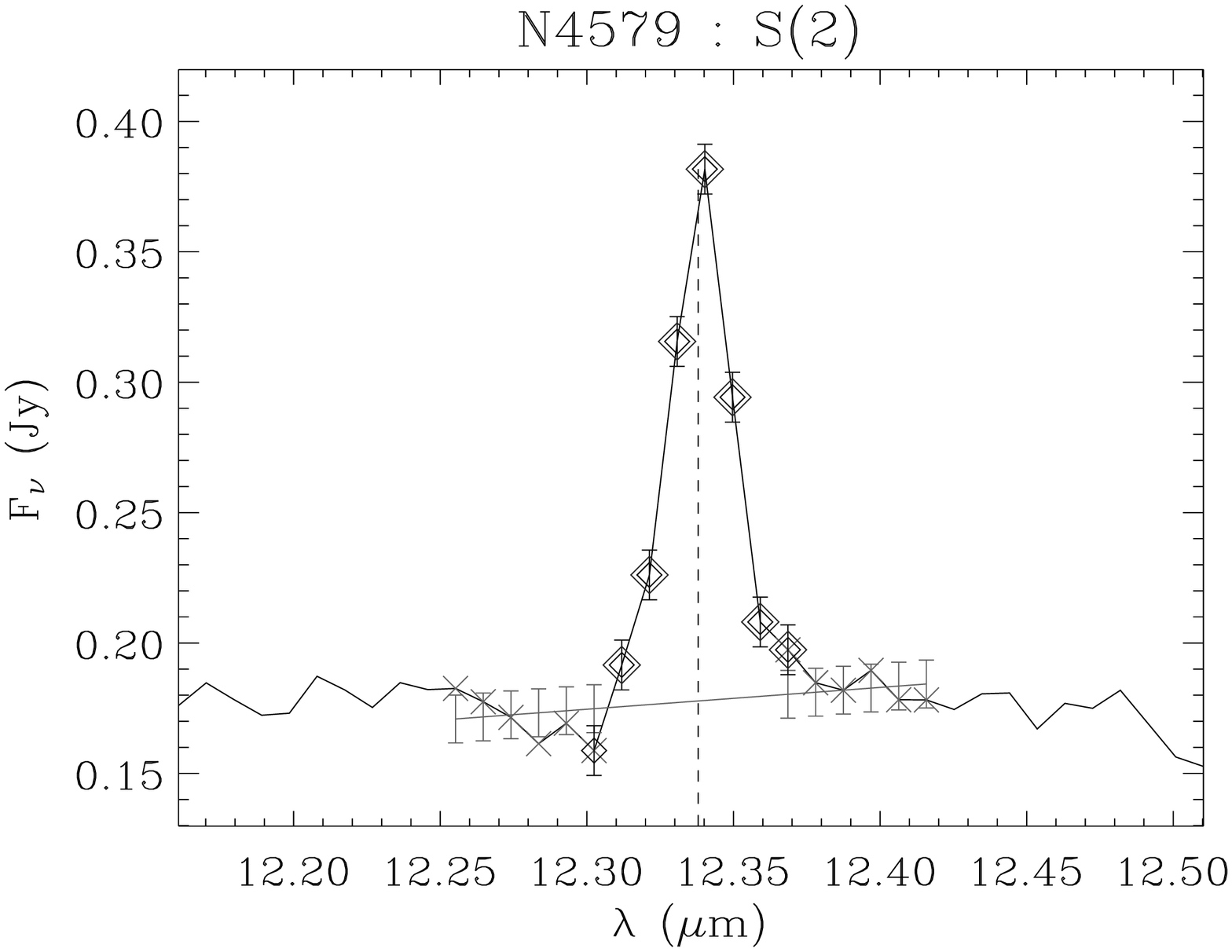}}}
\hspace*{-0.55cm}
\resizebox{5cm}{!}{\rotatebox{90}{\plotone{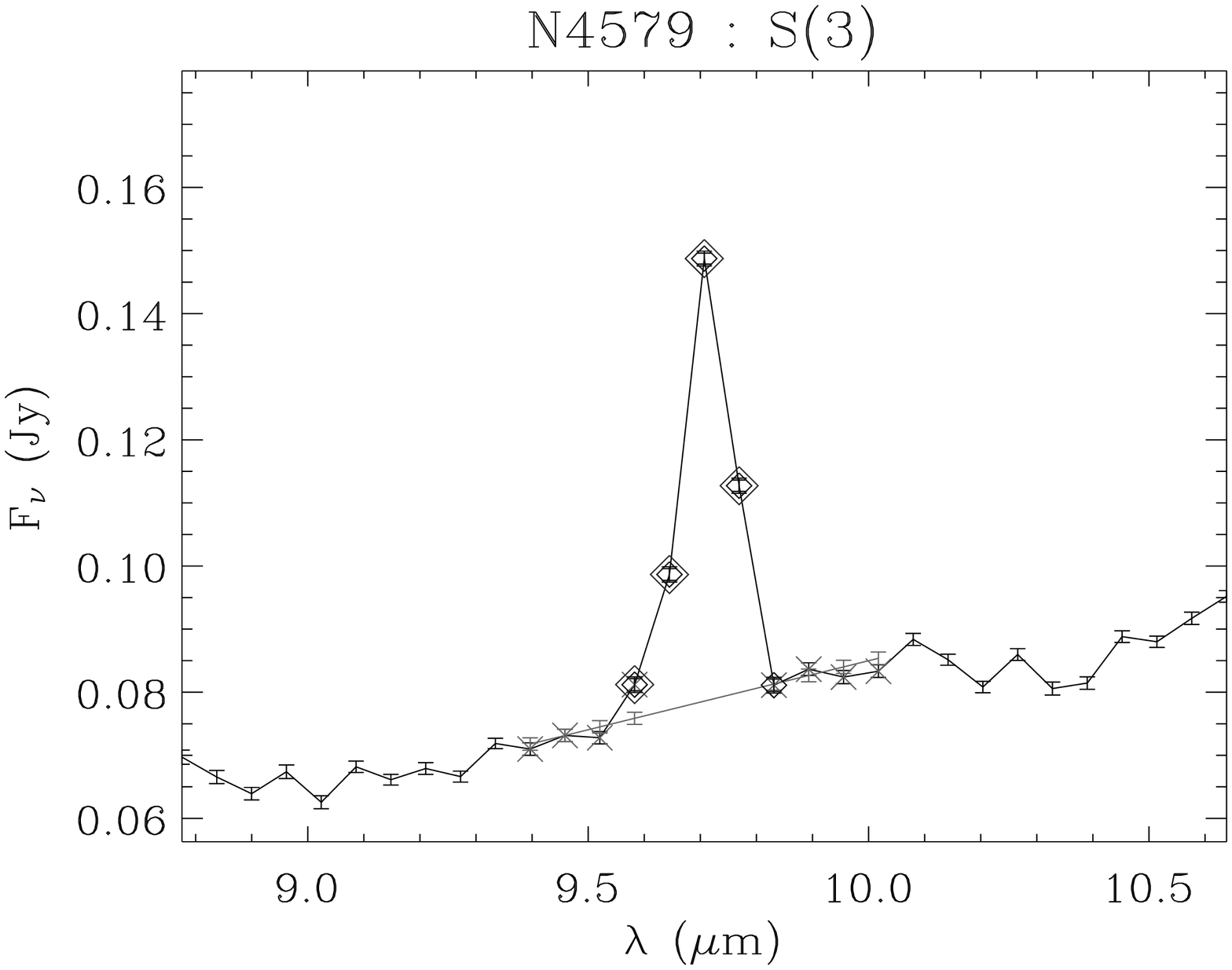}}} \\
\hspace*{-2.5cm}
\resizebox{5cm}{!}{\rotatebox{90}{\plotone{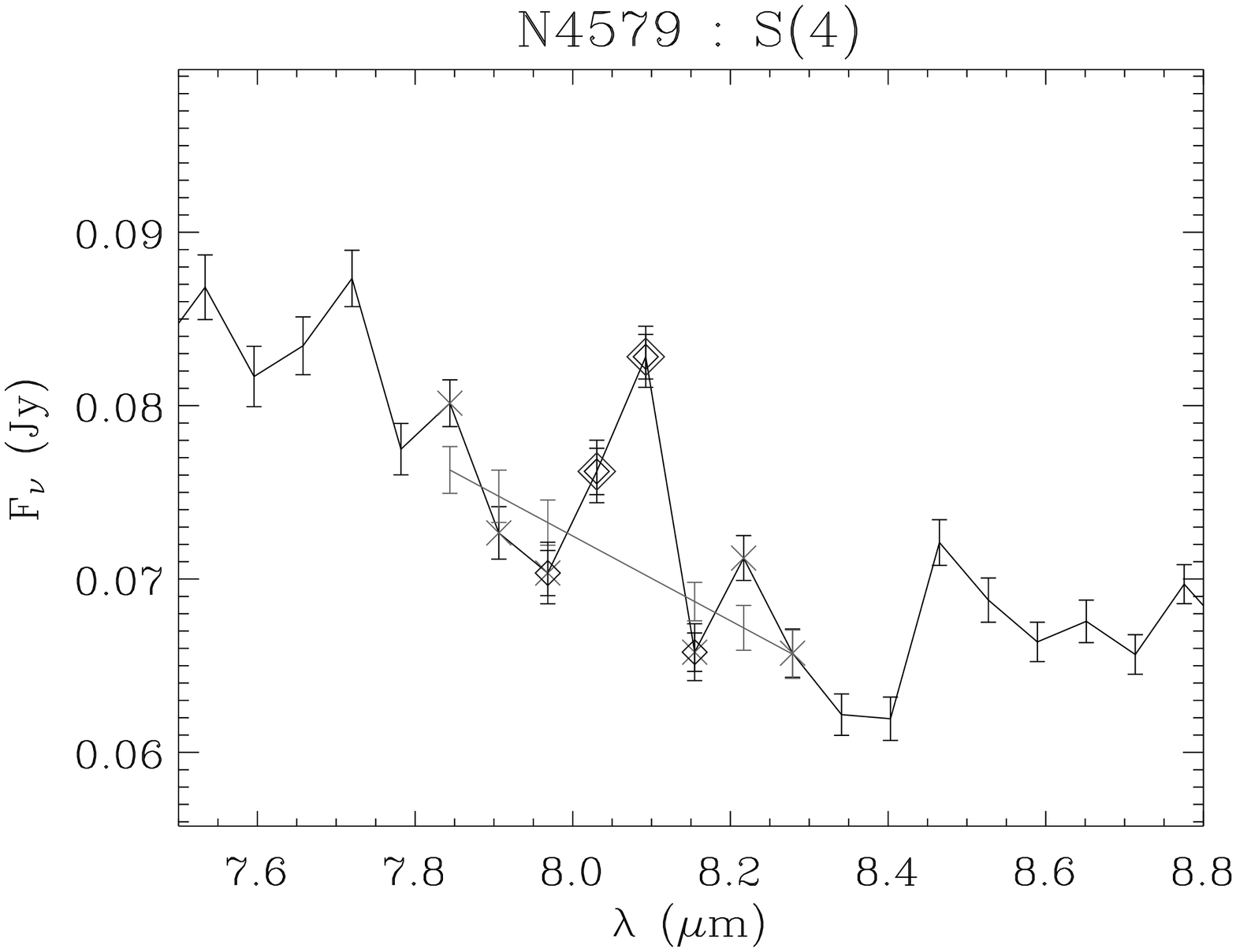}}}
\hspace*{-0.55cm}
\resizebox{5cm}{!}{\rotatebox{90}{\plotone{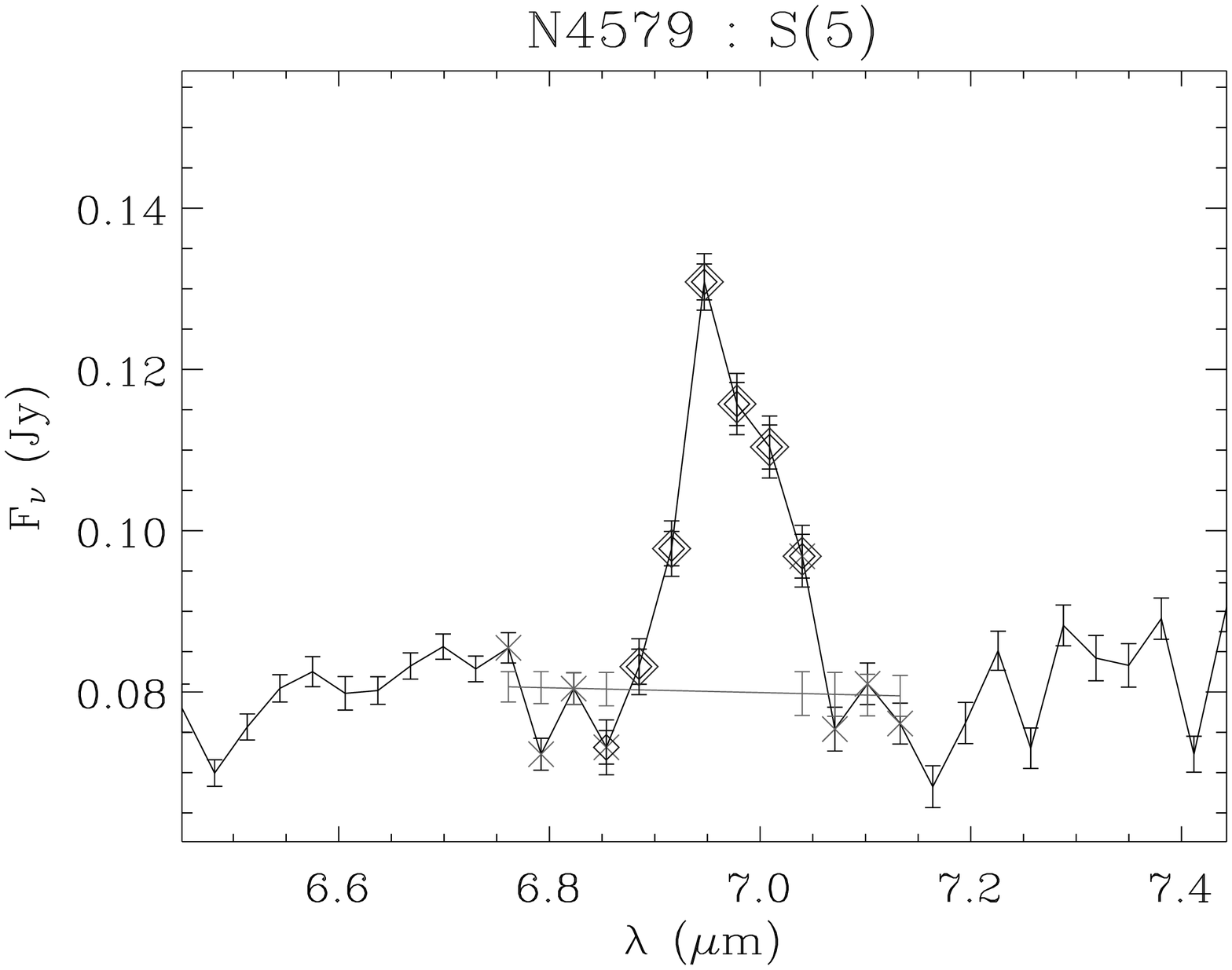}}}
\hspace*{-0.55cm}
\resizebox{5cm}{!}{\rotatebox{90}{\plotone{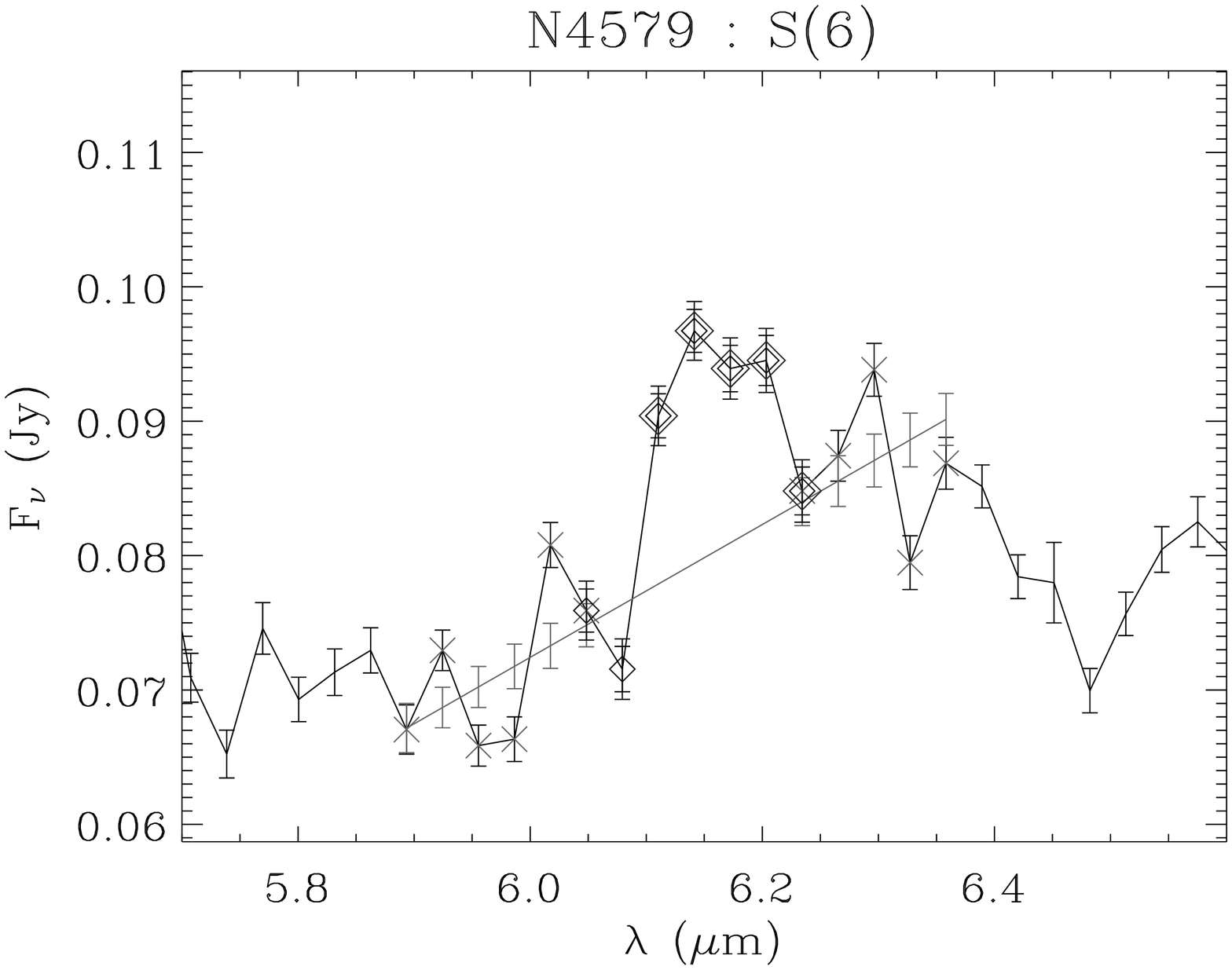}}}
\hspace*{-0.55cm}
\resizebox{5cm}{!}{\rotatebox{90}{\plotone{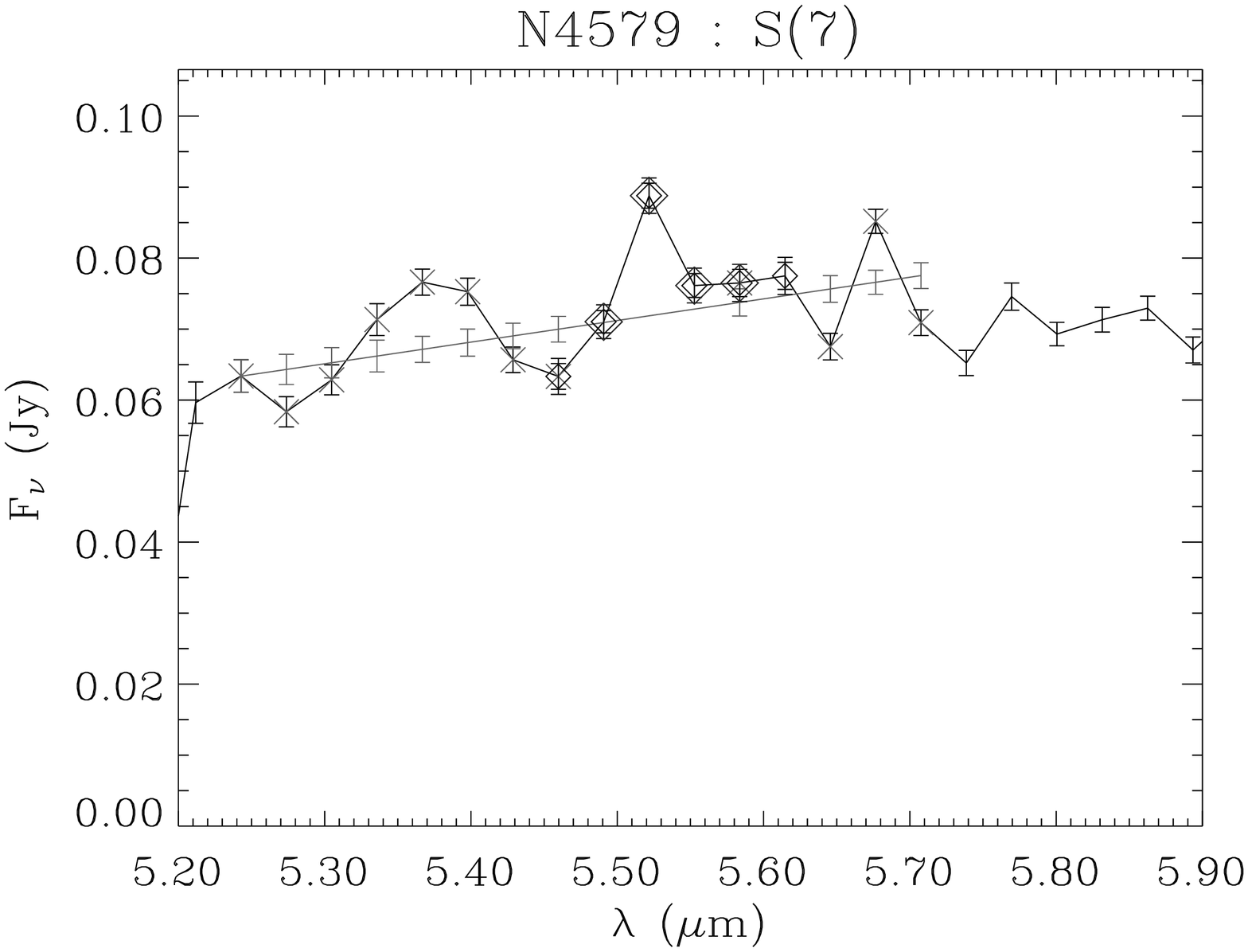}}}
\caption{(continued).
}
\end{figure}

\begin{figure}[!ht]
\resizebox{10cm}{!}{\rotatebox{90}{\plotone{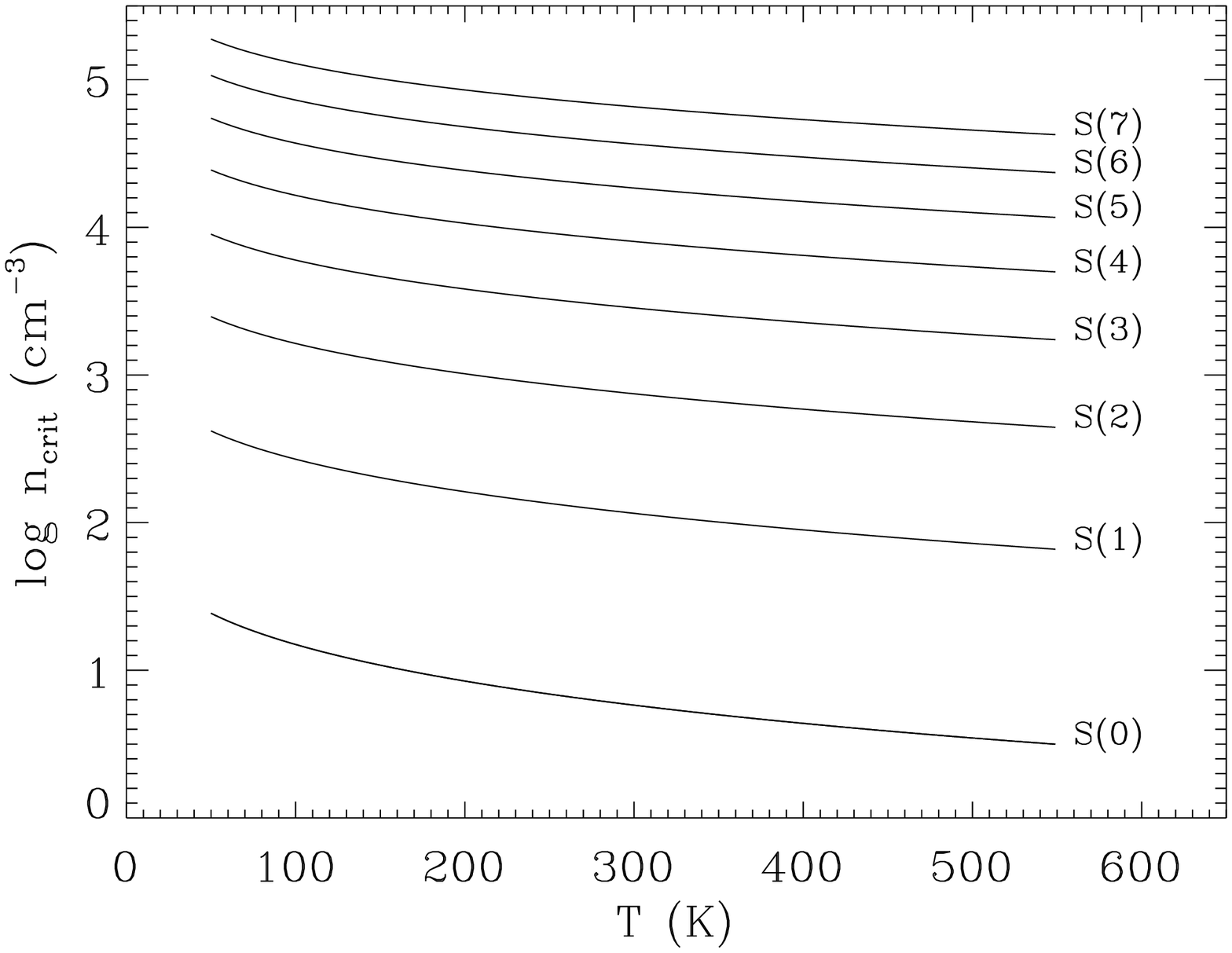}}}
\caption{Critical densities for collisional deexcitation by H$_2$ for the rotational transitions in the
wavelength range of the IRS instrument.
}
\label{fig:ncrit}
\end{figure}

\begin{figure}[!ht]
\hspace*{-2cm}
\resizebox{10cm}{!}{\rotatebox{90}{\plotone{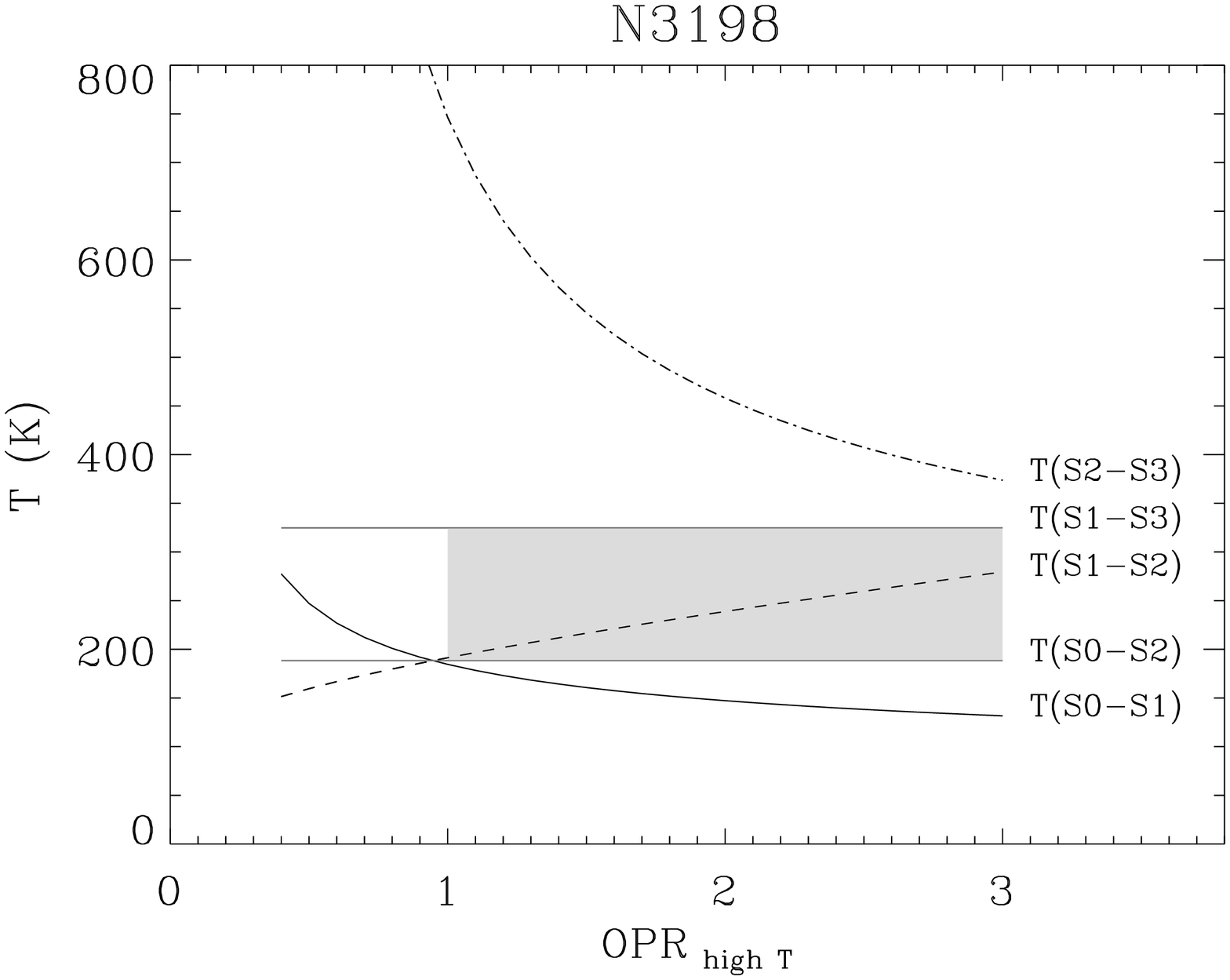}}}
\hspace*{-0.5cm}
\resizebox{10cm}{!}{\rotatebox{90}{\plotone{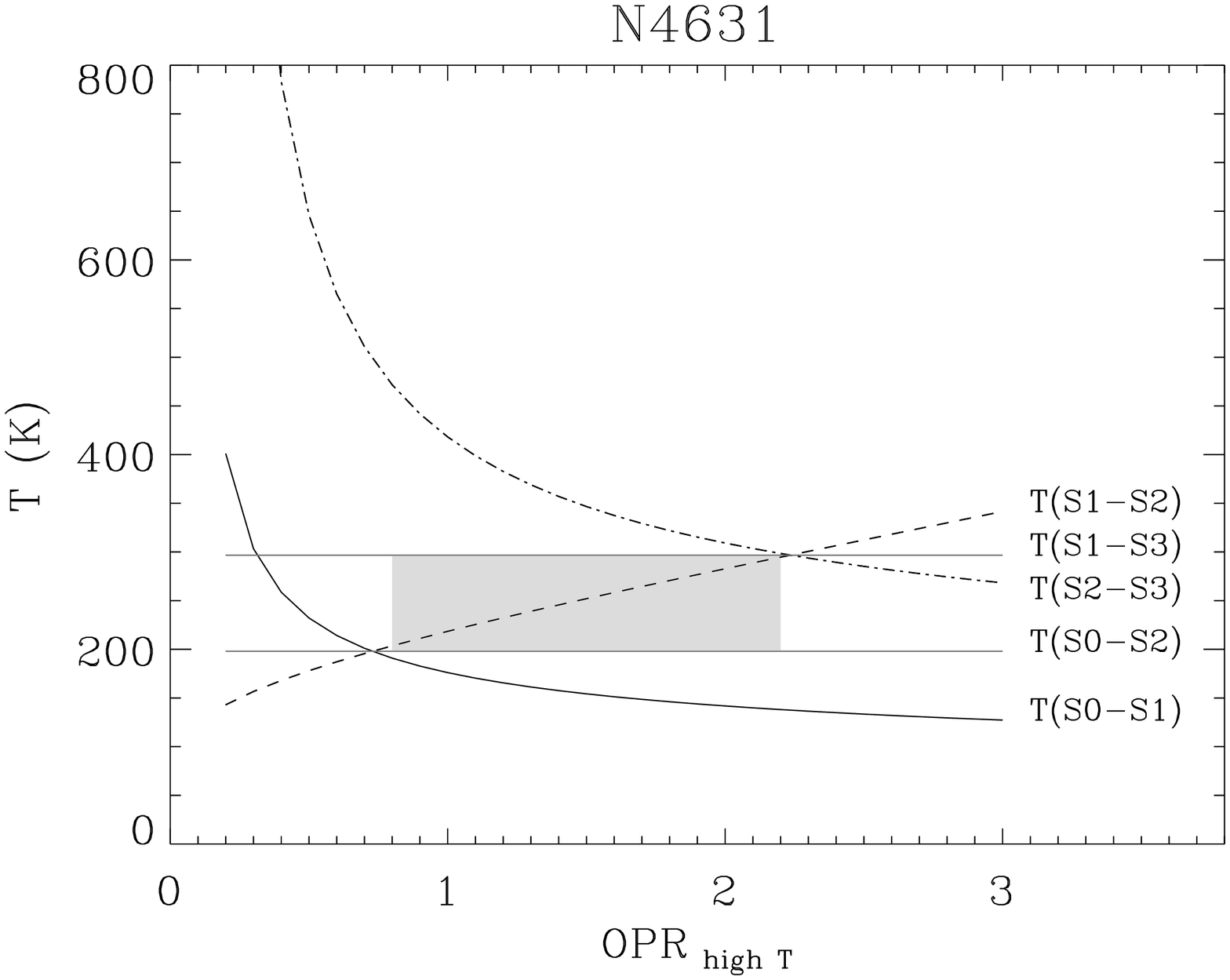}}}
\caption{Examples of how the possible range of apparent $OPR_{\rm \,high\,T}$ is determined
for each galaxy.
Thermalization requires the apparent temperatures derived from each pair of transitions
to be monotonic as a function of upper level energy; in particular, the conditions
$T(S1-S2) < T(S1-S3) < T(S2-S3)$ (indicated by the shaded areas) have to be satisfied.
The apparent temperatures derived for NGC\,3198 are compatible with $OPR_{\rm \,high\,T} = 3$,
whereas they require $OPR_{\rm \,high\,T} < 3$ for NGC\,4631 (see text).
}
\label{fig:diag_temp}
\end{figure}

\begin{figure}[!ht]
\vspace*{-1cm}
\hspace*{-2cm}
\resizebox{10cm}{!}{\rotatebox{90}{\plotone{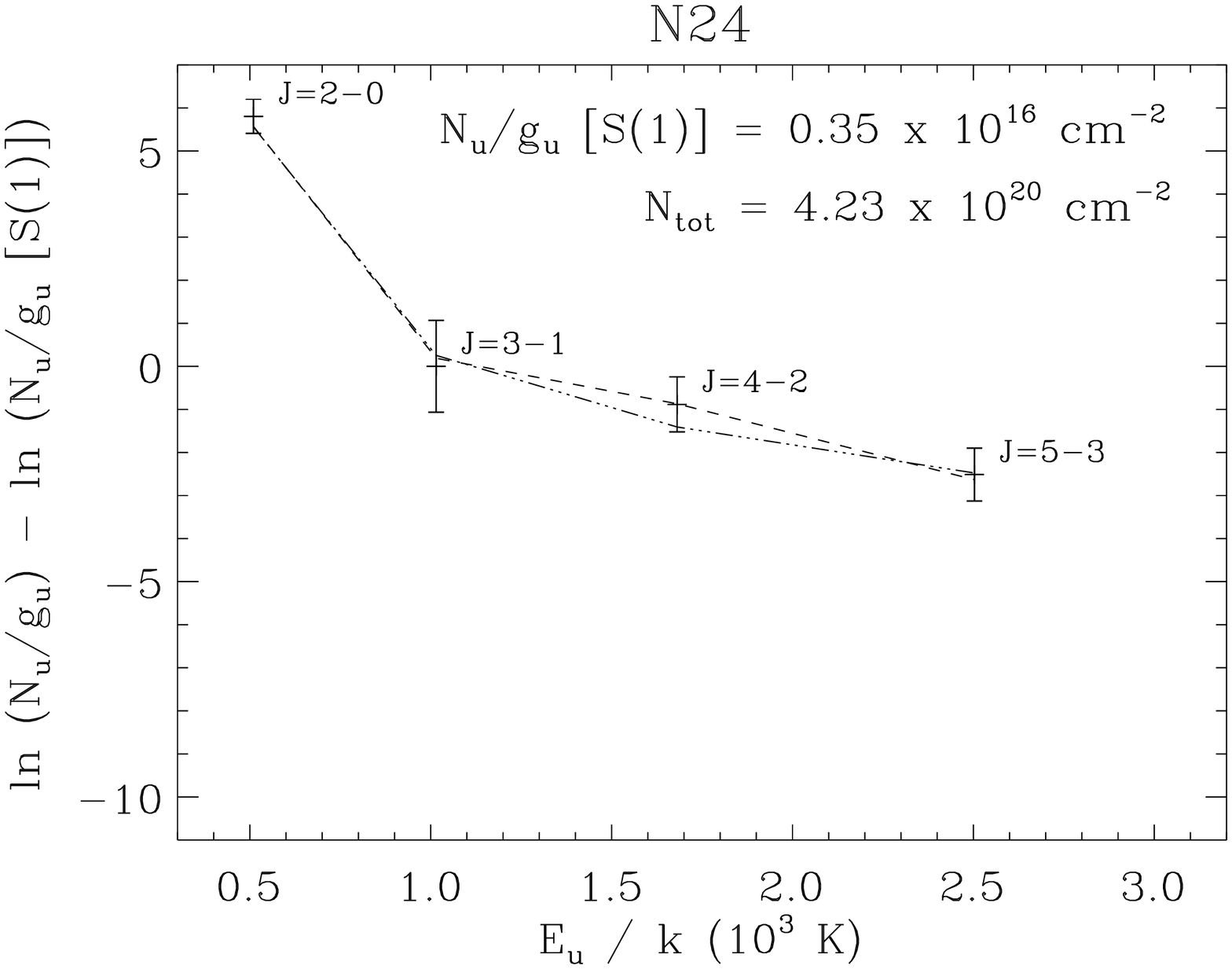}}}
\hspace*{-0.5cm}
\resizebox{10cm}{!}{\rotatebox{90}{\plotone{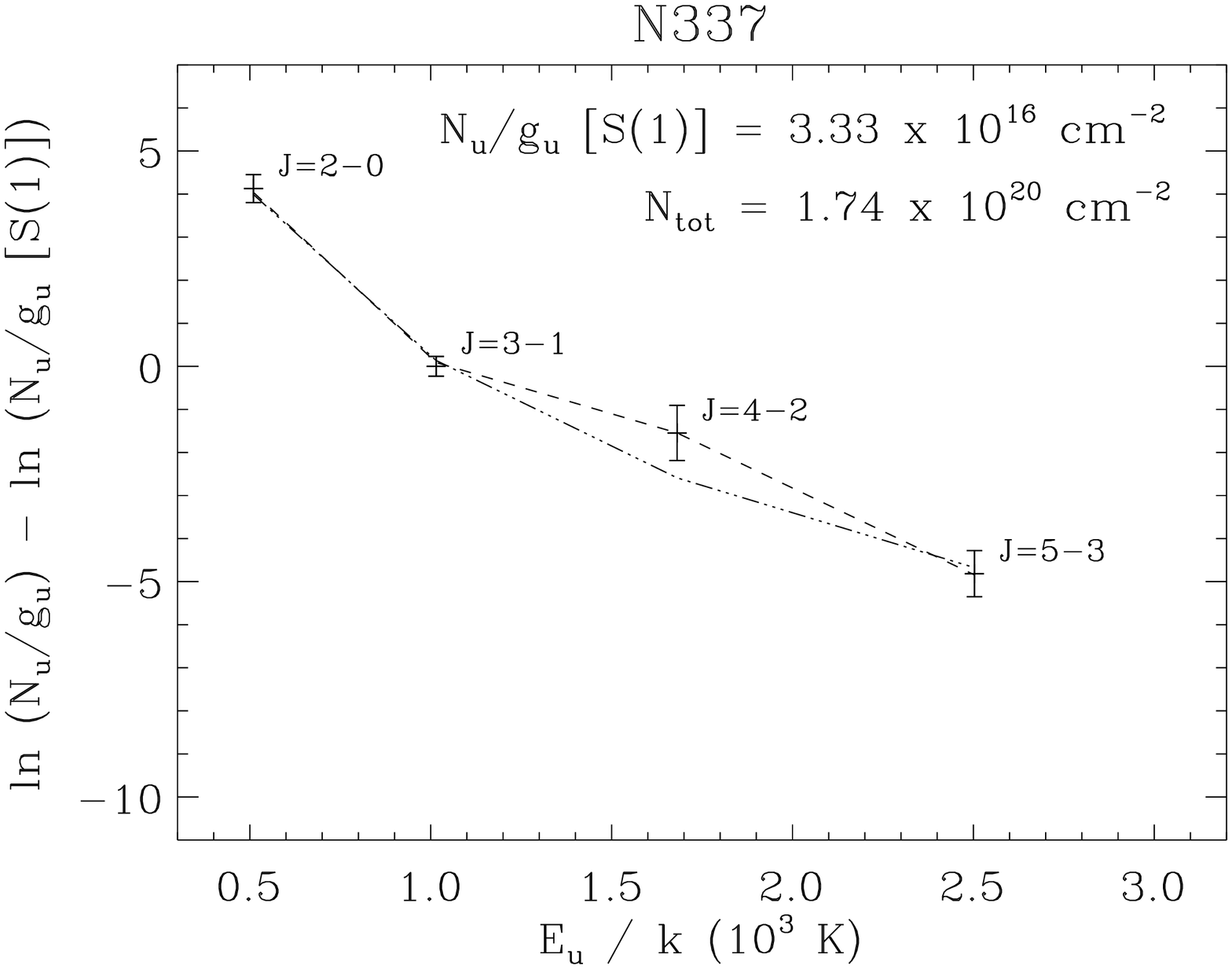}}}
\hspace*{-2cm}
\resizebox{10cm}{!}{\rotatebox{90}{\plotone{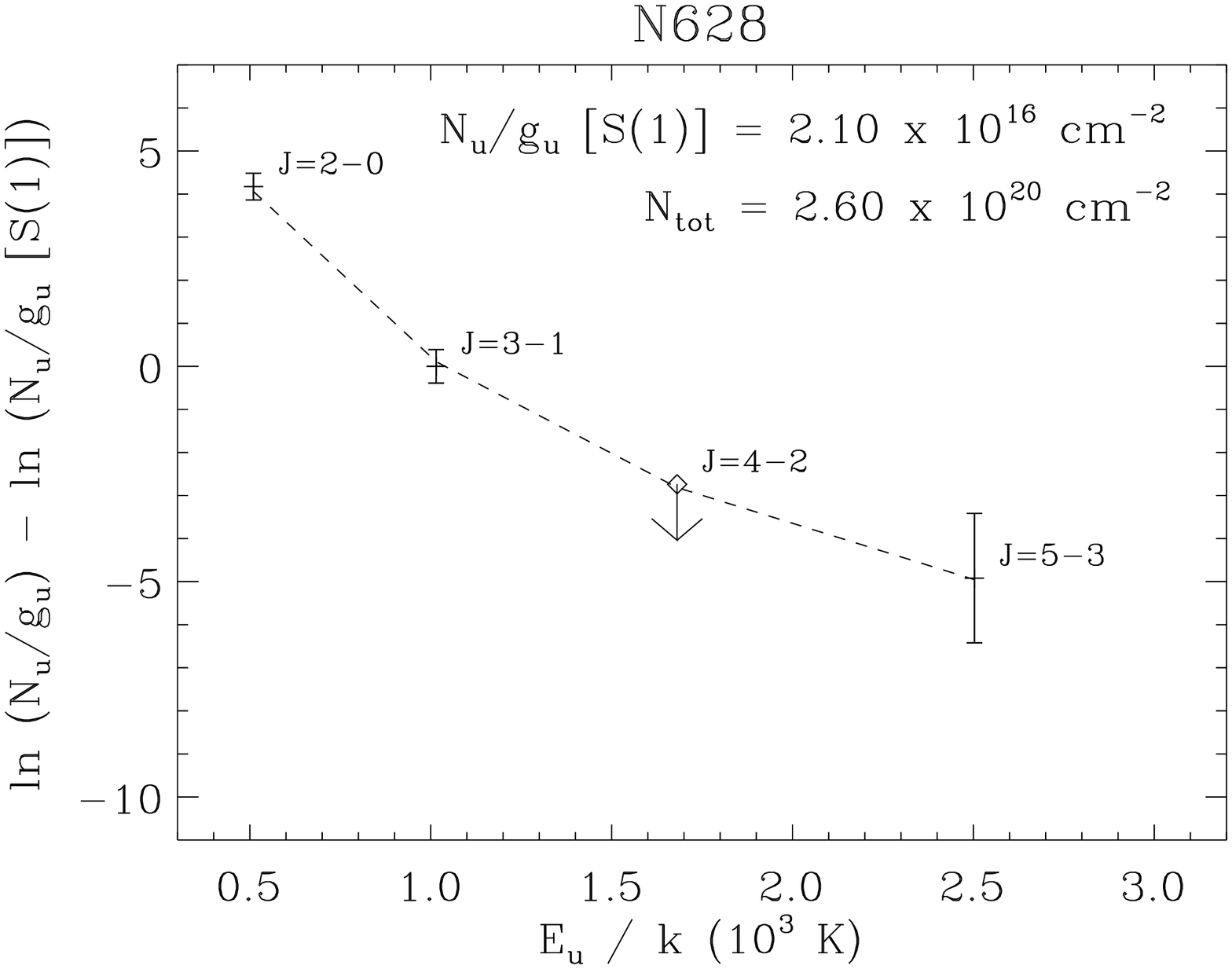}}}
\hspace*{-0.5cm}
\resizebox{10cm}{!}{\rotatebox{90}{\plotone{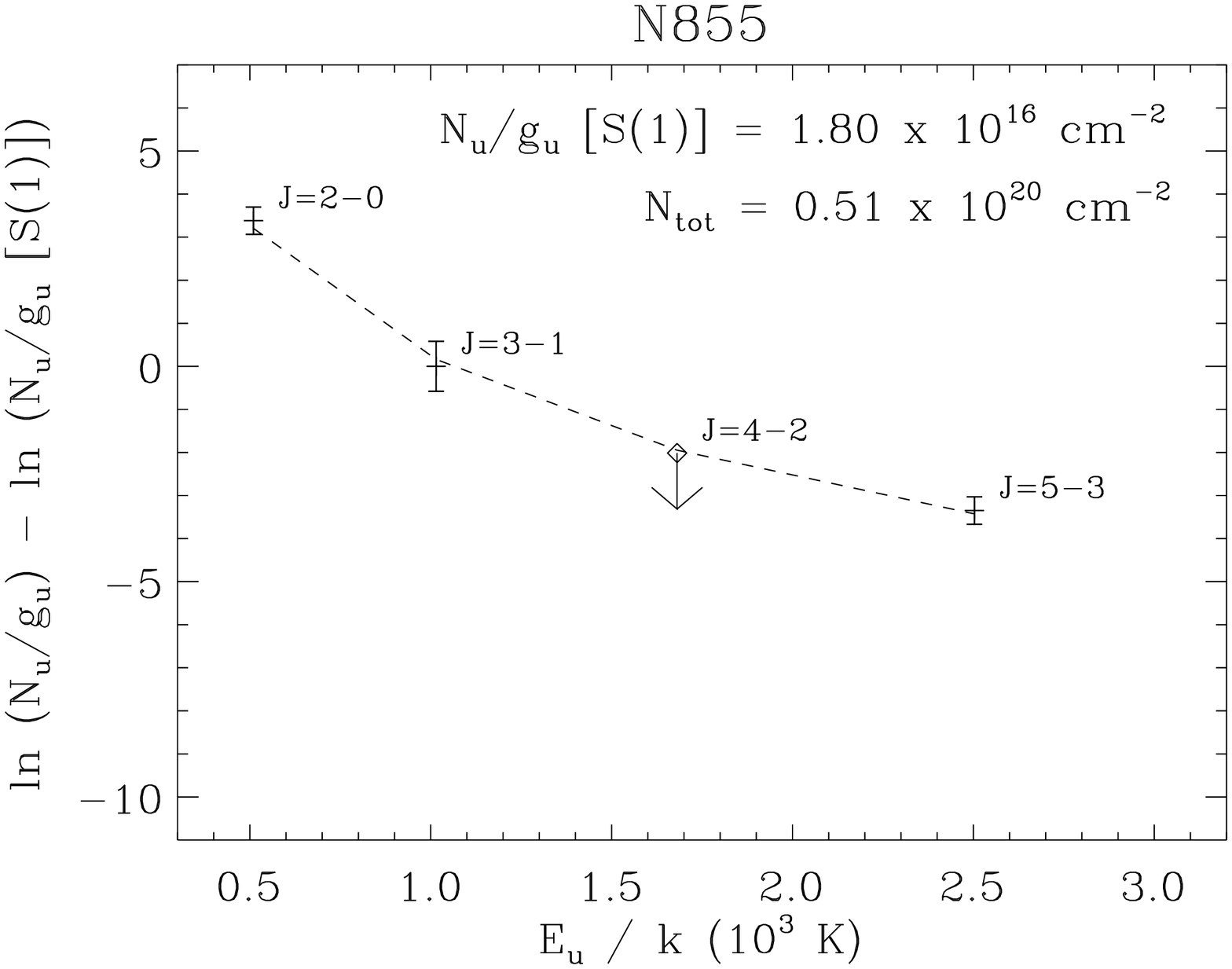}}}
\hspace*{-2cm}
\resizebox{10cm}{!}{\rotatebox{90}{\plotone{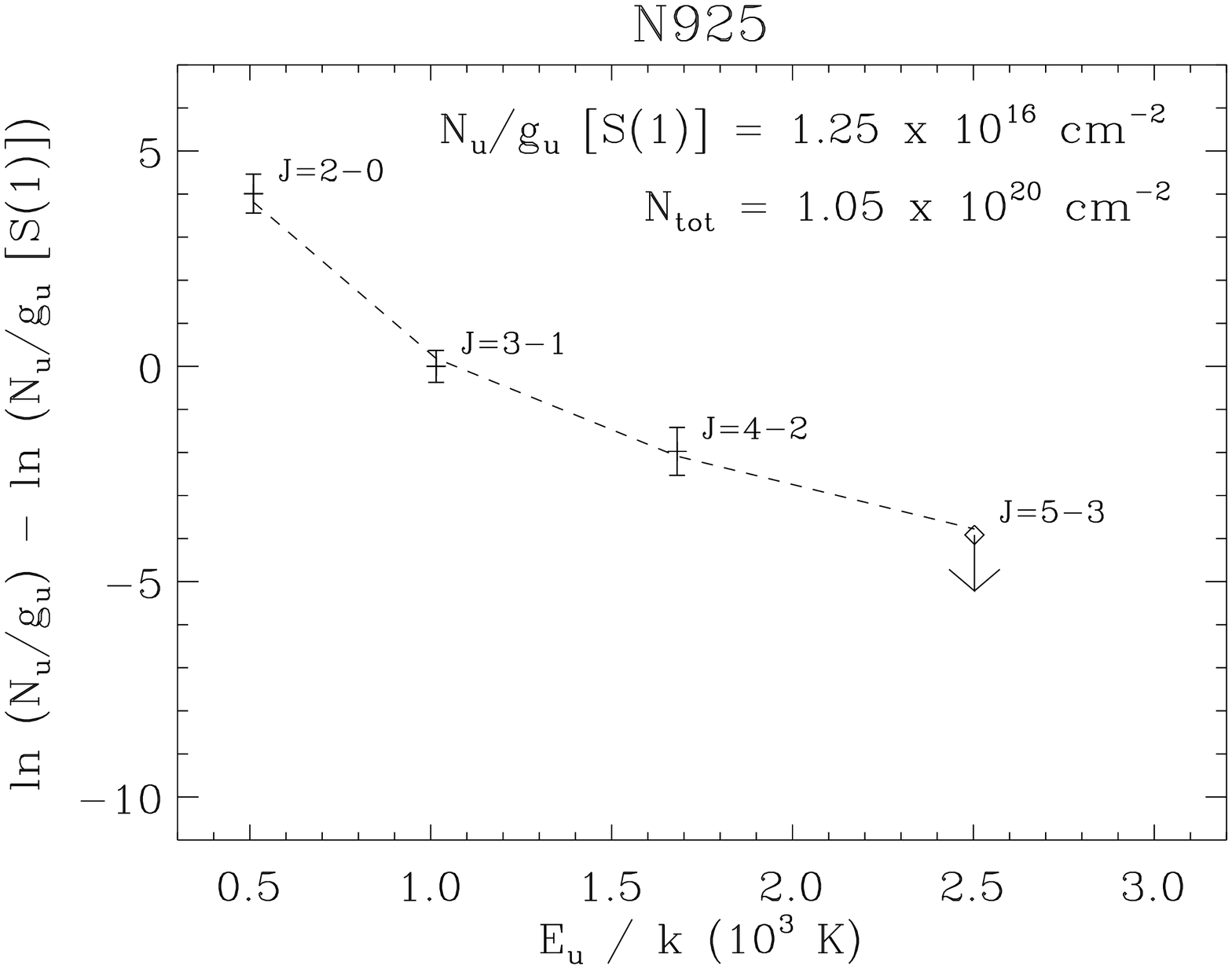}}}
\hspace*{-0.5cm}
\resizebox{10cm}{!}{\rotatebox{90}{\plotone{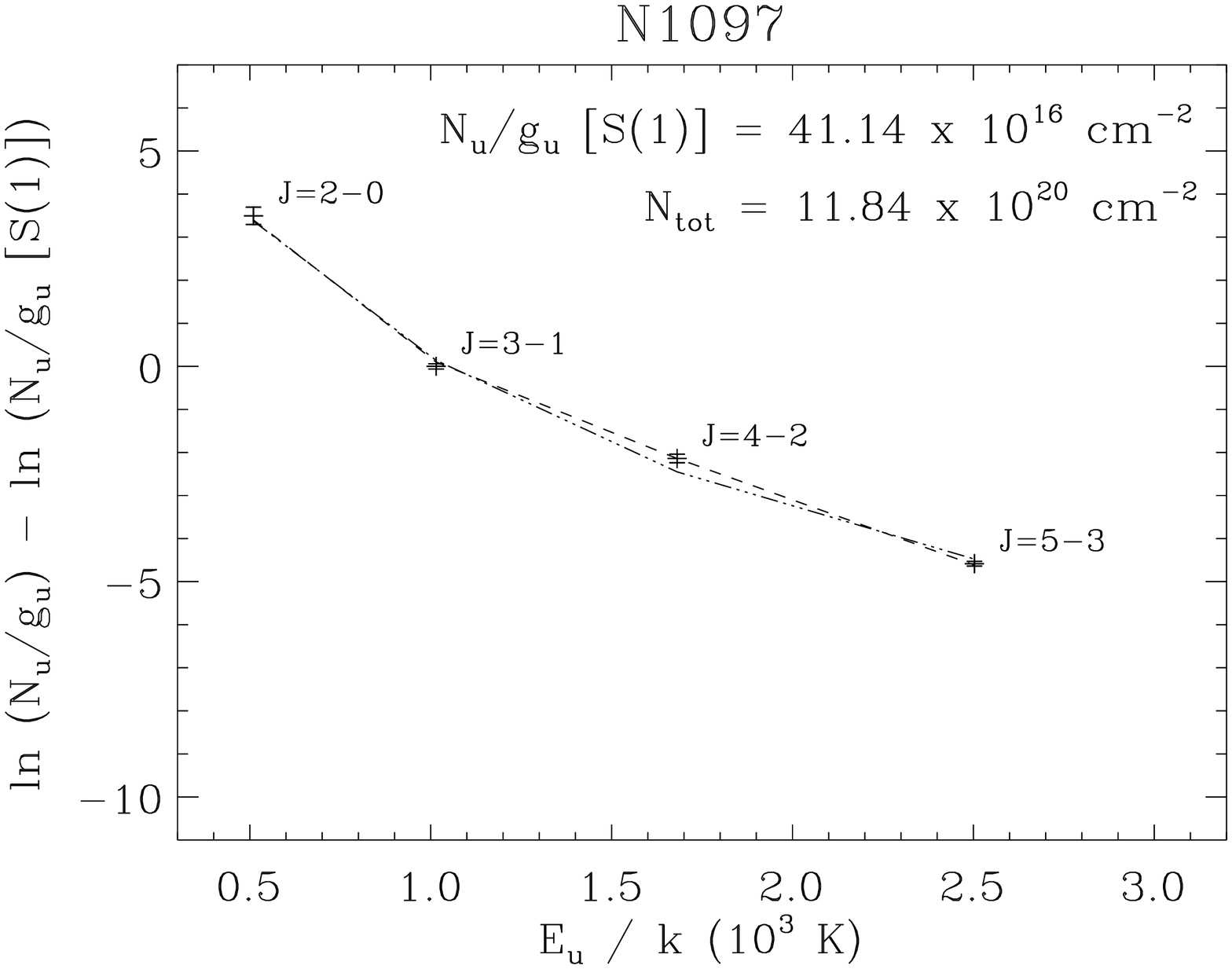}}}
\vspace*{-1cm}
\caption{Excitation diagrams. The $N_{\rm u} / g_{\rm u}$ ratios are normalized by the S(1)
transition. The dashed line indicates the best fit (see text). Whenever $OPR_{\rm \,high\,T} < 3$
is required by the temperature constraints, a dot-dash line shows for comparison
the fit obtained with $OPR_{\rm \,high\,T} = 3$.
}
\label{fig:diag_exc}
\end{figure}

\addtocounter{figure}{-1}
\begin{figure}[!ht]
\hspace*{-2cm}
\resizebox{10cm}{!}{\rotatebox{90}{\plotone{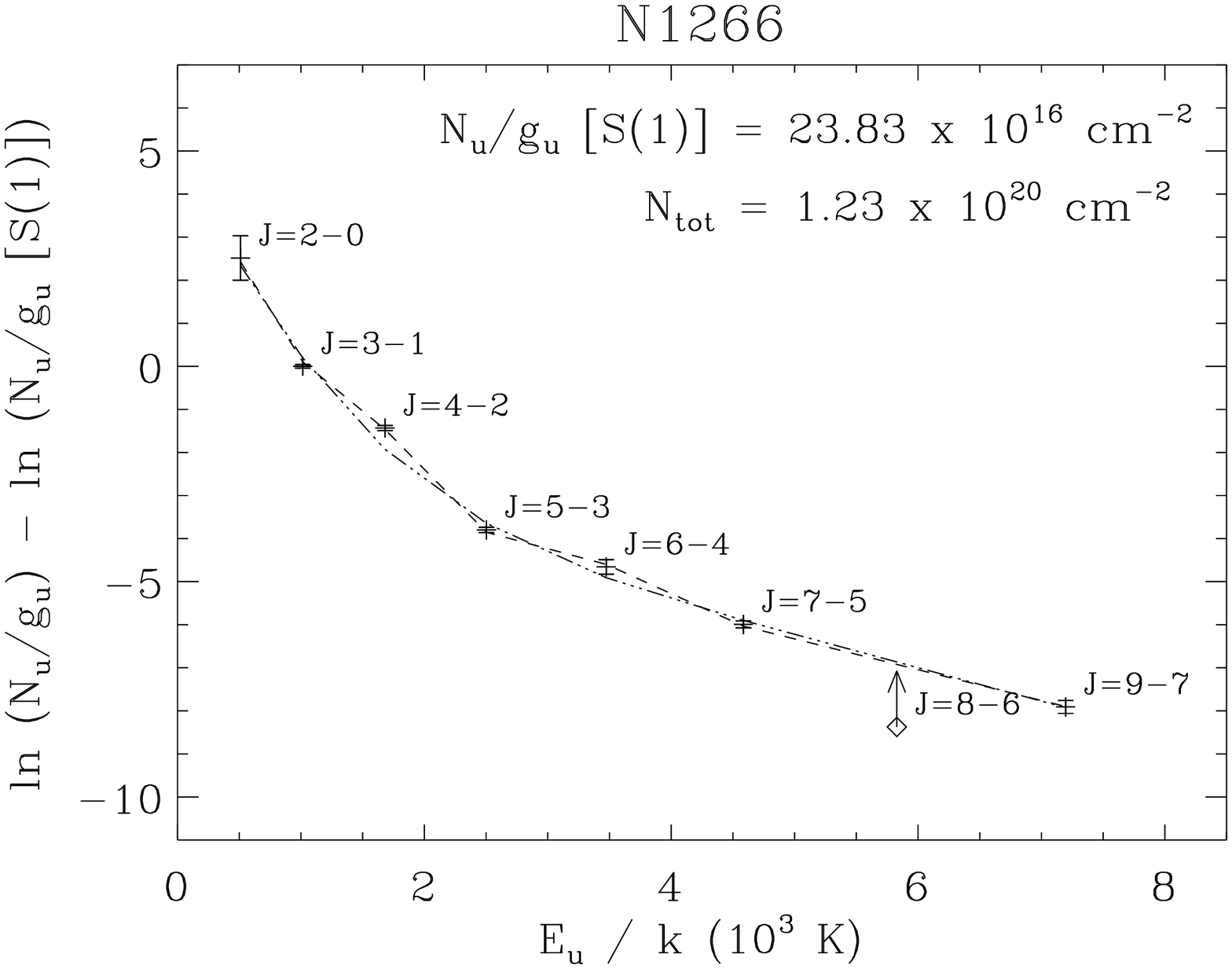}}}
\hspace*{-0.5cm}
\resizebox{10cm}{!}{\rotatebox{90}{\plotone{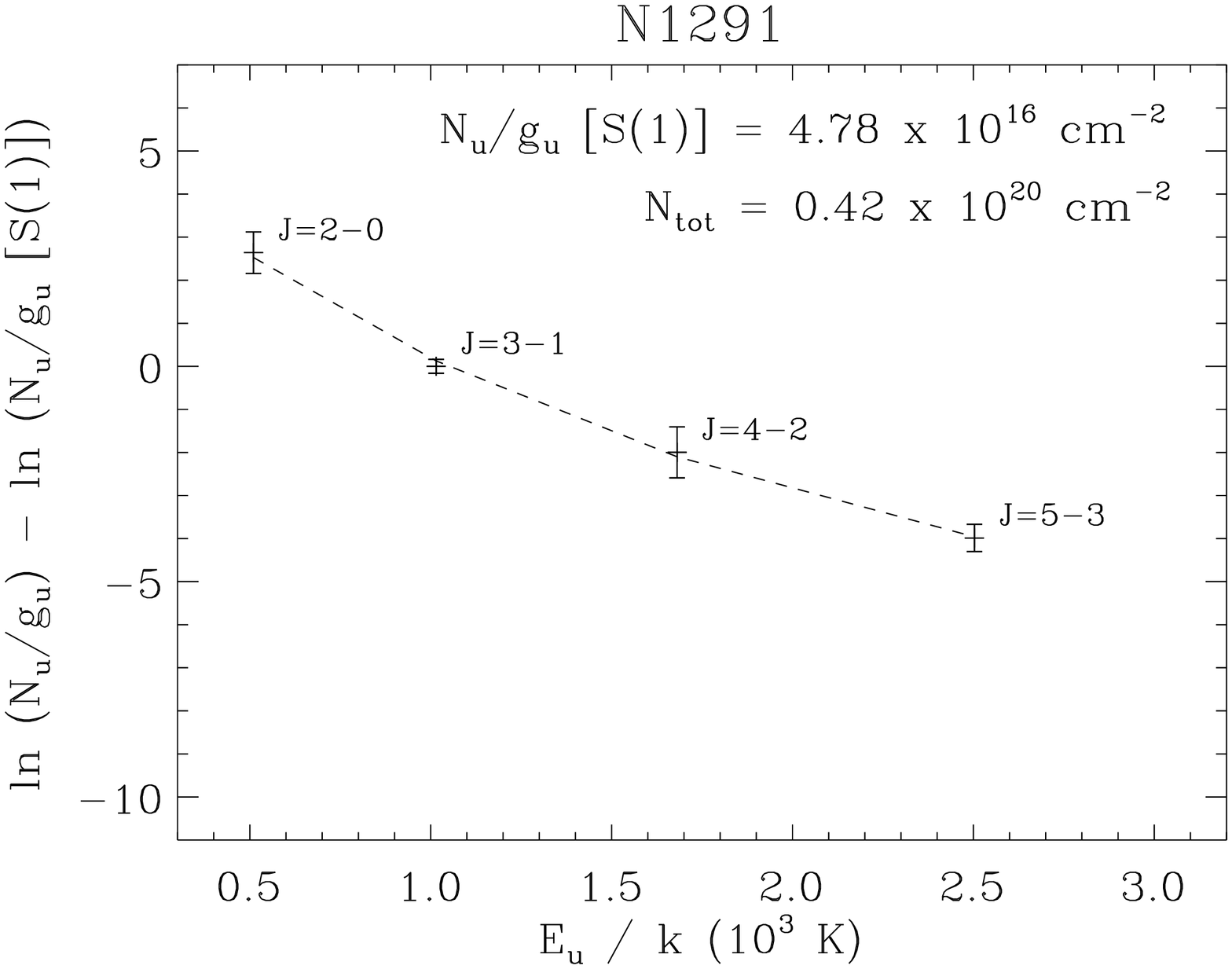}}}
\hspace*{-2cm}
\resizebox{10cm}{!}{\rotatebox{90}{\plotone{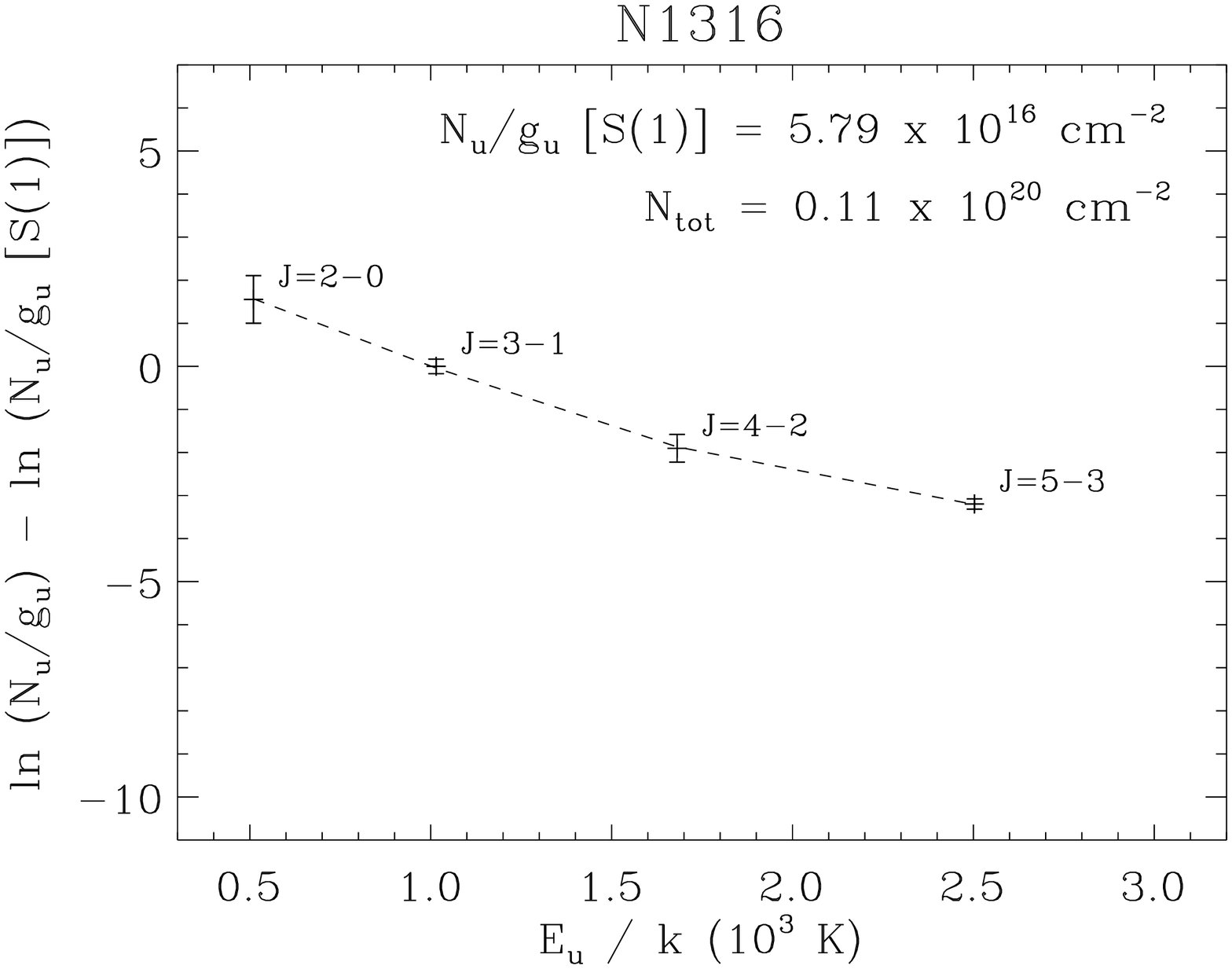}}}
\hspace*{-0.5cm}
\resizebox{10cm}{!}{\rotatebox{90}{\plotone{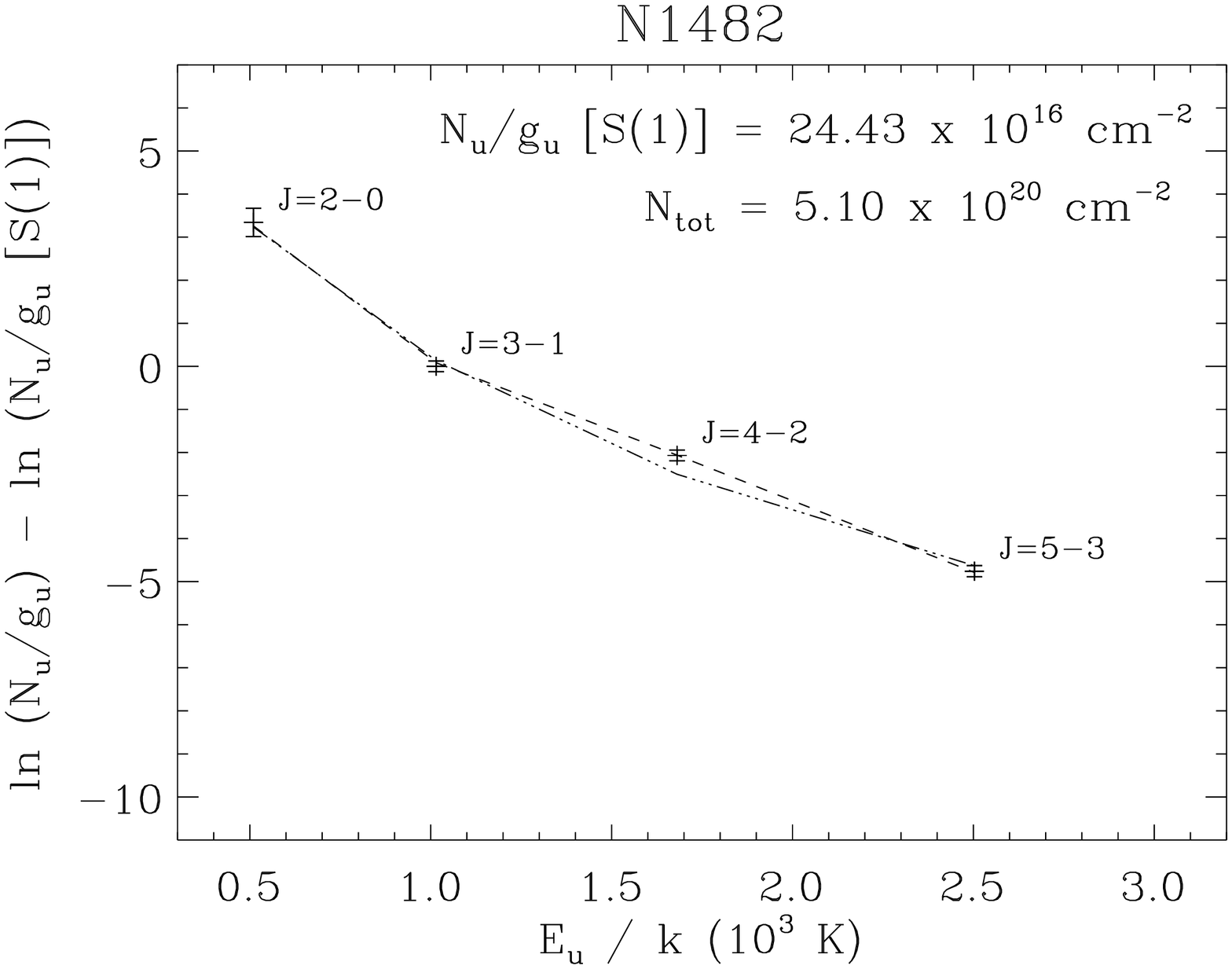}}}
\hspace*{-2cm}
\resizebox{10cm}{!}{\rotatebox{90}{\plotone{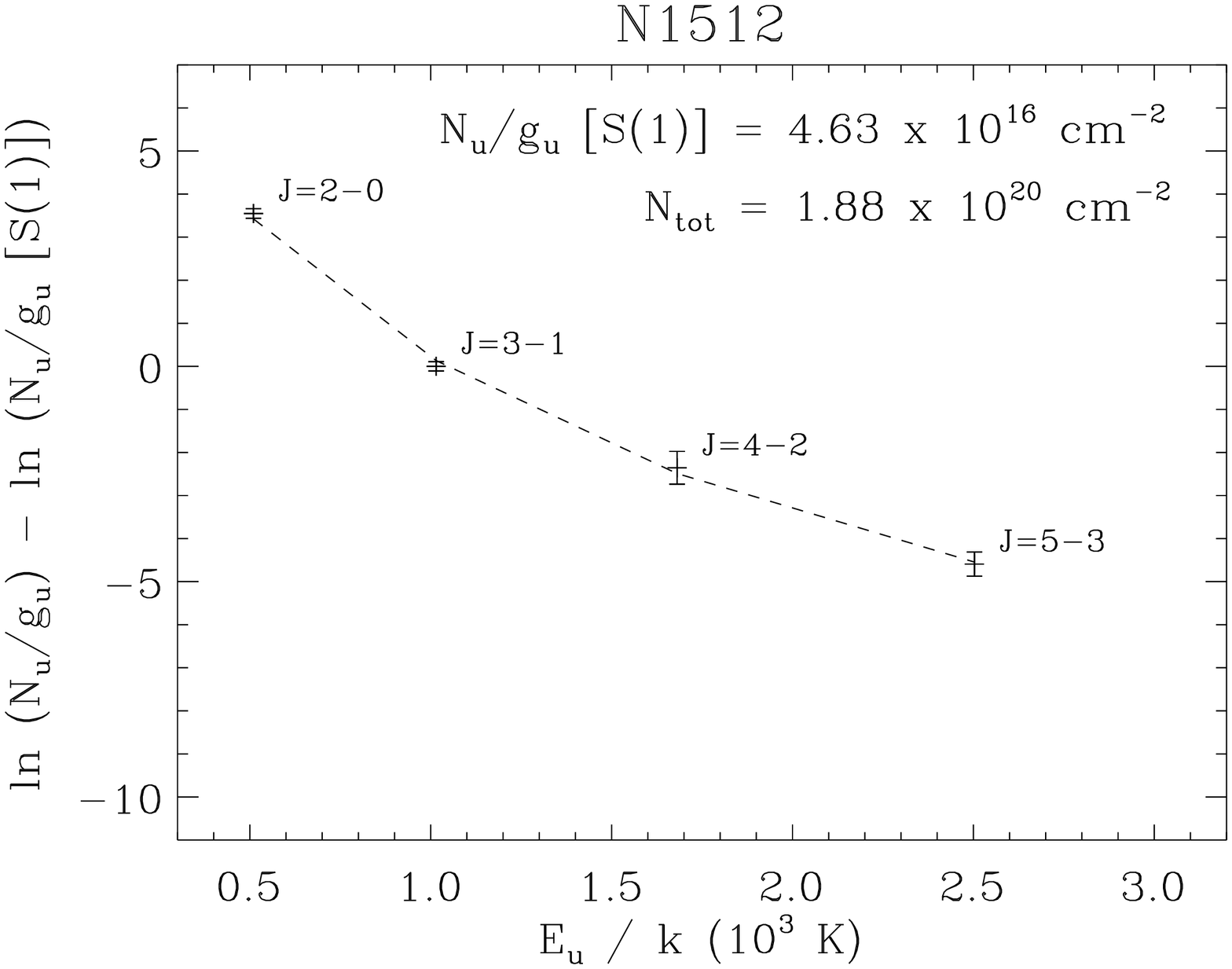}}}
\hspace*{-0.5cm}
\resizebox{10cm}{!}{\rotatebox{90}{\plotone{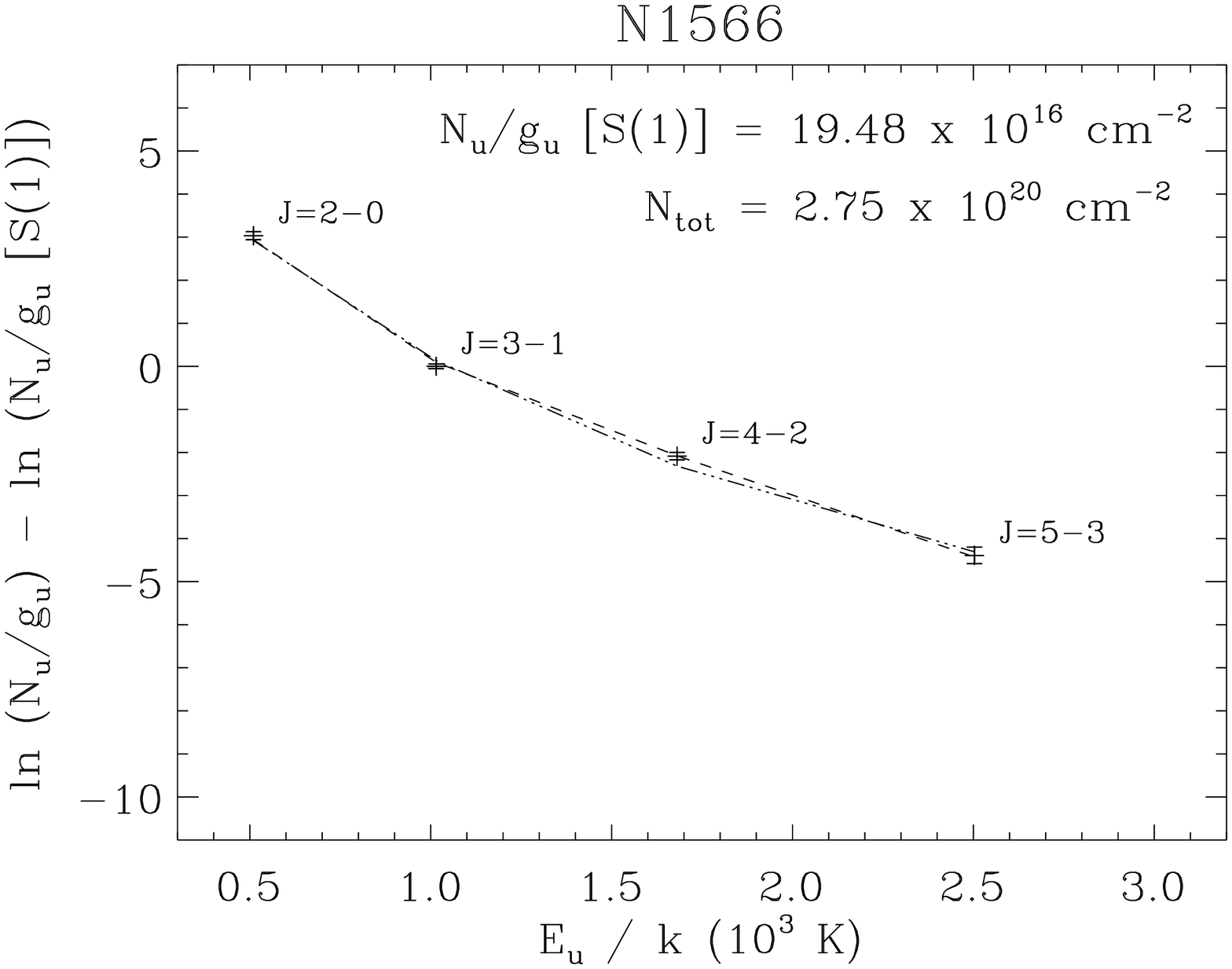}}}
\caption{(continued).
}
\end{figure}

\addtocounter{figure}{-1}
\begin{figure}[!ht]
\hspace*{-2cm}
\resizebox{10cm}{!}{\rotatebox{90}{\plotone{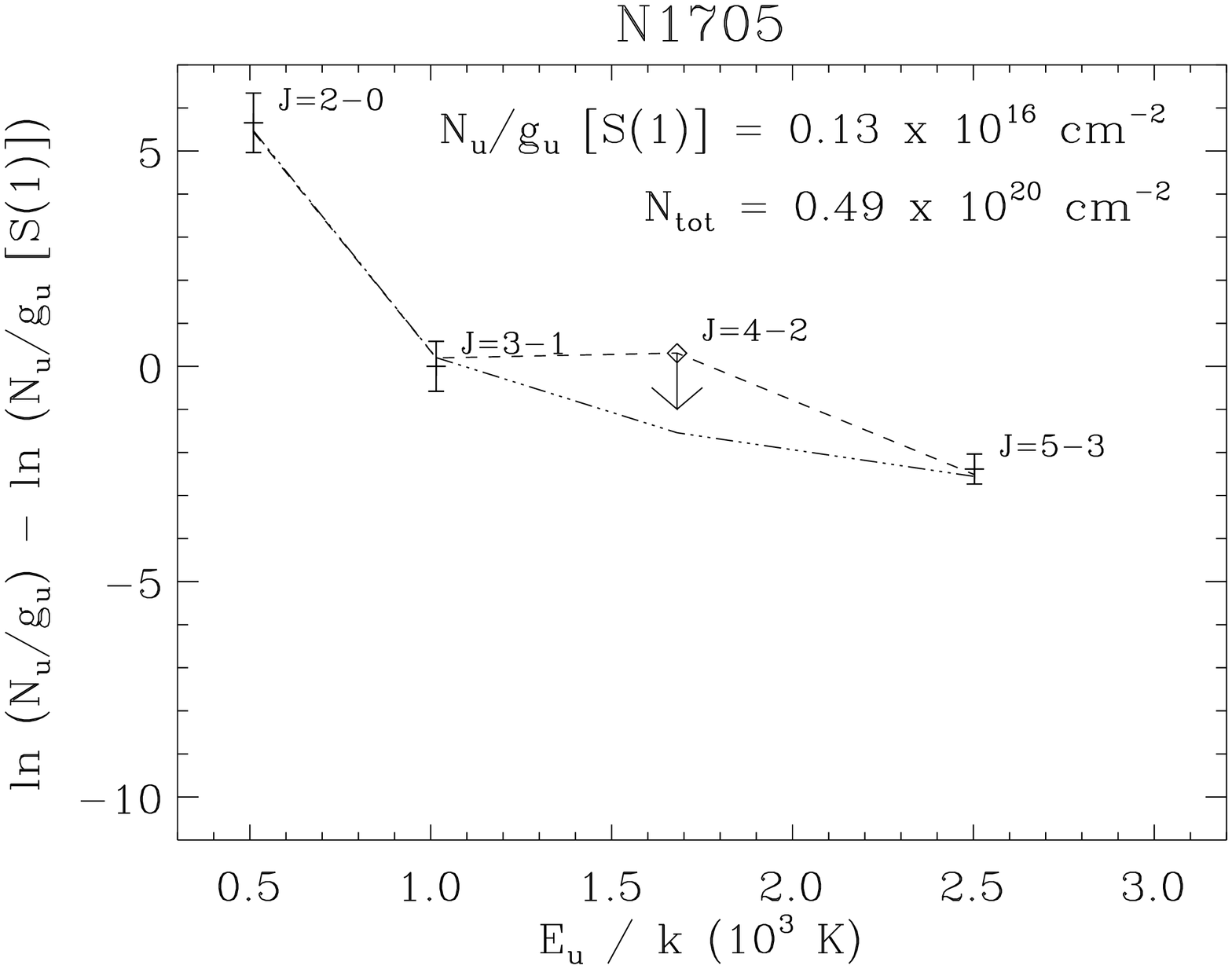}}}
\hspace*{-0.5cm}
\resizebox{10cm}{!}{\rotatebox{90}{\plotone{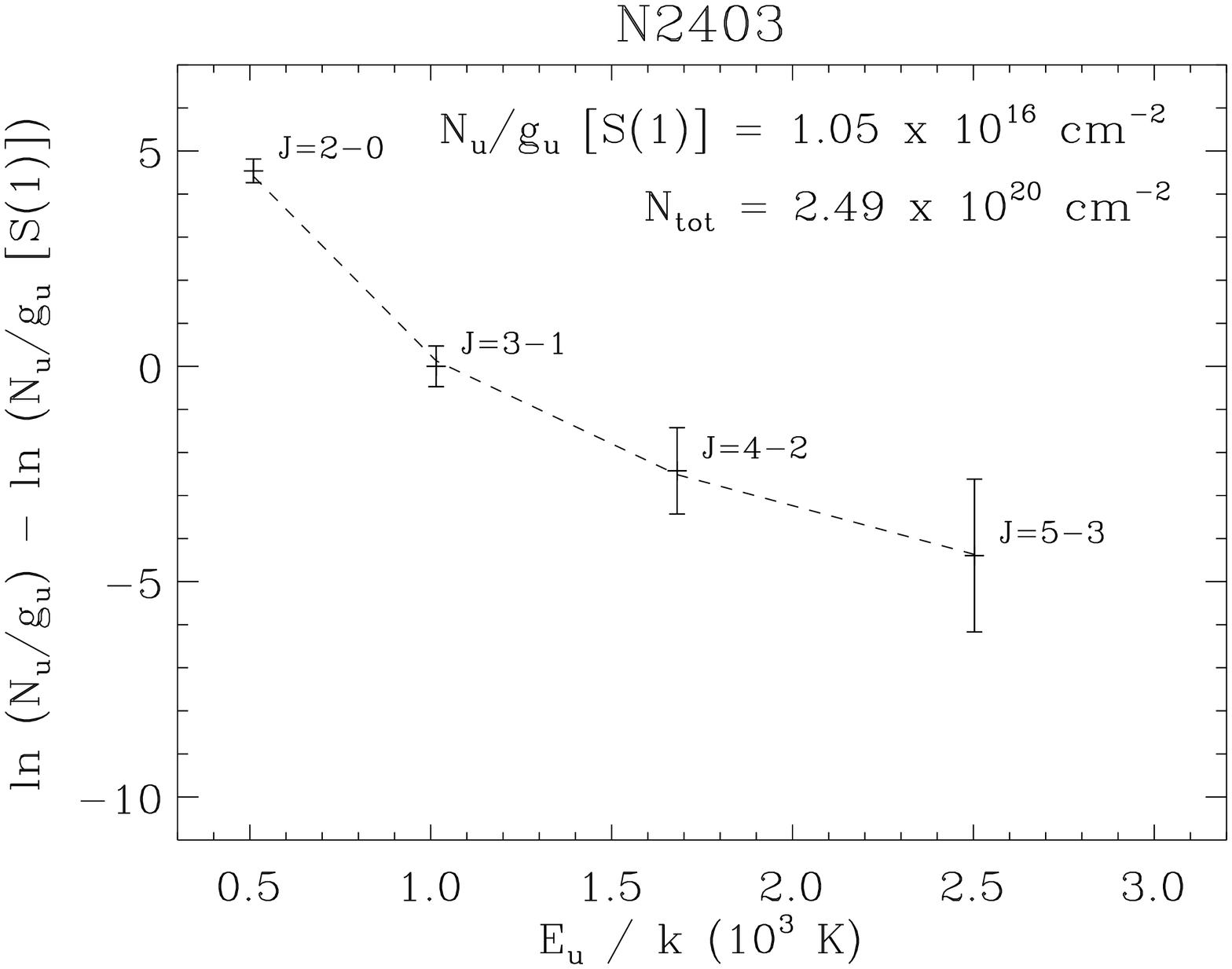}}}
\hspace*{-2cm}
\resizebox{10cm}{!}{\rotatebox{90}{\plotone{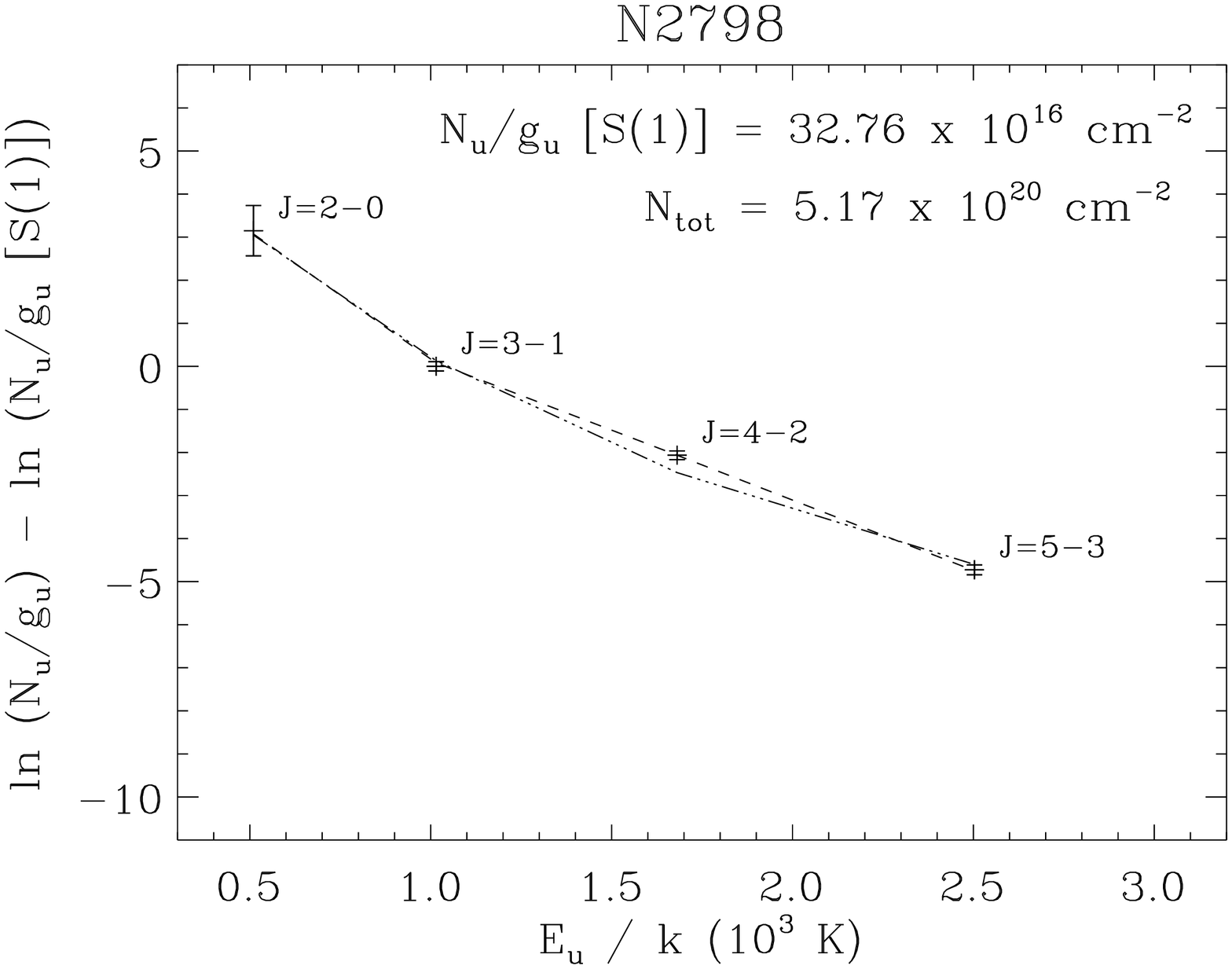}}}
\hspace*{-0.5cm}
\resizebox{10cm}{!}{\rotatebox{90}{\plotone{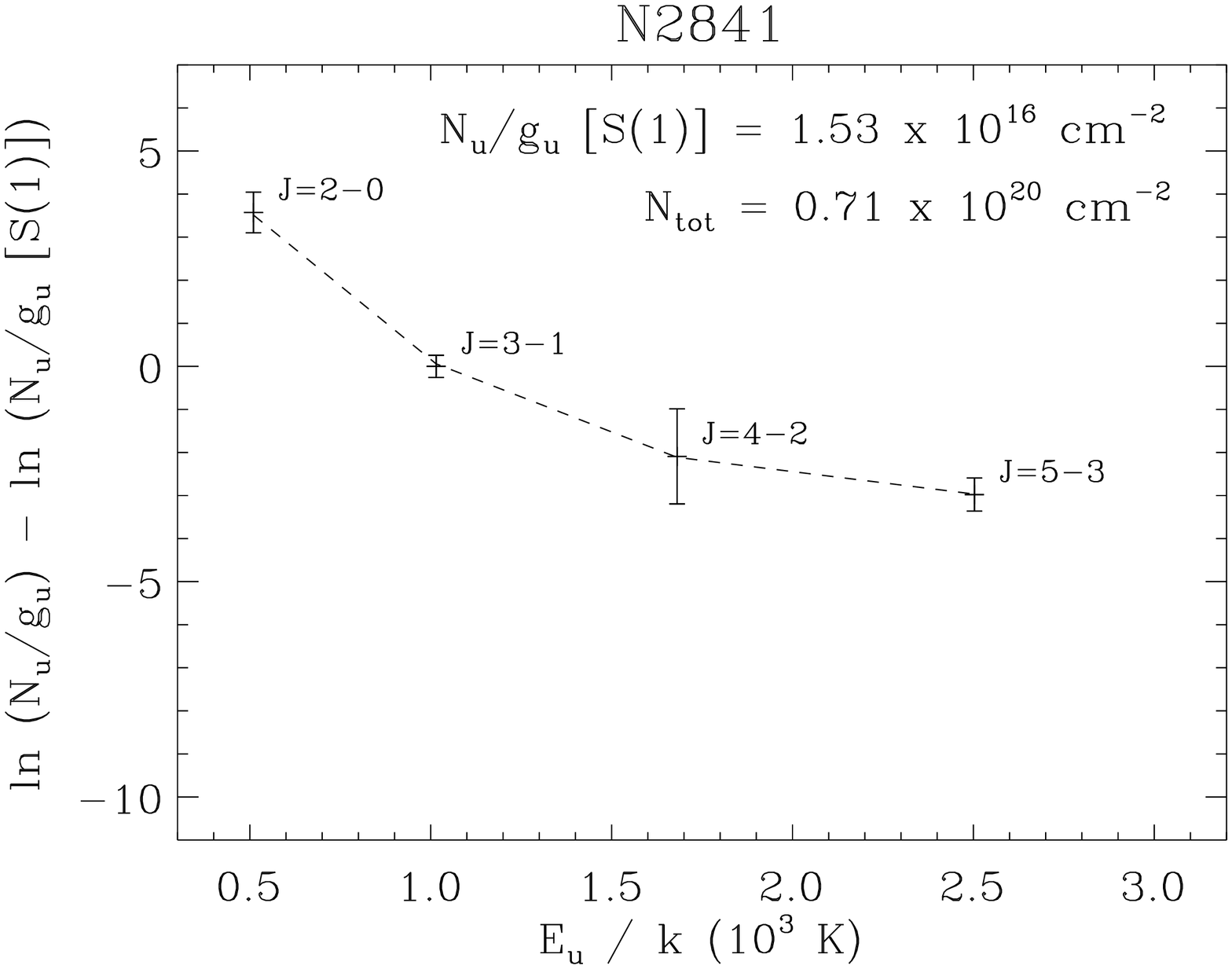}}}
\hspace*{-2cm}
\resizebox{10cm}{!}{\rotatebox{90}{\plotone{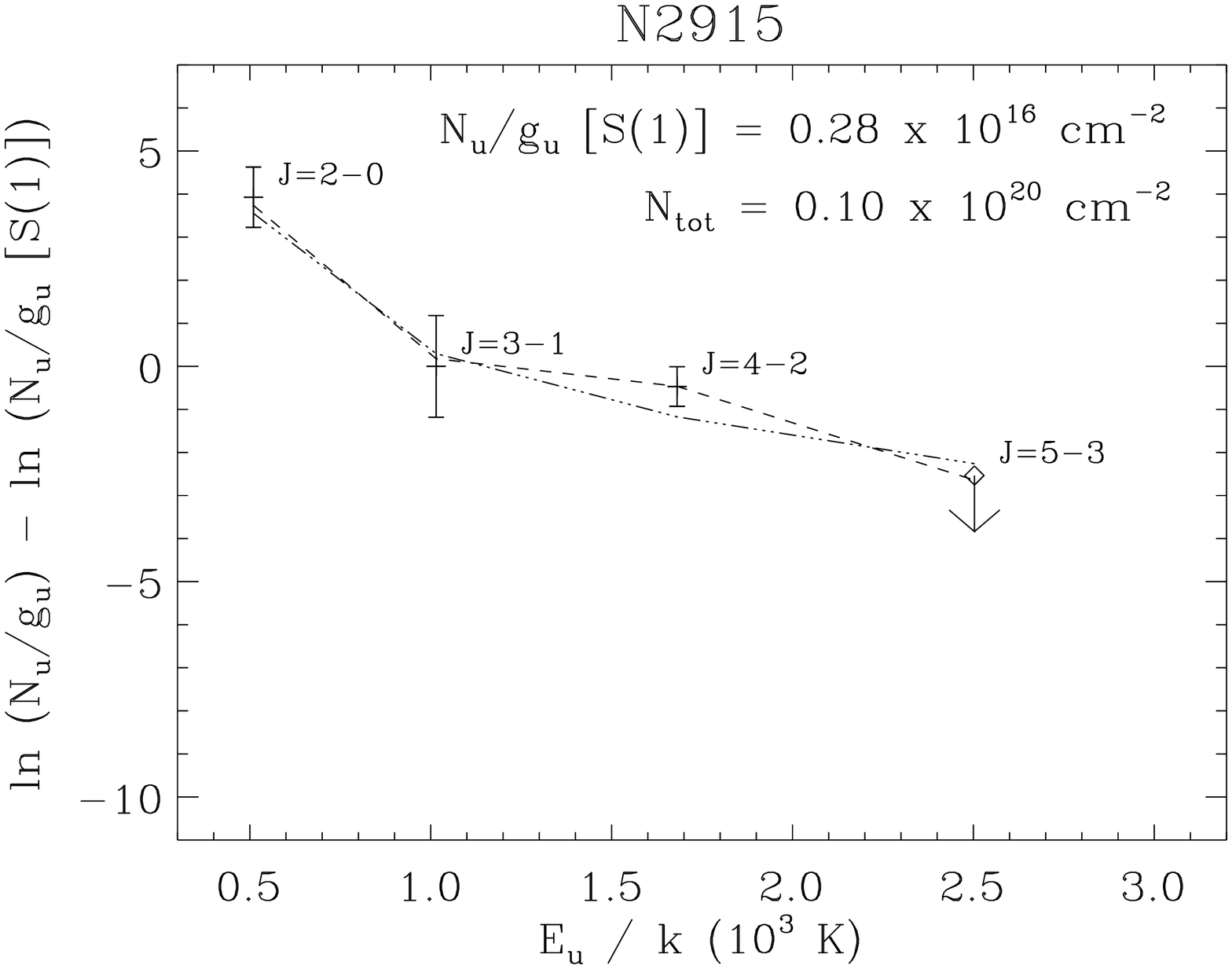}}}
\hspace*{-0.5cm}
\resizebox{10cm}{!}{\rotatebox{90}{\plotone{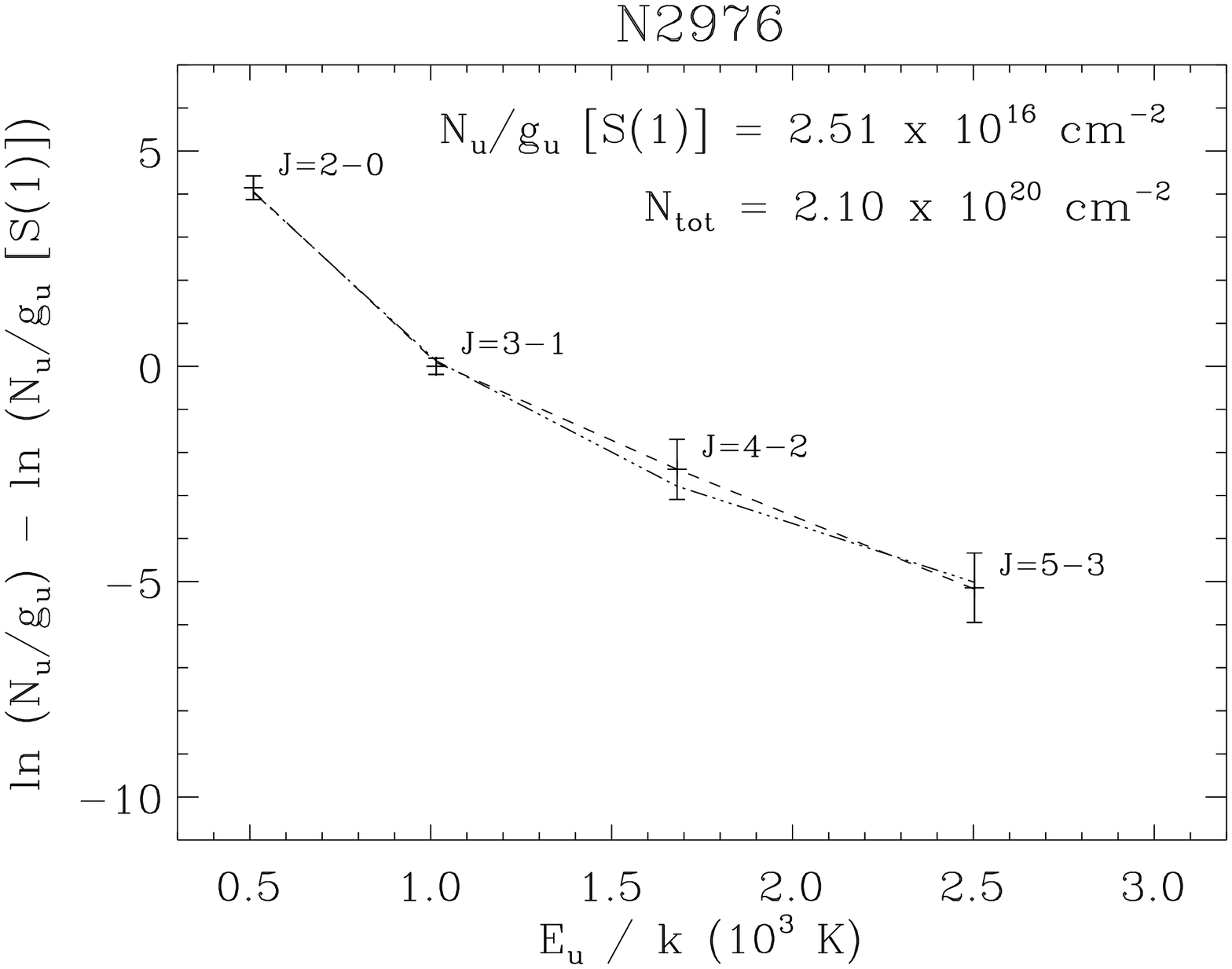}}}
\caption{(continued).
}
\end{figure}

\addtocounter{figure}{-1}
\begin{figure}[!ht]
\hspace*{-2cm}
\resizebox{10cm}{!}{\rotatebox{90}{\plotone{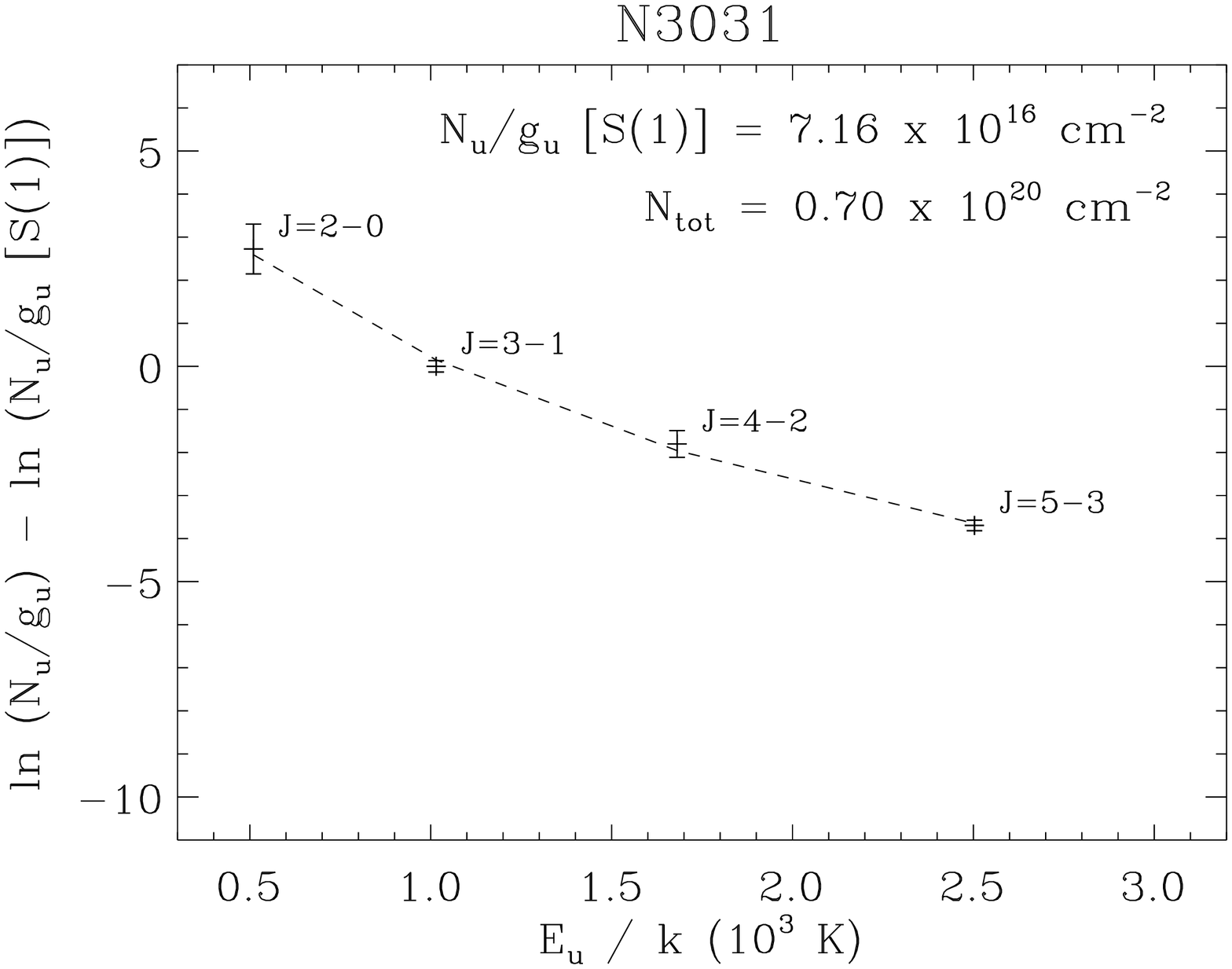}}}
\hspace*{-0.5cm}
\resizebox{10cm}{!}{\rotatebox{90}{\plotone{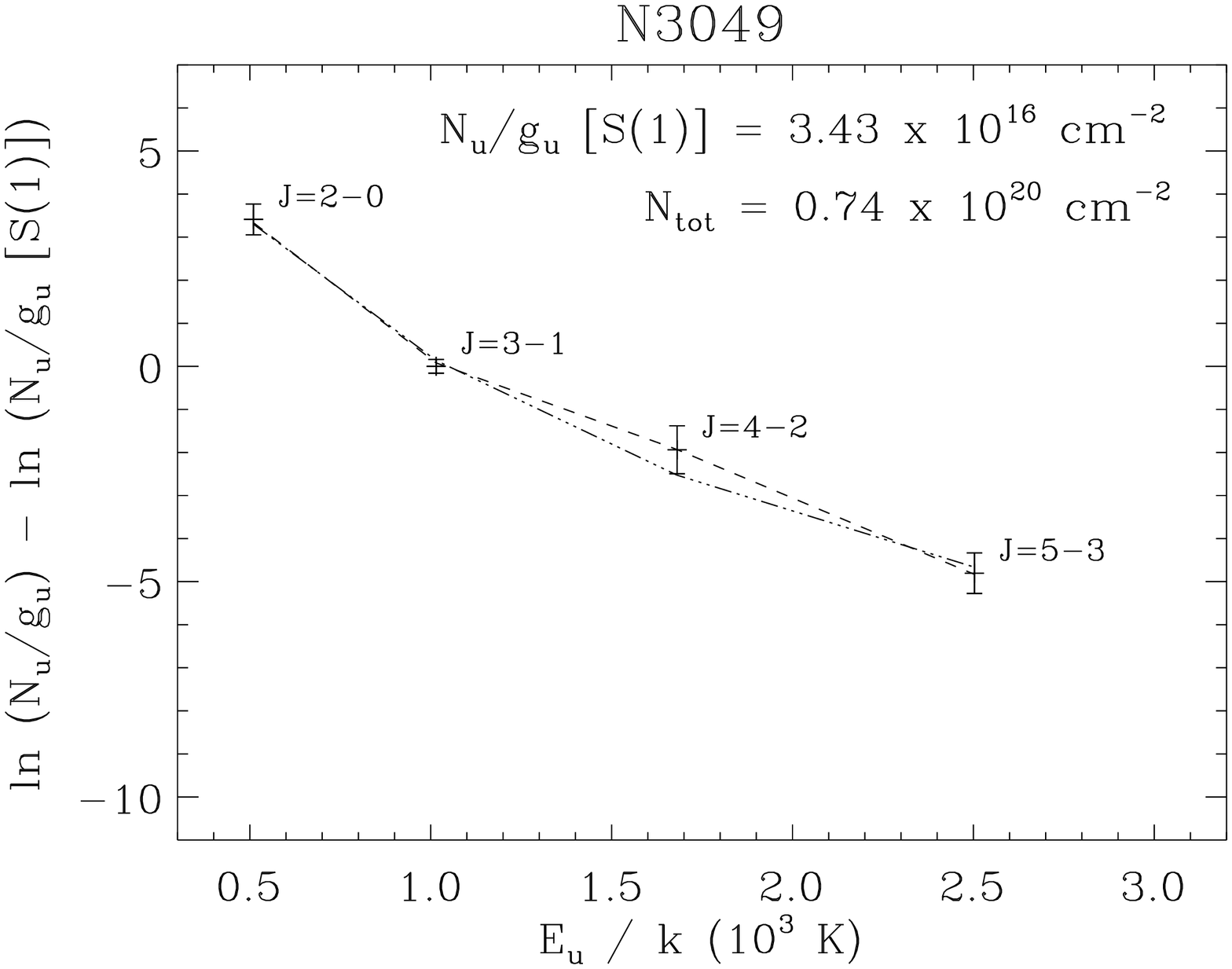}}}
\hspace*{-2cm}
\resizebox{10cm}{!}{\rotatebox{90}{\plotone{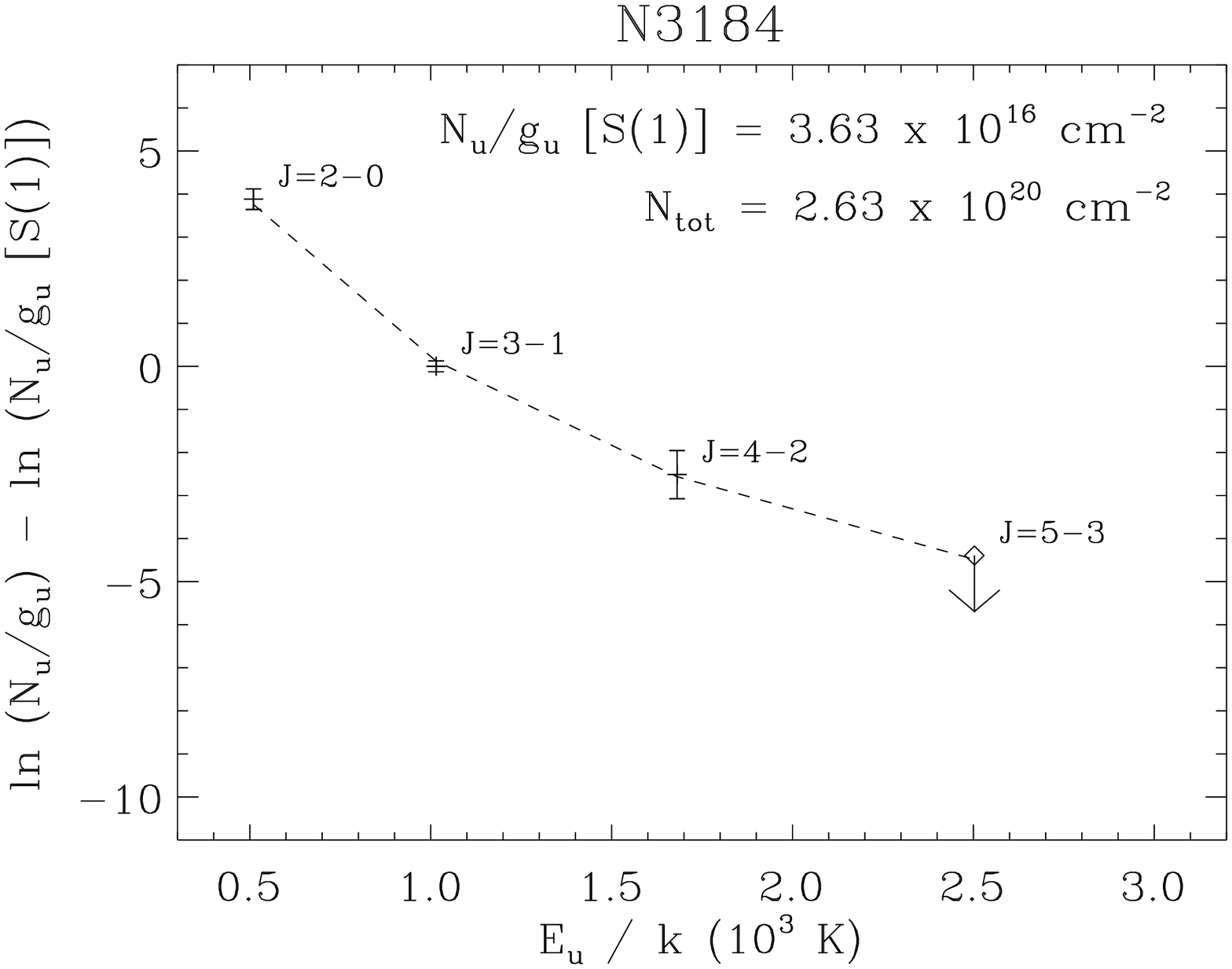}}}
\hspace*{-0.5cm}
\resizebox{10cm}{!}{\rotatebox{90}{\plotone{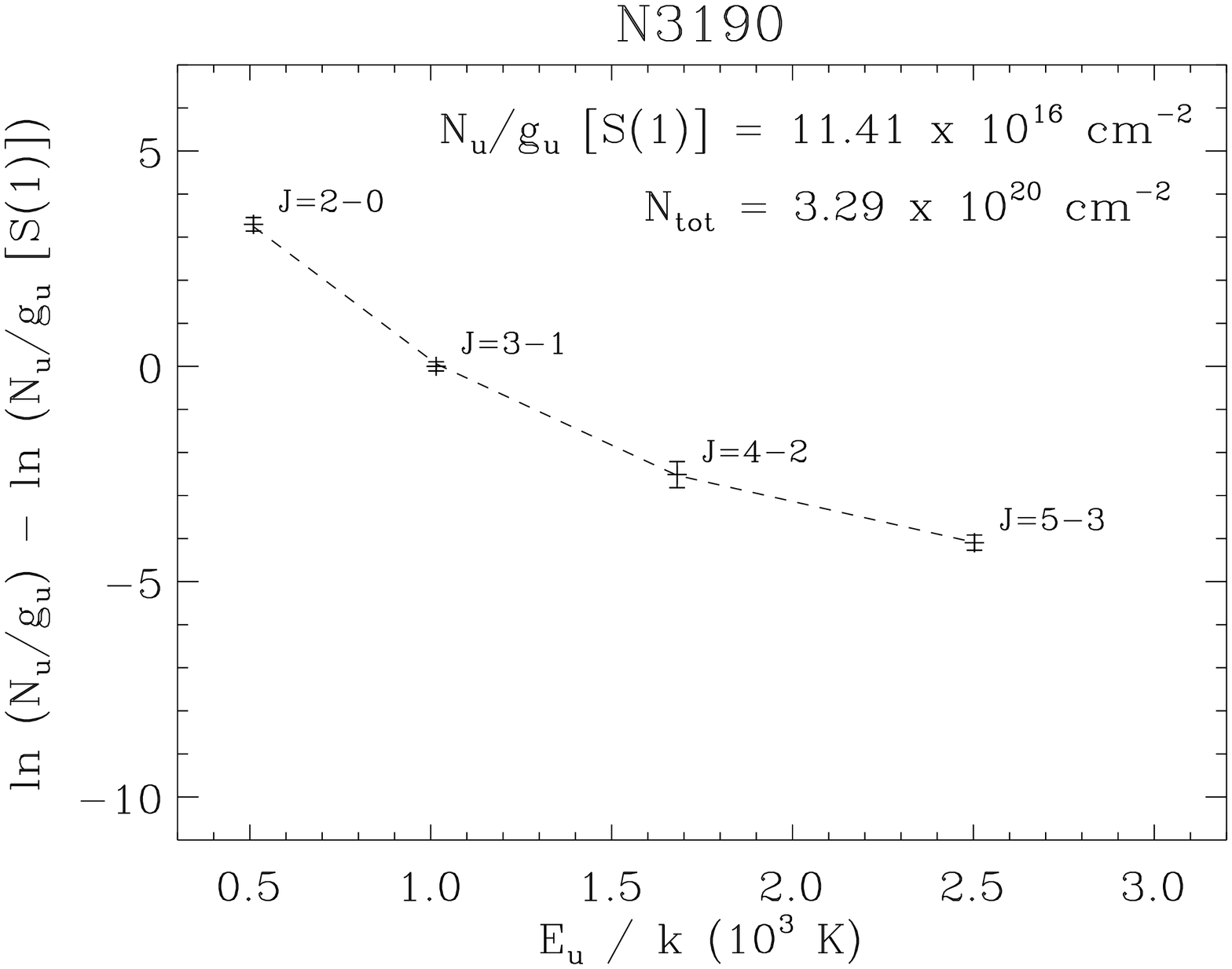}}}
\hspace*{-2cm}
\resizebox{10cm}{!}{\rotatebox{90}{\plotone{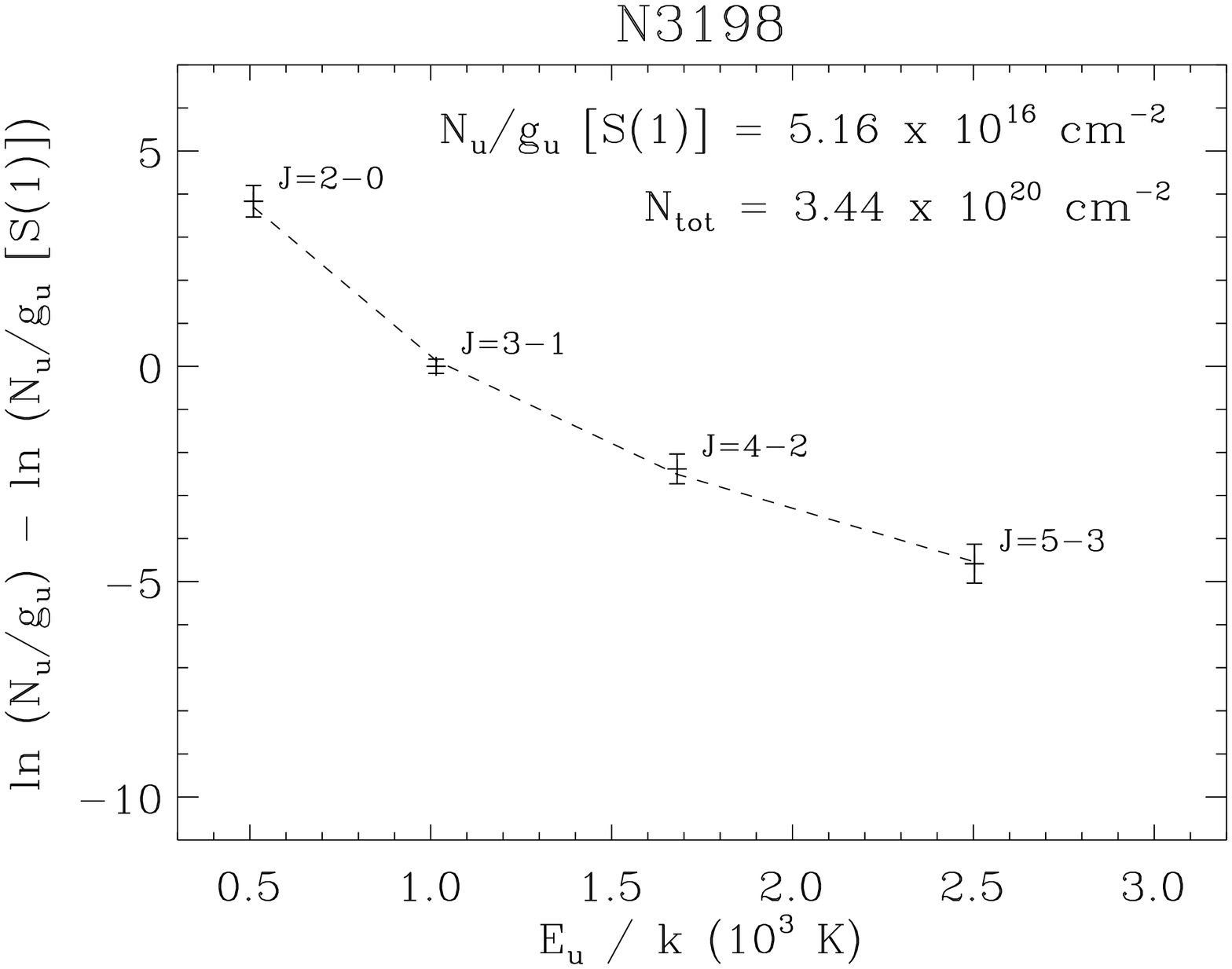}}}
\hspace*{-0.5cm}
\resizebox{10cm}{!}{\rotatebox{90}{\plotone{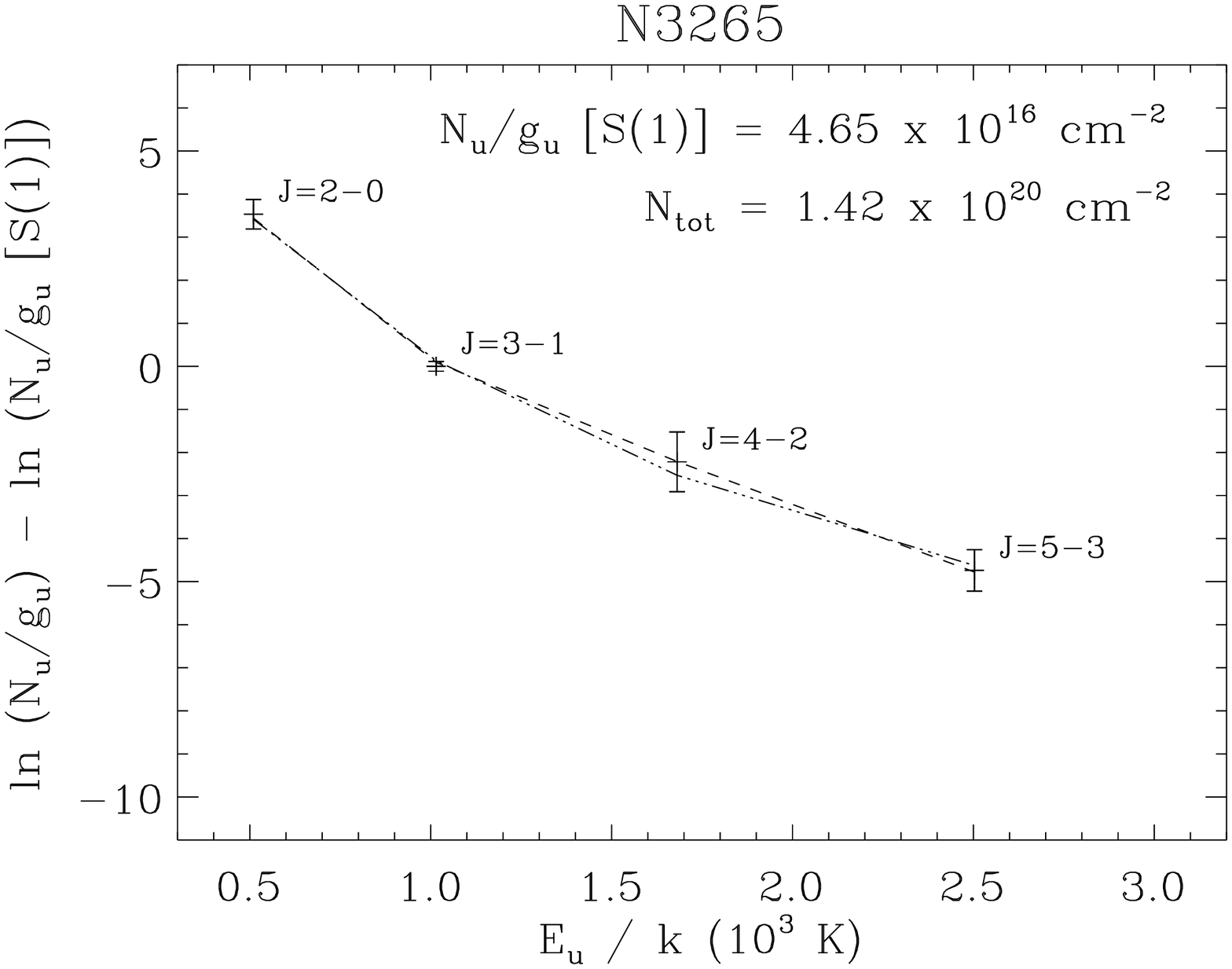}}}
\caption{(continued).
}
\end{figure}

\addtocounter{figure}{-1}
\begin{figure}[!ht]
\hspace*{-2cm}
\resizebox{10cm}{!}{\rotatebox{90}{\plotone{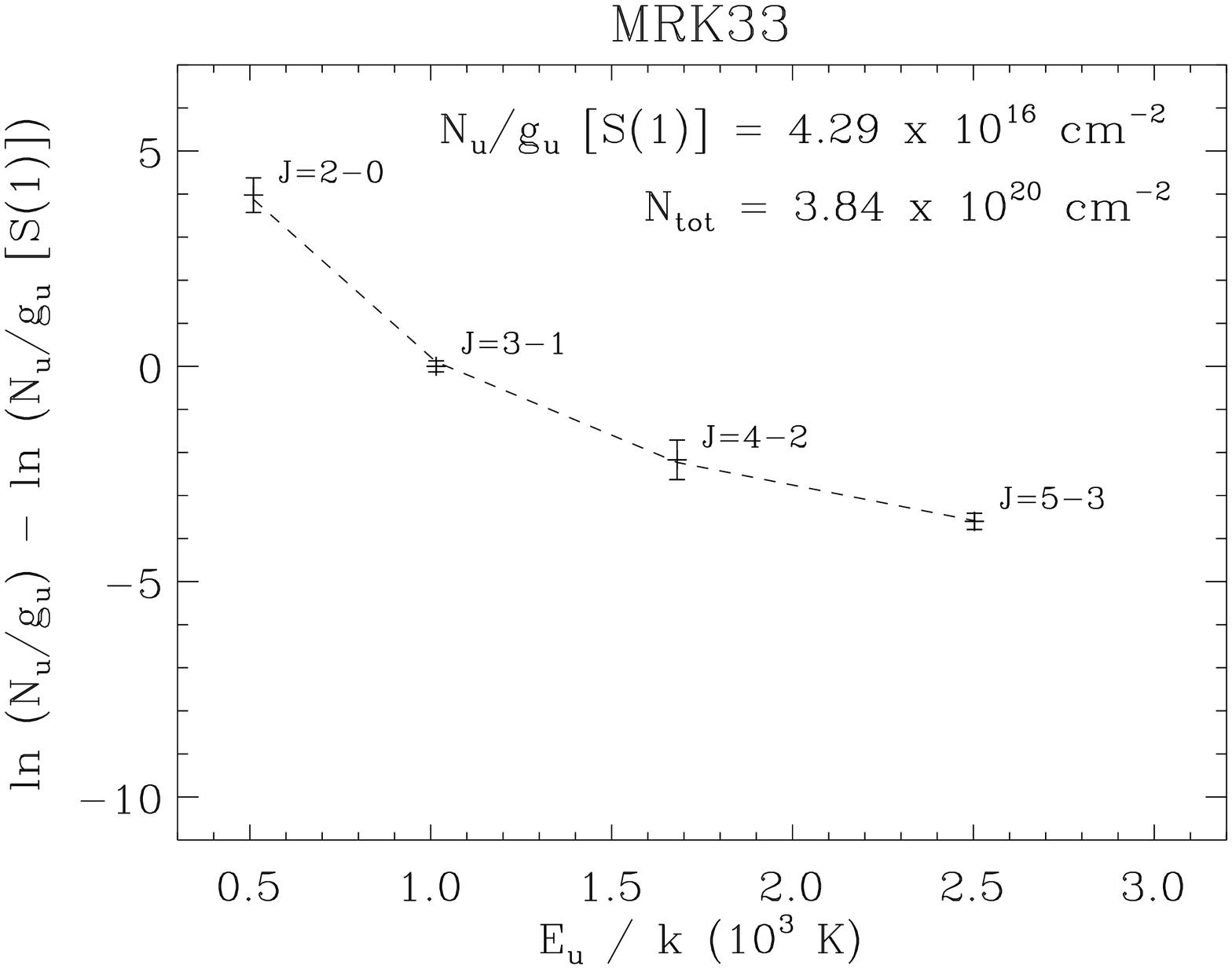}}}
\hspace*{-0.5cm}
\resizebox{10cm}{!}{\rotatebox{90}{\plotone{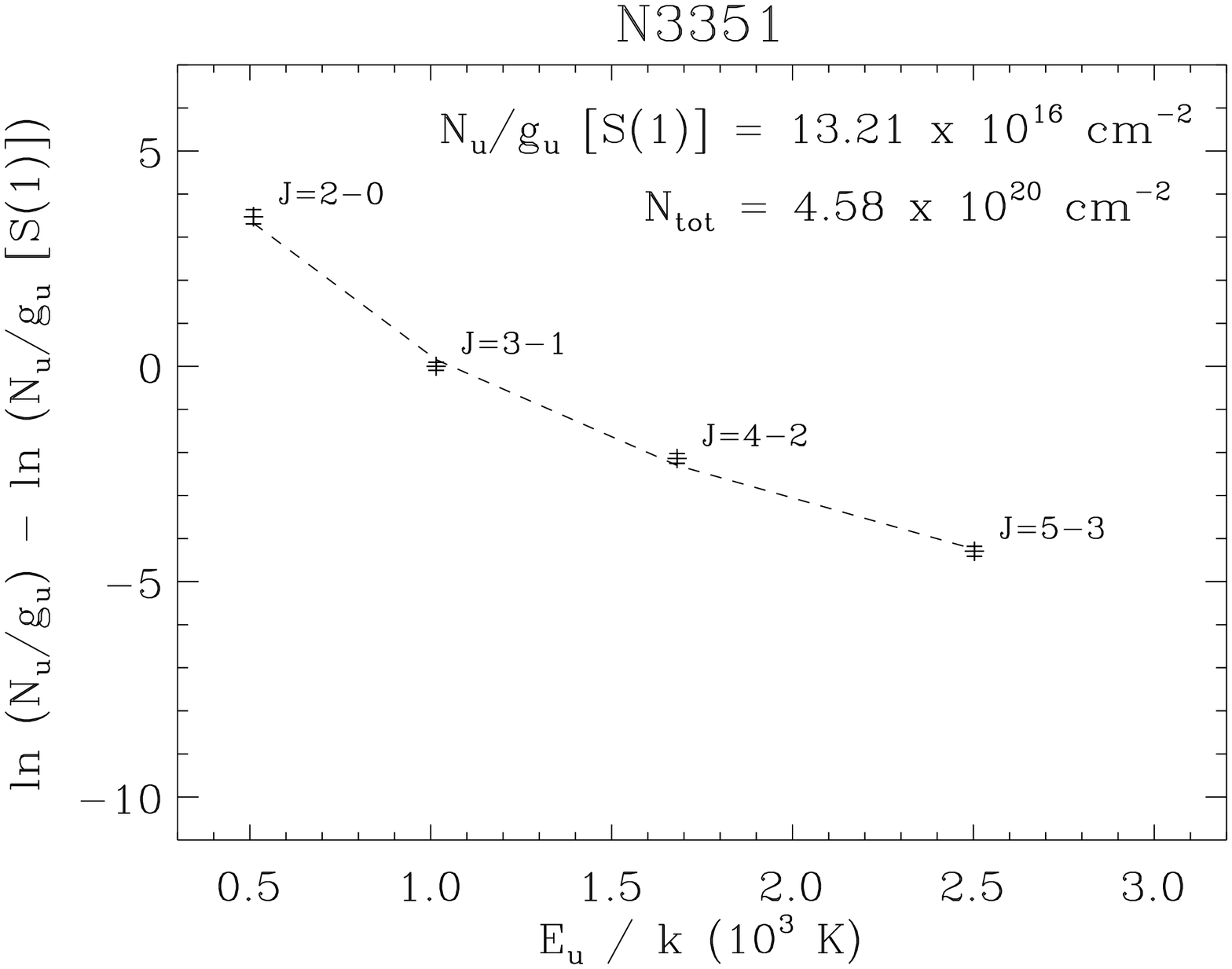}}}
\hspace*{-2cm}
\resizebox{10cm}{!}{\rotatebox{90}{\plotone{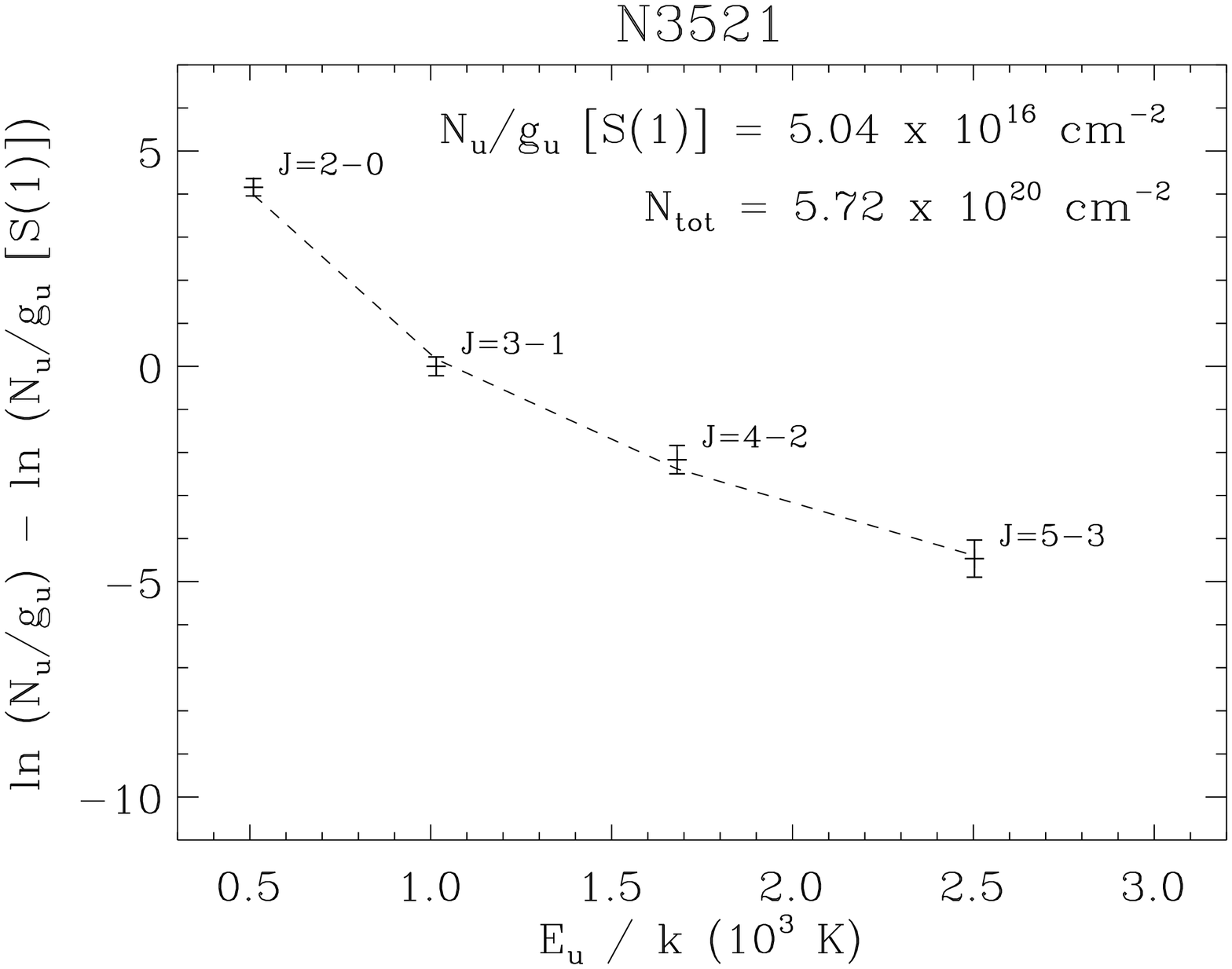}}}
\hspace*{-0.5cm}
\resizebox{10cm}{!}{\rotatebox{90}{\plotone{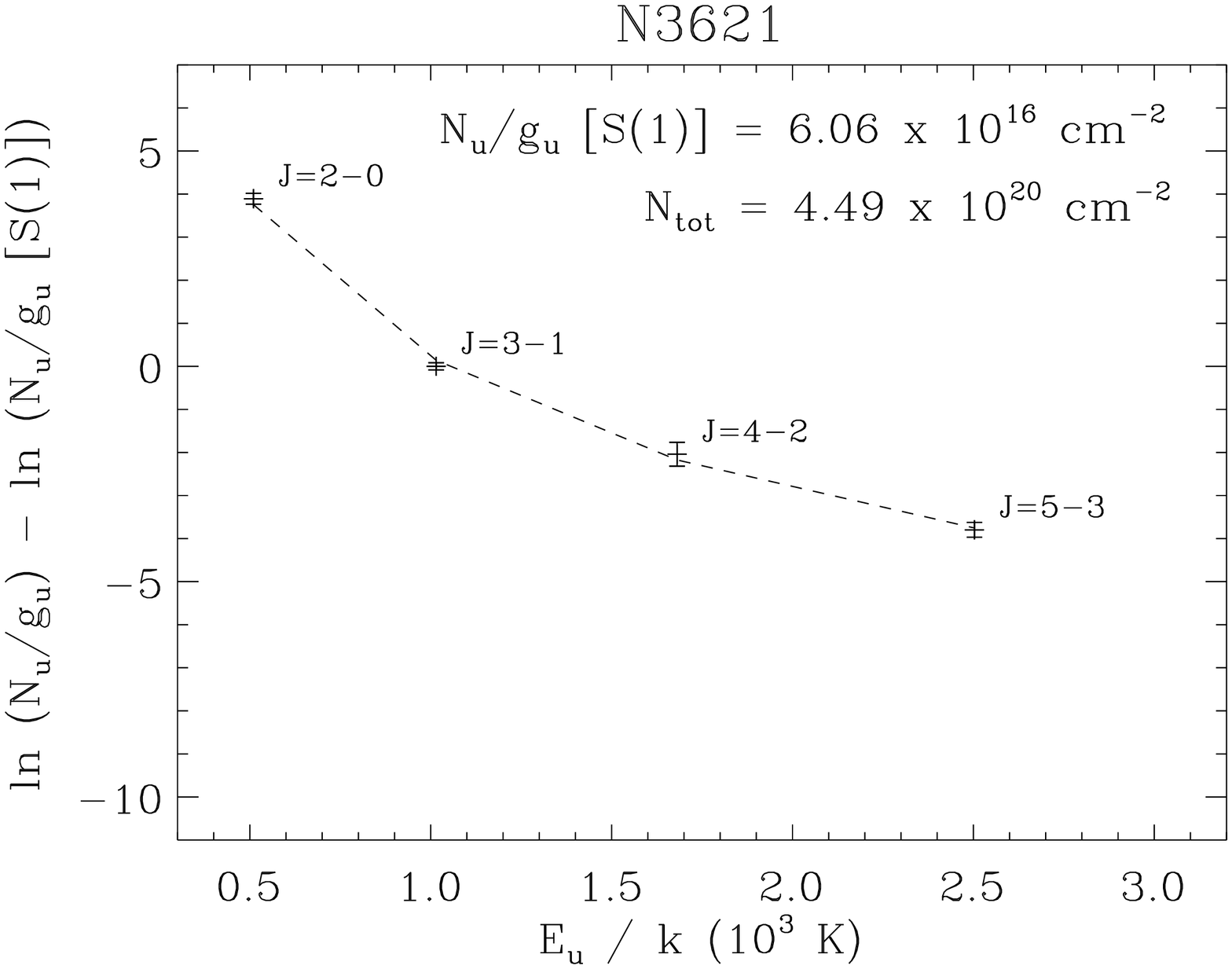}}}
\hspace*{-2cm}
\resizebox{10cm}{!}{\rotatebox{90}{\plotone{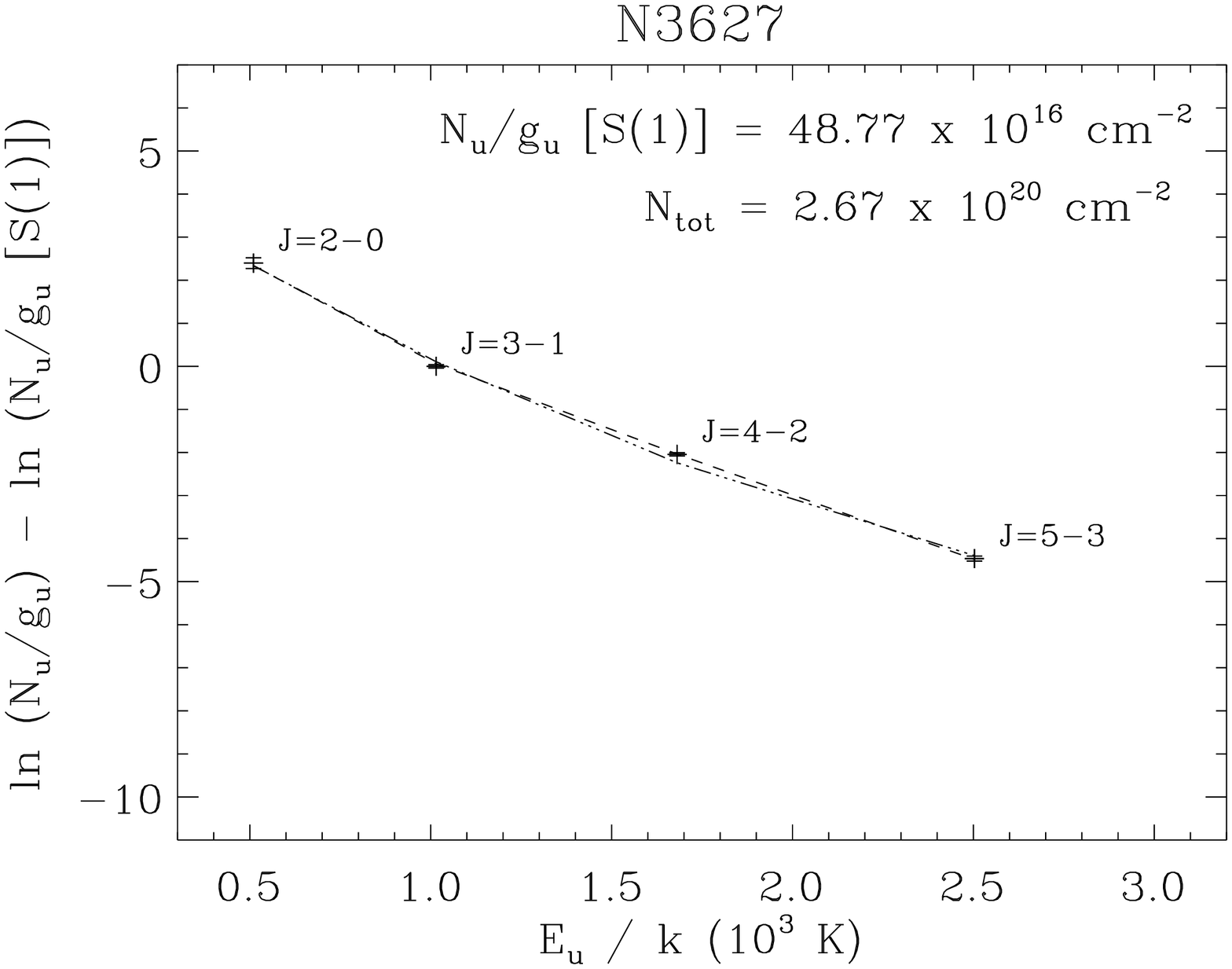}}}
\hspace*{-0.5cm}
\resizebox{10cm}{!}{\rotatebox{90}{\plotone{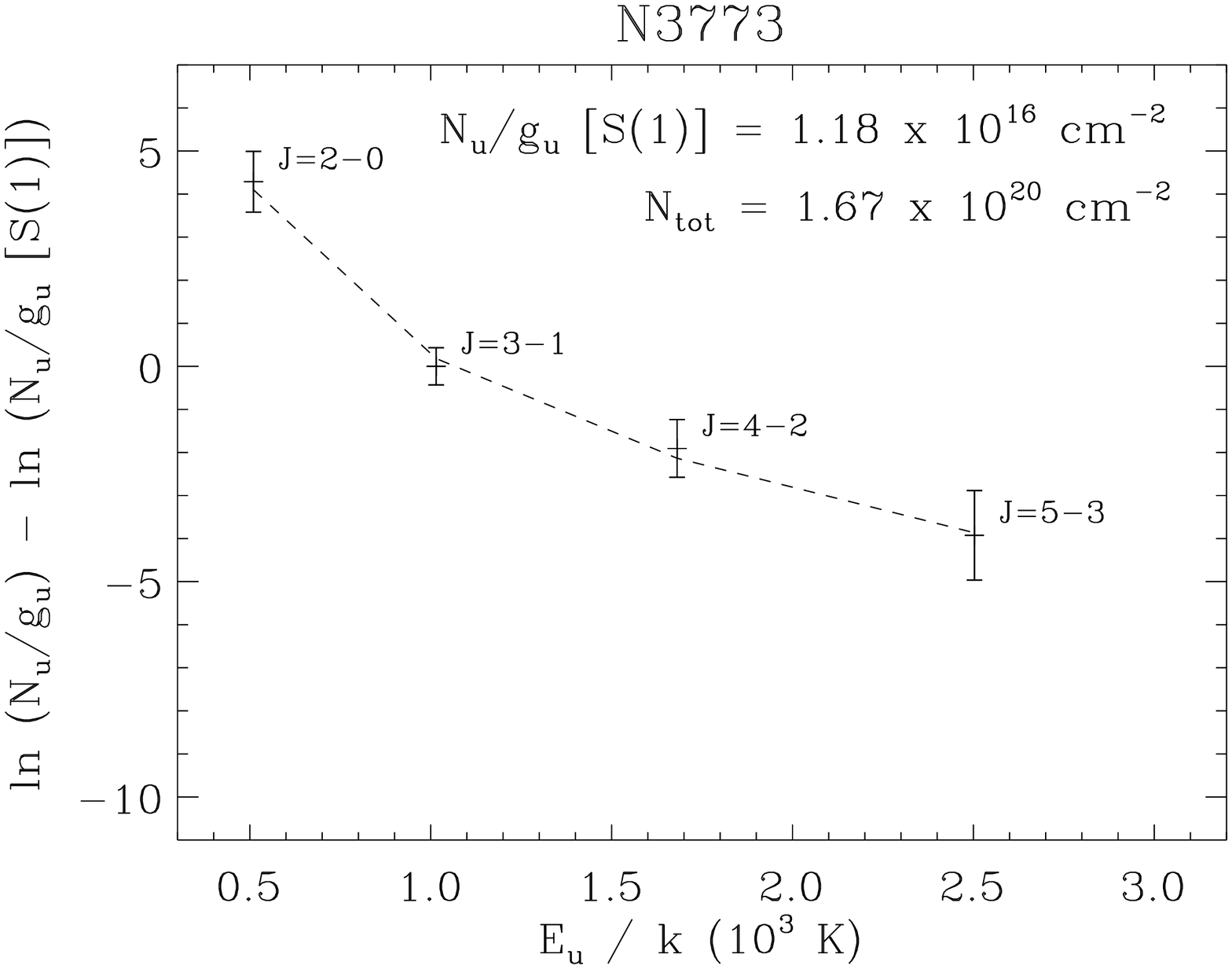}}}
\caption{(continued).
}
\end{figure}

\addtocounter{figure}{-1}
\begin{figure}[!ht]
\hspace*{-2cm}
\resizebox{10cm}{!}{\rotatebox{90}{\plotone{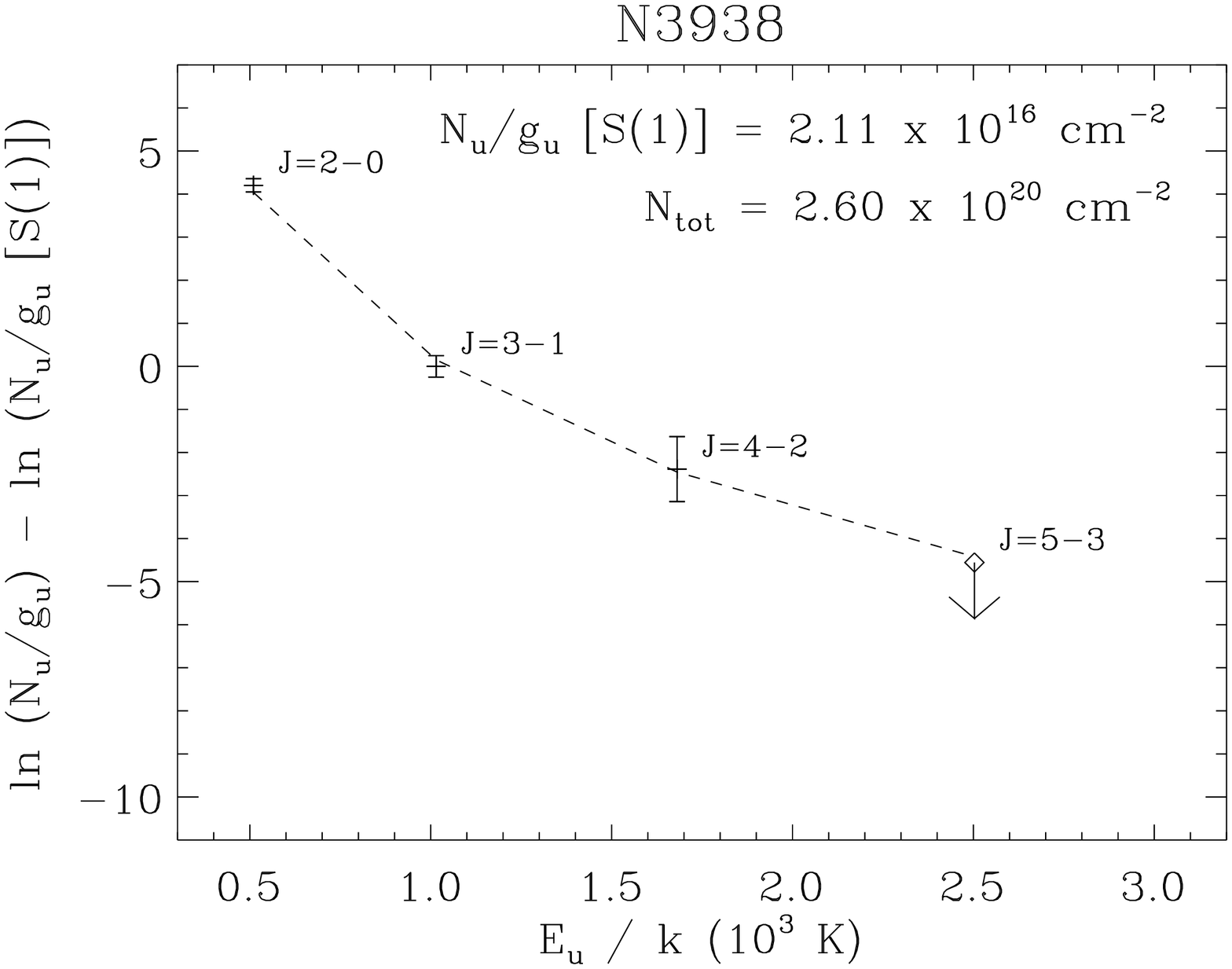}}}
\hspace*{-0.5cm}
\resizebox{10cm}{!}{\rotatebox{90}{\plotone{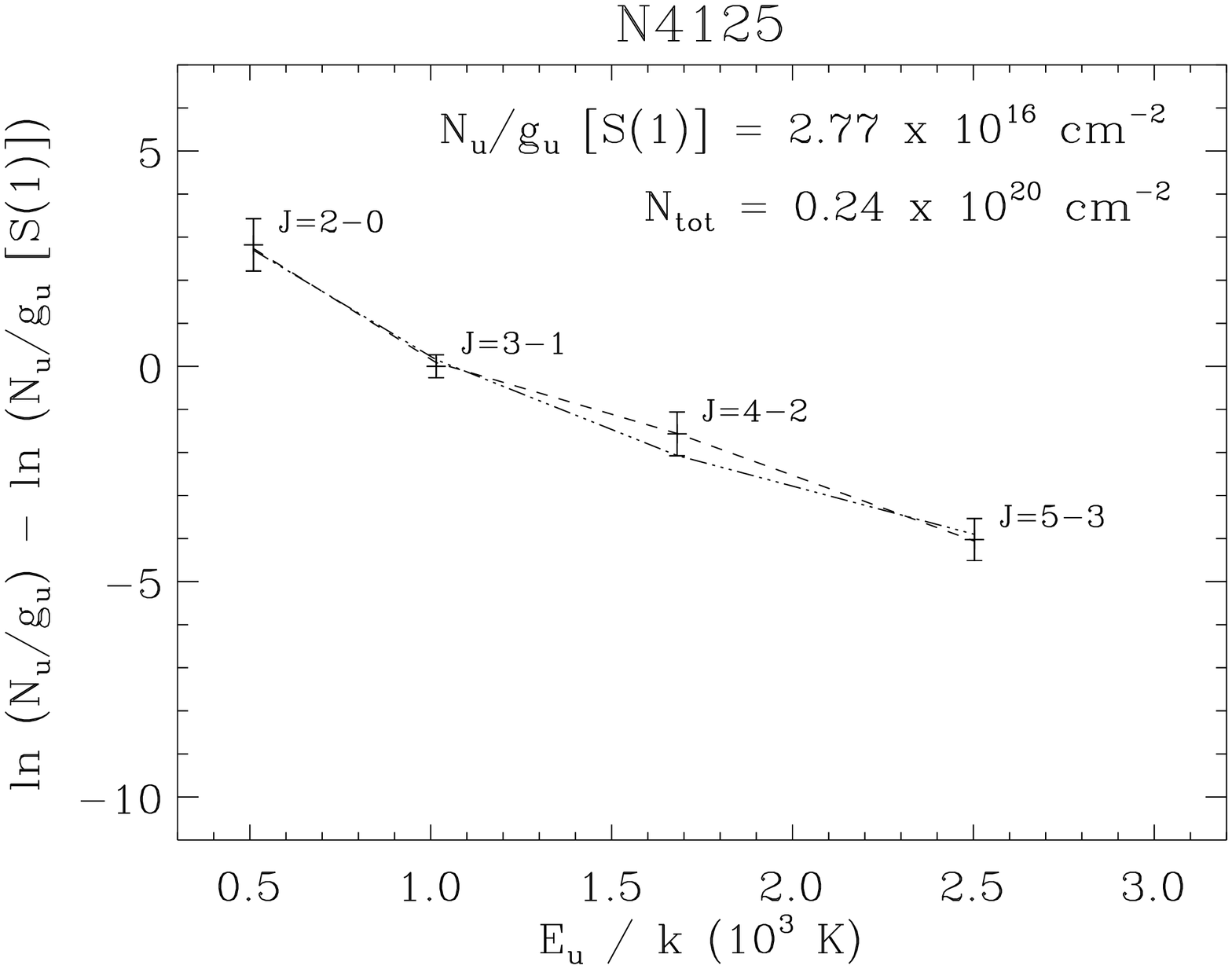}}}
\hspace*{-2cm}
\resizebox{10cm}{!}{\rotatebox{90}{\plotone{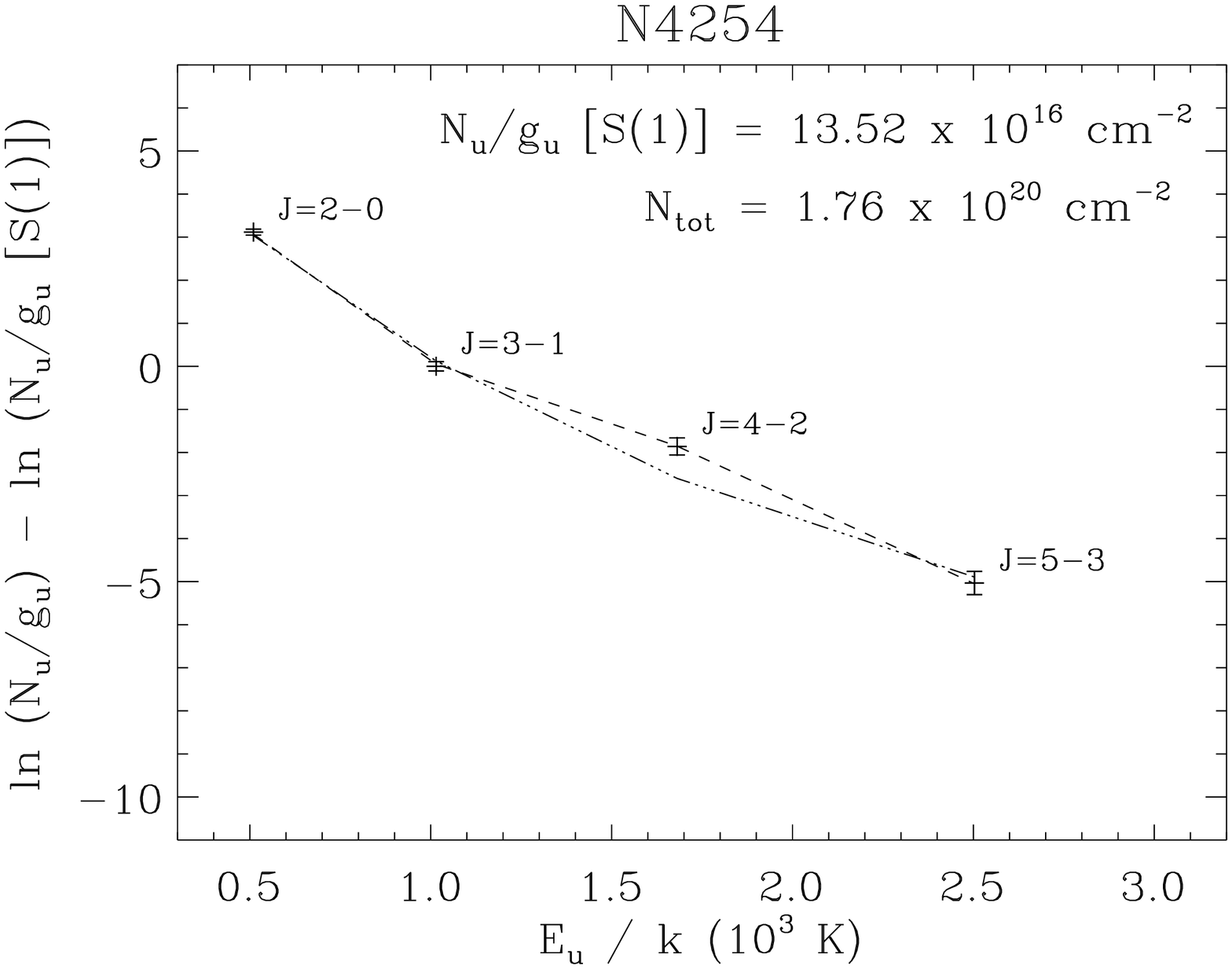}}}
\hspace*{-0.5cm}
\resizebox{10cm}{!}{\rotatebox{90}{\plotone{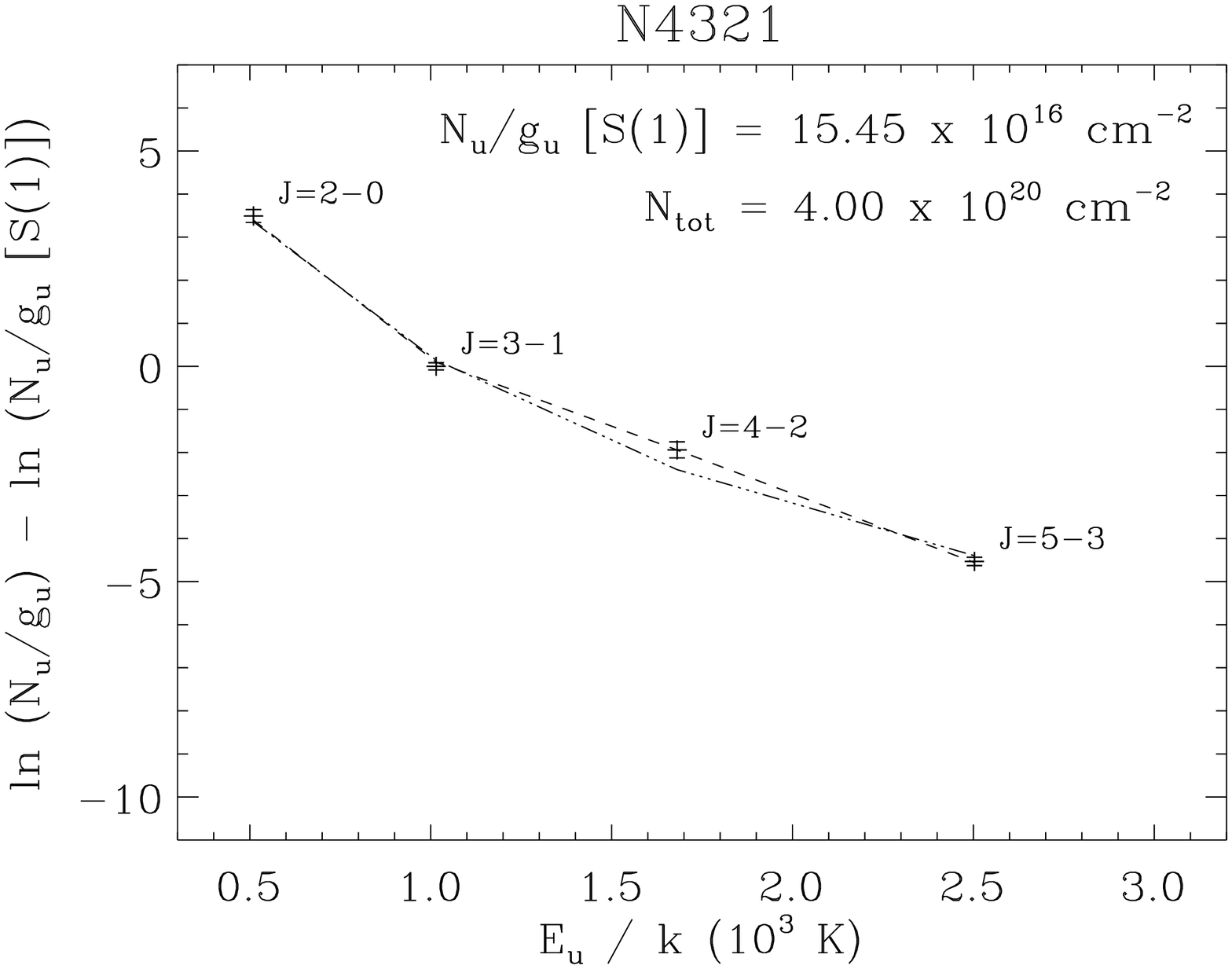}}}
\hspace*{-2cm}
\resizebox{10cm}{!}{\rotatebox{90}{\plotone{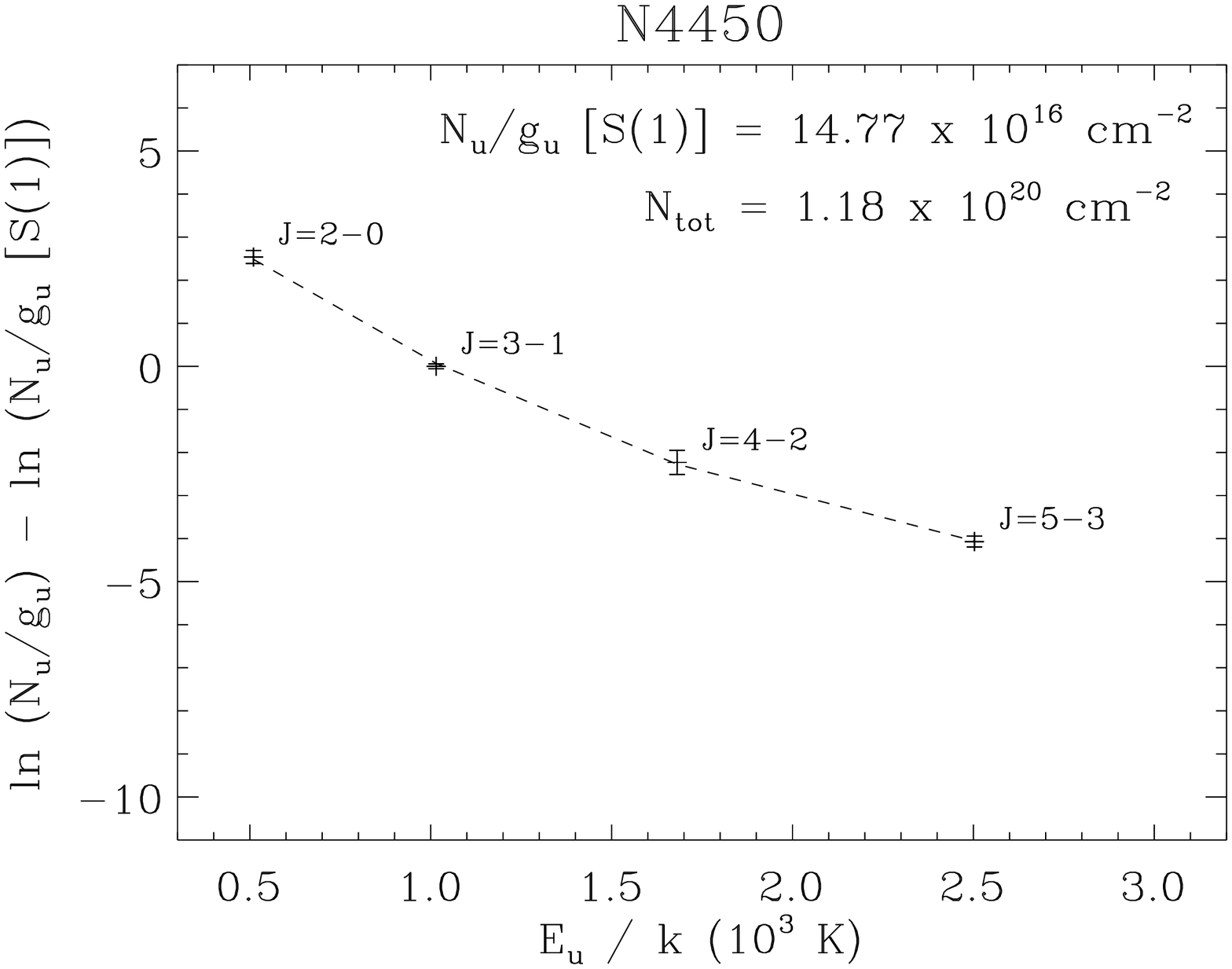}}}
\hspace*{-0.5cm}
\resizebox{10cm}{!}{\rotatebox{90}{\plotone{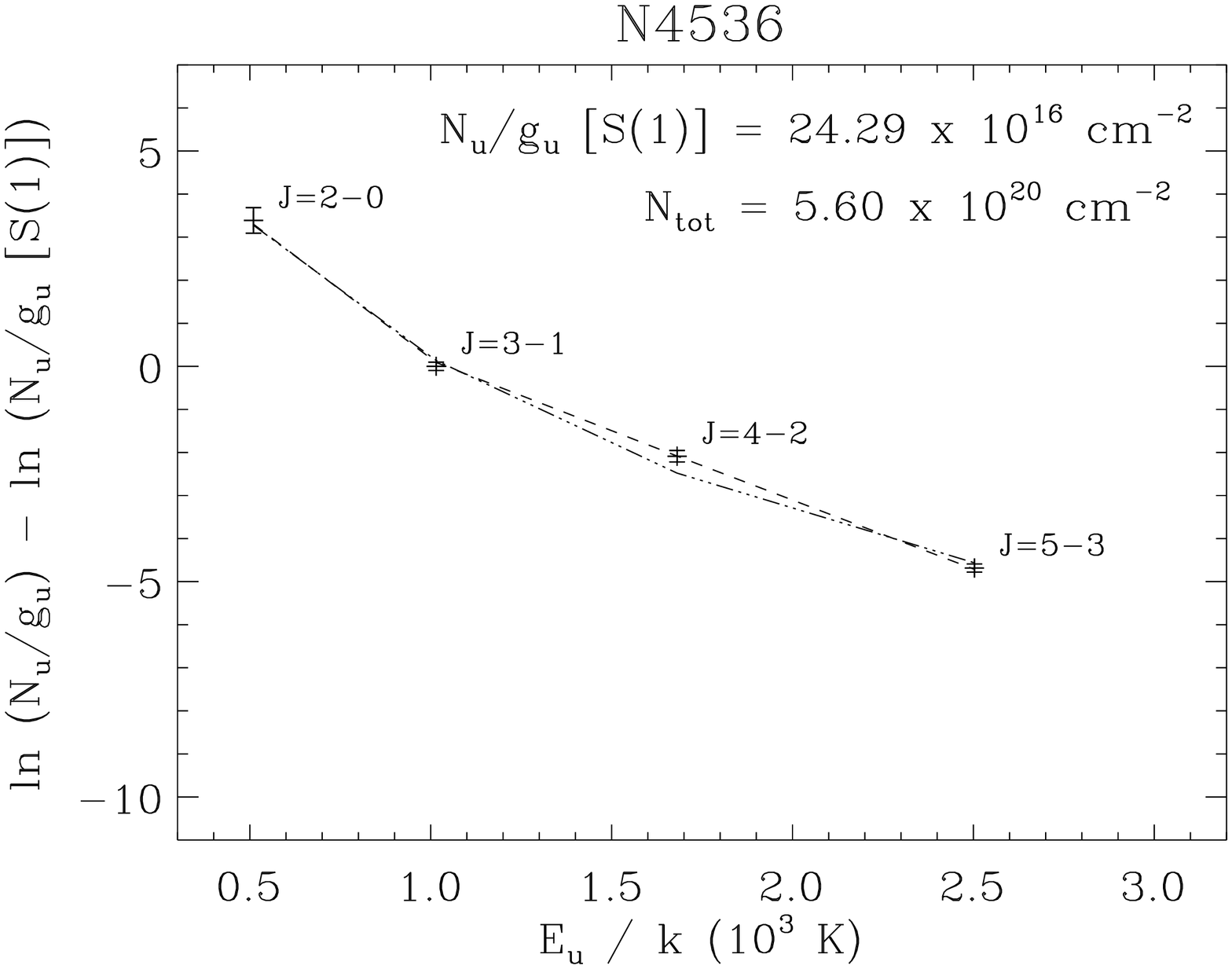}}}
\caption{(continued).
}
\end{figure}

\addtocounter{figure}{-1}
\begin{figure}[!ht]
\hspace*{-2cm}
\resizebox{10cm}{!}{\rotatebox{90}{\plotone{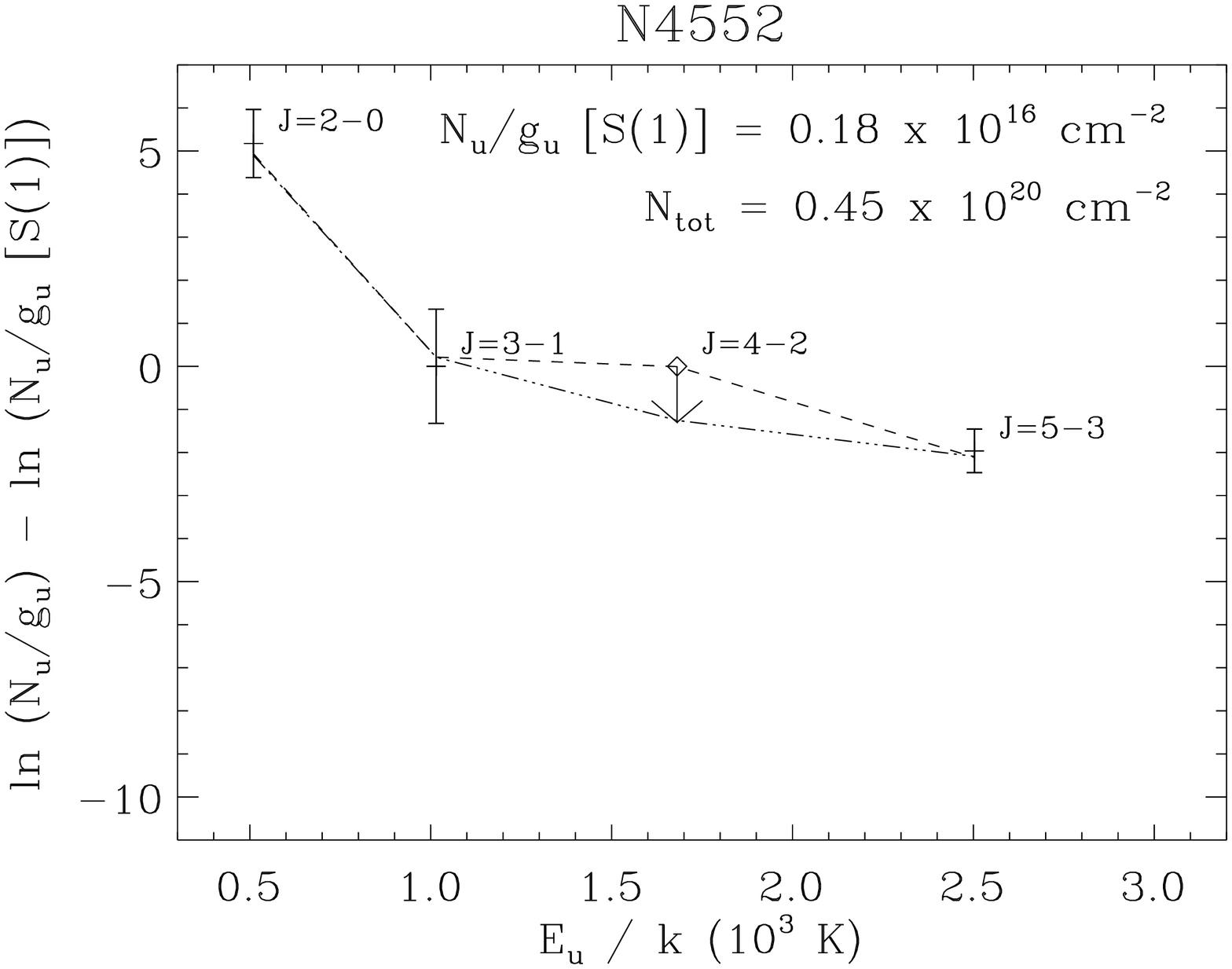}}}
\hspace*{-0.5cm}
\resizebox{10cm}{!}{\rotatebox{90}{\plotone{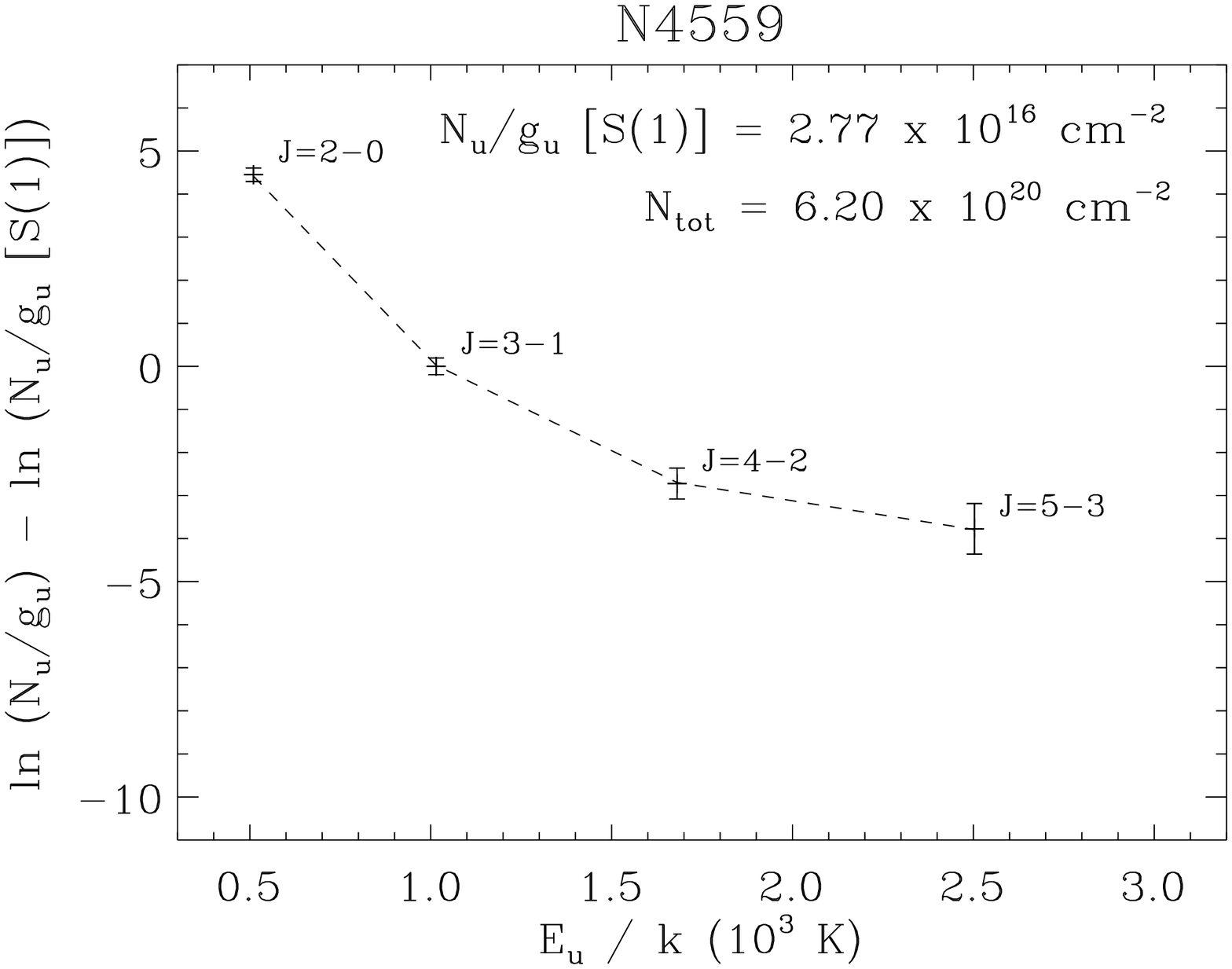}}}
\hspace*{-2cm}
\resizebox{10cm}{!}{\rotatebox{90}{\plotone{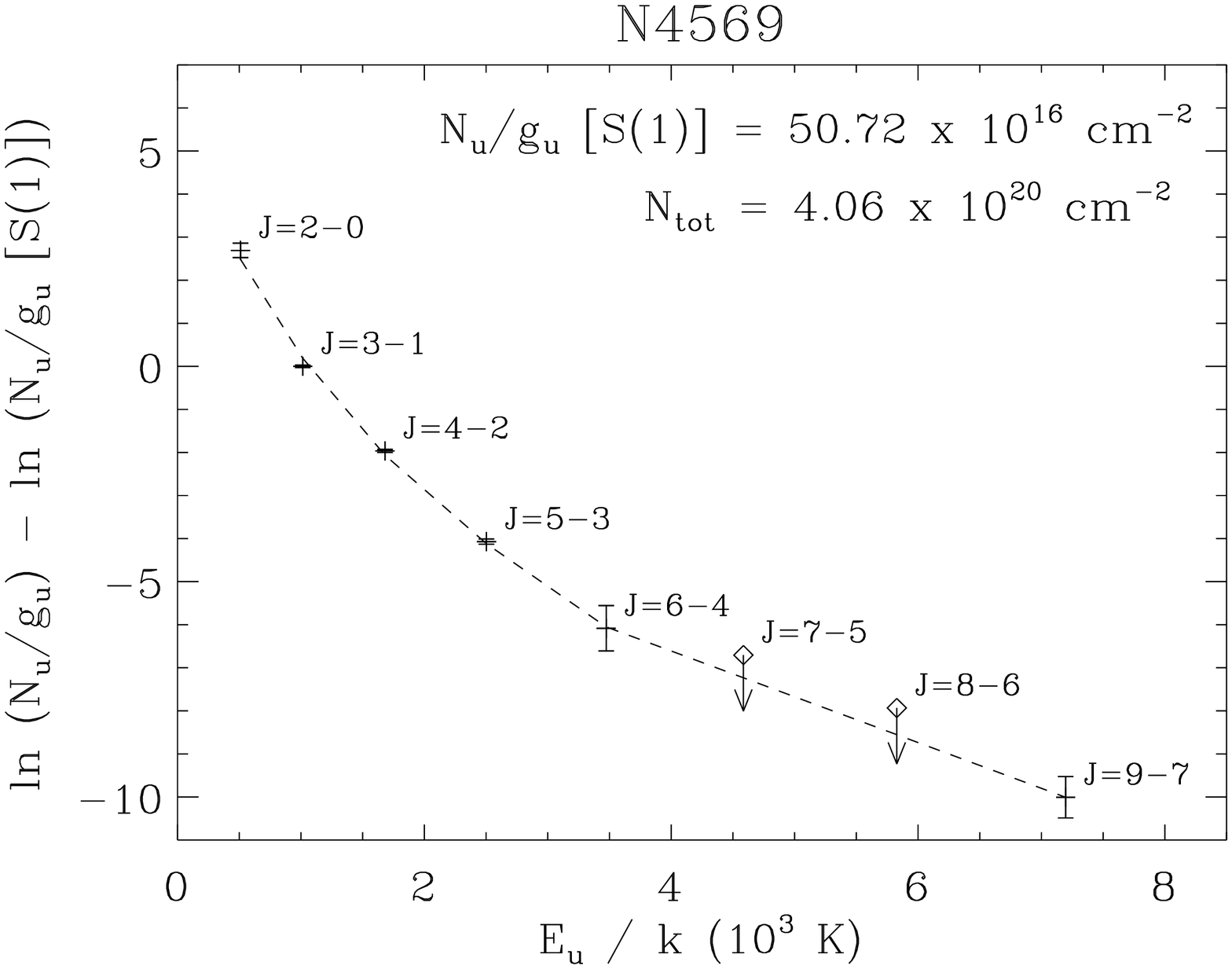}}}
\hspace*{-0.5cm}
\resizebox{10cm}{!}{\rotatebox{90}{\plotone{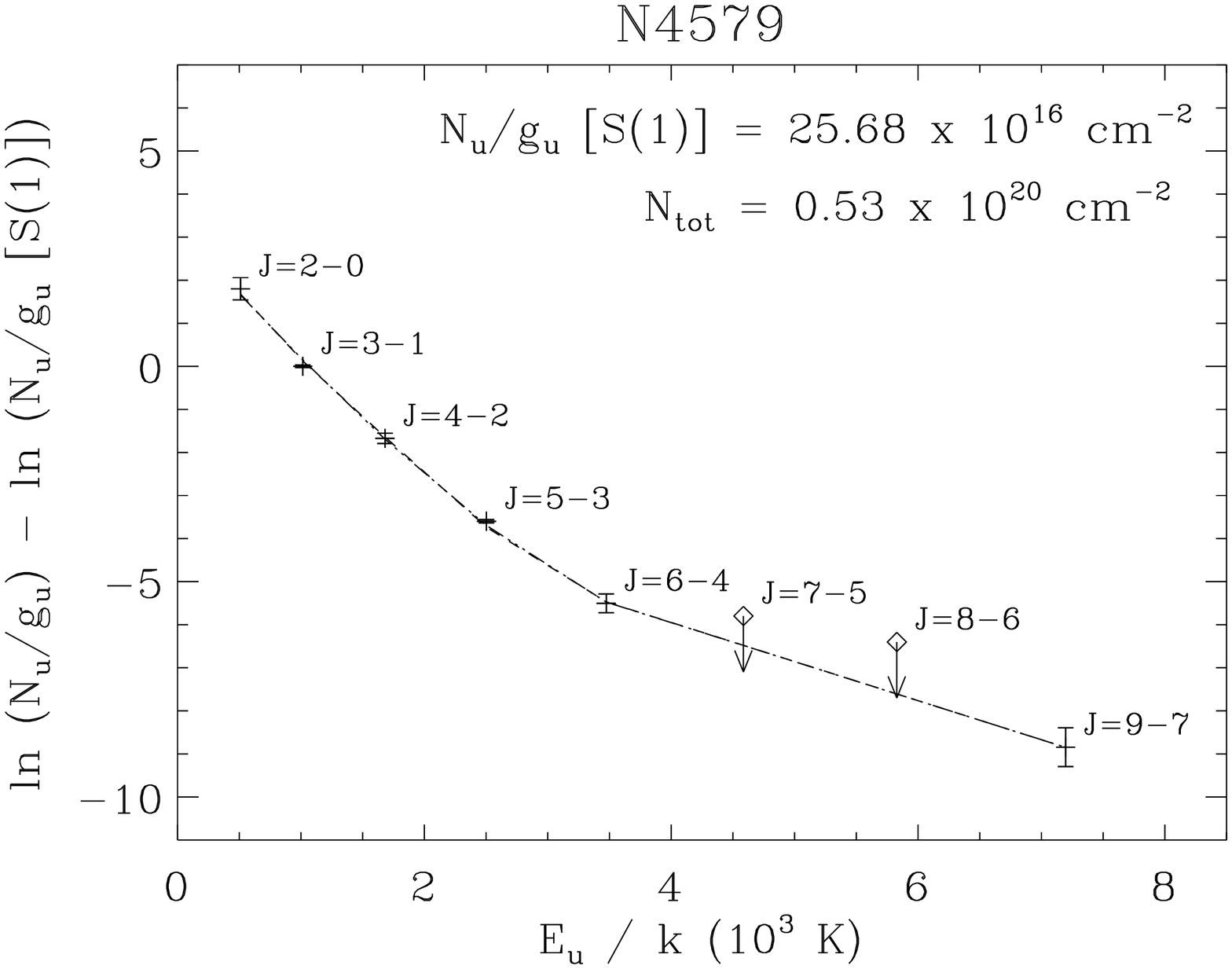}}}
\hspace*{-2cm}
\resizebox{10cm}{!}{\rotatebox{90}{\plotone{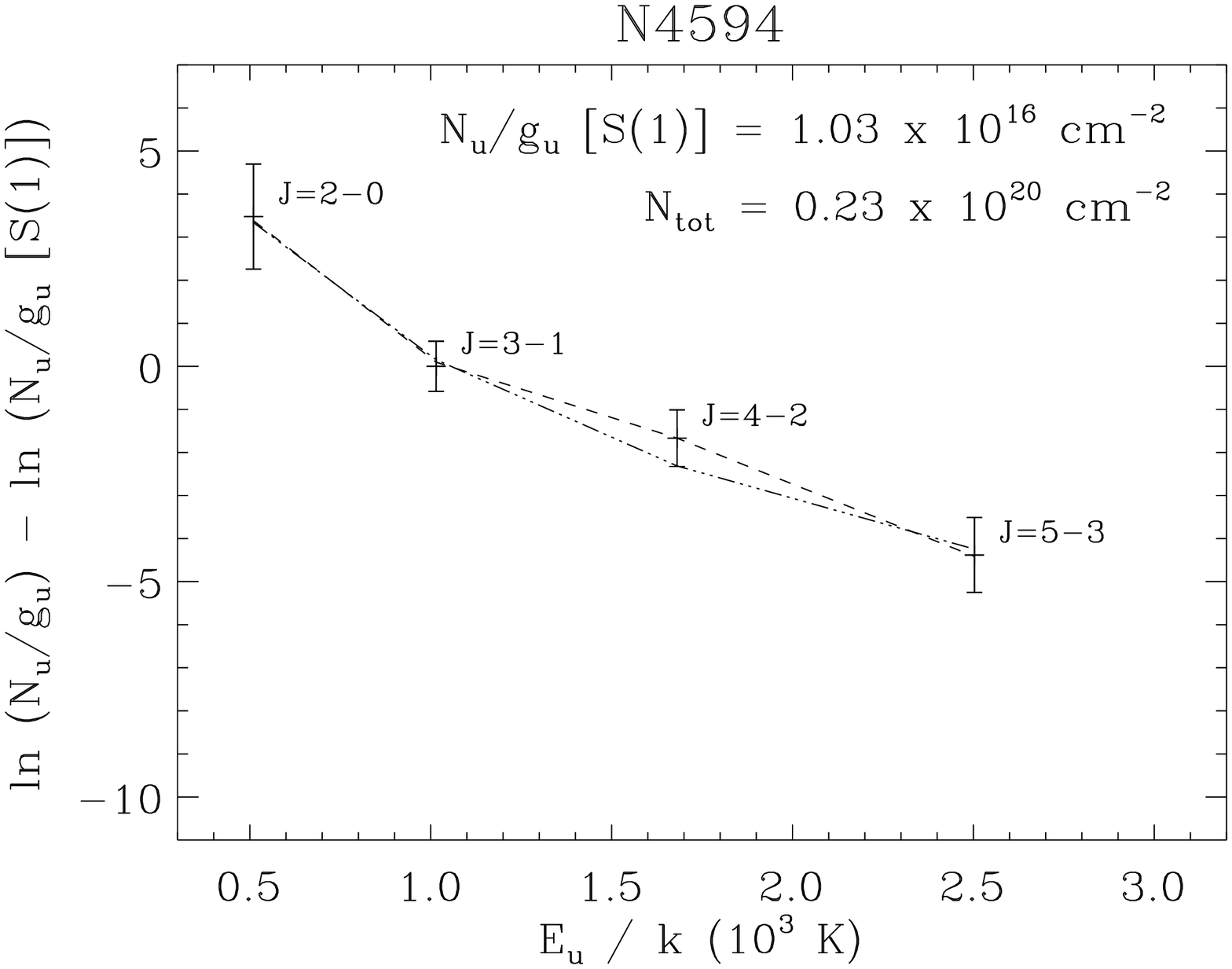}}}
\hspace*{-0.5cm}
\resizebox{10cm}{!}{\rotatebox{90}{\plotone{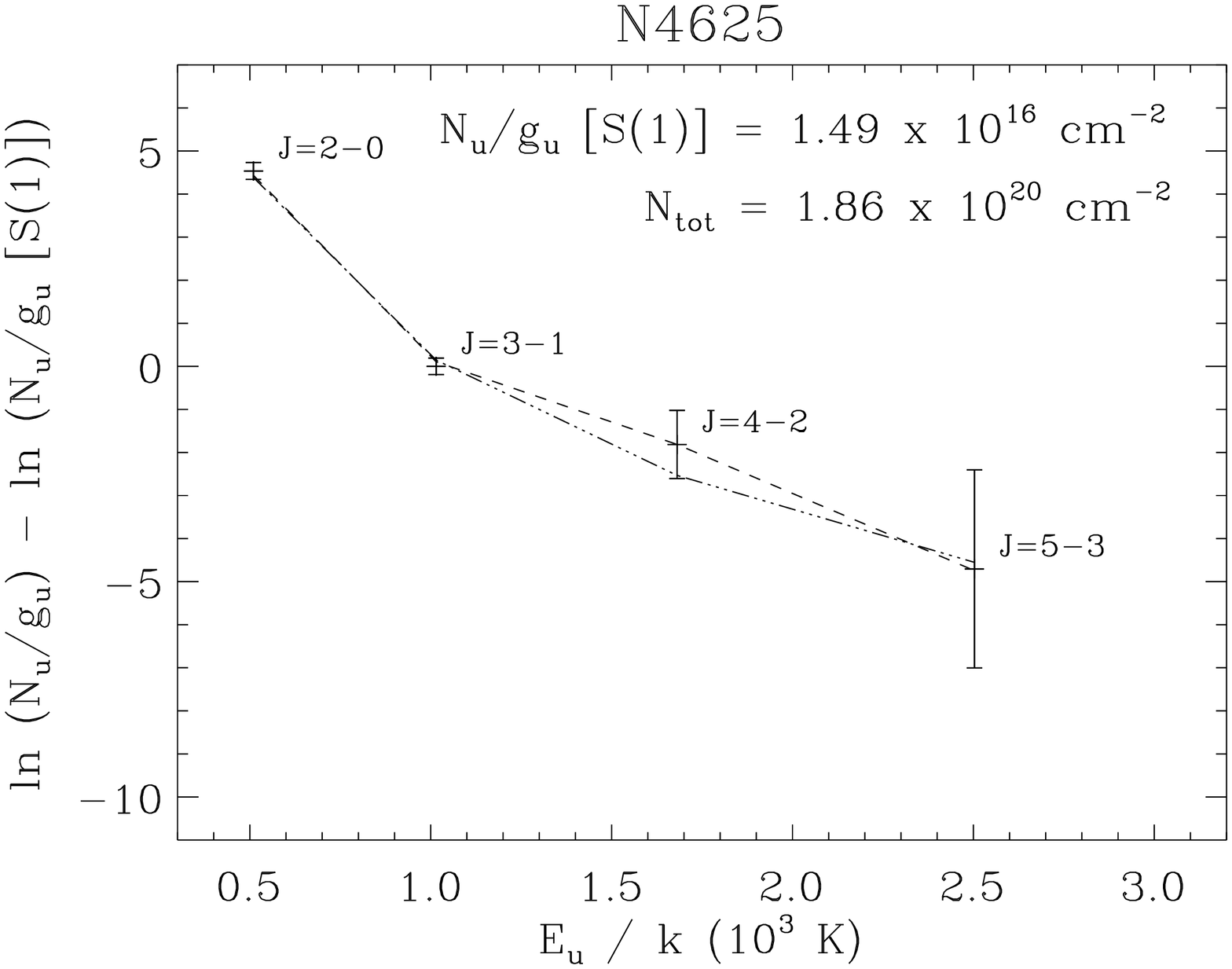}}}
\caption{(continued).
}
\end{figure}

\addtocounter{figure}{-1}
\begin{figure}[!ht]
\hspace*{-2cm}
\resizebox{10cm}{!}{\rotatebox{90}{\plotone{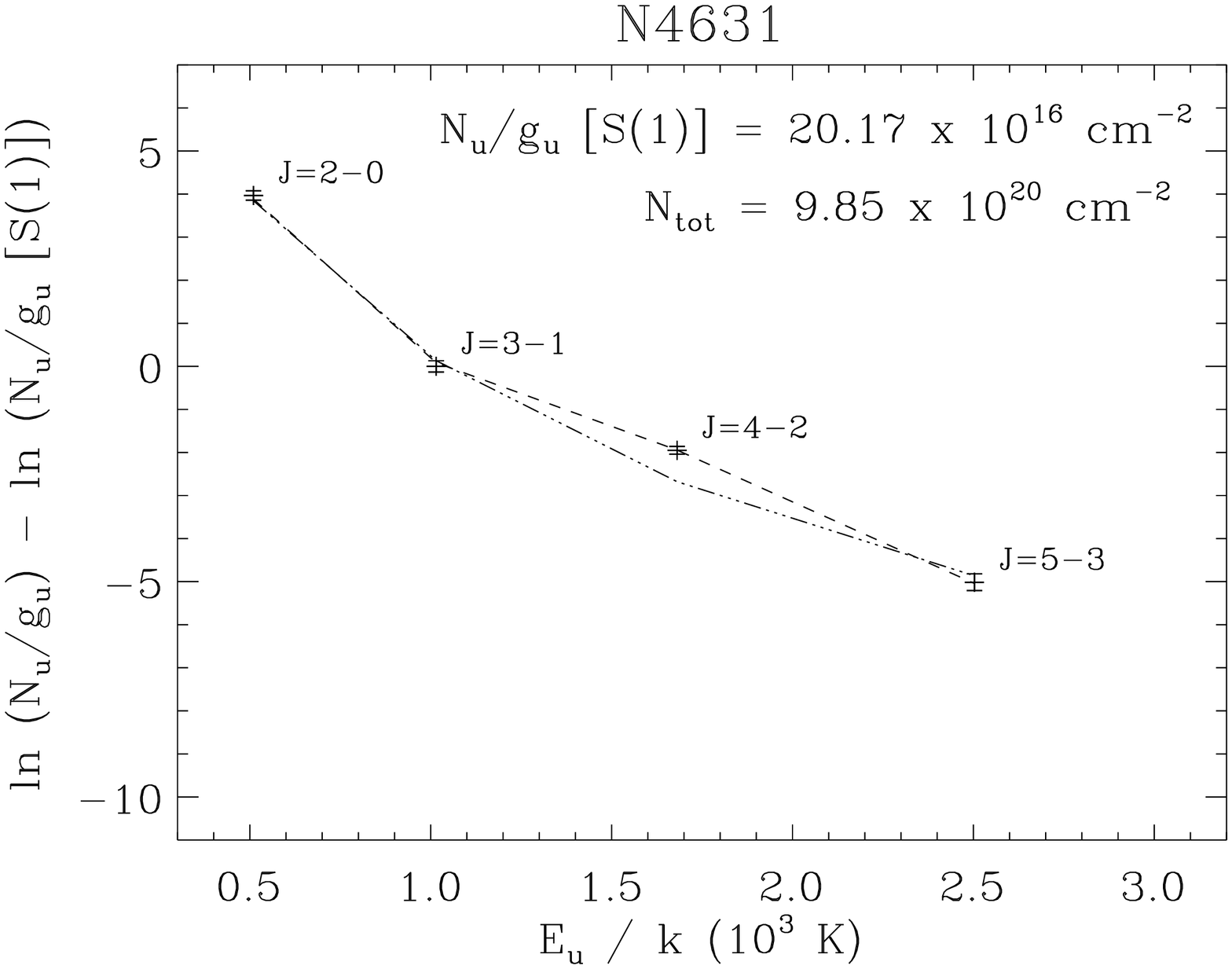}}}
\hspace*{-0.5cm}
\resizebox{10cm}{!}{\rotatebox{90}{\plotone{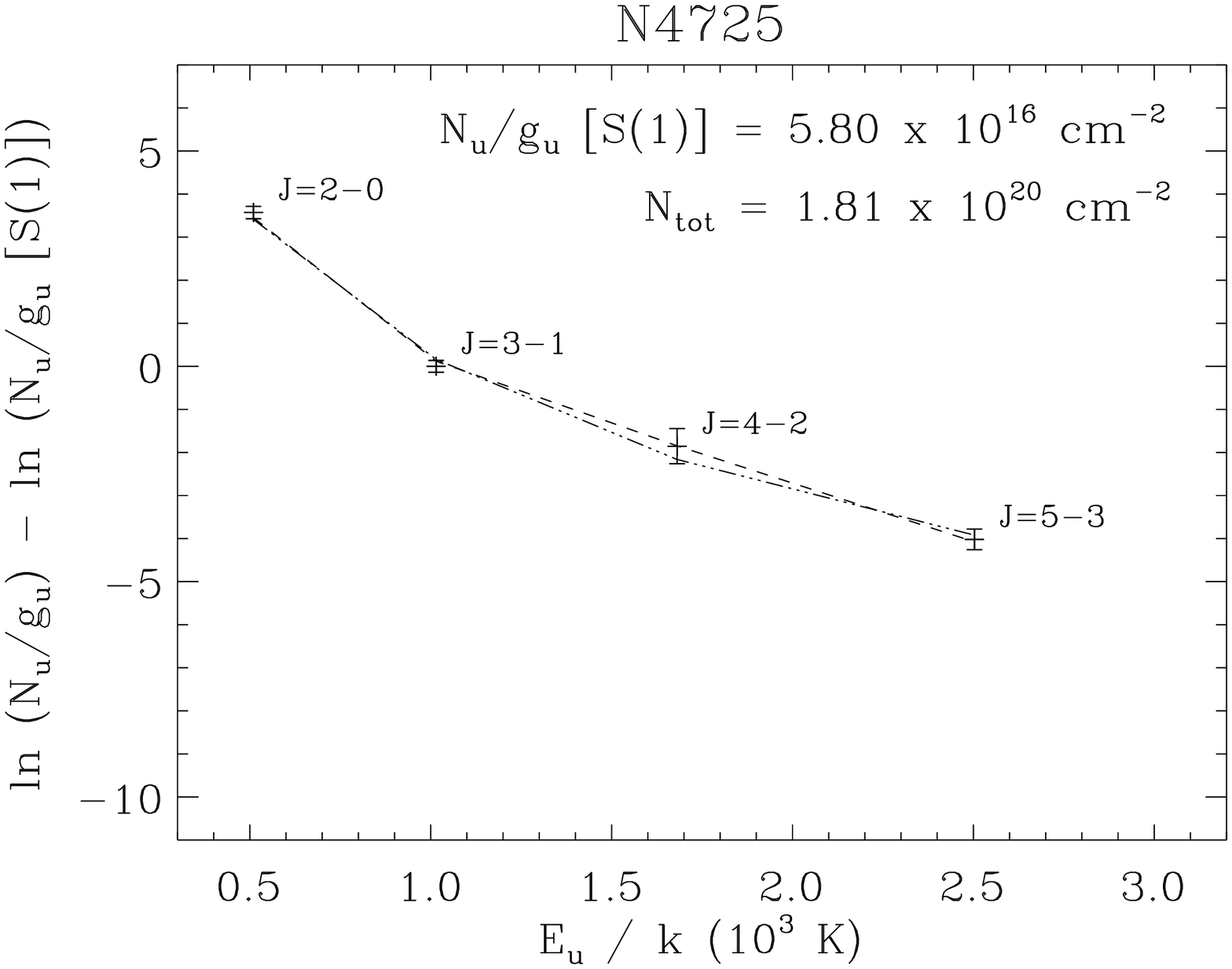}}}
\hspace*{-2cm}
\resizebox{10cm}{!}{\rotatebox{90}{\plotone{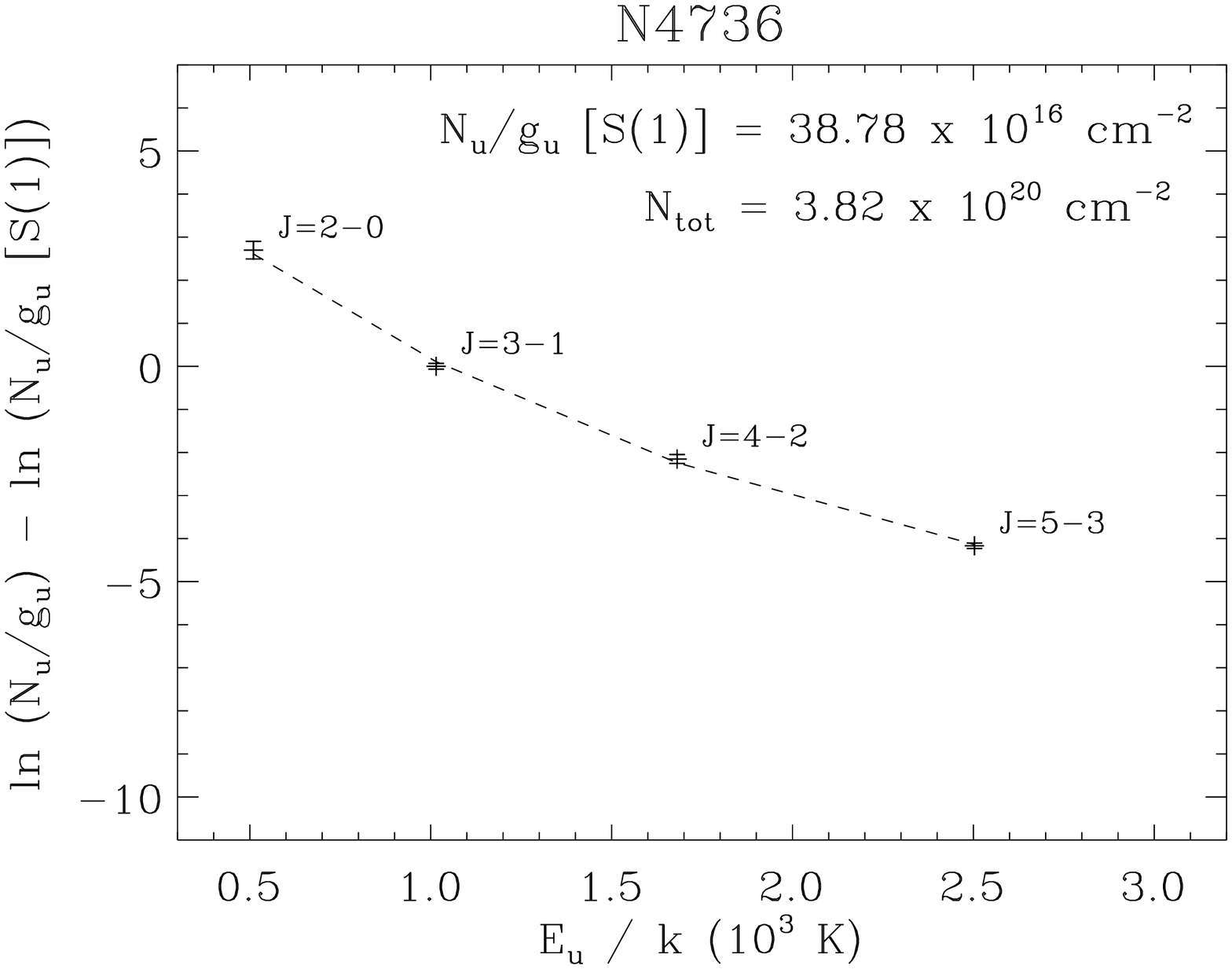}}}
\hspace*{-0.5cm}
\resizebox{10cm}{!}{\rotatebox{90}{\plotone{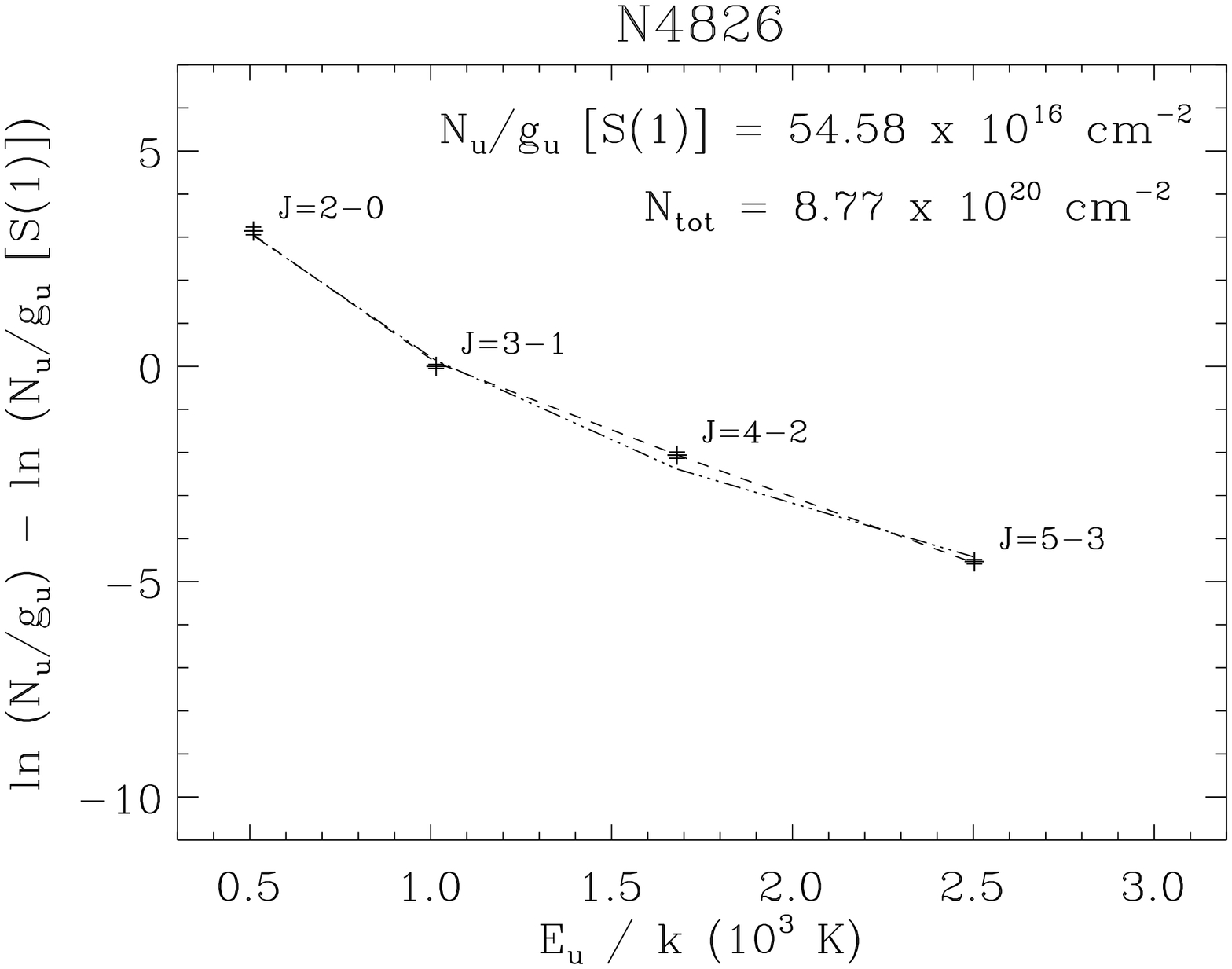}}}
\hspace*{-2cm}
\resizebox{10cm}{!}{\rotatebox{90}{\plotone{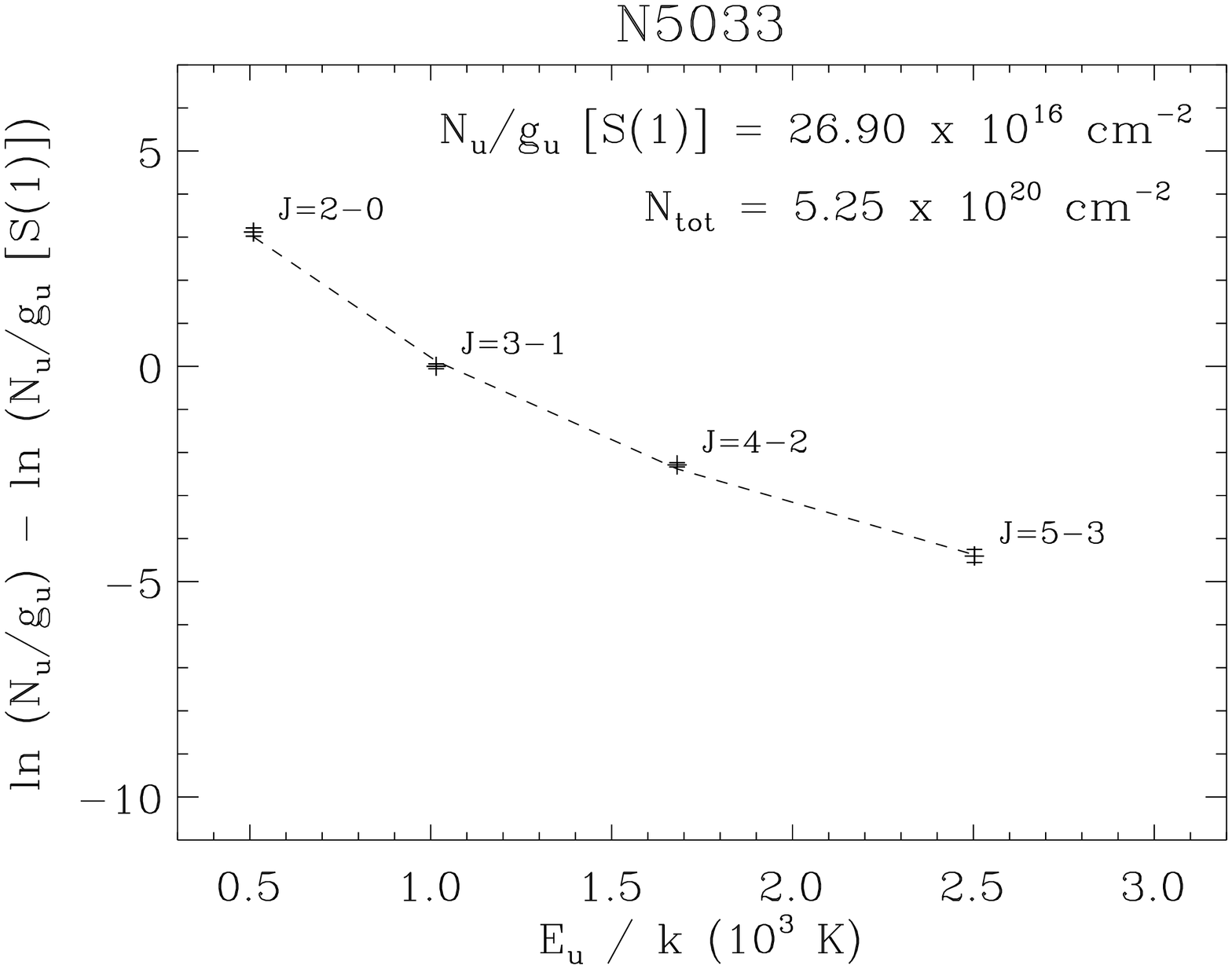}}}
\hspace*{-0.5cm}
\resizebox{10cm}{!}{\rotatebox{90}{\plotone{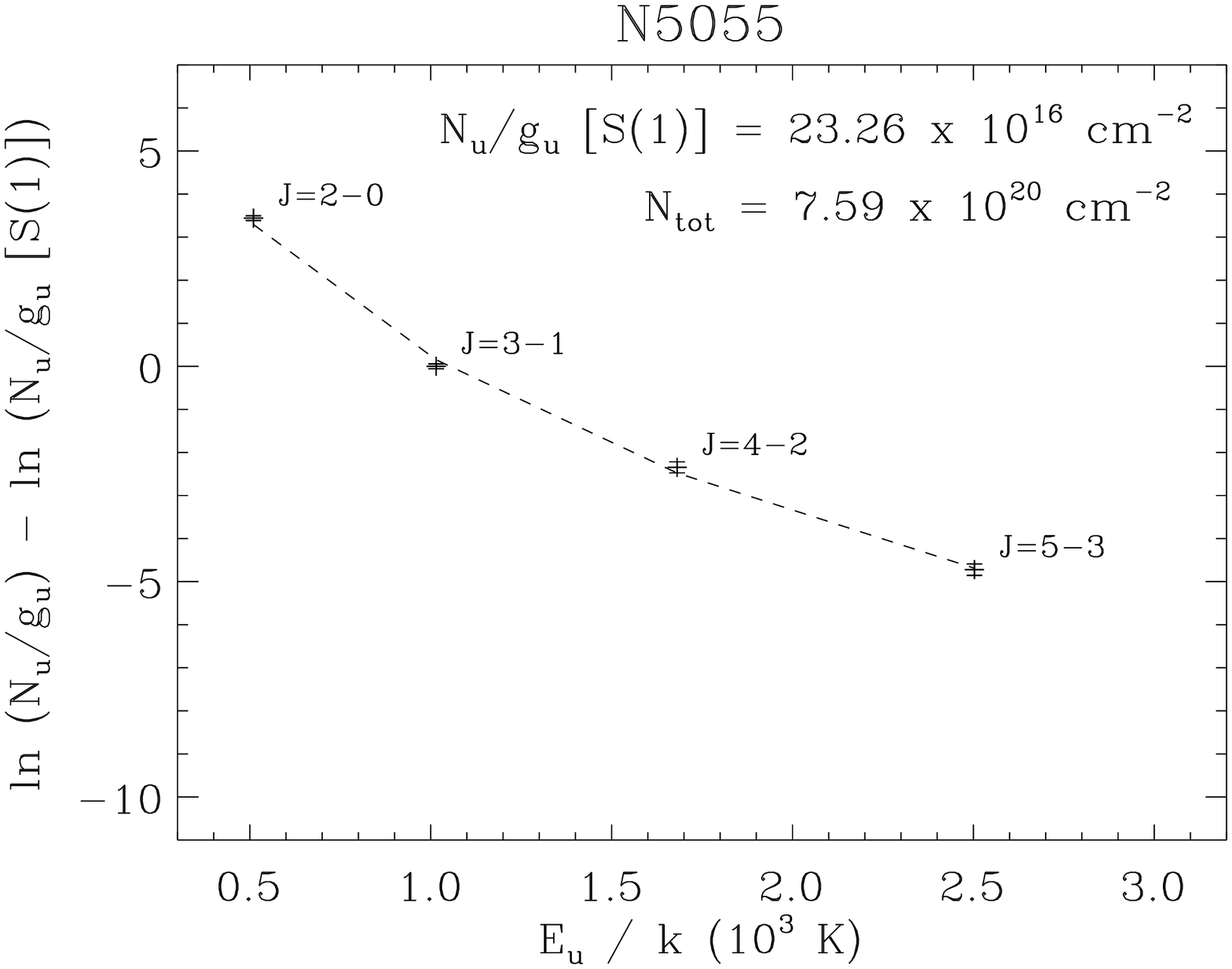}}}
\caption{(continued).
}
\end{figure}

\addtocounter{figure}{-1}
\begin{figure}[!ht]
\hspace*{-2cm}
\resizebox{10cm}{!}{\rotatebox{90}{\plotone{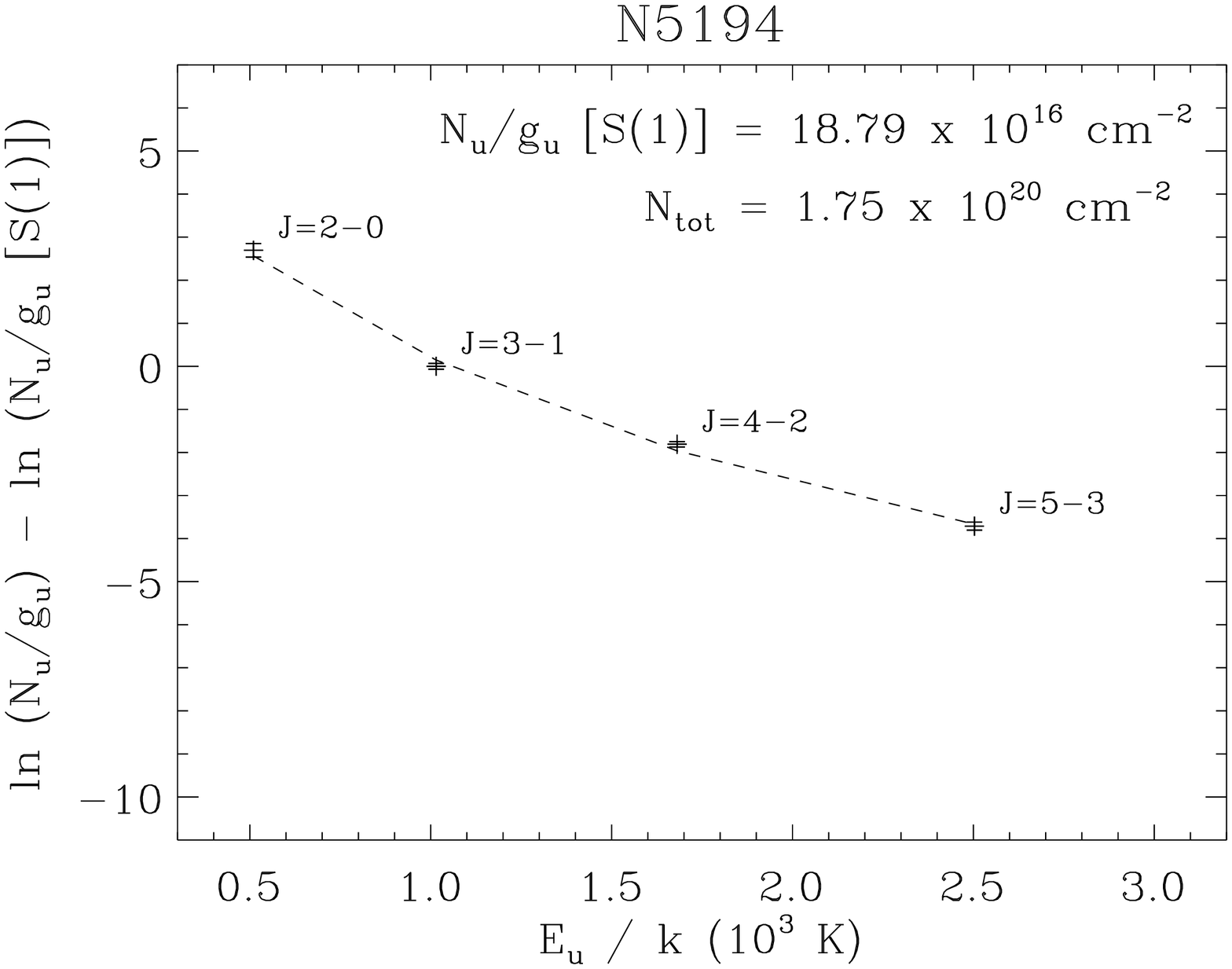}}}
\hspace*{-0.5cm}
\resizebox{10cm}{!}{\rotatebox{90}{\plotone{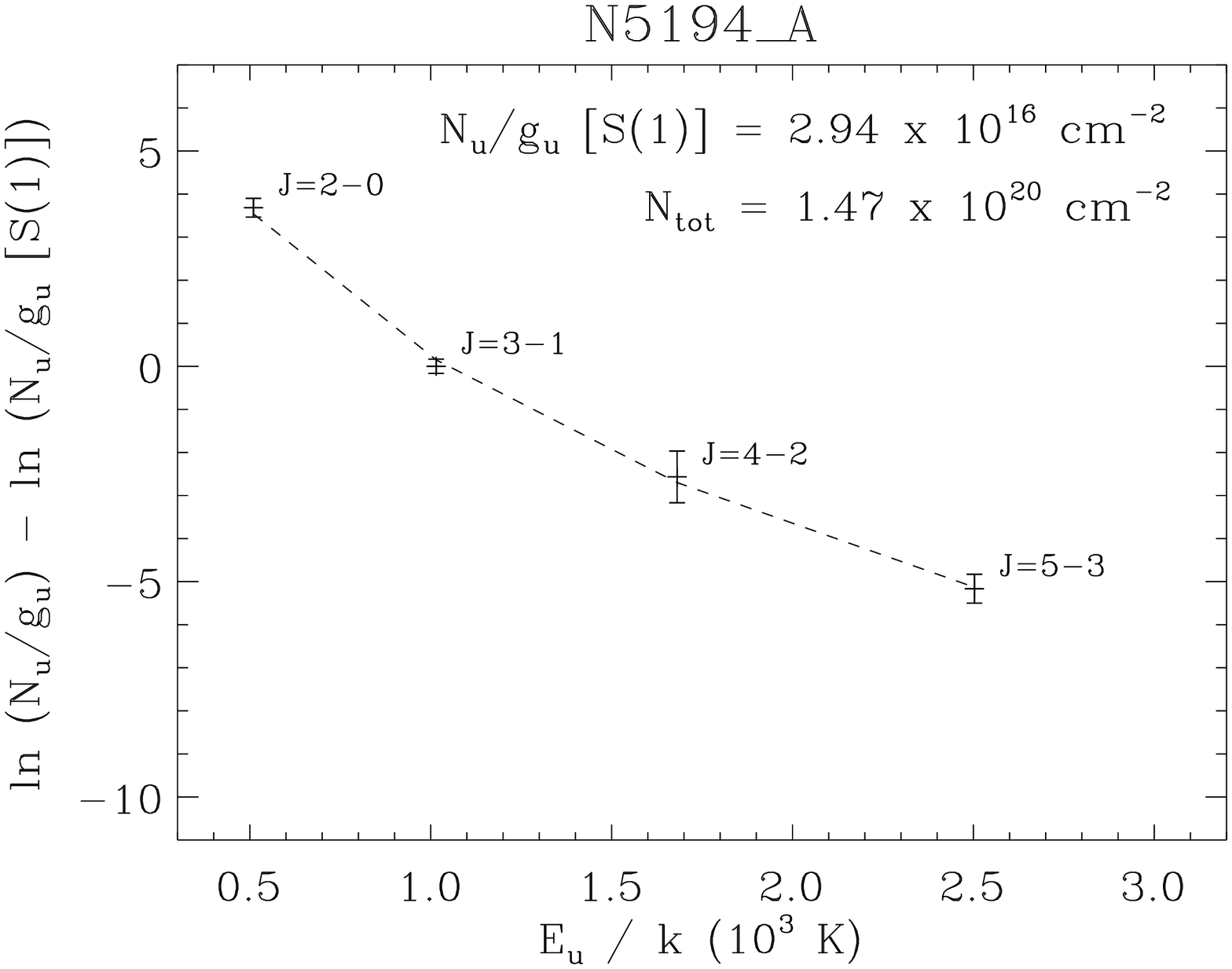}}}
\hspace*{-2cm}
\resizebox{10cm}{!}{\rotatebox{90}{\plotone{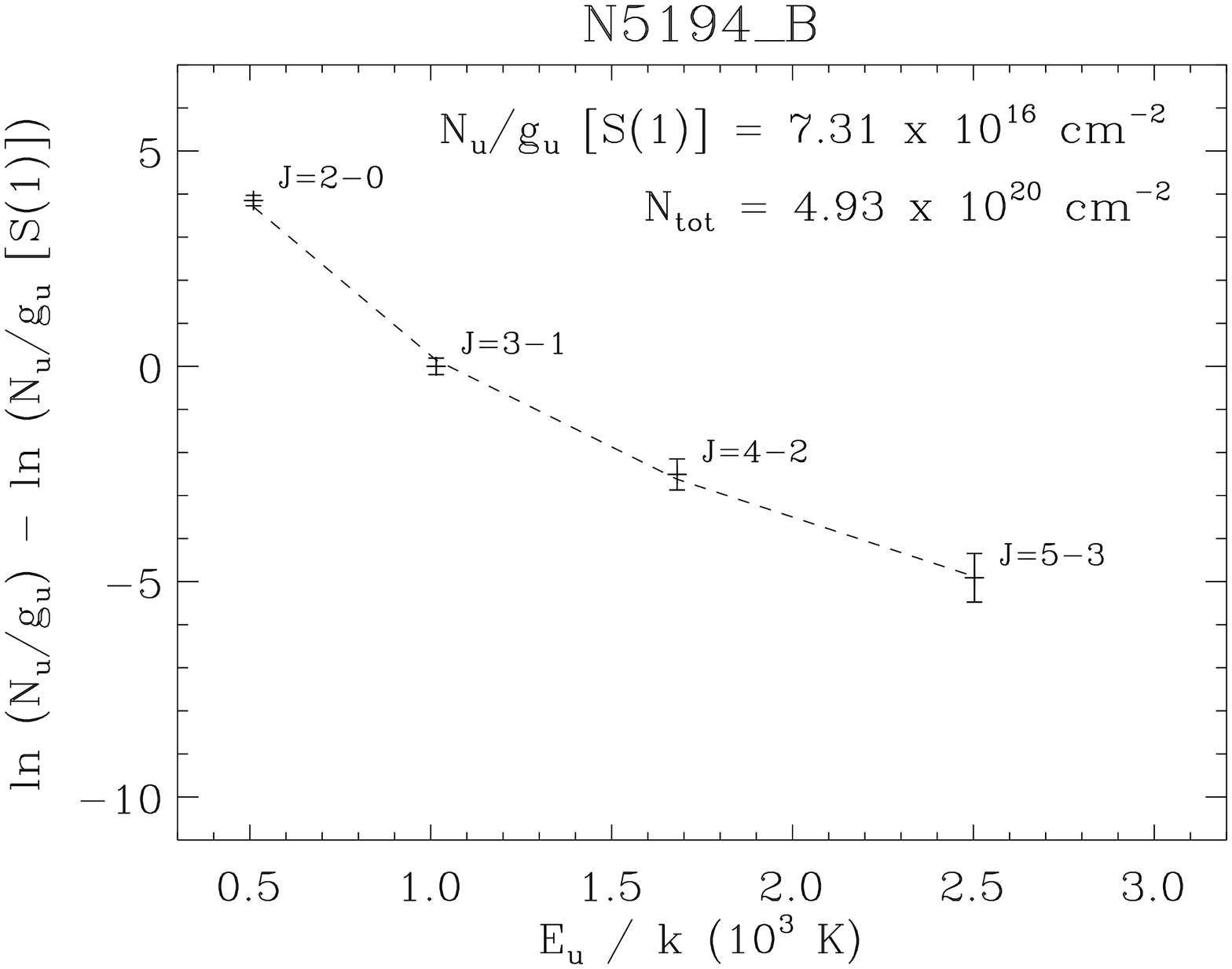}}}
\hspace*{-0.5cm}
\resizebox{10cm}{!}{\rotatebox{90}{\plotone{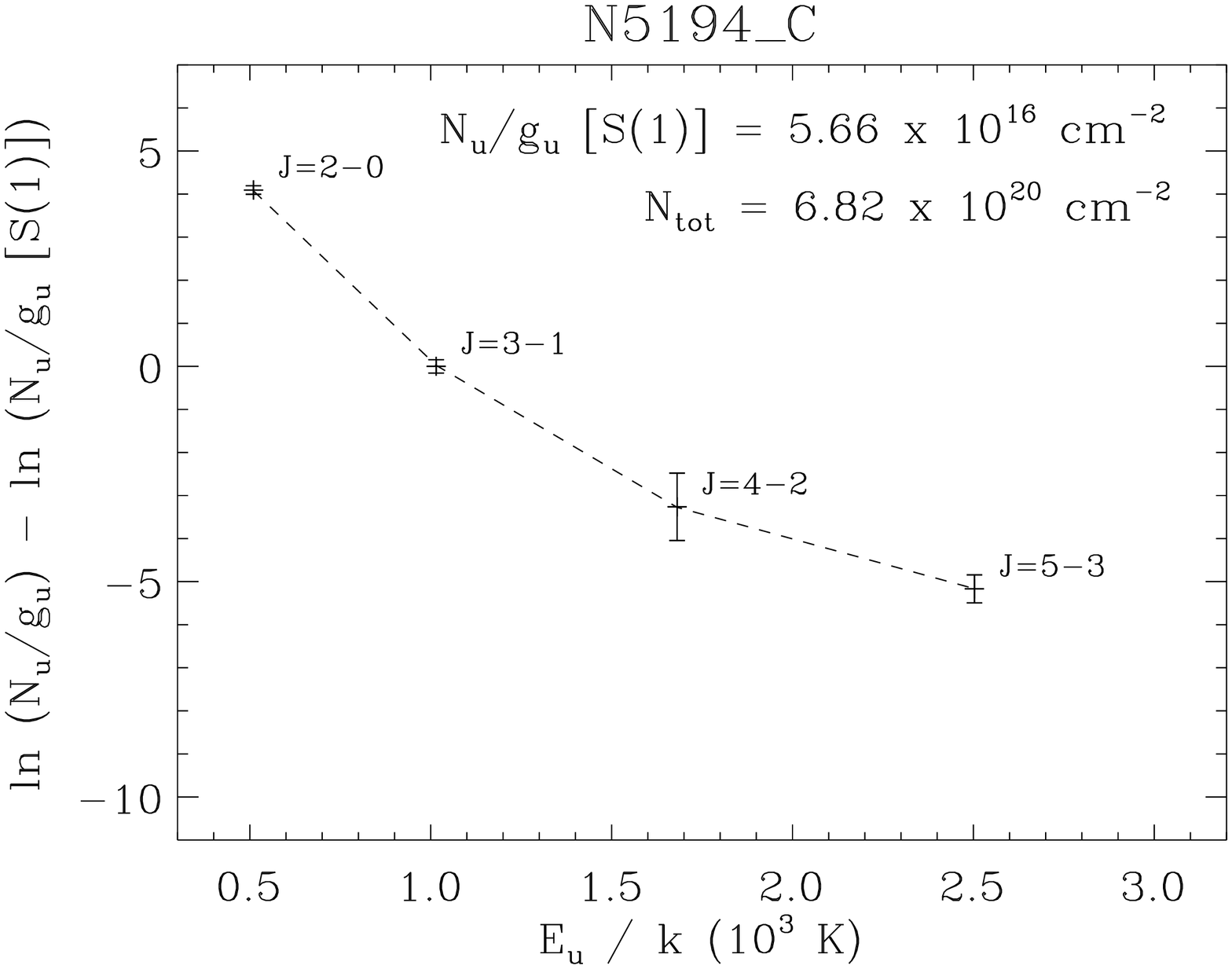}}}
\hspace*{-2cm}
\resizebox{10cm}{!}{\rotatebox{90}{\plotone{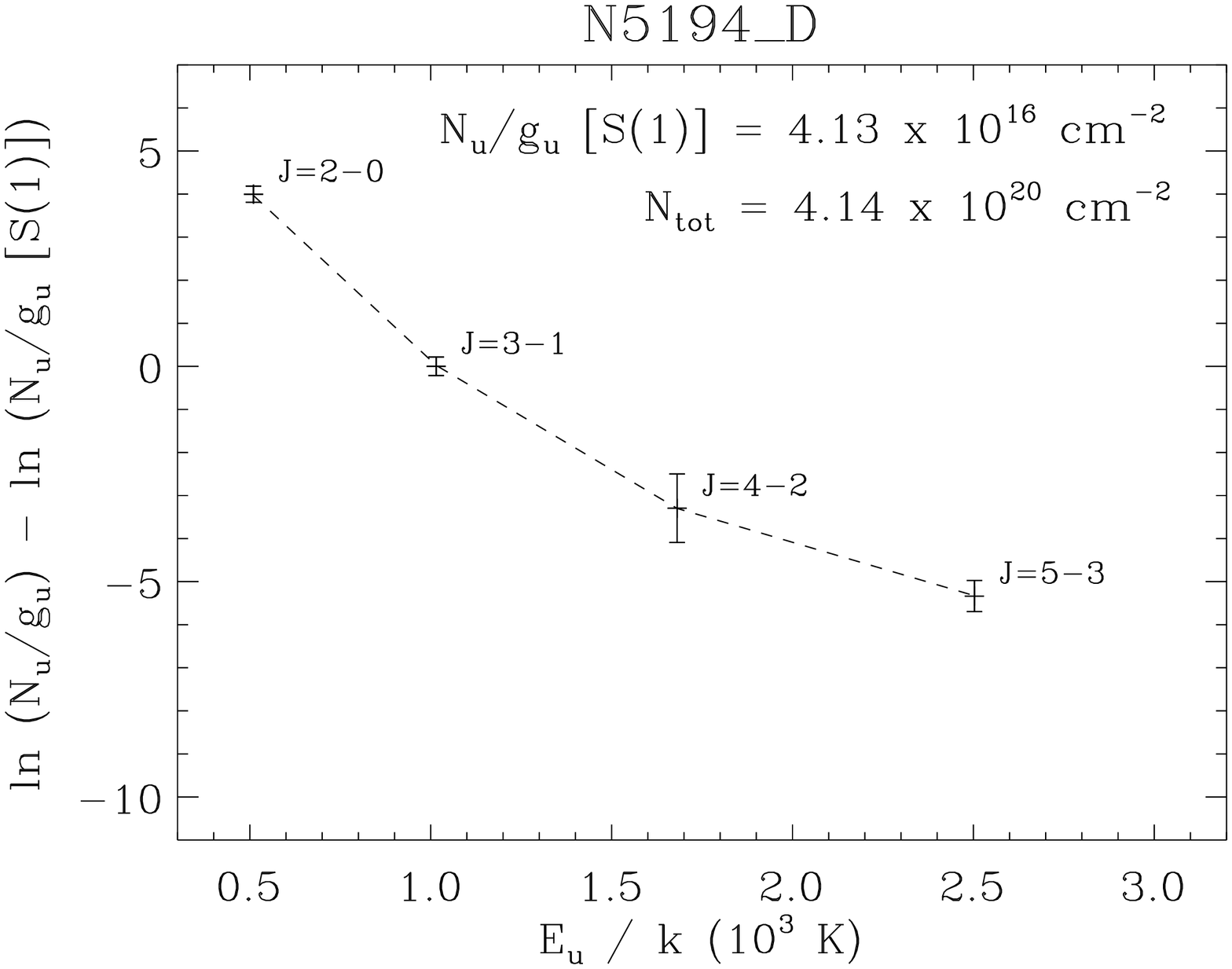}}}
\hspace*{-0.5cm}
\resizebox{10cm}{!}{\rotatebox{90}{\plotone{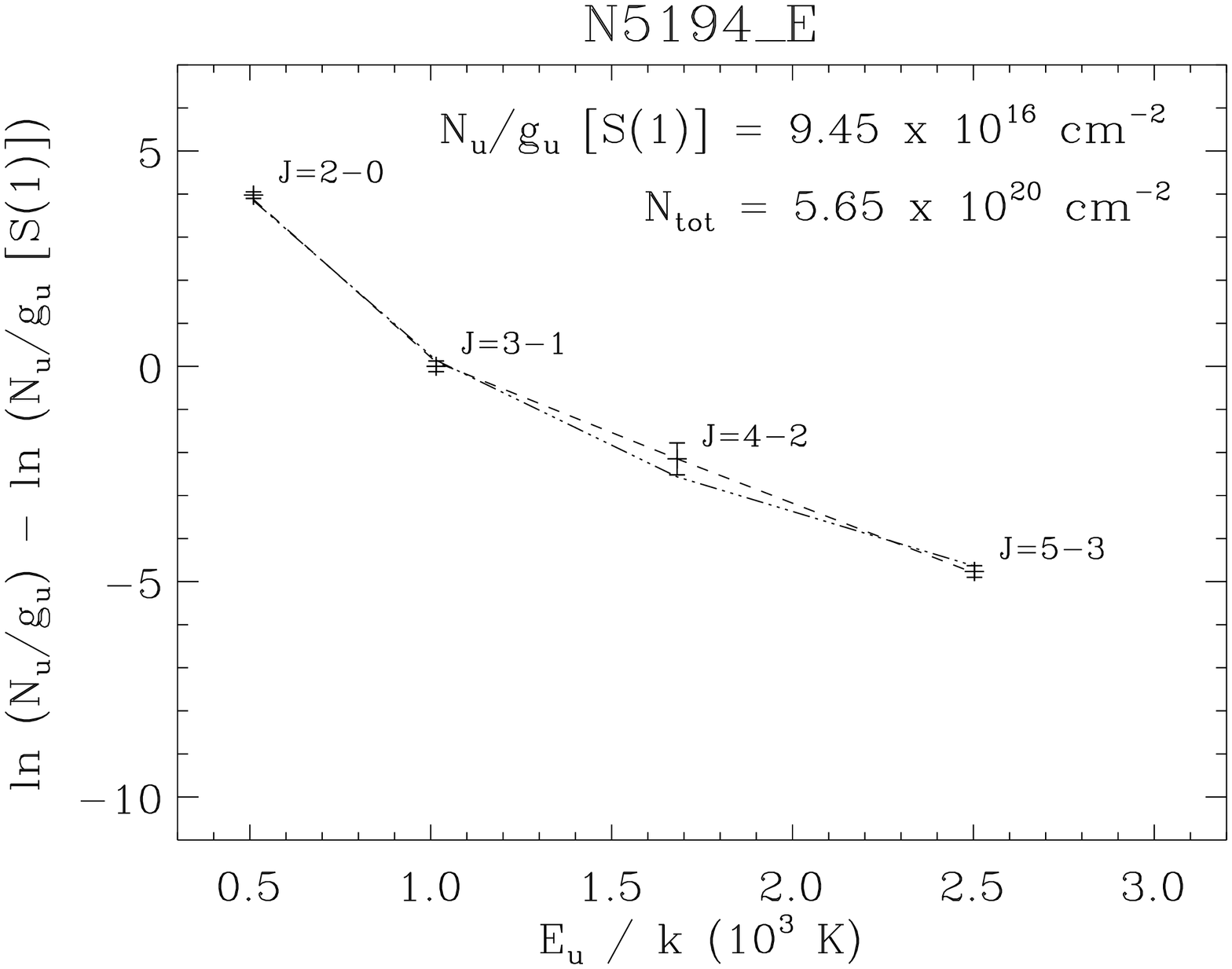}}}
\caption{(continued).
}
\end{figure}

\addtocounter{figure}{-1}
\begin{figure}[!ht]
\hspace*{-2cm}
\resizebox{10cm}{!}{\rotatebox{90}{\plotone{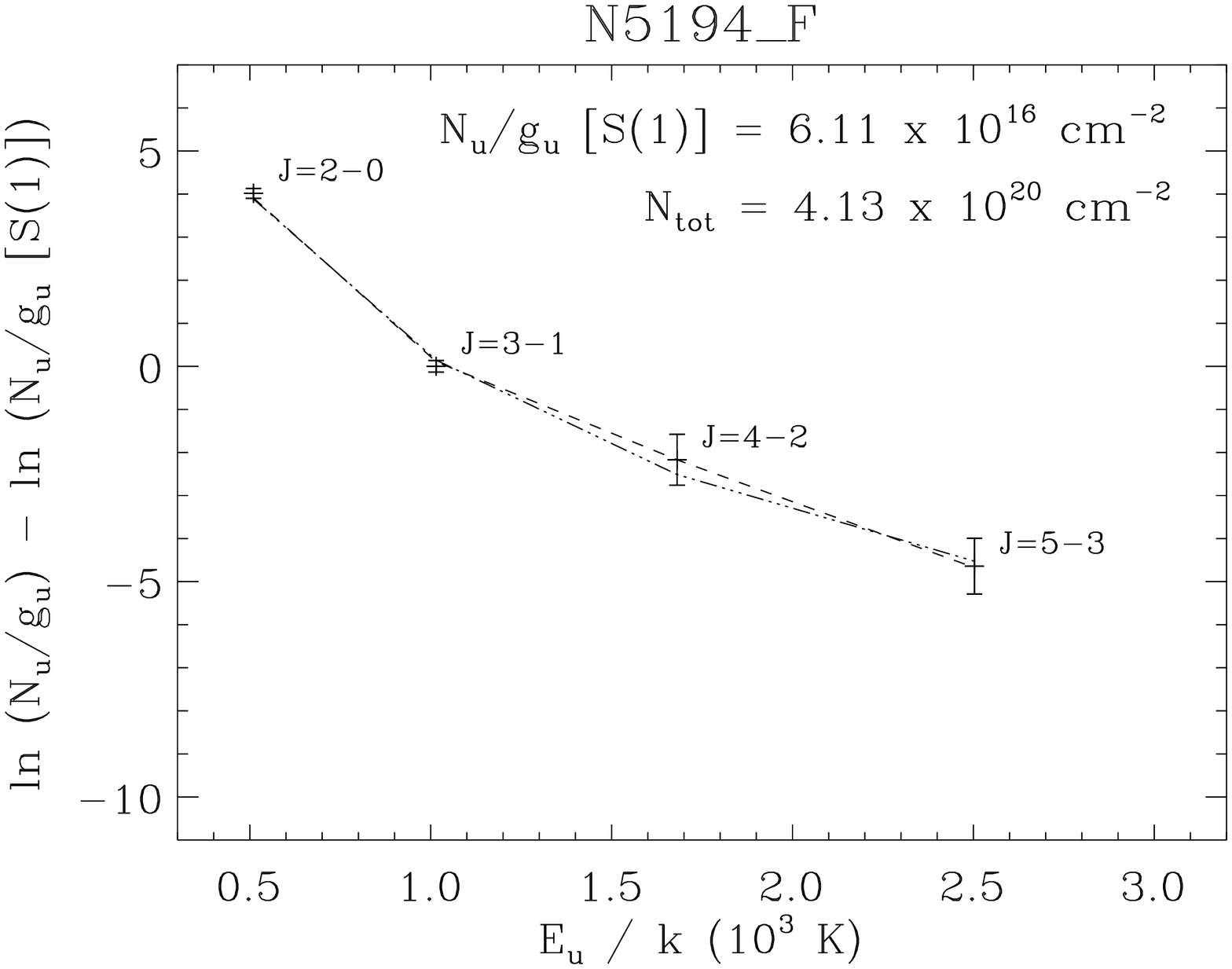}}}
\hspace*{-0.5cm}
\resizebox{10cm}{!}{\rotatebox{90}{\plotone{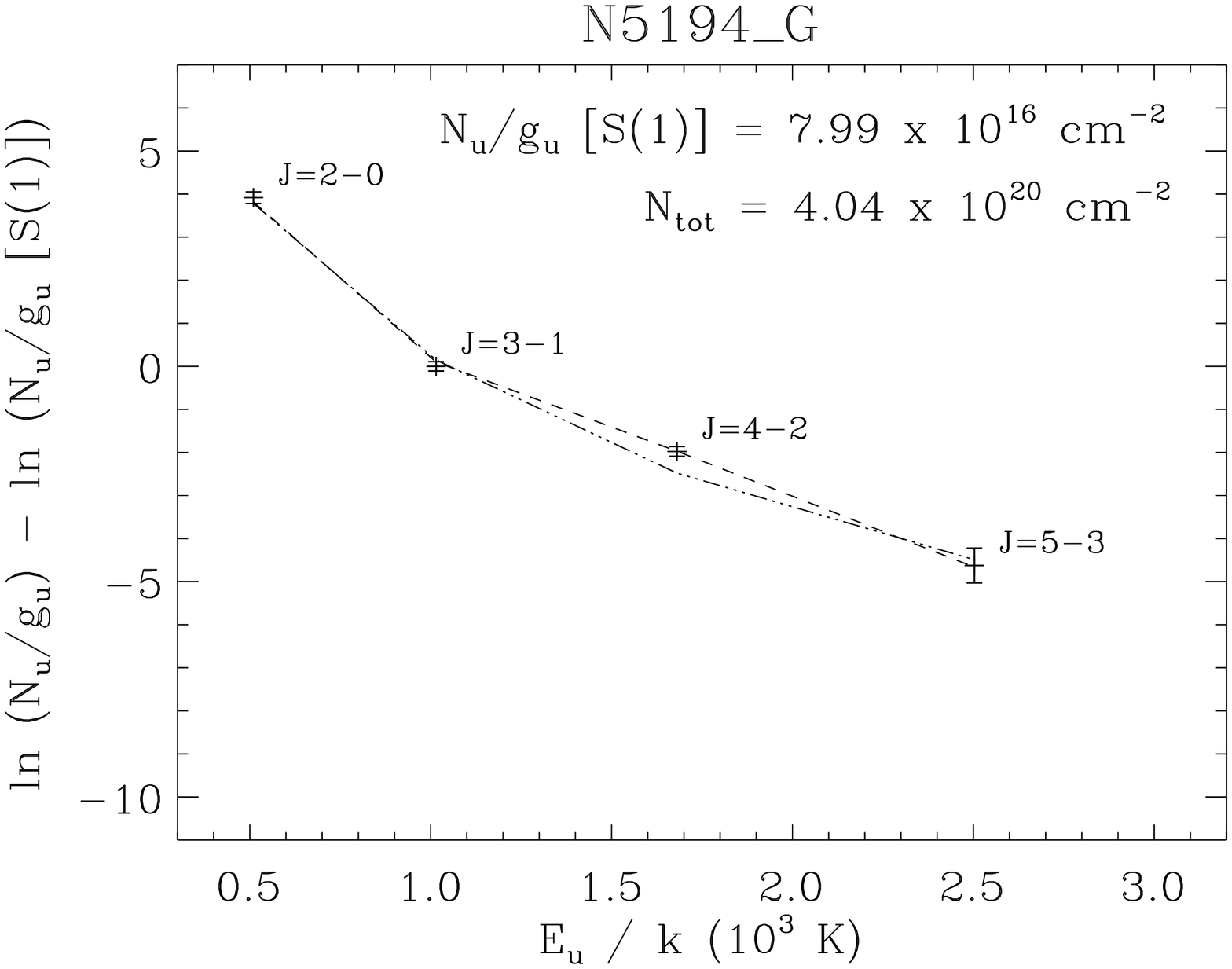}}}
\hspace*{-2cm}
\resizebox{10cm}{!}{\rotatebox{90}{\plotone{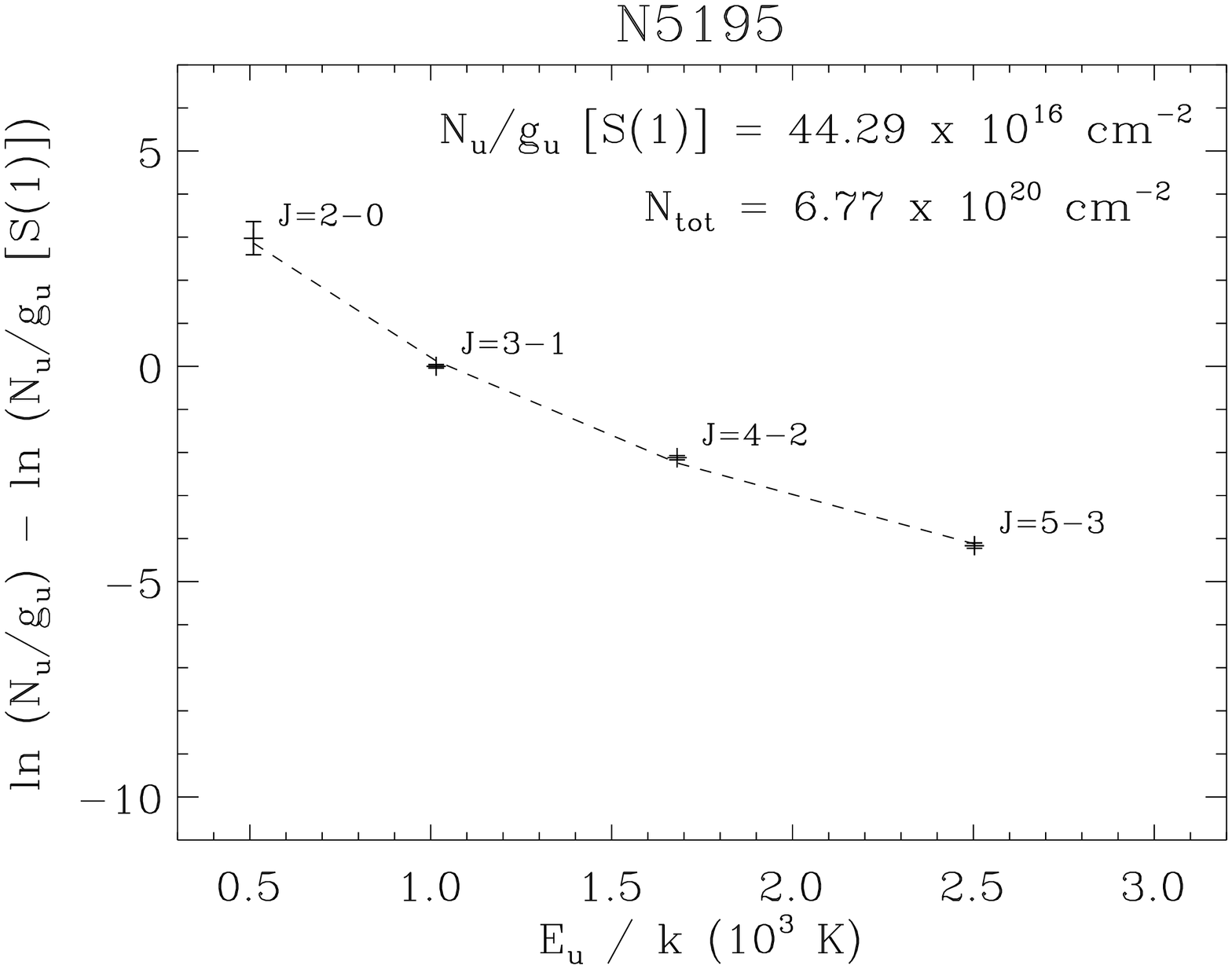}}}
\hspace*{-0.5cm}
\resizebox{10cm}{!}{\rotatebox{90}{\plotone{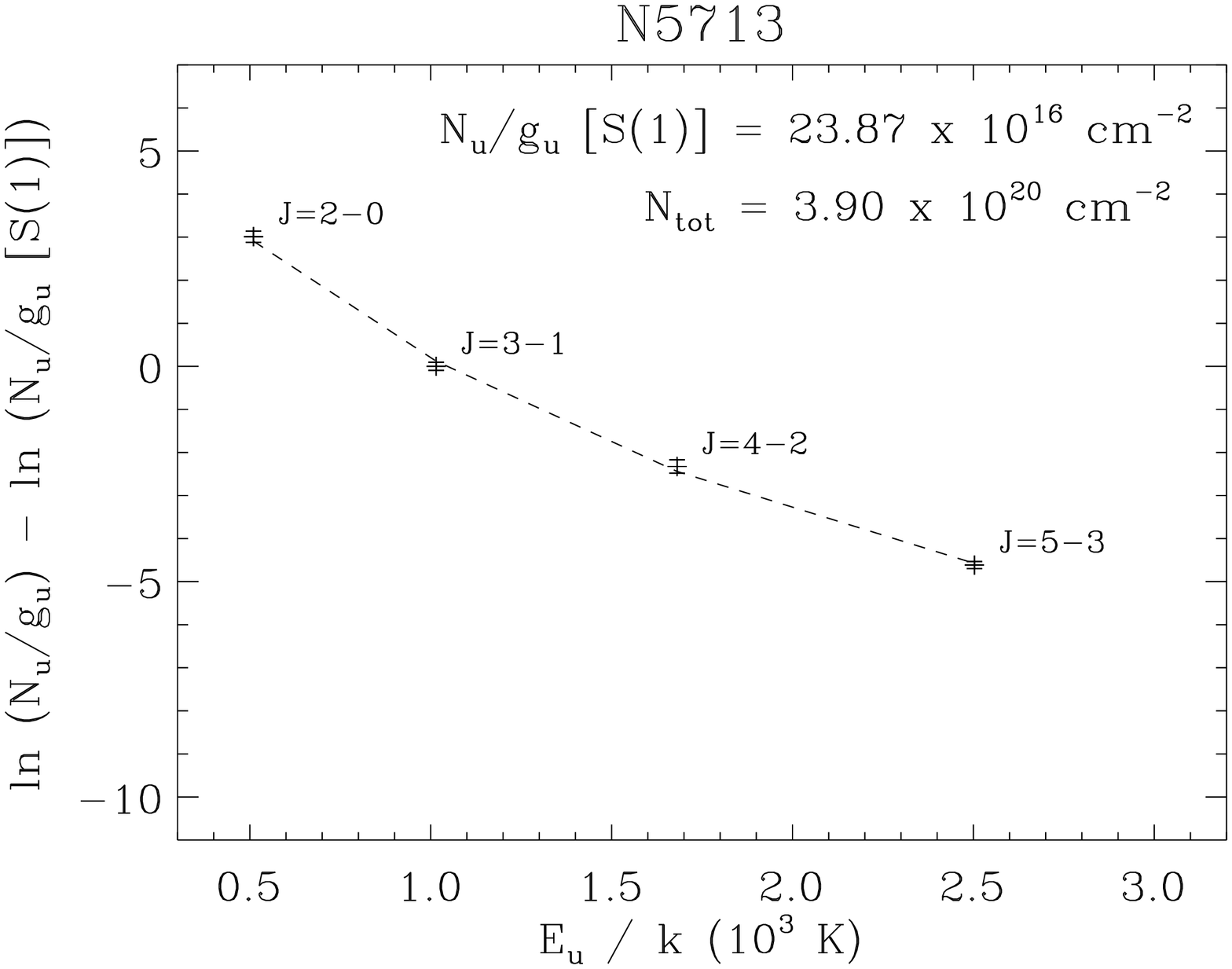}}}
\hspace*{-2cm}
\resizebox{10cm}{!}{\rotatebox{90}{\plotone{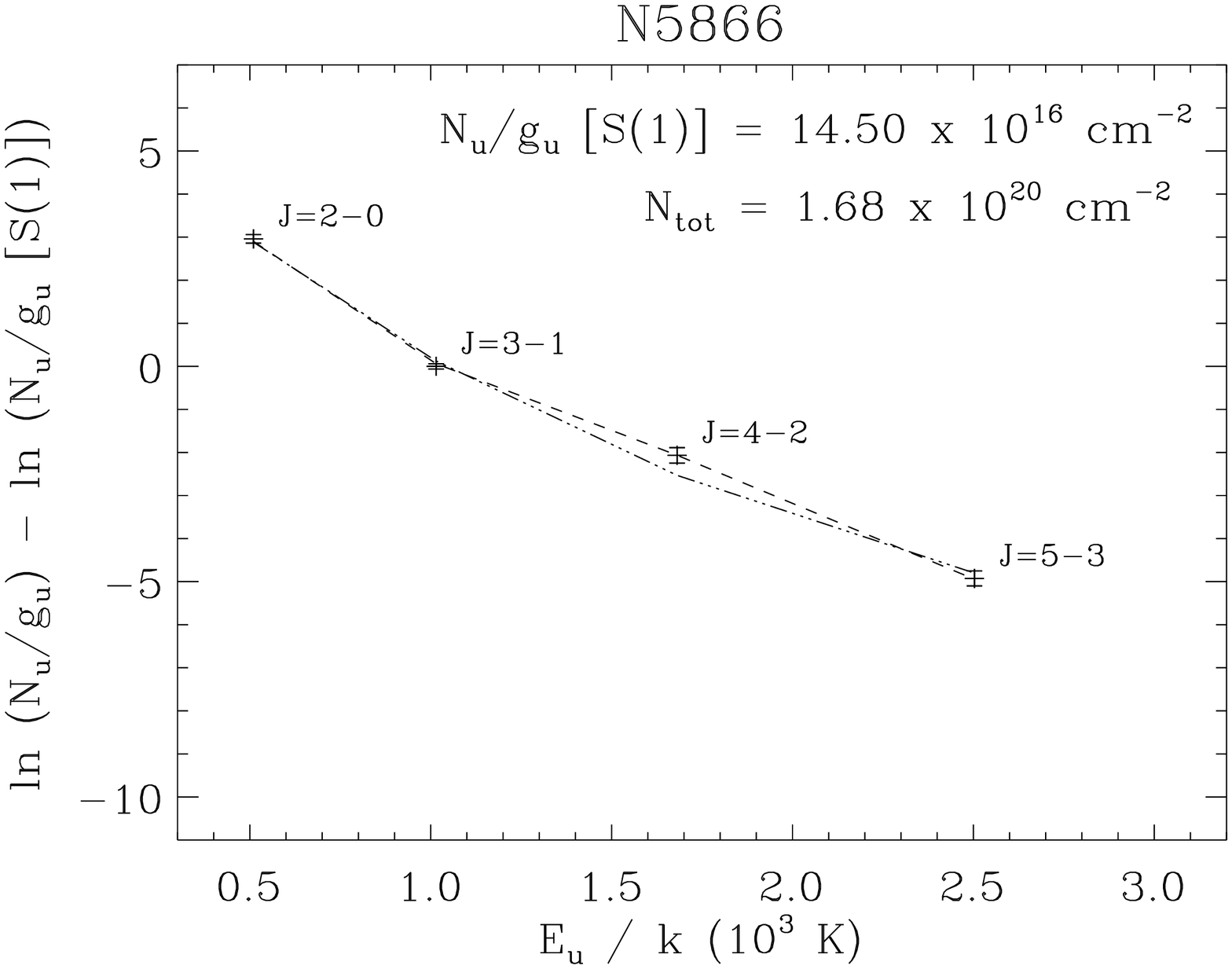}}}
\hspace*{-0.5cm}
\resizebox{10cm}{!}{\rotatebox{90}{\plotone{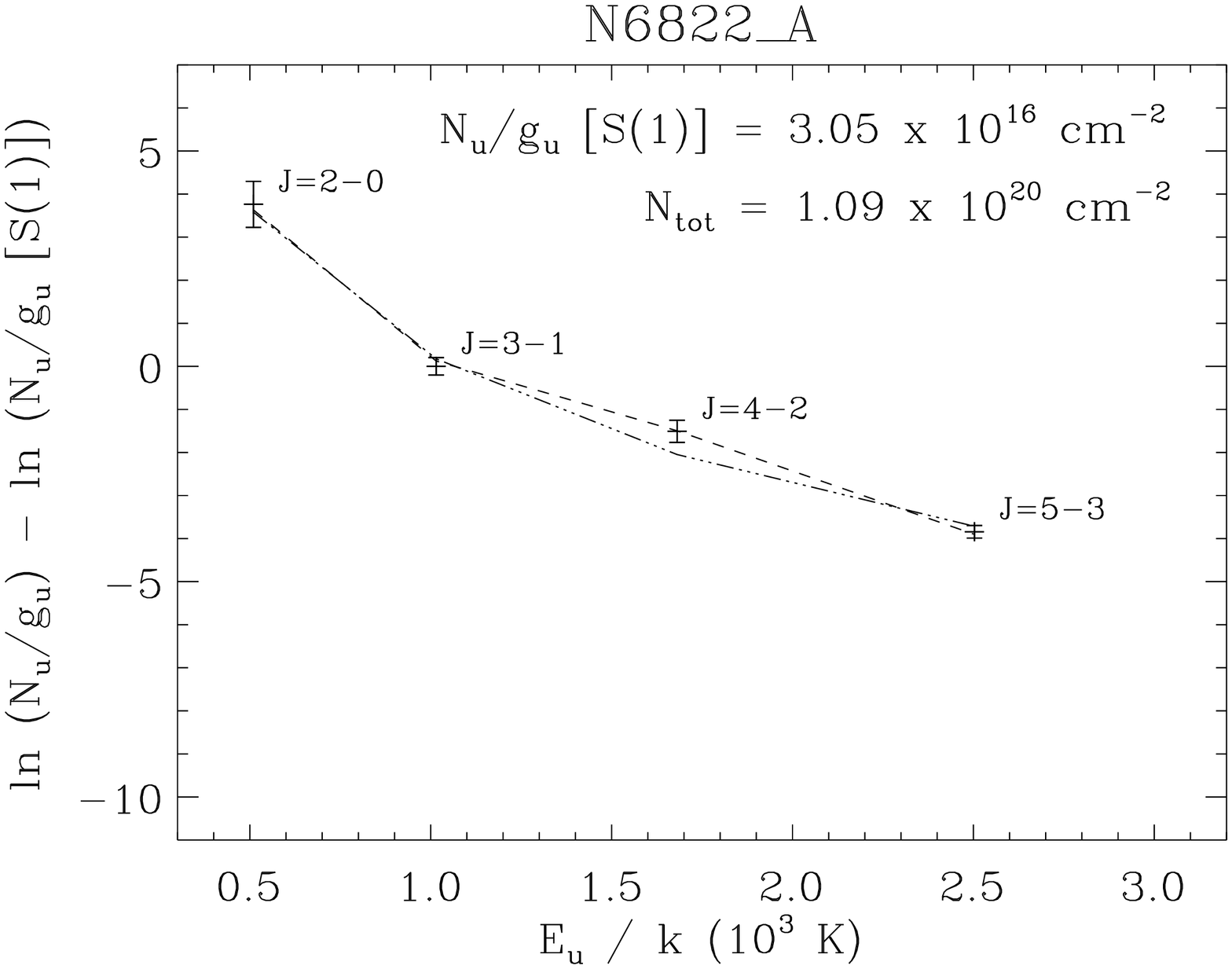}}}
\caption{(continued).
}
\end{figure}

\addtocounter{figure}{-1}
\begin{figure}[!ht]
\hspace*{-2cm}
\resizebox{10cm}{!}{\rotatebox{90}{\plotone{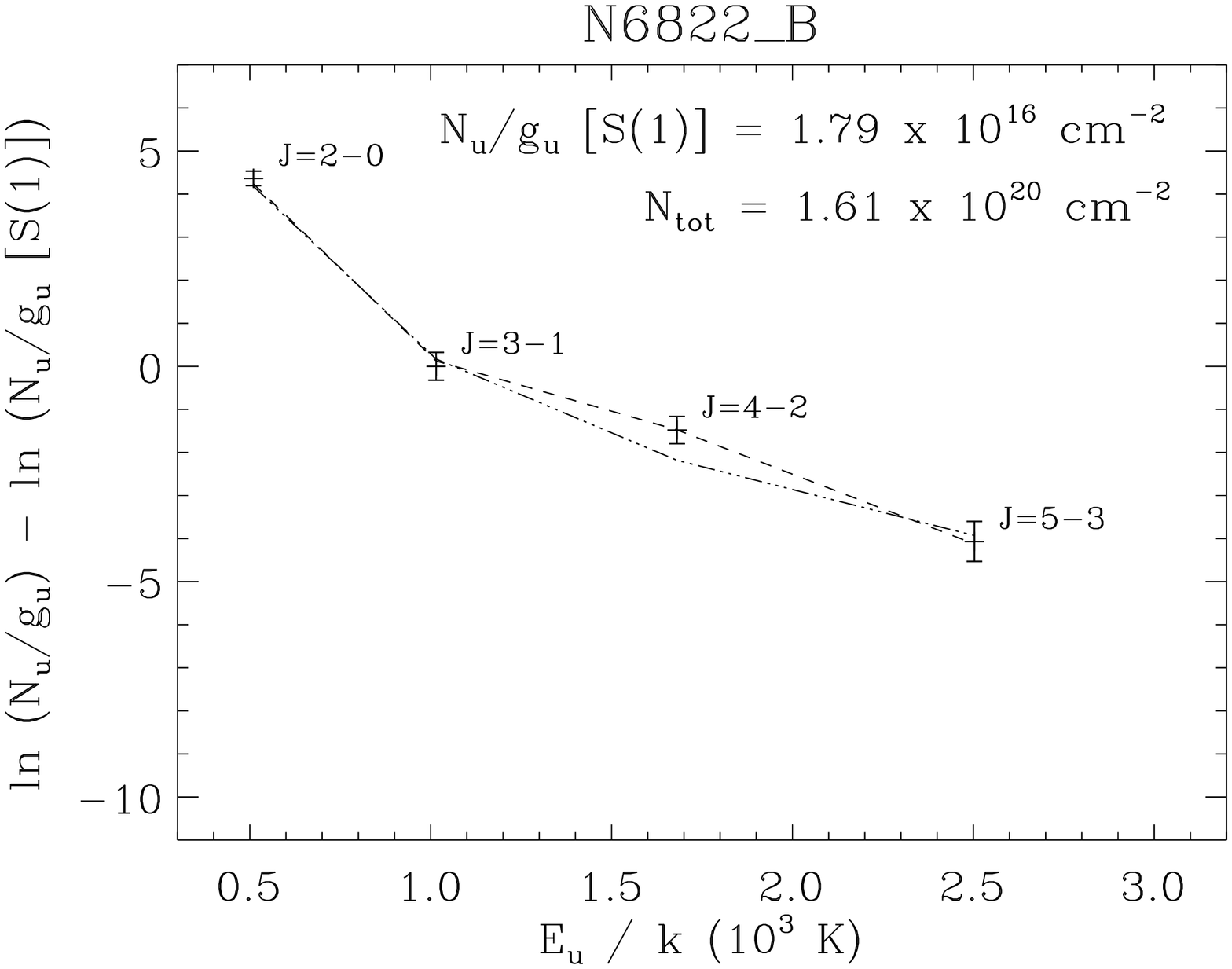}}}
\hspace*{-0.5cm}
\resizebox{10cm}{!}{\rotatebox{90}{\plotone{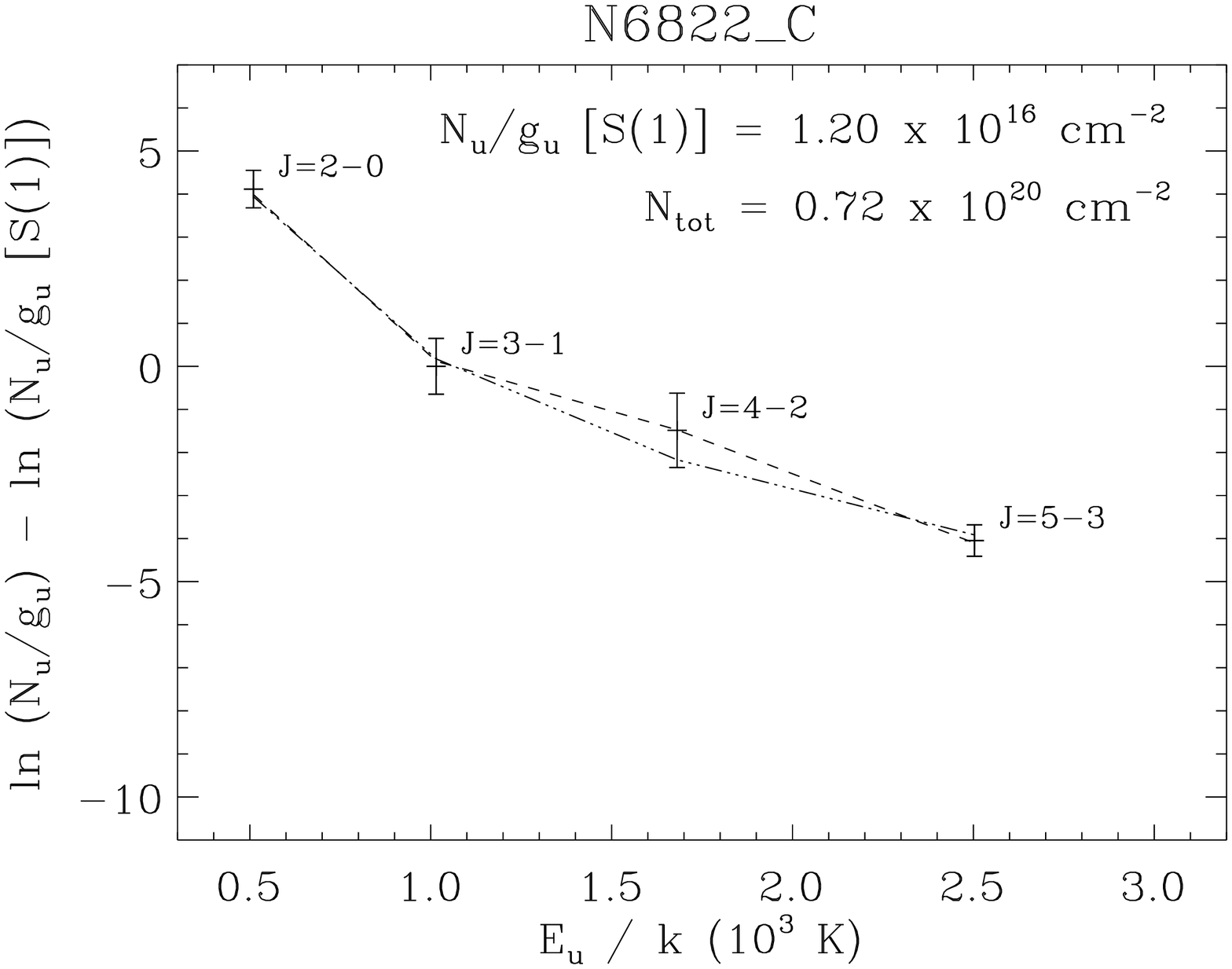}}}
\hspace*{-2cm}
\resizebox{10cm}{!}{\rotatebox{90}{\plotone{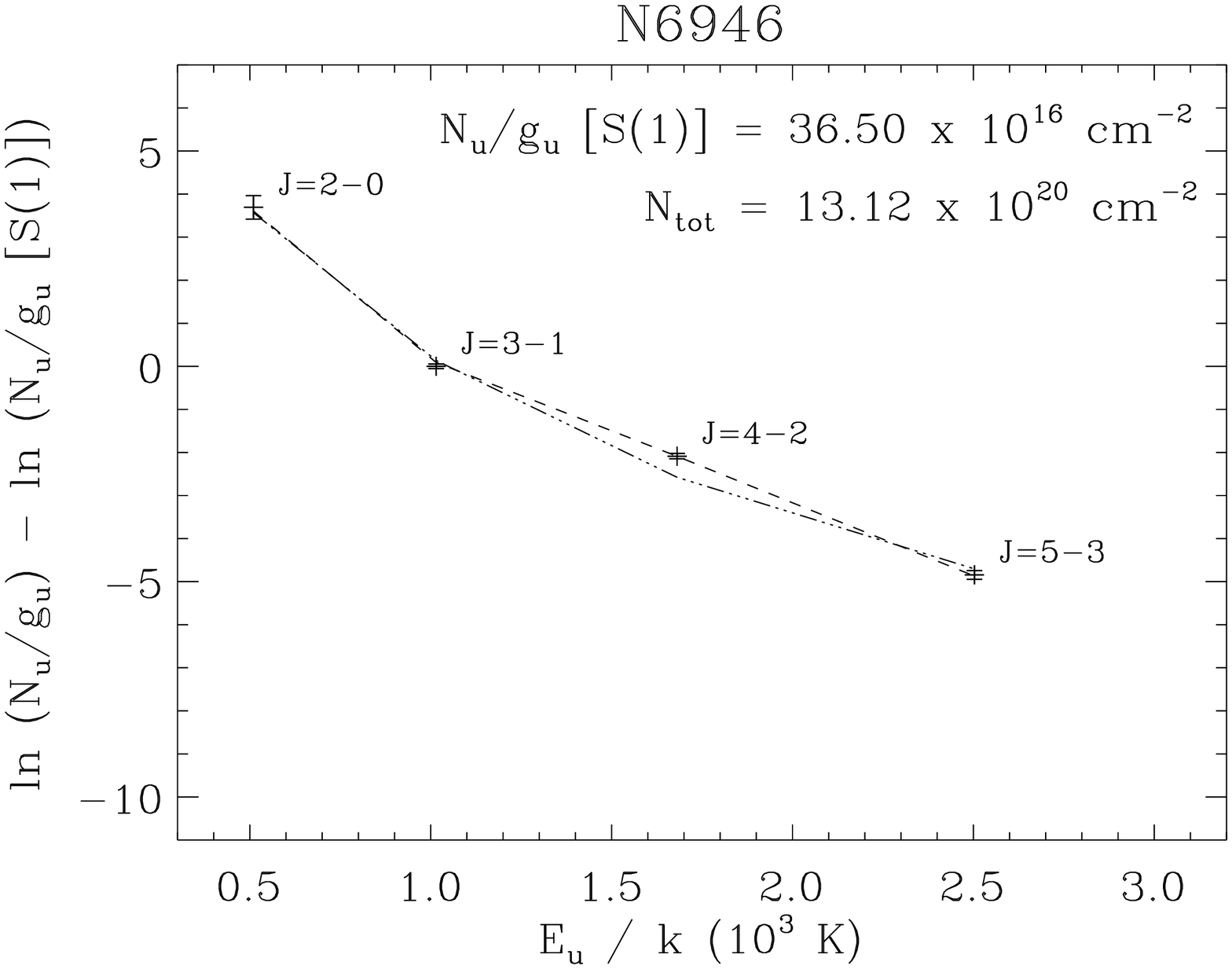}}}
\hspace*{-0.5cm}
\resizebox{10cm}{!}{\rotatebox{90}{\plotone{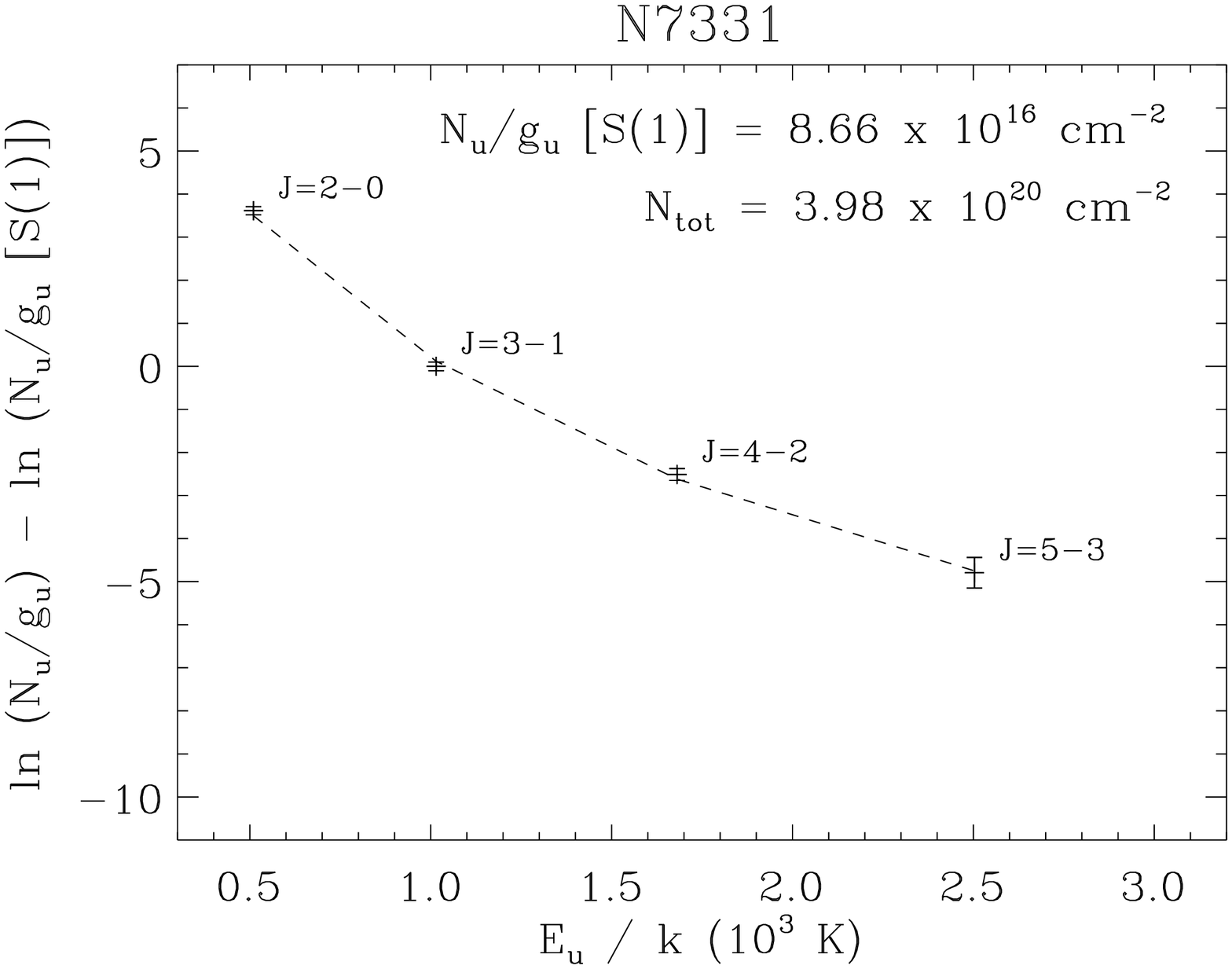}}}
\hspace*{-2cm}
\resizebox{10cm}{!}{\rotatebox{90}{\plotone{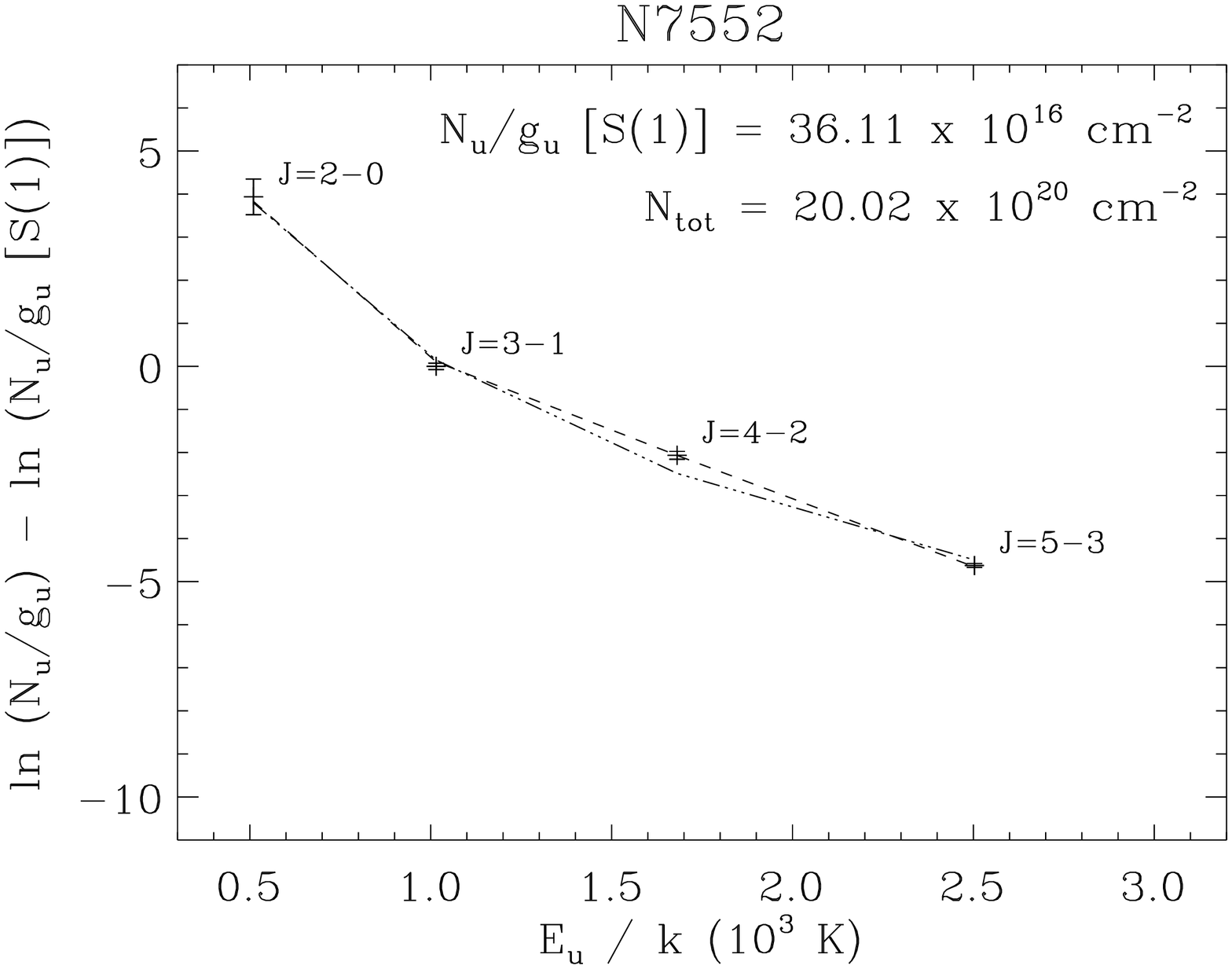}}}
\hspace*{-0.5cm}
\resizebox{10cm}{!}{\rotatebox{90}{\plotone{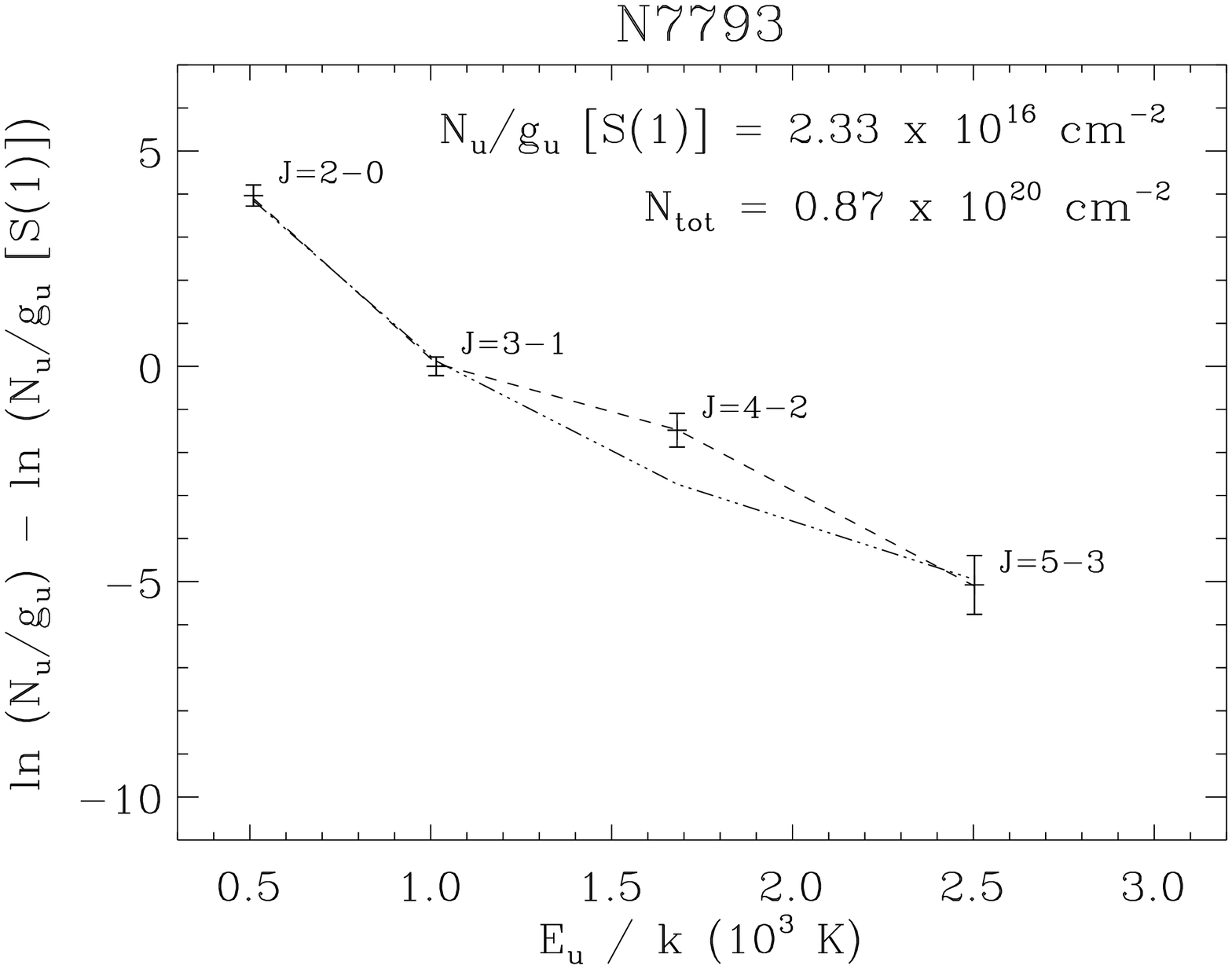}}}
\caption{(continued).
}
\end{figure}

\clearpage

\begin{figure}[!ht]
\hspace*{-2cm}
\resizebox{10cm}{!}{\rotatebox{90}{\plotone{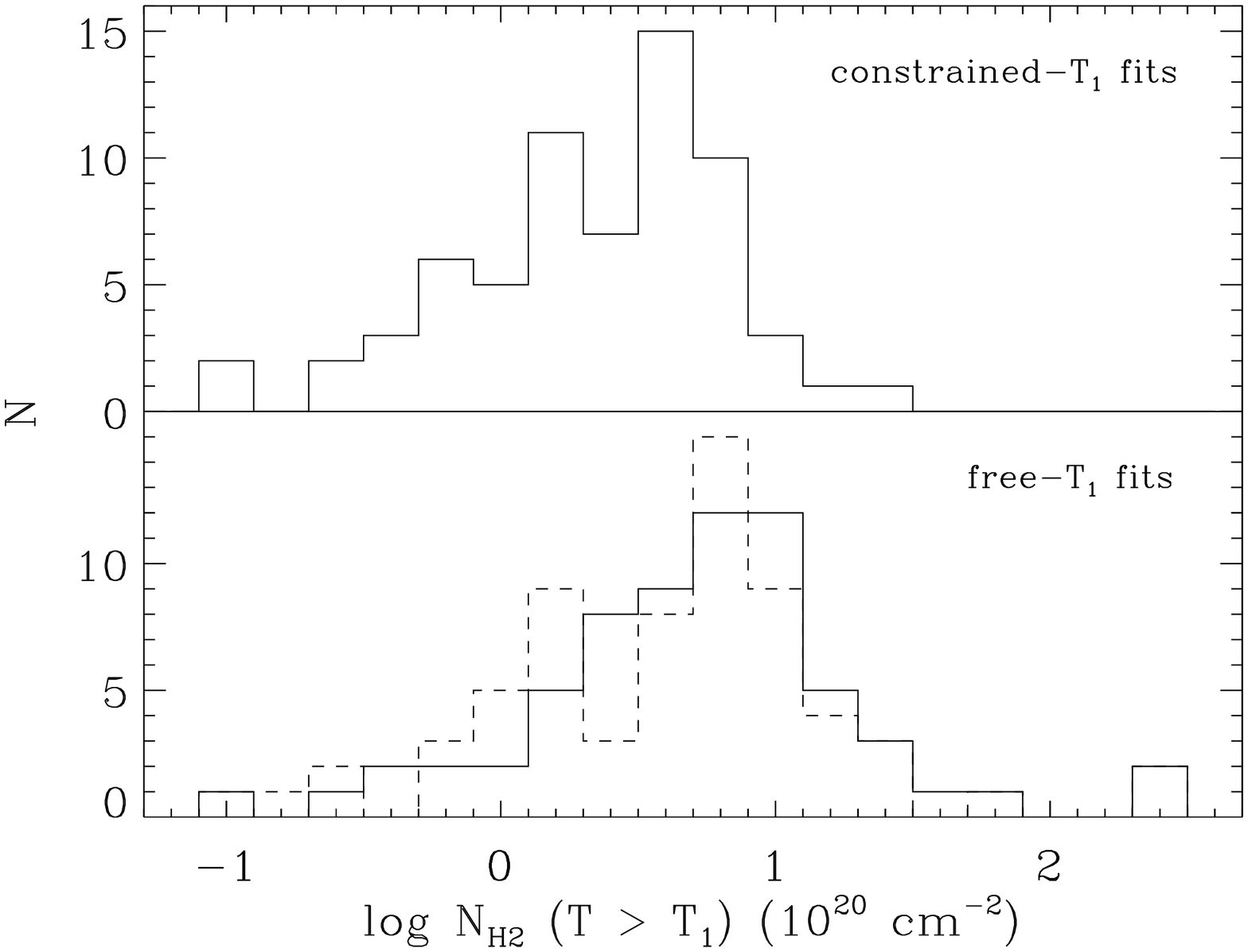}}}
\hspace*{-0.5cm}
\resizebox{10cm}{!}{\rotatebox{90}{\plotone{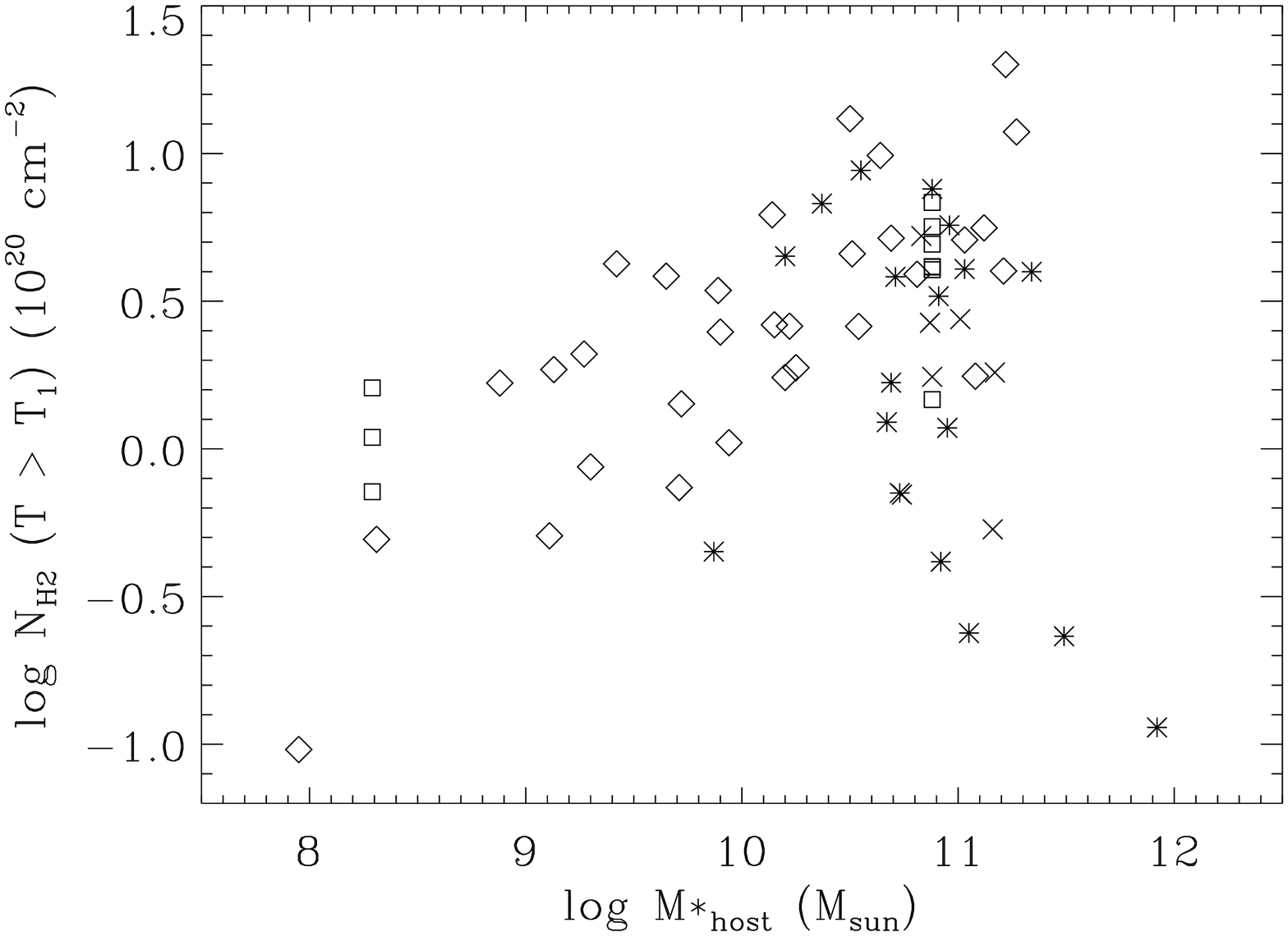}}}
\caption{
{\bf Left:} Histograms of the derived column densities in fits where the lower temperature is
constrained (top) and where it is free to vary (bottom). In the latter case, the solid line
indicates results with low $T_2$ values and the dashed line with high $T_2$ values, in
$OPR_{\rm \,high\,T} < 3$ fits (see text).
{\bf Right:} Column densities as a function of the total stellar mass of the host galaxy,
estimated as by \citet{Lee06}. Star-forming nuclei are represented as diamonds,
the extranuclear regions in NGC\,5194 and NGC\,6822 as squares, {\small LINER} nuclei
as star symbols and Sy nuclei as crosses. By definition, dwarf galaxies have
$M*_{\rm host} < 10^{9.7}$\,M$_{\sun}$.
}
\label{fig:coldens}
\end{figure}

\begin{figure}[!ht]
\resizebox{15cm}{!}{\plotone{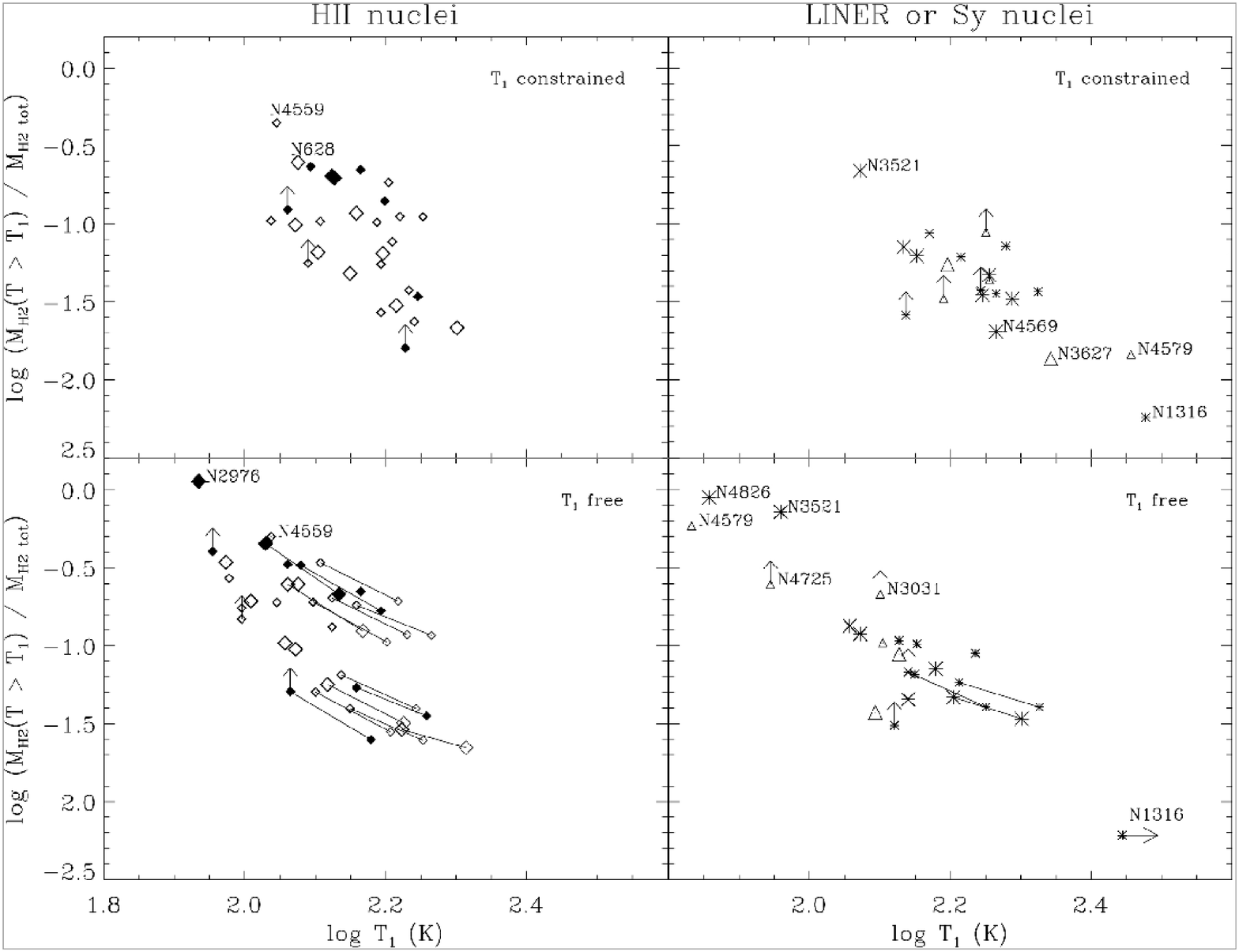}}
\caption{Fraction of H$_2$ in the warm phase ($T > T_1$) as a function of $T_1$, the lower temperature
of the two components fitted to the rotational lines. Nuclei classified as purely star-forming are
shown as diamonds, and regions in dwarf galaxies as filled diamonds; nuclei classified as
{\small LINER} or Sy are shown as star symbols or triangles, respectively.
The symbol size is increased for targets with the most robust CO flux estimates (see text).
In the free-$T_1$ fits with $OPR_{\rm \,high\,T} < 3$, results with the two values of $T_2$
considered here (see text) are connected by a line segment; the fits with
$T_2 = 1.5 \times T{\rm (S1-S3)}$ produce higher temperatures and lower warm H$_2$ masses
than the fits with $T_2 = 1.14 \times T{\rm (S1-S3)}$.
}
\label{fig:frac_warm}
\end{figure}

\begin{figure}[!ht]
\vspace*{-3.5cm}
\resizebox{15cm}{!}{\plotone{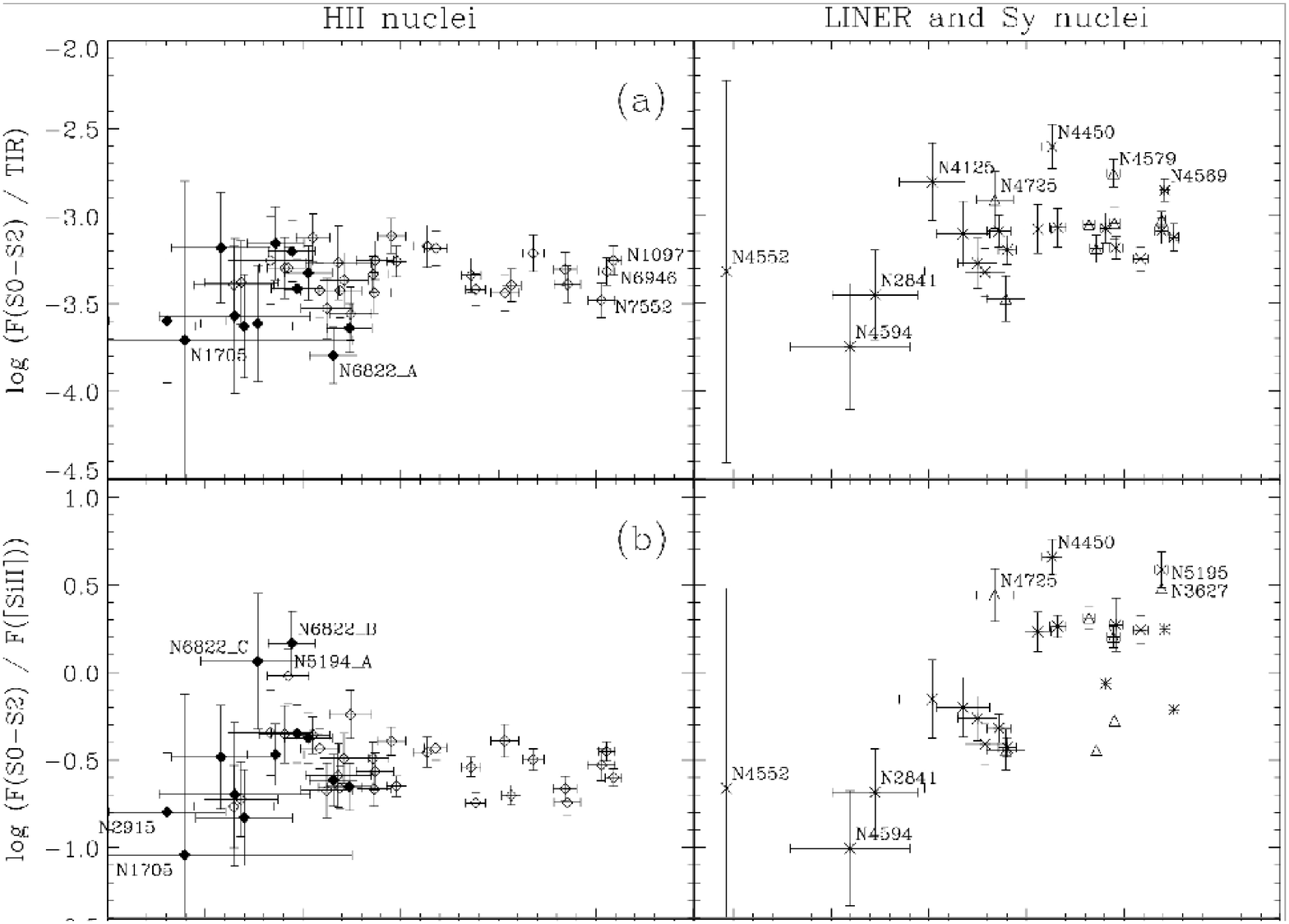}}
\vspace*{-0.12cm} \\
\resizebox{15cm}{!}{\plotone{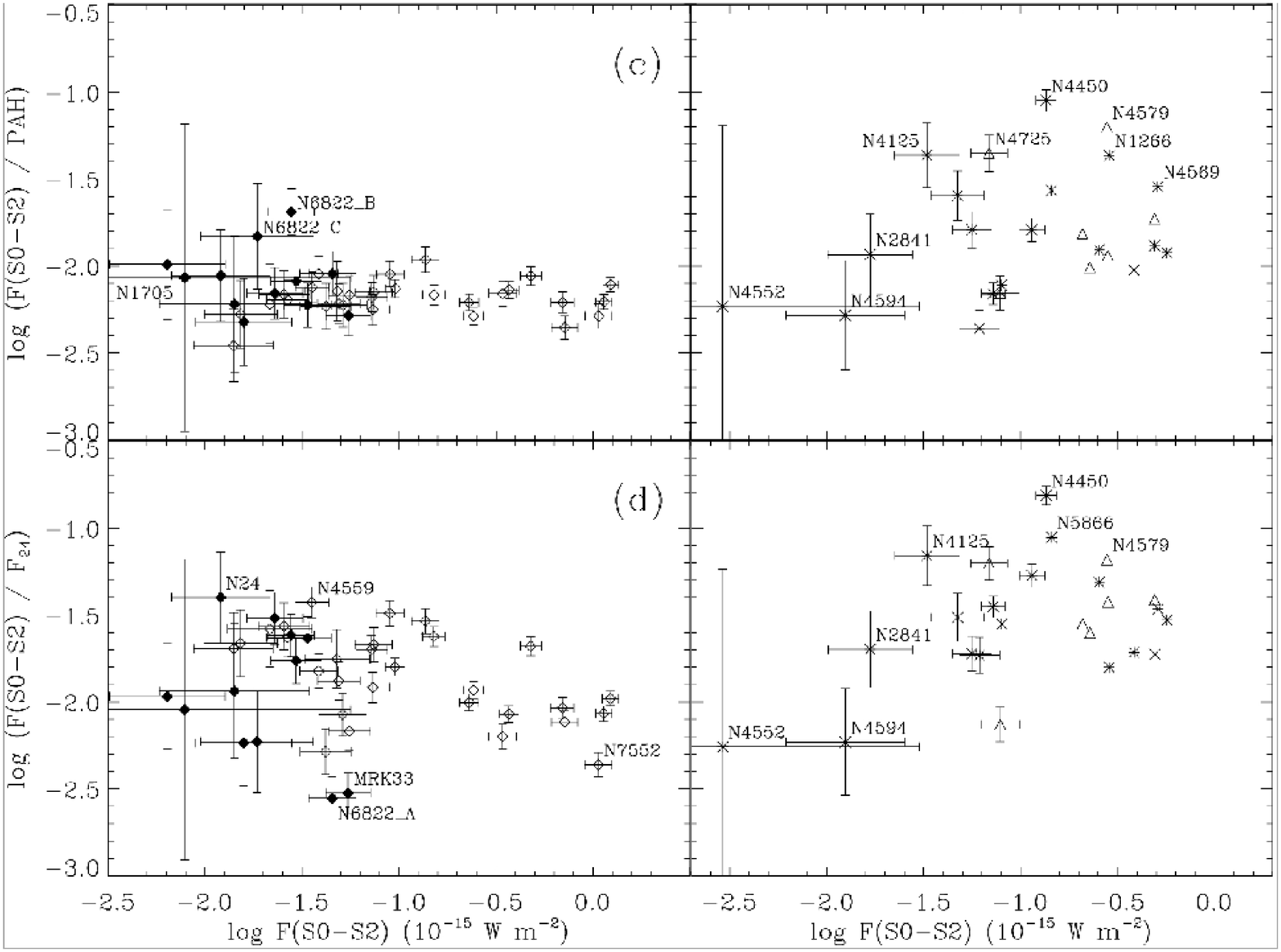}}
\vspace*{-0.5cm}
\caption{
Ratio of the power emitted in the sum of the S(0) to S(2) transitions to:
(a) the total infrared power; (b) the [Si{\small II}] line power; (c) the power emitted
in the aromatic bands within the IRAC4 filter (assuming a filter width of 13.9\,THz
and subtracting the stellar emission as mentioned in Sect.~\ref{images});
(d) the power emitted at 24\,$\mu$m within the MIPS1 filter (assuming a filter
width of 3.1\,THz). The symbol coding is as in Fig.~\ref{fig:frac_warm}.
Excluding the regions within NGC\,6822 (the targets with the smallest projected
aperture), as well as NGC\,1705 and NGC\,2915 (the galaxies with the smallest H$_2$
brightness), the average and dispersion of each logarithmic power ratio is:
(a) $-3.36 \pm 0.15$ for H{\small II} nuclei and complexes;
and $-3.12 \pm 0.24$ for {\small LINER} and Sy nuclei;
(b) $-0.53 \pm 0.17$ and $-0.04 \pm 0.42$, respectively;
(c) $-2.19 \pm 0.10$ and $-1.80 \pm 0.34$; (d) $-1.85 \pm 0.28$ and $-1.51 \pm 0.31$.
Note that the H$_2$ flux (abscissa) is almost proportional to the H$_2$ brightness,
because of the quasi-uniform beam.
}
\label{fig:frac_tir}
\end{figure}

\begin{figure}[!ht]
\resizebox{10cm}{!}{\plotone{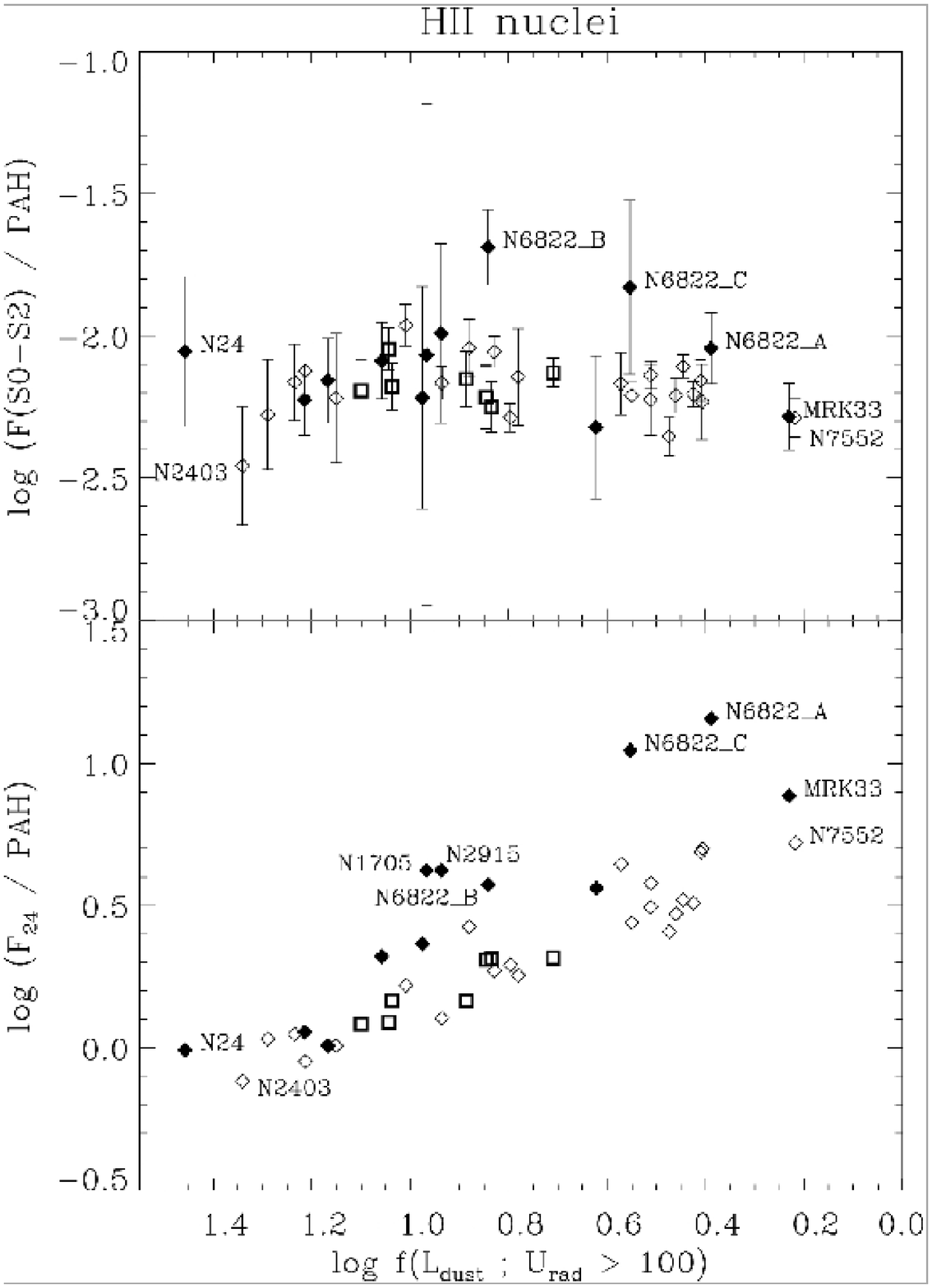}}
\caption{Power ratios of H$_2$ to aromatic bands (shown in Fig.~\ref{fig:frac_tir}c) and 24\,$\mu$m
emission to aromatic bands as a function of
$P_{24} = 1.05 ((\nu_{24} F_{24} - 0.14 \nu_{7.9} F_{\rm 7.9\,dust}) / (\nu_{71} F_{71} + \nu_{156} F_{156}) - 0.035)^{0.75}$,
a quantity closely related to $f(L_{\rm dust}~; U_{\rm rad} > 100)$ defined by
\citet{Draine07b}, which is the fraction of the total dust luminosity coming from
regions with radiation field intensities more than 100 times the average local field.
Only H{\small II} nuclei and complexes are shown. Dwarfs are represented by filled diamonds
and the extranuclear regions within NGC\,5194 by thick squares.
}
\label{fig:frac_pah_frac_highU}
\end{figure}

\begin{figure}[!ht]
\resizebox{15cm}{!}{\plotone{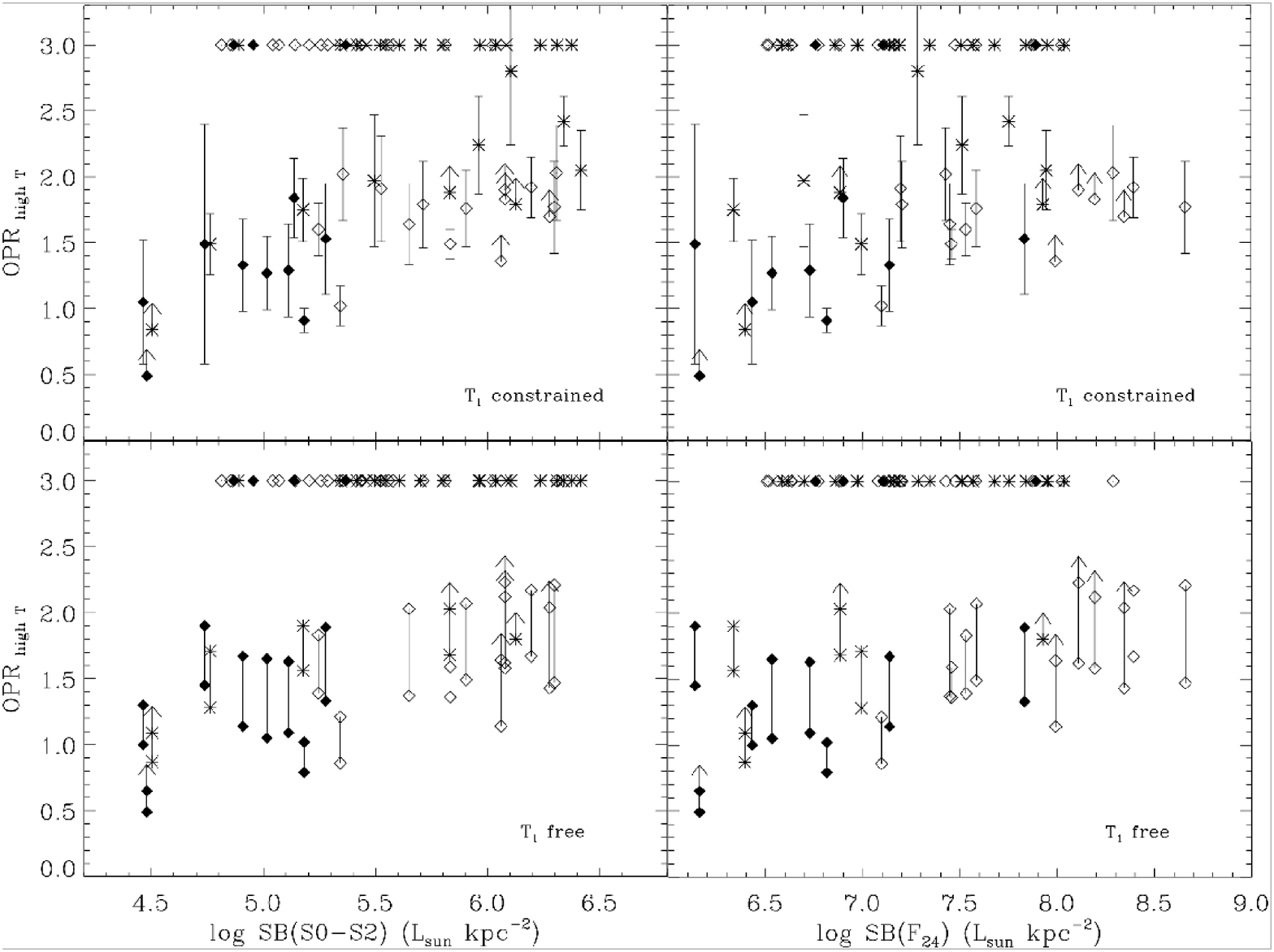}}
\caption{Ortho to para ratios as a function of the surface brightness in the sum of the
S(0) to S(2) transitions and as a function of the surface brightness in the 24\,$\mu$m band.
For a number of sources, based on the excitation diagram, we adopted a fixed
$OPR_{\rm \,high\,T} = 3$ (see text). Galaxies with upper limits in the S(2) line, or with
indications of possibly non-negligible optical depth at 10\,$\mu$m, are shown as lower
limits of $OPR_{\rm \,high\,T}$. For free-$T_1$ fits, results obtained from the two values
of $T_2$ considered here are connected by a line segment. The symbol coding is as
in Fig.~\ref{fig:frac_warm}, except that {\small LINER} and Sy nuclei are both shown
as star symbols.}
\label{fig:opr_sb}
\end{figure}

\begin{figure}[!ht]
\resizebox{10cm}{!}{\rotatebox{90}{\plotone{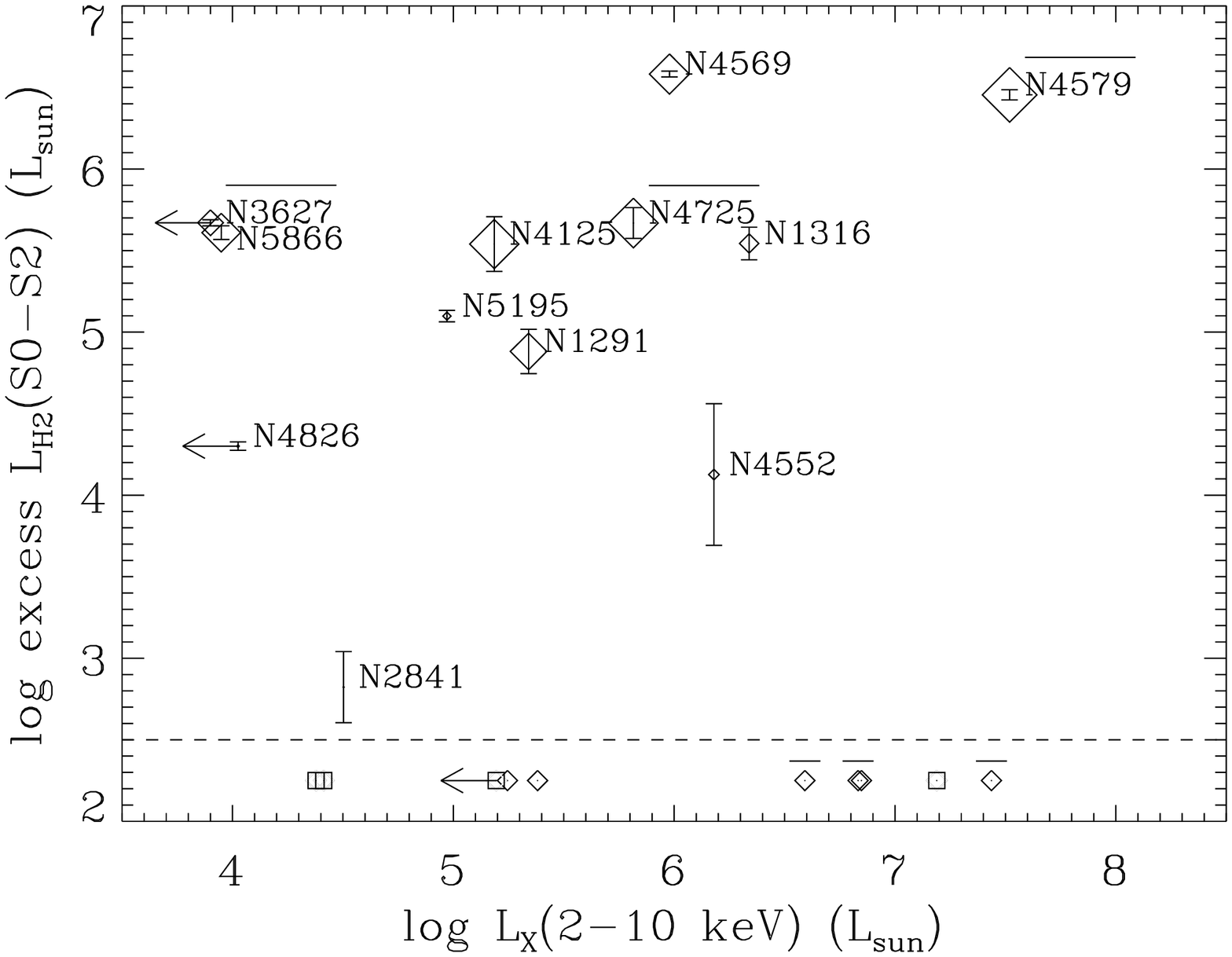}}}
\caption{Excess H$_2$ luminosity as a function of 2-10\,keV X-ray luminosity. The excess
H$_2$ emission is defined from the relation between H$_2$ and aromatic band power, shown
in Figure~\ref{fig:frac_tir}c, as the difference between the total H$_2$ emission and the
quantity $10^{-1.94} \times F_{\rm 7.9\,dust}$, which defines the upper envelope of H{\small II}
nuclei. Galaxies with no H$_2$ excess according to this definition were arbitrarily placed
at an ordinate of 2.25 below the dashed line. The size of the symbols is proportional to
the fraction of excess H$_2$ to total H$_2$ emission, coded as in Fig.~\ref{fig:snshocks}.
Sy nuclei are marked by an overlying horizontal bar, and star-forming nuclei by a square.
The large scatter and the high ratios of excess H$_2$ to X-ray luminosities argue against
X-rays playing a dominant role in the additional (nonstellar) excitation of H$_2$ in
{\small LINER}/Sy nuclei (see text).
}
\label{fig:xrays}
\end{figure}

\begin{figure}[!ht]
\resizebox{10cm}{!}{\rotatebox{90}{\plotone{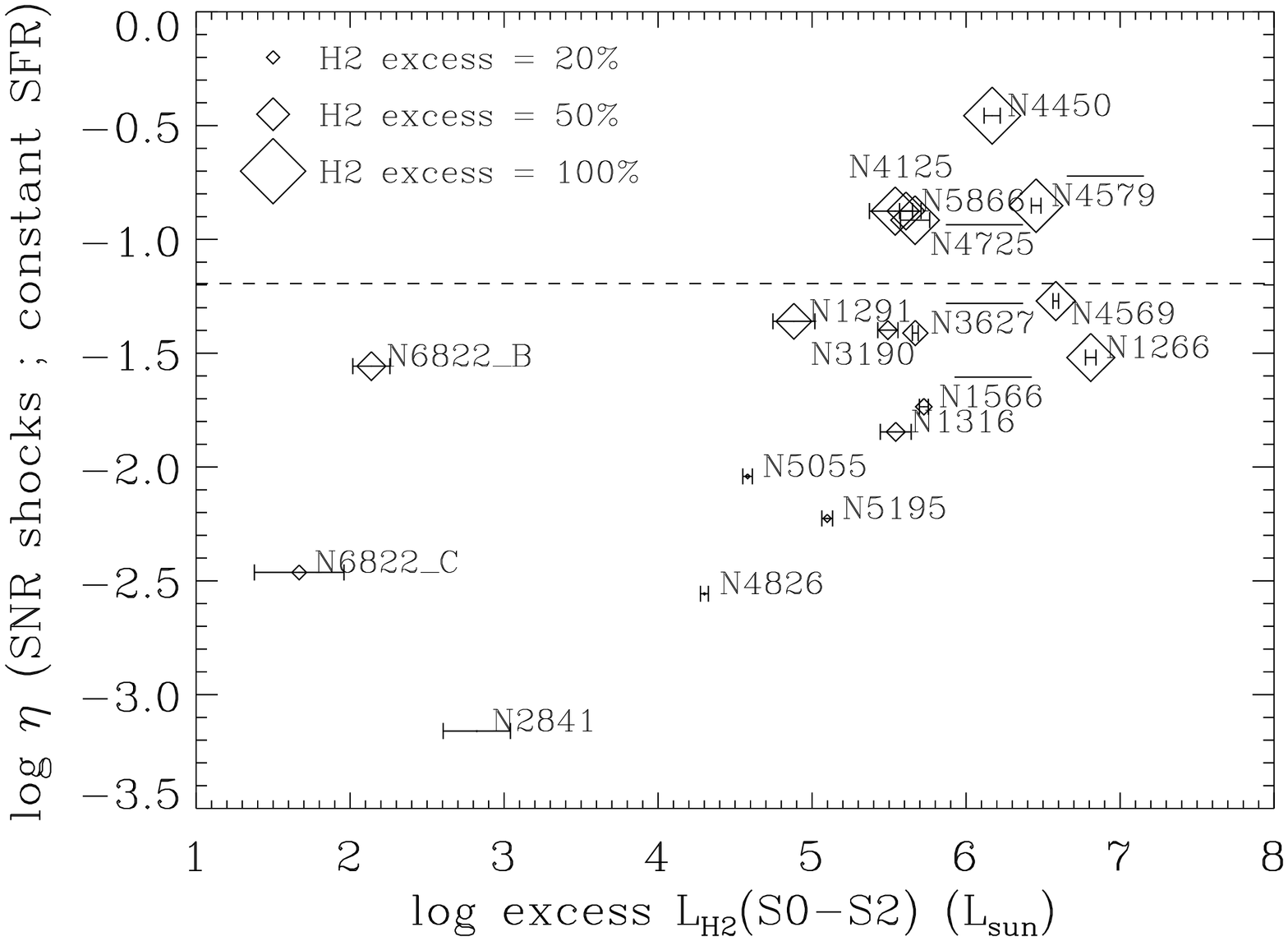}}}
\caption{Heating efficiency required to account for the excess H$_2$ emission by supernova
remnant shocks (ratio of the power in the sum of the S(0)--S(2) lines to the total
mechanical power), as a function of the excess H$_2$ luminosity, defined as in
Fig.~\ref{fig:xrays}. The dashed line indicates the maximum efficiency expected if all
the mechanical power produced by supernova remnants is absorbed in molecular clouds,
for the model parameters of \citet{Kaufman96}.
The size of the symbols is proportional to the fraction of excess H$_2$ to total H$_2$
emission. Sy nuclei are marked by an overlying horizontal bar. The regions B and C within
NGC\,6822 are the only non-AGN targets in this figure.
}
\label{fig:snshocks}
\end{figure}

\end{document}